\documentclass[aps,prd,superscriptaddress,onecolumn,10pt,url]{revtex4-1}
\usepackage[colorlinks=true,citecolor=green,linkcolor=red,urlcolor=blue]{hyperref}

\usepackage{amssymb}
\usepackage{graphicx,psfrag}
\usepackage{amsfonts}
\usepackage{amssymb}
\usepackage{amsmath}
\usepackage{epsfig, color}
\usepackage{setspace}
\usepackage{epstopdf}
\usepackage{bm}
\usepackage{mathtools}
\usepackage{braket}
 \usepackage{wasysym}
 \usepackage{dsfont}

\newtheorem{e-proposition}[theorem]{Proposition}

\newtheorem{e-definition}[theorem]{Definition\rm}

\def\va{{\boldsymbol{a}}}
\def\vk{{\boldsymbol{k}}}
\def\vj{{\boldsymbol{j}}}
\def\vr{{\boldsymbol{r}}}
\def\vq{{\boldsymbol{q}}}
\def\vp{{\boldsymbol{p}}}
\def\vb{{\boldsymbol{b}}}
\def\vd{{\boldsymbol{d}}}
\def\vl{{\boldsymbol{l}}}
\def\vn{{\boldsymbol{n}}}
\def\vn{{\boldsymbol{n}}}
\def\ve{{\boldsymbol{e}}}
\def\vm{{\boldsymbol{m}}}
\def\vx{{\boldsymbol{x}}}

\def\vK{{\boldsymbol{K}}}
\def\vG{{\boldsymbol{G}}}
\def\vR{{\boldsymbol{R}}}

\def\vA{{\boldsymbol{A}}}
\def\vE{{\boldsymbol{E}}}
\def\vB{{\boldsymbol{B}}}
\def\vF{{\boldsymbol{F}}}
\def\vS{{\boldsymbol{S}}}
\def\vM{{\boldsymbol{M}}}

\def\dhat{{\hat {\boldsymbol d}}}

\begin{document}
\title{Topological and geometrical aspects of band theory}

\author{J. Cayssol}
\email{jerome.cayssol@u-bordeaux.fr}
\affiliation{Univ. Bordeaux, CNRS, LOMA, UMR 5798, F-33405 Talence, France}
\author{J.N. Fuchs}
\email{jean-noel.fuchs@sorbonne-universite.fr}
\affiliation{Sorbonne Universit\'e, CNRS, Laboratoire de Physique Th\'eorique de la Mati\`ere Condens\'ee, LPTMC, F-75005 Paris, France}
\affiliation{Laboratoire de Physique des Solides, Universit\'e Paris Saclay, CNRS, F-91405 Orsay, France}



\begin{abstract}
This paper provides a pedagogical introduction to recent developments in geometrical and topological band theory following the discovery of graphene and topological insulators. Amusingly, many of these developments have a connection to contributions in high-energy physics by Dirac. The review starts by a presentation of the Dirac magnetic monopole, goes on with the Berry phase in a two-level system and the geometrical/topological band theory for Bloch electrons in crystals. Next, specific examples of tight-binding models giving rise to lattice versions of the Dirac equation in various space dimension are presented: in 1D (Su-Schrieffer-Heeger and Rice-Mele models), 2D (graphene, boron nitride, Haldane model) and 3D (Weyl semi-metals). The focus is on topological insulators and topological semi-metals. The latter have a Fermi surface that is characterized as a topological defect. For topological insulators, the two alternative view points of twisted fiber bundles and of topological textures are developed. The minimal mathematical background in topology (essentially on homotopy groups and fiber bundles) is provided when needed. Topics rarely reviewed include: periodic versus canonical Bloch Hamiltonian (basis I/II issue), Zak versus Berry phase, the vanishing electric polarization of the Su-Schrieffer-Heeger model and Dirac insulators.
\end{abstract}
\date{\today}
\maketitle
\tableofcontents
%
%
%
%
%


\section{Introduction}
Band theory was created in the 1930's right after the invention of quantum mechanics~\cite{Hoddeson:1987}. It relies mainly on Bloch's theorem, which rules the behavior of an electron in the periodic potential created by ions, and on Fermi-Dirac statistics, which governs the filling of energy bands by electrons. The coronation of band theory was the classification of crystals by Wilson~\cite{Wilson:1931} into insulators and metals depending on whether the band structure contains or not a partially filled band (i.e. a Fermi surface). The equations describing the semiclassical motion of a Bloch electron restricted to a single band were obtained by Bloch~\cite{Bloch:1928}, Peierls~\cite{Peierls:1929}, Jones and Zener~\cite{Jones:1934}. These equations have a form similar to that for a classical non-relativistic particle in the vacuum, except for the velocity which is replaced by the band's group velocity. Even at the quantum level, electrons in a crystal were believed to behave almost like electrons in the vacuum upon replacing the dispersion relation $E=(\hbar\vk)^2/(2m)$ by the band dispersion $E_n(\vk)$. This initial version of band theory could account for the electronic behavior of very many crystals~\cite{AshcroftMermin}. 

However, in the 1940's, 50's and 60's researchers started to realize that there may be more to band theory than simply individual energy bands. We should mention the work of pioneers such as Adams, Blount, Kohn, Luttinger, Roth, Slater, Wannier and others (see the review by Blount~\cite{Blount:1962}), who understood the importance of inter-band effects. For example, Karplus and Luttinger realized that in some materials, corrections to the group velocity could appear in the form of an anomalous transverse velocity~\cite{Karplus:1954}, which may explain the anomalous Hall effect. Also Kohn found that the effective Hamiltonian for a single band in a magnetic field was not only given by the band dispersion $E_n(\vk)$ but was shifted by an orbital magnetic moment (in addition to the Zeeman effect)~\cite{Kohn:1959}. But the situation remained obscure and, apart from band theory aficionados, nobody really paid attention. In hindsight, reading the review by Blount~\cite{Blount:1962}, one recognises already a lot of the ``modern'' concepts of geometrical band theory such as ``Berry'' connection, curvature, virtual inter-band transitions, analogy to electromagnetism but in reciprocal space, relation to the Dirac equation, etc. that we will encounter in this review. However, topological considerations were absent.

Everything changed with the experimental discovery of the integer quantum Hall effect~\cite{Klitzing1980} and its subsequent understanding as a topological effect by Thouless, Kohmoto, Nightingale and den Nijs~\cite{Laughlin1981,Thouless:1982} building on concepts that had emerged in the context of the Hofstadter butterfly~\cite{Hofstadter:1976}. It was soon recognised that this was a particular instance of a Berry phase effect~\cite{Berry:1984,Simon:1983}. It took some more years for Haldane~\cite{Haldane:1988} to clearly spell out that the quantum Hall effect (now known as the quantum anomalous Hall effect) could be understood as a pure band theory effect provided time-reversal symmetry was broken, but that one could dispense with Landau levels (no uniform magnetic field) and maintain the periodicity of the crystal. In particular, this showed that a filled band could indeed conduct electricity despite what had been written in solid-state physics textbooks for 50 years. Simultaneously, Volovik proposed a similar kind of topological effect in the framework of Bogoliubov-de Gennes (mean-field) description of superconductors and applied it to a film of superfluid helium 3~\cite{Volovik:1988}. It also involved coupling between bands but the origin of bands is in particle-hole coupling via the superfluid pairing and not in Bloch's theorem. This constitutes a first example of what is now called a topological superconductor~\cite{Bernevig:2013}. Around the same time, Zak realized that another type of Berry phase -- an open-path Berry phase along a non-contractible loop -- could be defined in the Brillouin zone using the torus geometry~\cite{Zak:1989}. A few years later, King-Smith, Vanderbilt and Resta understood that the Zak phase was related to the position operator and the key to settling the problematic issue of the proper definition of a bulk electric polarization for a crystal~\cite{KingSmith:1993,Vanderbilt:1993,Resta:1994}. In addition to this ``modern theory of electric polarization''~\cite{Resta:2007}, a parallel ``modern theory of orbital magnetization'' was developed and is reviewed in~\cite{XiaoBerryRMP:2010,Thonhauser:2011}. The semi-classical equations of motion for a Bloch electron restricted to a given band, and modified by Berry phase terms, were also obtained in final form by Chang, Sundaram and Niu~\cite{Chang:1996,Sundaram:1999} using a wave-packet approach, see~\cite{XiaoBerryRMP:2010} for review. In a few years, many long-standing and annoying problems of solid-state physics were solved by realizing that certain measurable quantities explicitly depend on the phase of the Bloch wave functions.

The next major step was taken by Kane and Mele, who realized that topology could also be present in systems that do not break time-reversal symmetry~\cite{Kane:2005a,Kane:2005b}. The original proposal was with graphene and could not be realized because of a too small intrinsic spin-orbit coupling of carbon but Bernevig, Hughes and Zhang proposed another system -- a HgTe/CdTe quantum well -- in which one could obtain a quantum spin Hall insulator~\cite{Bernevig:2006}. This was realized experimentally in the group of Molenkamp~\cite{Konig:2007}. This first example of a symmetry-protected topological insulator in two dimension was soon followed by its generalization to three dimensions (something not possible for the integer quantum Hall effect)~\cite{Fu:2007,Moore:2007,Roy:2009} and the subsequent experimental discovery, see~\cite{KaneRMP:2011} for review.

The proposal of Kane and Mele was concomitant with a major experimental discovery, that of graphene~\cite{Novoselov:2005,Zhang:2005}. This two-dimensional honeycomb carbon crystal has low-energy electrons that obey a massless Dirac equation rather than an effective single-band Schr\"odinger equation~\cite{Wallace:1947,DiVincenzo:1984}. This is another extension of band theory. A system that is neither a metal nor an insulator -- it is gapless but the density of states vanishes at the Fermi surface reduced to two points -- and whose description involves two bands that are strongly coupled. The vicinity of each contact point resembles a diabolo~\cite{Berry:2010} and is now called a Dirac cone. One characterization of these Dirac fermions is that they carry a $\pi$ Berry phase. Graphene is now considered as an example of a (symmetry-protected) topological semi-metal. The culmination in this modern version of band theory -- that may be summarized as Berry phase effects + graphene + topological insulators -- was in the periodic table (or ten-fold way) classification of topological insulators and superconductors by Schnyder, Ryu, Furusaki and Ludwig~\cite{Schnyder:2008} and by Kitaev~\cite{Kitaev:2009}. This classification was extended in several directions including topological semi-metals~\cite{Horava:2005,Zhao:2013}, of which Volovik should be mentioned as an early pioneer~\cite{Volovik:2003}.

Amusingly, in many of the above modern developments in band theory, one may see shadows of contributions by Dirac in high-energy physics. A first instance is the simultaneous invention of anti-matter (the positron) by Dirac and its solid-state version (the hole) by Peierls (this story is beautifully told in~\cite{Hoddeson:1987}). Obviously the Dirac equation~\cite{Dirac:1928} plays an important role as the simplest Hamiltonian describing the coupling between two (or four) bands. It was invented for relativistic electrons in the 3D vacuum but now serves to describe various crystals in the long-wavelength limit in 1D, 2D and 3D~\cite{Cayssol:2013b,Shen:2017,DiracMatter:2017}. The most prominent example is the honeycomb lattice of graphene, which gives rise at long wavelength to the emergence of two massless Dirac equations in 2D. Note also that in the review by Blount~\cite{Blount:1962}, it was already recognized that the Dirac equation was a model for band-coupling effects. At that time, it was mainly used as an effective description of bulk 3D bismuth. Another contribution of Dirac that has descendants in solid-state physics is the magnetic monopole~\cite{Dirac:1931}. It may be seen as a forerunner of the Aharonov-Bohm phase and more generally of Berry phases. Although elusive as magnetic charge in real space, the Dirac monopole actually exists in reciprocal space as a source of Berry flux and is related to band contact points. The well-known quantization of the magnetic strength of the Dirac monopole has a counterpart in the integer Chern numbers characterizing the bands. Later, Wu and Yang have shown that the Dirac monopole has topological significance and is related to the mathematical notion of fiber bundles~\cite{Wu:1975}.

In the present paper, we provide a pedagogical review of modern band theory focusing on geometrical and topological effects. These effects are all due to coupling between bands. The latter modify the effective description of an electron restricted to a given band, by the appearance of an emergent gauge field (also known as the Berry connection), that takes into account the possibility of virtual transitions to other bands. These are geometrical effects in band theory, i.e. effects in solid-state physics that do not only depend on the energy bands in the absence of external fields but also involve the cell-periodic Bloch eigenfunctions. In addition, and because the Brillouin zone is a compact manifold (a torus in $D$ dimensions), some of these geometrical effects turn topological. For geometrical band theory and Berry phase effects in solids, we recommend the reviews by Xiao, Chang and Niu~\cite{XiaoBerryRMP:2010}, Resta~\cite{Resta:2000,Resta:2011} and the book by Vanderbilt~\cite{Vanderbilt:2018}. On the topic of topological insulators, see Refs.~\cite{KaneRMP:2011,QiRMP:2011,KonigJPSJ:2008,Fruchart:2013,Witten:2016,Shankar:2018} and the books by Bernevig~\cite{Bernevig:2013} and by Asb\'oth, Oroszl\'any and P\'alyi~\cite{Asboth:2016}. On the subject of topological semi-metals and the classification of Fermi surfaces as topological defect, see the book by Volovik~\cite{Volovik:2003}. For topological superconductors, we recommend the chapters written by Hughes in~\cite{Bernevig:2013}. For the extension of these ideas to cold atoms or to photonics see Refs.~\cite{CooperRMP:2019,RMP_topophotonics:2019}.

The structure of our review is the following. In Sec.~\ref{sec:Diracmonopole}, we study the Dirac magnetic monopole. Then in Sec.~\ref{section0D}, we consider a quantum two-level system and show the appearance of a Berry phase, i.e. and emergent gauge structure in parameter space. Next, we turn to periodic crystals in Sec.~\ref{sec:bandtheory} and present geometrical and topological band theory. In particular, we give an introduction to the mathematical notion of fiber bundles. In Sec.~\ref{section1D}, we review the physics of one-dimensional non-interacting electrons on dimerized (or diatomic) chains, using the Su-Schrieffer-Heeger (SSH) and Rice-Mele (RM) models as examples. The following section~\ref{section2D} deals with two-dimensional band structures on the honeycomb lattice: we discuss the geometrical and topological aspects of graphene (Sec.~\ref{section2Dgraphene}), boron nitride (Sec.~\ref{section2Dbn}) and the Haldane model of a Chern insulator (Sec.~\ref{section2DHaldane}). In Sec.~\ref{sec:topinsulators}, we sketch a bigger picture of the notion of topological insulators. In Sec.~\ref{sec:topmetal}, we consider topological semi-metals (especially 3D Weyl semi-metals) in which the Fermi surface is treated as a topological defect and describe a connection between topological metals and insulators via the relation between topological defects and textures. Here, the mathematical notion of homotopy groups is outlined. In the general conclusion (Sec.~\ref{sec:conclusion}), we summarize the most important points and in an Appendix, we make a distinction between topological insulators (covered in this review) and topological order (not covered).

\section{Dirac magnetic monopole in real space}
\label{sec:Diracmonopole}

In 1931, Dirac investigated the compatibility of quantum mechanics with the presence of point-like magnetic charges~\cite{Dirac:1931}. At the level of classical electrodynamics, such a magnetic monopole is forbidden as a point charge but may exist as the termination of a semi-infinite solenoid: it is not only a point-like singularity in the magnetic field, but it also implies a line singularity in the vector potential $\vA(\vr)$ along the solenoid. Those latter singularities, called Dirac strings, are semi-infinite lines emanating from the monopole and extending to infinity. At first sight, such extended singularities in $\vA(\vr)$ look pretty harmful to quantum mechanics since $\vA(\vr)$ enters directly the Schr\"odinger equation. But Dirac realized that the framework of quantum mechanics can be kept perfectly coherent provided that the wave functions vanish along such Dirac strings (the Dirac veto). This condition leads to a relation between electrical charge and the monopole strength.
We start by presenting a quick argument for the Dirac quantization condition. Then we present a derivation or reinterpretation of the quantization condition due to Wu and Yang in the seventies~\cite{Wu:1975}, that allows one to avoid the concept of Dirac strings (and the related Dirac veto) and is important to realize the global topological nature of the Dirac monopole. As a general reference on the Dirac monopole, we recommend a chapter in the book by Ryder~\cite{Ryder}.

\subsection{Obstruction, Dirac string and quantization argument}

A magnetic monopole of strength $g$, located at the origin of real space $\mathbb{R}^3$, would produce a radial magnetic field given by : 
\begin{equation}
\vB = \frac{g}{r^2} \ve_r \, ,
\label{Bmonop}
\end{equation}
which is the solution of :
\begin{equation}
\boldsymbol{\nabla} \cdot \vB = 4 \pi g \, \delta(\vr) \, .
\label{divBmonop}
\end{equation}
The total magnetic flux piercing a closed surface (e.g. a sphere $S^2$) surrounding the origin is therefore : 
\begin{equation}
\oiint \,   \vB \, . \,  {\boldsymbol{dS} }  = 4 \pi g\, \label{Fluxmonop} .
\end{equation}

It is natural to search for a vector potential $\vA$ associated to the monopole radial magnetic field Eq. (\ref{Bmonop}) as $\boldsymbol{\nabla} . \vB =0$ in $\mathbb{R}^3-\{0\}\simeq S^2\times \mathbb{R}^+ \sim S^2$. It turns out that it is impossible to find a single regular/smooth vector potential expression (electromagnetic gauge) covering the whole space $\mathbb{R}^3-\{0\}$ \footnote{Mathematically this is expressed in the fact that the second cohomology group of the sphere is non trivial: $H^2(S^2)=\mathbb{Z}$.}. To understand this fact, let us assume the existence of such a smooth gauge and show some contradiction. If $\vA$ were smooth on the whole sphere $S^2$, for any closed line $\mathcal{C}$ on $S^2$, separating the sphere in two regions, it would be possible to apply Stokes' theorem for each region and get :
\begin{equation}
 \oiint_{S^2} \,   \vB \, . \,  {\boldsymbol{dS} }= \oint_{\mathcal{C}} \vd\vl \, \, . \,  \vA -  \oint_{\mathcal{C}} \vd\vl \, \, . \,  \vA  =0  \, .
\label{Stokesmonop}
\end{equation}
Then the flux of $\vB=\boldsymbol{\nabla}\times \vA$ through $S^2$ (or any closed surface) would always be zero in contradiction with Eq. (\ref{Fluxmonop}). The conclusion is therefore that a magnetic monopole is only possible if there is an obstruction to finding such a smooth global gauge for the vector potential. Therefore there should be at least a point on $S^2$ where the potential $\vA$ is singular. By connecting this singularity for continuously varying radius of the sphere, one gets a line singularity called a Dirac string, which is a not necessarily straight, starts from the monopole and extends to infinity (see Fig.~\ref{fig:diracstrings}). 

The Dirac string can be used to partially (in a restricted space) bypass the obstruction discussed above. The idea is to see the monopole as the free end of an infinite line of magnetic dipoles (or a semi-infinite thin solenoid), the other end of the line being sent to infinity. This is equivalent to attaching a thin solenoid to the magnetic pole. The role of this solenoid is to feed some flux $(-g)$ into the sphere to compensate the radial flux $g$ created by the monopole itself. In this way the total magnetic flux through the sphere is exactly zero and the total magnetic field (of the system pole + solenoid) can be written as the curl of a vector potential. Then the magnetic field of the monopole alone reads $\boldsymbol{\nabla}\times \vA - \vB_{sol}$, where $\vB_{sol}$ is the highly singular magnetic field located inside the thin solenoid.    

By solving the problem of a scalar wave function in the background of the monopole, Dirac has shown that the wave function must vanish along the Dirac string (i.e. a nodal line), which implies a quantization condition. A heuristic argument to get this quantization condition is to anticipate that the obstruction originates from the finite magnetic flux $4 \pi g$. One may suspect that when this magnetic flux is a multiple of the quantum of flux $h/e$, its effect is undetectable, and the phases of the wave functions can be defined in a consistent way. This lead to :
\begin{equation}
4 \pi g =  n \frac{h}{ e}      \hspace{5mm} {\rm with}  \hspace{5mm}    n \in   \,  \mathbb{Z}   \, ,
\end{equation}
which is Dirac's quantization condition.

\subsection{Wu and Yang's construction with two patches}
\begin{figure}[h!]
\begin{center}
\includegraphics[width=14cm]{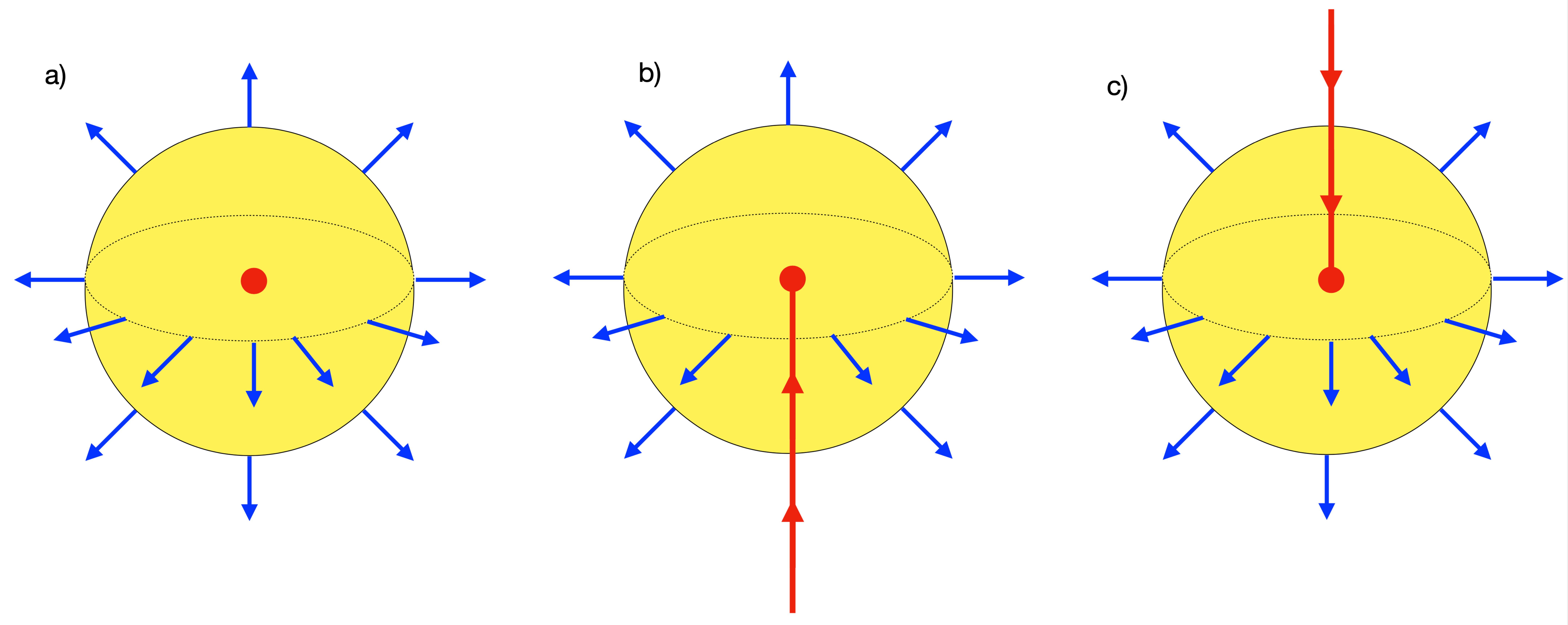}
\caption{\label{fig:diracstrings} The Dirac monopole is represented as a red point. (a) The monopole alone creates a radial magnetic field (blue arrows) with a total flux $4\pi g$ through the sphere $S^2$ oriented outward if $g>0$. This magnetic field cannot be written as the curl of a vector potential. (b) The magnetic monopole plus an attached semi-infinite thin solenoid carrying a magnetic flux (solenoid in red) $-g$ that compensates the flux from the monopole itself. The magnetic field of this composite system "monopole+solenoid" can be expressed as the curl of a vector potential $\vA^{(n)}$. The latter is regular everywhere on the north cap except along the negative $z$ axis. (c) The location of the radial Dirac string/solenoid is changed (with respect to b)) and therefore the magnetic field of the "monopole+solenoid" is the curl of the vector potential  $\vA^{(s)}$ defined everywhere except along the north radial line (positive $z$ axis). The analytic expressions of $\vA^{(n)}$ and $\vA^{(s)}$ are given in the text, see Eqs.~(\ref{ANem},\ref{ASem}).}
\end{center}
\end{figure}
In 1975, T. Wu and C.N. Yang proposed an alternative way to handle the Dirac monopole problem which circumvents the use of the Dirac strings~\cite{Wu:1975}. The idea consists in working with two-distinct gauges, $\vA^{(n)}$ and $\vA^{(s)}$, each of them well-defined in a restricted subset of space, respectively $R_n$ and $R_s$ (see Fig.~\ref{fig:diracstrings}). The two regions $R_n$ and $R_s$ are such that their union covers all space, and that their intersection is not empty. Using this procedure, the magnetic flux : 
\begin{equation}
 \oiint_{S^2} \,   \vB \, . \,  {\boldsymbol{dS} }= \oint_{\mathcal{C}} \vd\vl \, \, . \, (  \vA^{(n)}   - \vA^{(s)}  ) \neq 0  \, ,
\label{StokesmonopWuYang}
\end{equation}
is not necessary zero because $\vA^{(n)}$ and $\vA^{(s)}$ are distinct (they differ by the gradient of a scalar function).

Let us now find expressions for the different vector potentials $\vA^{(n,s)}$ using a specific path $\mathcal{C}=\mathcal{C}_\theta$, which is the parallel-type circle defined by the constant value $\theta$ of the polar angle, separating the sphere in a north cap $\mathcal{N}$ and a south cap $\mathcal{S}$. Applying Stokes' theorem to the north cap :
\begin{equation}
\iint_{\mathcal{N}} \vB \, . \,  dS \ve_r  = 2 \pi (1-\cos \theta ) g  =   \int \vA^{(n)} \, . \, \vd \vl = 2 \pi r \sin \theta \, \vA^{(n)}.\ve_\varphi \,  ,
\end{equation}
yields the following expression for the electromagnetic vector potential :
\begin{equation}
\vA^{(n)}=  \frac{g(1 - \cos \theta)}{ r \sin \theta}  \ve_\varphi    \, ,
\label{ANem}
\end{equation}
which is singular when $\theta = \pi$. The negative $Oz^{-}$ axis corresponds to the Dirac string singularity. The potential $\vA^{(n)}$ is well-defined everywhere else, namely in the set $R_n=\mathbb{R}^3-(Oz^{-})$. Note that any closed surface within $R_n$ does not contain the monopole, so the total flux is zero. 

Applying similarly Stokes' theorem to the south cap :
\begin{equation}
\iint_{\mathcal{S}} \vB \, . \,  dS \ve_r  = 2 \pi (1+\cos \theta ) g  =  - \int \vA^{(s)} \, . \, \vd \vl = - 2 \pi r \sin \theta \, \vA^{(s)}.\ve_\varphi \, ,
\end{equation}
provides another expression for the vector potential :
\begin{equation}
\vA^{(s)}=  -\frac{g(1 + \cos \theta)}{ r \sin \theta}  \ve_\varphi    \, ,
\label{ASem}
\end{equation}
which is singular along the positive $Oz$ axis ($\theta \rightarrow 0$). 

Along any parallel $\mathcal{C}_\theta$, both vector potentials are well-defined and we can therefore compare them. The difference of the two vector potentials is given by a gradient :
\begin{equation}
\vA^{(n)} - \vA^{(s)}=  \frac{2 g}{ r \sin \theta}  \ve_\varphi  =   \boldsymbol{\nabla} \left( 2g \varphi \right)  \, .
\end{equation}

Up to this point, everything was derived within classical electrodynamics. Quantum mechanics enters via the concept of gauge invariance/connection which states that the wave functions along a parallel $\mathcal{C}_\theta$ in the gauge $(n)$ and $(s)$ differ by a phase factor as
\begin{equation}
\Psi^{(n)}(\vr) = \Psi^{(s)}(\vr) \, e^{i \frac{e}{\hbar} \int^{\vr} (\vA^{(n)} -\vA^{(s)} )  \, \vd \vl }
= \Psi^{(s)}(\vr) \, e^{i \frac{e}{\hbar}2 g \varphi}\, ,
\end{equation}
where $\varphi$ is the azimuthal angle. Hence, in order to ensure the single-valuedness of the wave function when comparing $\varphi$ and $\varphi+2\pi$, one  must impose that
\begin{equation}
 \frac{e}{\hbar} 2g 2 \pi = n 2 \pi \, ,
\end{equation}
where $n$ is an integer. This can be rewritten in the following way : 
\begin{equation}
4 \pi g =  n \frac{h}{ e}      \hspace{5mm} {\rm with}  \hspace{5mm}    n \in   \,  \mathbb{Z}   \, ,
\end{equation}
meaning that the total flux of the monopole has to be a multiple of the flux quantum $h/e$, or that the monopole strength $g$ has to be an integer multiple of $g_0\equiv \hbar/(2e)$. The latter plays the role of a quantum of magnetic strength, i.e. of the ``smallest magnetic pole''. The number of such quanta (or the ``charge'' of the magnetic monopole) is therefore
\begin{equation}
\frac{g}{g_0} = \frac{1}{2\pi} \oiint_{S^2} \,   \frac{e}{\hbar}\vB \, . \,  {\boldsymbol{dS} } = n \, \,  \in \,  \mathbb{Z}  \, .
\end{equation}
The above ``quantization'' of the magnetic strength is unusual and is known as topological quantization (see the book by Thouless~\cite{Thouless:1998}). Its origin is different from usual ``quantum numbers'' that arise because of the eigen-spectrum of a Hermitian operator (an observable). Here it is the result of a topological constraint, i.e. that the gluing condition for the wave function on the equator of the sphere is related to mappings from the circle to the circle and therefore to the winding number $n$ as the fundamental group of the circle is $\Pi_1(S^1)=\mathbb{Z}$ (an introduction to homotopy groups is given in section \ref{homotopy}). Dirac noticed that if one magnetic monopole is present in the universe, then all charges have to be quantized to preserve the single-valued character of wave functions. 

In summary, Dirac extends the realm of electromagnetism. In classical Maxwell electromagnetism, magnetic monopoles do not exist $\boldsymbol{\nabla}\cdot \vB=0$. With quantum mechanics, magnetic monopoles can exist $\boldsymbol{\nabla}\cdot \vB=4\pi g \delta(\vr)$ but only if their flux $4\pi g$ is a multiple of the flux quantum $h/e$.

\subsection{Dirac monopole as a a fiber bundle}
[This section may be omitted in a first reading by readers not familiar with the mathematical notion of a fiber bundle, to which we give a brief introduction later in section \ref{sec:introfiberbundles}]

In 1931, Dirac started its seminal paper on monopoles~\cite{Dirac:1931} by a philosophical discussion about the evolution of mathematics that occurs in parallel to physics and shifts towards always more abstract concepts, citing examples such as Riemannian geometry and non-commutative algebra. Amusingly, the mathematician Hopf published the very same year his work on the higher homotopy groups of the 3-sphere $S^3$ which is somewhat related to the Dirac monopole issue. The Hopf fibration relies on the fact that $S^3$ can be seen as being a nontrivial fiber bundle with base space $S^2$ and fiber $S^1$, ``non trivial'' meaning that globally $S^3\neq S^2\times S^1$ although the equality holds true locally. Dirac was probably not aware of this work and it took more than 40 years to physicists and mathematicians to realize that the mathematical structure behind the Dirac monopole (with unit charge) is indeed the Hopf fiber bundle~\cite{Ryder:1980,Minami:1979}. Wu and Yang are actually the ones who realized that the mathematical structure behind the Dirac monopole was that of fiber bundles. Here, the base space is the total space minus the position of the monopole i.e. $\mathbb{R}^3-\{0\}\sim S^2\times \mathbb{R}^+ \sim S^2$ and the fiber corresponds to the phase of the wave function i.e. $U(1)\sim S^1$. As the total space $S^3$ is not globally the direct product of the base space $S^2$ and the fiber $S^1$, the fiber bundle is said to be non-trivial or twisted. A twisted fiber bundle can be characterized by a topological invariant. When the fiber is a complex vector space (here a one-dimensional Hilbert space), this invariant is known as the first Chern number and reads
\begin{equation}
n = \frac{1}{h/e} \oiint_{S^2} \,  \vB \, . \,  {\boldsymbol{dS} }\, ,
\end{equation}
which we recognize again as the ``charge'' of the magnetic monopole or the number of flux quanta piercing the sphere. Actually, for a twisted fiber bundle, the ``wave function'' is no longer a function but becomes a more general object (as understood by Dirac) and now known as a ``wave-section'', building on the notion of a section of a fiber bundle. The main difference with an ordinary function is that a wave-section or generalized wave function can have a non-integrable phase.

\section{Emergent Berry monopole for a two-level system \label{section0D}}

In this section, we consider a two-level system (TLS) driven by some external parameters, a typical example being a spin coupled to an external magnetic field. We use this fundamental system to introduce the concept of Berry phase, also called geometric phase \cite{Berry:1984,WilczekShapere:1989,XiaoBerryRMP:2010,Bernevig:2013,Vanderbilt:2018}. The Berry phase is a phase angle (a number defined modulo $2 \pi$) that quantifies the global phase evolution of a quantum state when this state is transported along a closed loop in the external parameter space. The related concepts of Berry connection, curvature and flux are also introduced in this minimal context. Finally, we describe the analogy between two apparently unrelated situations : the Berry connection of a TLS driven by two external parameters on the one hand~\cite{Berry:1984,WilczekShapere:1989,XiaoBerryRMP:2010}, and the electromagnetic vector potential of a charge moving in the background field of a Dirac monopole on the other hand~\cite{Dirac:1931}. The main idea is the existence of a topological number which is the flux of the Berry curvature in the TLS case, and the flux of magnetic field for the Dirac monopole.   

\subsection{Two-level system \label{subsec:BerryTLS}}

Before discussing $D$-dimensional lattice systems ($D=1,2,3$), we start by introducing briefly the fundamental topological aspects of the $0$-dimensionnal quantum TLS, whose Hamiltonian generically reads :
\begin{equation}
\label{H_spin}
H  =  \vd \cdot {\boldsymbol \sigma}  = \begin{pmatrix} 
d_z & d_x - i d_y \\
d_x + i d_y  & -d_z 
\end{pmatrix} \, .
\end{equation}
The TLS consists in some isospin degree of freedom described by standard Pauli matrices ${\boldsymbol \sigma}=(\sigma_x,\sigma_y,\sigma_z)$, coupled to its environment via the external parameters $\vd=(d_x,d_y,d_z)$. In the example of a spin 1/2 in a magnetic field, the vector $\vd$ corresponds to the external magnetic field. Alternatively, Eq. (\ref{H_spin}) may also describe a superconducting qubit, like the Cooper pair box, the fluxonium, or the transmon. In this latter example, the isospin ${\boldsymbol \sigma}$ would describe some charge, phase or flux degrees of freedom, and the vector $\vd=(d_x,d_y,d_z)$ would be a set of control parameters depending on gate voltages, bias fluxes, etc... Whatever the 
physical implementation is, the energy levels are generically given by :
\begin{equation}
\label{E_spin}
E_{\pm} = \pm |\vd | \, ,
\end{equation}
meaning that the spectrum is determined by the norm of the vector of external parameters $\vd$ solely. In contrast, the corresponding spinor wave functions of the excited state $\ket{\Psi _+^{(n)}} $ and ground state $\ket{\Psi _-^{(n)}} $ depend solely on the direction of the vector $\vd$, and can be written :
\begin{equation}
\ket{\Psi _+^{(n)}} = \begin{pmatrix} \cos{\frac{\theta}{2}}  \\ \sin{\frac{\theta}{2}} e^{i \varphi}  \end{pmatrix}  \hspace{5mm} {\rm and}  \hspace{5mm} \ket{\Psi _-^{(n)}}=\begin{pmatrix} \sin{\frac{\theta}{2}} e^{-i \varphi} \\ -\cos{\frac{\theta}{2}}   \end{pmatrix} \, ,
\label{spinorsGaugeNorth}
\end{equation}
where $\theta$ and $\varphi$ are respectively the polar (colatitude) and azimuthal (longitude) angles of the vector $ \vd$. The explicit forms of the spinors given in Eq. (\ref{spinorsGaugeNorth}) are not unique because they are defined up to a global phase. The choice made in Eq. (\ref{spinorsGaugeNorth}) defines a gauge where the spinor $\ket{\Psi _-^{(n)}}$ is not well-defined when $\theta = \pi$. Indeed at the south pole of the Bloch sphere this spinor reads 
\begin{equation}
\ket{\Psi _-^{(n)}} \rightarrow   \begin{pmatrix}  e^{-i \varphi} \\ 0 \end{pmatrix} \, ,
\end{equation}
and $\varphi$ is not defined at the poles. In contrast, $\ket{\Psi _-^{(n)}} \rightarrow (0,-1)^T$ is well-defined at the north pole, and in fact everywhere except at the south pole, hence the superscript $(n)$. This gauge is characterized by the fact that the phase $\varphi$ disappears from the spinors $\ket{\Psi _\pm^{(n)}} $ at the north pole, where $\theta=0$, but not at the south pole.

 In order to cure the fact that the ground state spinor is not defined unambiguously at the south pole of $S^2$, it is possible to choose a different gauge simply by multiplying the spinors Eq. (\ref{spinorsGaugeNorth}) by an overall $e^{\pm i \varphi}$ factor, leading to :      
\begin{equation}
\label{spinors0D}
\ket{\Psi _+^{(s)}} = \begin{pmatrix} \cos{\frac{\theta}{2}} e^{-i \varphi}  \\  \sin{\frac{\theta}{2}}   \end{pmatrix}  \hspace{5mm} {\rm and}  \hspace{5mm} \ket{\Psi _-^{(s)}}=\begin{pmatrix} \sin{\frac{\theta}{2}}  \\ - \cos{\frac{\theta}{2}} e^{i \varphi}  \end{pmatrix} \, .
\end{equation}
Note that the phase factor $e^{\pm i \varphi}$ is now multiplied by $\cos{\frac{\theta}{2}}$ instead of $\sin{\frac{\theta}{2}}$, compare with Eq.~(\ref{spinorsGaugeNorth}). Within this new gauge, denoted by the superscript $(s)$, the ground state wave function is now well-defined at the south pole, but at the expense of being ill-defined at the north pole because when $\theta \rightarrow 0$, then $\ket{\Psi _-^{S}} \rightarrow (0,e^{i \varphi})^T$~\footnote{One may think that it would be a good idea to have a more symmetric phase by having $e^{\pm i\varphi/2}$. Actually in that case the wave function is no longer single-valued, which is a problem.}. There is always a singularity remaining:  changing from gauge $(n)$ towards gauge $(s)$ only moves away this singularity but cannot remove it completely. We will see that this is related to the fact that the total Berry flux is non zero. 

We have exhibited two distinct gauges, namely $(s)$ and $(n)$, and we will keep on comparing them below for pedagogical purposes, but there are of course an infinity of other possible gauges. They are all perfectly equivalent and useful to describe the system at given values of the parameters. In a time-independent problem, this huge gauge freedom is merely a matter of fixing a working convention for representing states at once, and of course this initial choice will not 
alter the final results. The stationary states of the Hamiltonian are fixed up to a global phase. Once this phase is chosen, it is possible to keep this fixed basis to study the unitary evolution of the state of the system.

In a time-dependent problem, the situation is more subtle. Even in the adiabatic limit, the parameters $\vd$ of the Hamiltonian change and thus one needs to diagonalize a different Hamiltonian at each value of the parameters, and therefore pick a different choice of global phase for each values of these parameters. There is a huge amount of gauge freedom, and clearly the physical observables cannot depend on this arbitrariness.

\subsection{Gauge freedom and Berry connections}

We  now consider a driven-TLS described by a parameter-dependent Hamiltonian $H(\theta,\varphi)$. If we are interested in the variation of the spinors, it is convenient to use a parametrization in terms of the spherical angles $\theta$ and $\varphi$ of the vector $\vd$. A spinor state is attached to each point of the unit sphere $S^2$, which is called the Bloch sphere in this context (the Riemann sphere for mathematicians). The Bloch sphere representation is extensively used to monitor the evolution of a spin state in nuclear magnetic resonance (NMR), or a qubit state in quantum electronic circuits. Mathematically, the mapping from the unit vector $\hat{\vd}$ to the normalized spinor $\ket{\Psi_+^{(n)}}$ is the stereographic projection from $S^2$ to the complex projective plane $\mathbb{C}P^1$, performed from the south pole. 

The next step is to focus on the evolution of the groundstate $\ket{\Psi }=\ket{\Psi _-}$ as the angular parameters $\theta$ and $\varphi$ are varied, so we drop the subscript in the following. This is essentially the idea of the adiabatic following of a single level. At this point, we have projected on a single band (the lowest level). We call it a band because we consider that the Hamiltonian depends on two parameters $(\theta,\varphi)$. As long as only the specific properties of the spinorial wave functions are investigated, the specific dispersion of the band $E_{-}(\theta,\varphi)=-|\vd(\theta,\varphi)|$ upon $(\theta,\varphi)$ is not relevant. Provided the level do not cross, i.e. $|\vd(\theta,\varphi)| \neq 0$ for all parameter values, it is even possible to do a band flattening procedure which leads to a constant groundstate energy $E_-$. To follow the evolution of an eigenstate $\ket{\Psi }$, it is natural to compute overlaps such as 
\begin{equation}
 \braket{\Psi(\theta,\varphi)  \mid  \Psi(\theta+d\theta,\varphi)  }        \hspace{5mm} {\rm and}  \hspace{5mm}    \braket{\Psi(\theta,\varphi)  \mid  \Psi(\theta,\varphi+d\varphi)  }      \, .
\end{equation}
Extracting the phase of these inner products leads to define the Berry connection of a ket $\ket{\Psi}$ as the inner products~\cite{XiaoBerryRMP:2010}:
\begin{equation}
\label{berryconnec}
A_\theta=  i  \braket{\Psi  \mid \partial_\theta \Psi }        \hspace{5mm} {\rm and}  \hspace{5mm}   A_\varphi =  i  \braket{\Psi  \mid \partial_\varphi \Psi }       \, ,
\end{equation}
where $\partial_\theta =\partial / \partial \theta $ and $\partial_\varphi =\partial / \partial \varphi $. The components $A_\theta$ and $ A_\varphi$ are real quantities, because $\braket{\Psi  \mid \partial_\theta \Psi }$ and $\braket{\Psi  \mid \partial_\varphi \Psi }$ are purely imaginary. Indeed $\partial_\theta (\langle \Psi | \Psi \rangle)=0=\langle \partial_\theta  \Psi | \Psi \rangle+ \langle   \Psi | \partial_\theta \Psi \rangle$ and therefore $(\langle \partial_\theta  \Psi | \Psi \rangle)^*=\langle \Psi | \partial_\theta  \Psi \rangle = -\langle \partial_\theta  \Psi | \Psi \rangle$.    

The Berry connections are gauge-dependent objects. For instance, the Berry connection associated to the ground state $|\Psi_-\rangle$ is then given by :
\begin{equation}
   A^{(n)}_\theta=  0     \hspace{5mm} {\rm and}  \hspace{5mm}    A^{(n)}_\varphi = \sin^2 \frac{\theta}{2}  \, .
\label{berryconnecSPINgaugeNorth}
\end{equation}
within the $(n)$-gauge, and by :
\begin{equation}
 A^{(s)}_\theta=  0     \hspace{5mm} {\rm and}  \hspace{5mm}  A^{(s)}_\varphi= - \cos^2 \frac{\theta}{2}  \, ,
\label{berryconnecSPINsouth}
\end{equation}
within the $(s)$-gauge. These two different expressions for $A^{(s)}_\varphi$ differ by a constant one, which is the gradient $\partial_\varphi \varphi =1$.

Berry noticed that integrals of these connections around closed loops in parameter space are gauge-independent. Let us consider the circulations of $A^{(s)}_\varphi$ and $A^{(n)}_\varphi$ along a specific path $\mathcal{C}_\theta$, which is defined as the parallel-type circle at constant $\theta$, and oriented from $\varphi=0$ to $\varphi=2 \pi$. Those two circulations read :
\begin{equation}
\Phi^{(n)}  =  \oint_{\mathcal{C}_\theta} d\varphi  \, A^{(n)}_\varphi   = 2 \pi \sin^2 \frac{\theta}{2}    \hspace{5mm} {\rm and}  \hspace{5mm} \Phi^{(s)}  =  \oint_{\mathcal{C}_\theta} d\varphi  \, A^{(s)}_\varphi   = - 2 \pi \cos^2 \frac{\theta}{2}  \, .
\label{BerryPhases}
\end{equation}
Clearly these circulations differ by $\Phi^{N}-\Phi^{S}=2 \pi$, and therefore describe the same phase. The relevant gauge-invariant quantity is not the Berry phase (except in the modulo $2\pi$ sense), but rather its exponential, i.e. the Berry phase factor also known as an abelian Wilson loop (more on Wilson loops in Sec.~\ref{sec:nabp}):
\begin{equation}
W(\mathcal{C}_\theta)=e^{i \Phi^{(n)} } =  e^{i \Phi^{(s)} }  \, .
\label{ExpoBerryPhases}
\end{equation}
In conclusion, the Berry phase accumulated along a closed path is gauge independent modulo $2 \pi$, and therefore may be observable in some interference experiments. In contrast, the Berry phase accumulated by a quantum state along an open path of the parameter space typically/usually depends on the gauge, except if one takes special care by defining a closing procedure, see Ref.~\cite{Resta:2000}. We will see one such example of open-path Berry phase when discussing the Zak phase, see Sec.~\ref{sec:zak}.

The Berry phase is an example of anholonomy, i.e. the failure to come back to the exact same initial state after performing parallel transport along a closed path in a curved parameter space. It is actually a quantum version of a well-known geometrical effect. An elementary example, not in quantum mechanics, is that of the parallel transport of a stick on the surface of earth (the globe). Imagine a walker starting from the north pole and holding a stick in a given direction. The walker now moves to the south along a meridian trying to maintain the stick parallel at each moment (that's the notion of parallel transport). The walker next reaches the equator, makes a left turn and walks along the equator for a quarter of its length, before turning left again to move along a meridian towards the north and finally reaches the north pole again. In this closed path, trying to parallel transport a stick, the surprise of the walker is that the final direction of the stick makes an angle (90 degrees in our example) with the original direction. The angle between the initial and final direction is equal to the solid angle covered on the globe (namely $1/8$ of the total solid angle $4\pi$ in our example). The Berry phase is a quantum version of such a classical anholonomy.

\subsection{Berry curvature and Chern number \label{subsec:ChernTLS}}

It is important to define physical quantities that are independent of the gauge choice. By taking the curl of the Berry connection Eq. (\ref{berryconnec}), it is possible to get rid of the gradients and obtain such a gauge-invariant quantity, the so-called Berry curvature. In a 2D parameter space, the curl has only one component which is a pseudo-scalar :
\begin{equation}
F_{\theta \varphi}   =\partial_{\theta} A_\varphi - \partial_{\varphi} A_{\theta} =  i  \braket{\partial_\theta \Psi  \mid \partial_\varphi \Psi }    -i \braket{\partial_\varphi \Psi  \mid \partial_\theta \Psi } =i  \braket{\partial_\theta \Psi  \mid \partial_\varphi \Psi }  + c.c. \,  .
\label{berrycurvatureSPIN}
\end{equation}
For the TLS, the Berry curvature reads :
\begin{equation}
F_{\theta \varphi}   =\frac{1}{2} \sin \theta \,  ,
\label{berrycurvatureSPINexplicit}
\end{equation}
and its total flux integrated over the whole parameter space is finite :
\begin{equation}
 \iint_{S^2}   d\varphi d\theta \, F_{\theta \varphi} =\frac{1}{2}  \int_0^{2\pi} d\varphi  \int_0^{\pi} d\theta \,  \sin \theta  = 2 \pi \, . 
\label{berryfluxSPIN}
\end{equation}
This can be seen as i) the integral of the function $(\sin \theta)/2$ over the square $[0,2\pi]\times[0,\pi]$, or alternatively as ii) the flux of a radial vector of constant length $1/2$ through the unit sphere whose surface element is $d\theta \, d\varphi \, \sin \theta $. The nonzero value of the total flux through the parameter sphere signals a topological feature, that we will interpret in the next section as originating from the presence of a monopole of unit strength at the origin.  

\begin{figure}
\begin{center}
\includegraphics[width=11cm]{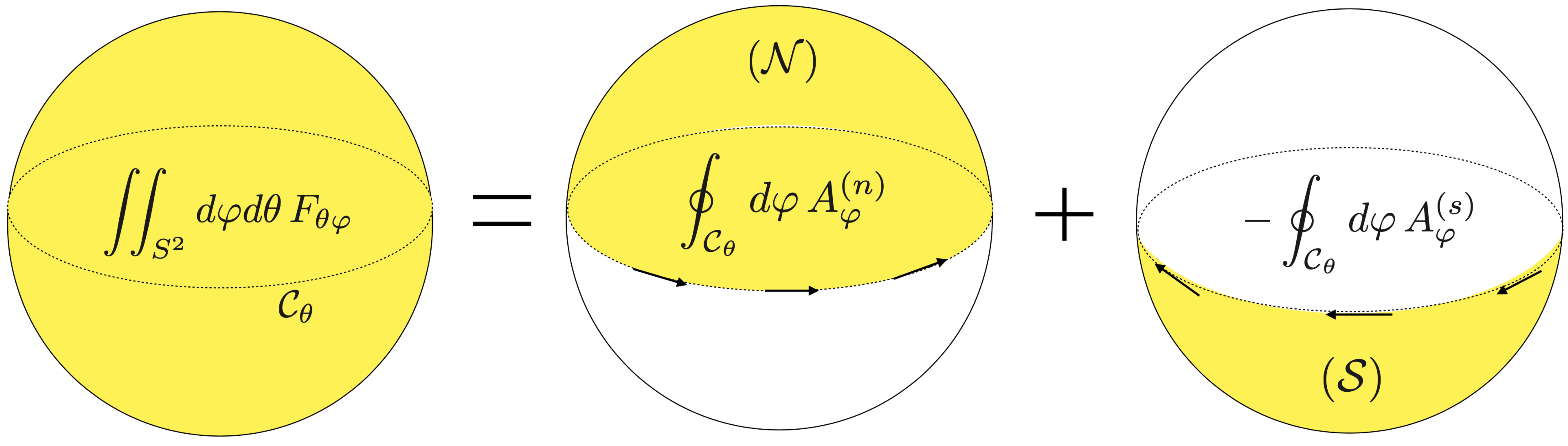}
\caption{The total flux of the Berry curvature $F_{\theta \varphi}$ through the whole Bloch sphere $S^2$ is equal to the difference of circulations of the Berry connection along the equator between two distinct gauges $(n)$ and $(s)$.}
\label{FigChernSpheres}
\end{center}
\end{figure}

One can prove that the total Berry flux is always a multiple of $2 \pi$ for a single band, using Stokes' theorem on appropriate domains of the sphere. Let us define two submanifolds realizing a partition of the sphere :  the cap ($\mathcal{N}$) gathering the regions located at the north of the parallel $\mathcal{C}_\theta$, and the cap ($\mathcal{S}$) at the south of $\mathcal{C}_\theta$ (Fig. \ref{FigChernSpheres}).

Within the northern cap ($\mathcal{N}$), one may safely apply Stokes' theorem using the gauge $(n)$:   
\begin{equation}
\iint_{\mathcal{N}} d\varphi d\theta \, F_{\theta \varphi}  =  \oint_{\mathcal{C}_\theta} d\varphi  \, A^{(n)}_\varphi   = 2 \pi \sin^2 \frac{\theta}{2}  \, ,
\label{StokesNorth}
\end{equation}
because $A^{(n)}_\varphi$ is well-defined over ($\mathcal{N}$). 

Within the southern cap ($\mathcal{S}$), one may similarly apply Stokes' theorem but using the gauge $(s)$:  
\begin{equation}
\iint_{\mathcal{S}} d\varphi d\theta \, F_{\theta \varphi} = - \oint_{\mathcal{C}_\theta} d\varphi \, A^{(s)}_\varphi     =  2 \pi \cos^2 \frac{\theta}{2}  \, ,
\label{StokesSouth}
\end{equation}
where the minus sign is due to the orientation of the circle $\mathcal{C}_\theta$, which should be reversed to apply Stokes' theorem to the south cap $\mathcal{S}$. 
Finally, the total flux through the 2-sphere is :
\begin{equation}
 \iint_{S^2} d\varphi d\theta \, F_{\theta \varphi} = \int_{\mathcal{N}} d\varphi d\theta \, F_{\theta \varphi} + \int_{\mathcal{S}} d\varphi d\theta \, F_{\theta \varphi} = \oint_{\mathcal{C}_\theta} d\varphi \,  ( A_\varphi^{(n)}  - A_\varphi^{(s)} )   = 2 \pi \, ,
\label{berryfluxPartition}
\end{equation}
where the last equality is consistent with the direct evaluation Eq. (\ref{berryfluxSPIN}). More generally, this shows that the total Berry flux is the circulation of the difference between Berry connections written in two distinct gauges and therefore it is the circulation of a gradient over a closed path, which has to be a multiple of $2 \pi$, here simply $2 \pi$. This allows one to define the Chern number as the total flux of the Berry curvature in units of $2 \pi$. This quantization of the total Berry flux is also valid for any closed 2D-manifold, like a torus $T^2$, because the same demonstration can be done by covering the parameter manifold by patches. A smooth connection is defined over each patch by a proper choice of gauge, the Berry phase accumulation along the closed path separating the patches has to be unique modulo $2 \pi$. 
\medskip

In conclusion, the Berry curvature is a local gauge-independent field. The Berry flux through the whole parameter space is a global gauge-independent quantity, which is a multiple of $2 \pi$. The Chern number is an integer, which is the total Berry flux in units of $2 \pi$. Recently, this quantization of the Berry flux has been measured for an individual superconducting qubit \cite{Schroer:2014}. It is actually possible to simulate the physics of a complete band structure with a single TLS that is driven in time. For example, the physics of the Haldane model on the honeycomb lattice, that we discuss below in Sec.~\ref{section2DHaldane}, was simulated in~\cite{Roushan:2014}.

In lattice systems of space dimensions two (see Sec.~\ref{section2D}), this Chern number is very important because it is related to physical observables and to their topological robustness.

\subsection{Berry flux monopole in parameter space}

In the previous paragraph, we have presented a justification of the integer character of the Chern number which is very reminiscent of the Wu-Yang construction of the flux quantization for a Dirac magnetic monopole. There is indeed a strong analogy between the structure of the two problems although they might seem very different at first sight.  

To facilitate the analogy, let us perform a change of parameter space from the spherical angular parameters $(\theta,\varphi) \in S^2$ to the Euclidian space $\mathbb{R}^3$ spanned by the cartesian parameters $(d_x,d_y,d_z)$ of the driven TLS. The Berry connection introduced previously appears in a new guise, because it is defined here with respect to the cartesian components of the parameter field $\vd=(d_x,d_y,d_z)$, rather than in terms of the spherical angles $(\theta, \varphi)$ of its direction :
\begin{equation}
\widetilde{\vA}=  i  \braket{\Psi  \mid {\boldmath \nabla}_\vd \Psi }    =    i  \braket{\Psi  \mid \left(   \ve_\theta \, \frac{1}{d}  \frac{\partial}{\partial \theta}   + \ve_\varphi \, \frac{1}{d \sin \theta}  \frac{\partial}{\partial \varphi}      \right) \mid \Psi }       \, .
\end{equation}
Here $\widetilde{\vA}=(\widetilde{A}_x,\widetilde{A}_y,\widetilde{A}_z)$ instead of $\vA=(A_\theta,A_\varphi)$. The radial component of the gradient could have been written, but it is actually vanishing because the eigenkets are independent of the norm $d=\mid \vd \mid$. This ``change of variables" leads to the following expressions. In the ``north-gauge" $(n)$, from Eq.~(\ref{spinorsGaugeNorth}), we immediately obtain for the new vector potential 
\begin{equation}
\widetilde{\vA}^{(n)}= \frac{ \sin^2 (\theta / 2) }{d \sin \theta}  \ve_\varphi =  \frac{1 - \cos \theta}{2 d \sin \theta}  \ve_\varphi    \, ,
\label{AN}
\end{equation}
corresponding to Eq.~(\ref{berryconnecSPINgaugeNorth}).  Similarly, in the ``south gauge" $(S)$ the new connection reads :
\begin{equation}
\widetilde{\vA}^{(s)}= -\frac{ \cos^2 (\theta / 2) }{d \sin \theta}  \ve_\varphi =  -\frac{1 + \cos \theta}{2 d  \sin \theta}  \ve_\varphi    \, ,
\label{AS}
\end{equation}
replacing Eq.~(\ref{berryconnecSPINsouth}). In this representation, the singularities at the poles can be seen even more explicitly in the expressions of the connections. For instance when $\theta \rightarrow 0$, the norm of  $\widetilde{\vA}^{(n)}$ is regular, while the norm of $\widetilde{\vA}^{(s)}$ diverges. We recognize that the Berry connections Eqs.~(\ref{AN},\ref{AS}) map exactly to the electromagnetic vector potentials Eqs.~(\ref{ANem},\ref{ASem}) provided one sets $g=1/2$ (i.e. $n=1$ with $\hbar=1$, $e=1$) and $d \rightarrow r$. The parameter space, spanned by the components of $\vd$, replaces the real space $\vr$ of the original Dirac monopole problem. The location of the Berry monopole at $\vd=0$ corresponds to a level degeneracy as $E_\pm =\pm |\vd|=0$.

The Berry curvature is obtained by taking the curl of the connection, in either gauge (so we drop the $(n),(s)$ indexes below) :
\begin{equation}
\vF = {\boldmath \nabla}_\vd \times \widetilde{\vA}= \frac{1 }{d \sin \theta}  \frac{\partial}{\partial \theta} (\sin \theta \widetilde{A}_\varphi ) \, \ve_r   = \frac{1 }{2 d^2}  \, \ve_r  \, ,
\label{AS2}
\end{equation}
which is a radial vector field. It is worth noticing that the Berry curvature is a 3-component vector in this 
definition, while it was a pseudoscalar in Sec.~(\ref{subsec:BerryTLS}). This Berry curvature vector is a local and gauge-invariant quantity, and it is therefore observable in principle. It is important to notice than the dimension of the parameter space is not related to the dimension of real space. 
As in the previous section, one can built a global gauge-invariant quantity by evaluating the flux of the Berry curvature through a surface. For instance, the flux of $\vF$ through a 
sphere, with radius $d$, surrounding the origin  :
\begin{equation}
 \oiint \, \vF \, . \,  {\boldsymbol{dS} }  = 4 \pi d^2/2 d^2=2 \pi  \,  .
\end{equation}
The geometric (or Berry) phase structure of the TLS is in fact related to the existence of a monopole in parameter space $(d_x,d_y,d_z)$. Within an electromagnetic analogy, the Berry phases can be interpreted as quantum mechanical phases accumulated by a charge coupled to a fictitious vector potential. This analogy between the driven-TLS and the Dirac monopole can be spelled out in detail. In the first case we have a quantum TLS described by a spinor wave function and no orbital coupling to a magnetic field. If we describe it in an approximate manner as a scalar -- one-component instead of two for the spinor -- wave function upon projection on a single band, i.e. adiabatic following, then we are forced to introduce an emergent gauge field (the Berry connection). The latter corresponds to a magnetic monopole in parameter space and accounts for the effect of virtual transitions to the other band (the band that we got rid of upon projection). Afterwards, we realize that we study a scalar wave function in the field of a magnetic monopole in parameter space. This is nothing but the situation considered by Dirac in real space. Hence we see that the problem of a scalar wave function in the field of a monopole is an adiabatic approximation to the quantum evolution of a spinor wave function. We also note that both the Dirac monopole and the TLS share the mathematical structure of the Hopf fibration~\cite{Urbantke:2003}.

\section{Geometrical and topological band theory}
\label{sec:bandtheory}

In the previous section, we outlined the topological features of a driven two-level system (TLS) controlled by two independent external parameters, essentially the angles $\theta$ and $\varphi$ that determine its spinor ground state. We now move to electrons on $D$-dimensional lattices. These particles may carry spin and/or some other internal isospin (orbital, sublattice,...). We neglect electron-electron interactions and concentrate on the geometrical and topological features of the band structure. Independent electrons (or cold atoms \cite{CooperRMP:2019}) in a periodic $D$-dimensional system occupy bands of Bloch states separated by gaps. In the absence of disorder, these Bloch states are labelled by a crystal momentum (or Bloch wave vector) $\vk$ living in $D$-dimensional periodic Brillouin zone (BZ), which is a torus $T^D$. Inter-band effects may occur when at least two bands are coupled. Then for each value of the crystal momentum $\vk$ one has essentially an effective TLS whose two states are generically described by the spinors Eq. (\ref{spinors0D}). In this context, the angles $\theta$ and $\varphi$ become functions of the crystal momentum $\vk$, and it is not necessary to drive externally the parameters. Indeed, the physical quantities are naturally expressed in terms of sums over the occupied Bloch states. The physics depends on the dimension $D$ and also on how the point $(\theta,\varphi)$ covers the Bloch sphere as $\vk$ spans the whole BZ $T^D$, which is constrained by the symmetries of the Bloch Hamiltonian. The Berry phase, connection, curvature and Chern number concepts (defined in the previous section on a $0$D TLS) have been extended very fruitfully to electrons in crystals, or atoms in optical lattices. 

In this section, we first introduce the mathematical concept of fiber bundle, and then show its implementation in band theory. An important difference with the previous sections (Dirac monopole and two-level system) is that the parameter space in band theory is a torus (the BZ) instead of a sphere. When relevant, we will point out the consequences of this difference. Next, we study geometrical phases and review the semi-classical formalism describing the dynamics of a particle in a given band in the presence of the Berry curvature field which accounts for the influence of inter-band transitions on the intra-band motion. This leads to the anomalous velocity concept and the related Hall effects. 

\subsection{Introduction to fiber bundles}
\label{sec:introfiberbundles}

\begin{figure}
\begin{center}
\includegraphics[width=7cm]{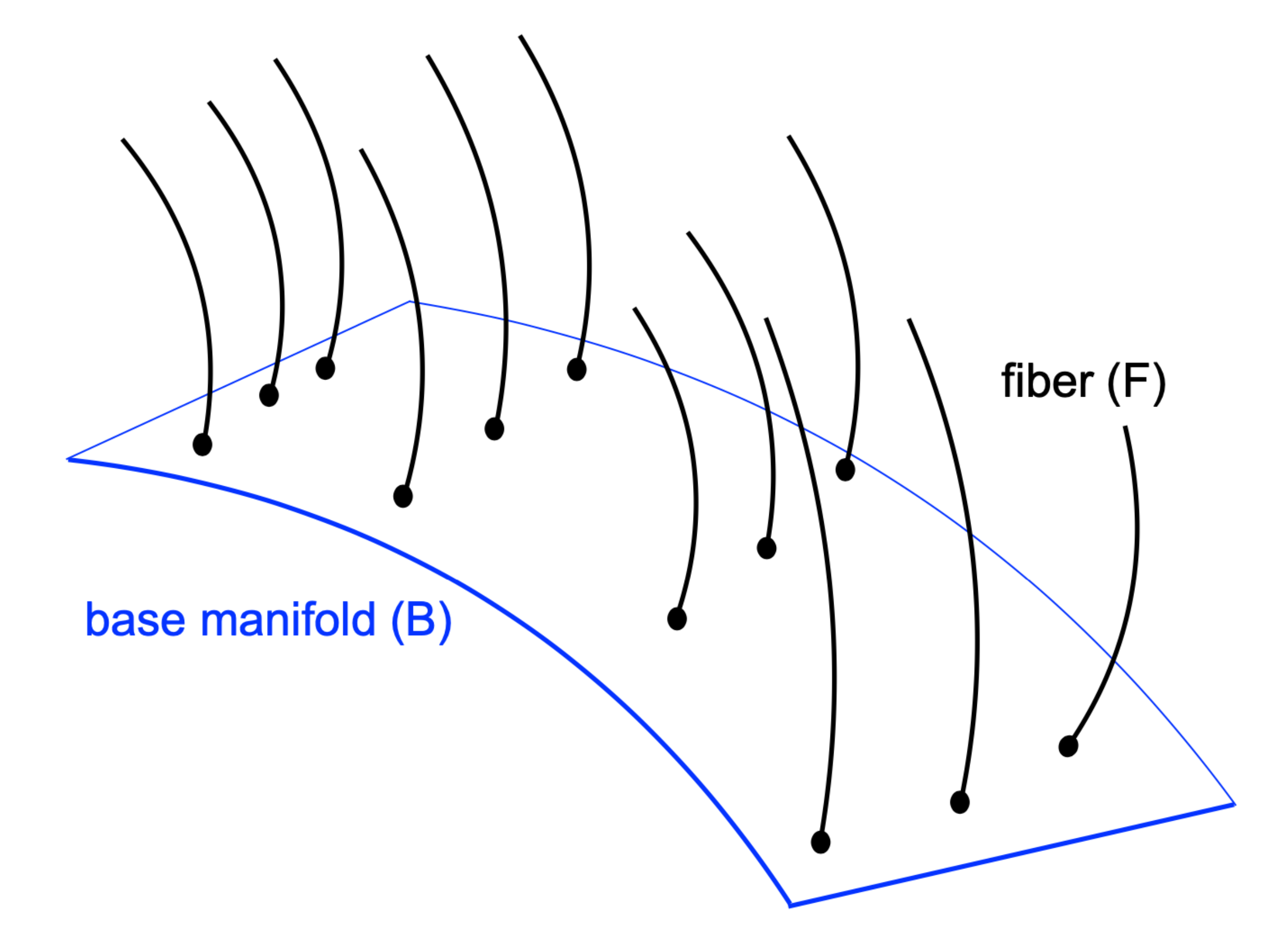}
\caption{A fiber bundle $E$ is a geometric object made of a base space $B$ at each point of which a fiber $F$ is attached. Here the base space is two-dimensional and the fibers are one-dimensional. Figure adapted from~\cite{Rowland}.}
\label{FigFB}
\end{center}
\end{figure}
The mathematical objects hiding behind the geometrization of band theory are fiber bundles (see Nakahara for a reference accessible to physicists~\cite{Nakahara} and \cite{Fruchart:2013}). A fiber bundle $E$ is a geometrical object made of a base manifold $B$, at each point of which a fiber $F$ is attached (see Fig.~\ref{FigFB}). A fiber is itself a manifold that may be, e.g., a real or a complex vector space. Locally, a fiber bundle $E$ resembles the direct product $B\times F$. A very simple example is $E=\mathbb{R}^3$ that may be described as a fiber bundle of base $B=\mathbb{R}^2$ and fiber $F=\mathbb{R}$ (alternatively it can also be described as a fiber bundle with base $B=\mathbb{R}$ and fiber $F=\mathbb{R}^2$). Even if locally, a fiber bundle $E$ resembles the direct product $B\times F$, this needs not be the case globally over the complete base space. When a fiber bundle is simply the direct product of a base space and a fiber $E=B\times F$, it is said to be topologically trivial. This is the case of the above example, $\mathbb{R}^3=\mathbb{R}^2\times \mathbb{R}$. When it is not, it is said to be non-trivial or twisted. The (local) geometry of a fiber bundle is described by objects such as connections and curvature, whereas its (global) topology is characterized by topological invariants called characteristic classes (for example, a Chern number). 

One also defines a map $p$ that projects from $E$ to $B$ and a structure group $G$ that acts on the fiber. A standard notation for a fiber bundle is
\begin{equation}
F \to E \xrightarrow{\text{p}} B  \, \, .
\end{equation}
An important notion about fiber bundles is that of a section $s$. It is a continuous map from $B$ to $E$. Naively, it could be thought as being the inverse of the projection map $p$. This is the case for a trivial fiber bundle but not for a twisted fiber bundle. Actually, the existence of a global and non-vanishing section is equivalent to the fiber bundle being trivial. A section can also be thought of as a generalization of a function defined over the base space. If the fiber bundle is trivial, a global section is simply an ordinary function. If it is twisted, the section is a function that is defined over different patches that together cover the base space. In other words, a section appears as a multi-valued (and hence ill-defined) function. Alternatively, a section can be seen as an extension of the notion of a function. One definition of a twister fiber bundle is that there is an obstruction in finding a global section that is non vanishing.

When the fiber is itself the structure group $G$, then the fiber bundle is called a principal $G$-bundle. This will turn out to be the relevant type of fiber bundles in band theory.

\begin{figure}
\begin{center}
\includegraphics[width=5cm]{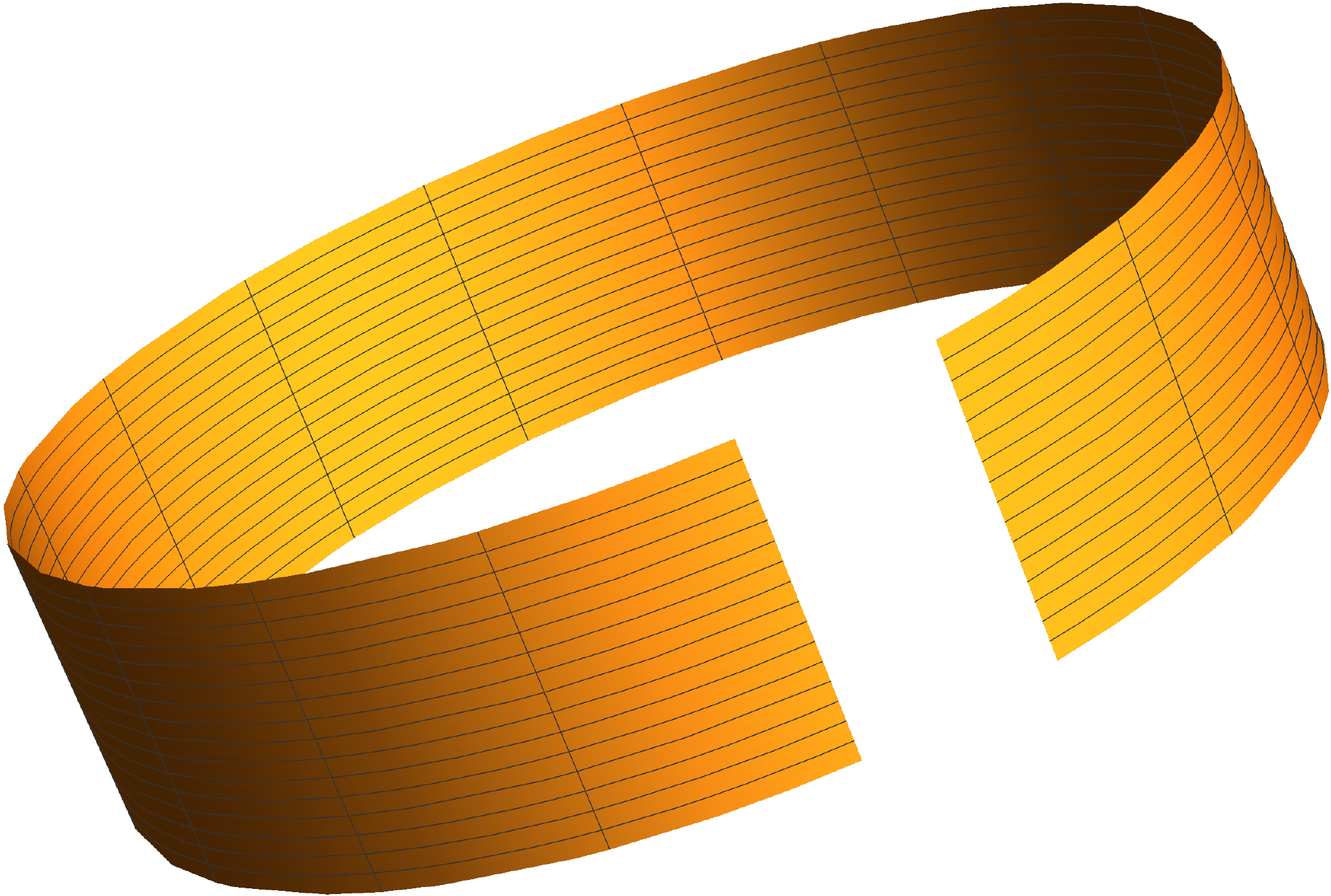}
\includegraphics[width=5cm]{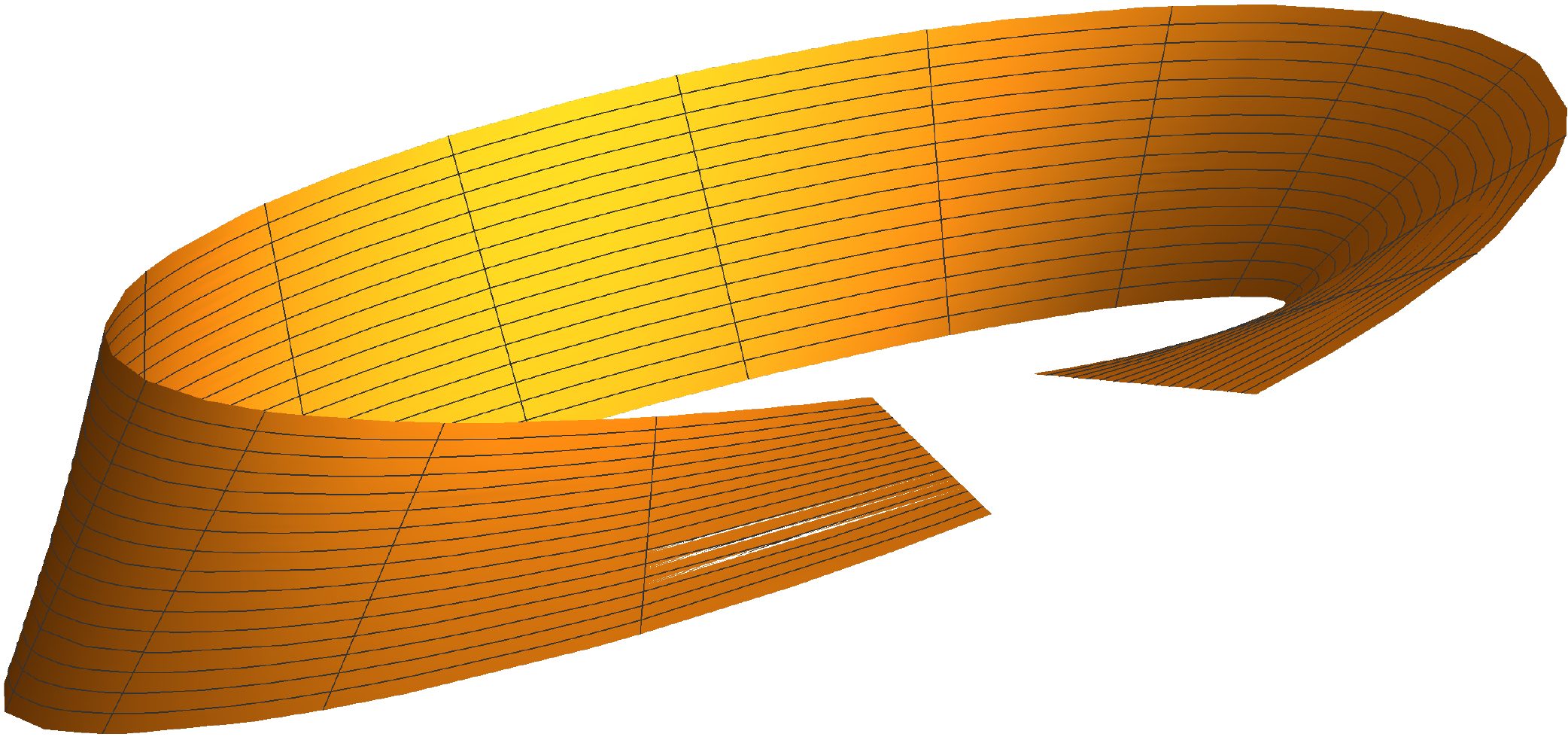}
\includegraphics[width=5cm]{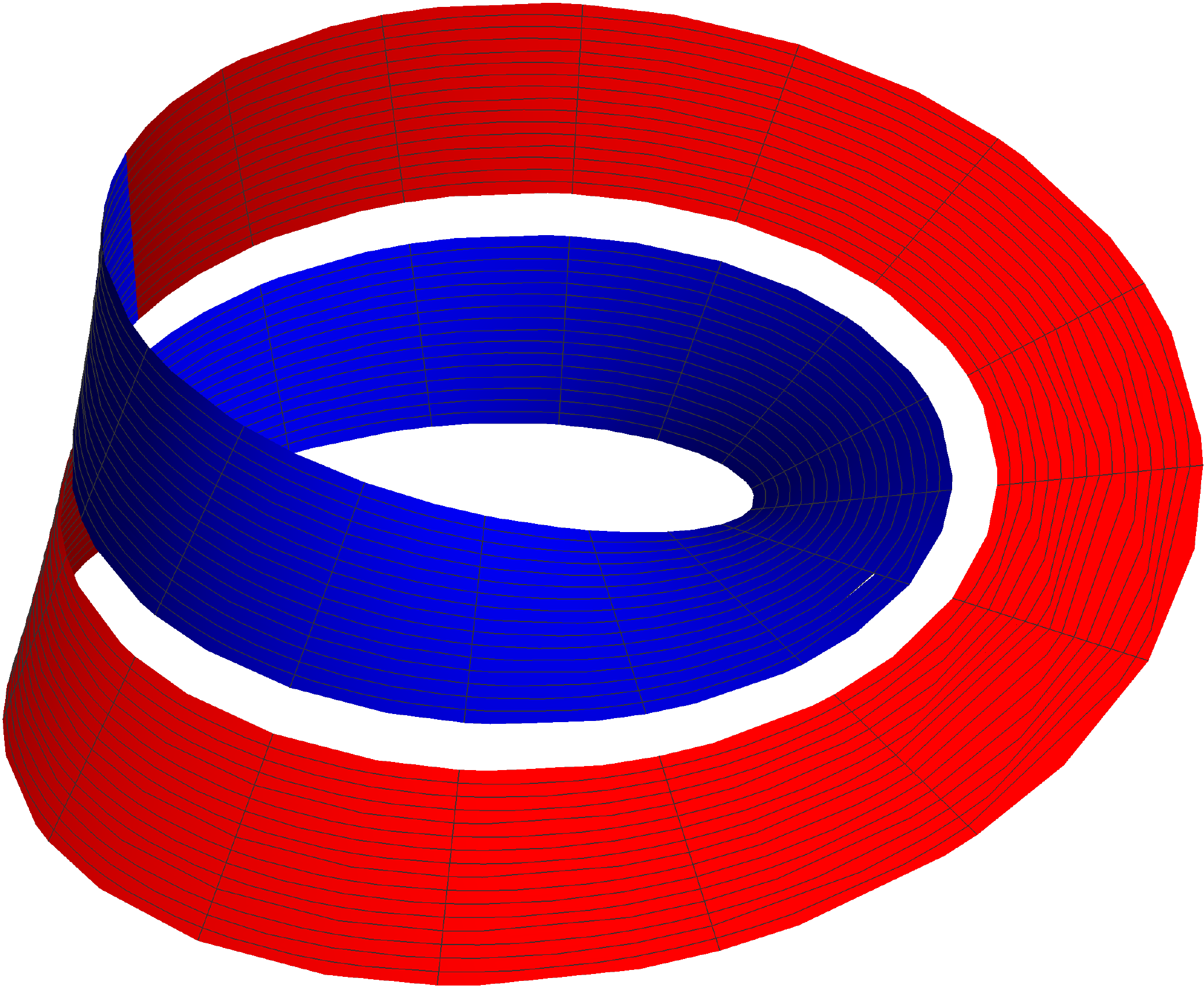}
\caption{\label{FigMoebius} Gluing of an open ribbon into either a cylinder (Left) or a M\"obius strip (Middle). Right: cutting a M\"obius strip in the middle of each fiber yields a single non-M\"obius strip (neither a cylinder nor a M\"obius strip).}
\end{center}
\end{figure}

The simplest example of a non-trivial fiber bundle is the well-known M\"obius strip. In this case, the base space is a circle $B=S^1$ and the fiber is a line segment $F=[-1,1]$. The structure group is $G=\mathbb{Z}_2=\{1,-1;\times\}$, whose non-trivial element ($-1$) flips the fiber $[-1,1]$ into $[1,-1]$. Locally, at each point of the circle $B$, one places a line segment $F$ perpendicularly. The global subtlety is in how these fibers are arranged around the circle. Let us call $\theta$ the parameter that spans the base space ($0\leq \theta \leq 2\pi$). If the gluing of the last fiber (the one at $\theta=2\pi^-$) to the first one (at $\theta=0^+$) is done naively, then one ends up with a regular cylinder, which is a trivial fiber bundle: the $\text{cylinder }=S^1 \times [-1,1]$ (see Fig.~\ref{FigMoebius} Left). If the gluing is done after twisting or flipping the last fiber (i.e. by gluing the upper end of the fiber at $\theta=2\pi^-$ to the lower end of the fiber at $\theta=0^+$), then one ends up with a M\"obius strip (see Fig.~\ref{FigMoebius} Middle). The latter is a non-trivial fiber bundle: the $\text{M\"obius strip } \neq S^1 \times [-1,1]$. The two fiber bundles (cylinder and M\"obius strip) are locally identical but are globally very different. For example, the M\"obius strip has a single edge and a single side, whereas the cylinder has two different edges and two different sides (an inside and an outside). For a twisted bundle, there is no global non-vanishing section, which translates into the following funny fact for the M\"obius strip: when cutting it in its middle (i.e. along the value $0$ in each fiber $F=[-1,1]$), one still obtains a connected object (made of a single piece but twice longer) that is neither a cylinder nor a M\"obius strip (see Fig.~\ref{FigMoebius} Right). It is similar to a M\"obius strip but with two twists instead of a single. 

Another example of a fiber bundle is that of the torus $T^2$ that can be described as a trivial fiber bundle of base space $S^1$ and fiber $S^1$ (very similar to the cylinder with $F=[-1,1]$ replaced by $F=S^1$): $T^2=S^1\times S^1$. The structure group is now $G=U(1)$ instead of $\mathbb{Z}_2$. When twisting such a fiber bundle in the same way as the M\"obius strip, one obtains the Klein bottle. Locally the torus and the Klein bottle are very similar, but globally one is a trivial and the other one a twisted fiber bundle. 

As a last example of a fiber bundle, we go back to our favorite Dirac monopole (of unit charge) which corresponds to the Hopf fiber bundle $E=S^3$ with base space $B=S^2$, fibers $F=S^1$ and structure group $G=U(1)$. Why $S^3$? Because $S^3$ is the Hilbert space for a spinor wave function $|\psi\rangle=(z_1,z_2)^T$. Normalization of the spinor means that its two complex components $z_1=x_1+i y_1$ and $z_2=x_2+i y_2$ satisfy $1=|z_1|^2+|z_2|^2=x_1^2+y_1^2+x_2^2+y_2^2$ which is indeed the equation of a unit sphere in 4-dimensional space, i.e. $S^3$. In the previous section on the Berry monopole, we have seen that there is a TLS hiding in the back of the Dirac monopole. The topology of such a fiber bundle is characterized by the Chern number, which can also be interpreted as the number of monopoles enclosed by the base space. The projection $p$ from $E=S^3$ to $B=S^2$ is known as the Hopf map.

\subsection{Band structure and Bloch fiber bundle}

In the context of band theory for a single electron in a periodic potential (a crystal), the relevant fiber bundles are called Bloch bundles. The base (parameter) space is the BZ torus ($\vk \in T^D$) and the fibers are complex vectors spaces (Hilbert spaces) corresponding to band space (with band index $n$), the dimension of which depends on the number $N_b$ of bands: $n=1,..,N_b$. For example, in a two-band system, one may define a Bloch bundle with a two-dimensional Hilbert space as a fiber. In this case, the complete Hilbert space is seen as a fiber bundle and the latter is not very interesting as it can be shown to be always trivial, see e.g.~\cite{Fruchart:2013}. A more interesting fiber bundle is obtained when one only takes a sub-set of bands, e.g. a fiber obtained by keeping only the one-dimensional Hilbert space corresponding to the lower band (imagine the case of a two-band insulator: the lower band is filled and the upper band empty and they are separated by a band gap). For a nice review on fiber bundles in the context of band theory, see Ref.~\cite{Fruchart:2013}.

\subsubsection{Bloch Hamiltonian and cell-periodic Bloch states}
Let us consider a single-particle Hamiltonian $H$ having translation invariance under a Bravais lattice (e.g. a tight-binding Hamiltonian). Most of this section is valid in any space dimension $D$, but at the end we will focus on space dimension $D=2$ which is special for topology as we have seen in the TLS section. A parameter-dependent Bloch Hamiltonian is defined via a unitary transformation as
\begin{equation}
H(\vk)=e^{- i \vk \cdot \vr} H e^{i \vk \cdot \vr}\, ,
\label{eq:blochham}
\end{equation}
where $\vr$ is the position operator and $\vk$ is a parameter with the dimension of a wave vector. In the following, we will refer to it as the canonical Bloch Hamiltonian or simply \emph{the} Bloch Hamiltonian. Bloch's theorem allows us to diagonalize the Hamiltonian as
\begin{equation}
H |\psi_{n,\vk}\rangle = E_n(\vk) |\psi_{n,\vk}\rangle \, ,
\label{bev}
\end{equation}
where $\vk$ is the crystal momentum and $n$ is a discrete band index. The Bloch eigenvectors are
\begin{equation}
|\psi_{n,\vk}\rangle = e^{i \vk \cdot \vr} |u_{n}(\vk)\rangle \, ,
\end{equation}
where in coordinate representation the ``cell-periodic Bloch state'' $u_{n,\vk}(\vr) = \braket{\vr | u_{n} (\vk) } $ obeys  
\begin{equation}
u_{n,\vk}(\vr+\vR)=u_{n,\vk}(\vr) \, ,
\label{eq:cpbs}
\end{equation}
namely it is periodic in real space with the periodicity of the unit cell ($\vR$ belongs to the Bravais lattice). In the following, in order to emphasize the fact that the Bloch wavevector $\vk$ now plays the role of a parameter (unlike the band index $n$ which remains a quantum number), we have chosen to write the cell-periodic Bloch eigenvector as $| u_{n} (\vk) \rangle $ rather than $| u_{n,\vk} \rangle $. Let us now examine the $\vk$-dependence of $| u_{n} (\vk) \rangle $ in the reciprocal space. One must have $|\psi_{n,\vk+\vG}\rangle \propto |\psi_{n,\vk}\rangle$ up to a global phase, which shows that the crystal momentum $\vk$ can be restricted to the first BZ. A common choice of phase is to ask that $|\psi_{n,\vk+\vG}\rangle=|\psi_{n,\vk}\rangle$ so that
\begin{equation}
|u_{n}(\vk+\vG)\rangle = e^{-i \vG \cdot \vr} |u_{n}(\vk)\rangle \, ,
\label{eq:pgc}
\end{equation}
which is known as the ``periodic gauge choice''~\cite{Resta:2000}. It does not fully fix the gauge but restricts the possible gauge choices. This choice is not always possible: in 2D there is a famous obstruction to it (known as a non-zero Chern number), that we discuss below. The cell-periodic Bloch eigenvectors $|u_{n}(\vk)\rangle$ will be the main players in the following~\footnote{Another reason for choosing $u_{n,k}(x)$ over $\psi_{n,k}(x)$ is that the $u_{n,k}(x)$'s all belong to the same Hilbert space as they have the same boundary conditions in $x=0$ and $x=a$~\cite{XiaoBerryRMP:2010,Vanderbilt:2018}.}. In general, they do not have the periodicity of the reciprocal lattice, see Eq.~(\ref{eq:pgc}). However, the energy bands do have the periodicity of the reciprocal lattice $E_n(\vk+\vG)=E_n(\vk)$.

The reason for performing the unitary transformation (\ref{eq:blochham}) is that we want a \emph{parameter-dependent} Hamiltonian in order to be able to separate two different dynamics: that associated with changing the wave vector $\vk$ (slow) and that associated with changing the band index $n$ (fast). The goal is to obtain an effective description for the dynamics of an electron restricted to a single band (this will be done by \emph{projecting} on a single band) nevertheless taking into account the coupling to other bands (this will occur via an emergent gauge field). To make this discussion more concrete, let us consider a one-dimensional tight-binding model (but with several bands) with hopping amplitude $t$, lattice spacing $a$ and crystal size $L$. The timescale for the intra-band dynamics can be estimated as $\sim \hbar L/(t a)$ (because the spacing between allowed $k$ values is $\Delta k =2\pi/L$ and the typical band velocity is $t a/\hbar$ so that the typical energy change $\Delta E\sim t a \Delta k \sim t a/L$), and that for the inter-band dynamics as $\sim \hbar/\text{bandgap}\sim \hbar/t$. For a macroscopic crystal $L\gg a$, the first timescale is much larger than the second. The idea that the emergent (Berry) gauge structure generally appears via a separation of two timescales such that the effective dynamics of the slow (or ``heavy'') degrees of freedom gets modified by the integration over the fast (or ``light'') degrees of freedom is well explained in Ref.~\cite{Berry:1989}. Paraphrasing Michael Berry, the reaction of the fast degrees of freedom (``light system'') onto the slow degrees of freedom (``heavy system'') occurs via the appearance of an emergent gauge field, as we will see. 

\subsubsection{Berry curvature and quantum metric}
The $|u_{n}(\vk)\rangle$'s are the parameter-dependent eigenvectors of the Bloch Hamiltonian:
\begin{equation}
H(\vk) |u_{n}(\vk)\rangle = E_n(\vk) |u_{n}(\vk)\rangle \, .
\end{equation}
This last equation appears very similar to (\ref{bev}) but is actually quite different. Whereas $\{|\psi_{n,\vk}\rangle,n,\vk\}$ form an orthonormal basis in Hilbert space, it is not so for $\{|u_{n}(\vk)\rangle,n,\vk\}$. Indeed $|u_{n}(\vk)\rangle$ and $|u_{n}(\vk')\rangle$ are eigenvectors of two different Hamiltonians $H(\vk)$ and $H(\vk')$ and therefore need not be orthogonal (however $\langle u_n(\vk)|u_{n'}(\vk)\rangle=\delta_{n,n'}$ as $|u_{n}(\vk)\rangle$ and $|u_{n'}(\vk)\rangle$ are eigenvectors of the same Hamiltonian $H(\vk)$). Their overlap 
\begin{equation}
\langle u_n(\vk)|u_n(\vk')\rangle \neq \delta_{\vk,\vk'} 
\label{overlap}
\end{equation}
is a non-zero complex number in general. Note that, precisely at this point, we have restricted the discussion to a single band (the $n^\text{th}$ band). This is the moment, where we stop discussing the dynamics in the full Hilbert space and project on a single band of interest. The corresponding fiber $F$ is a one dimensional complex vector space, i.e. essentially a $U(1)$ phase, while the first BZ torus $T^D$ plays the role of the base space $B$. 

When $\vk'=\vk +d \vk$ is close to $\vk$, one may study the deviation of this overlap (\ref{overlap}) from unity and define the evolution (i) of its phase and (ii) of its norm: 

\medskip

(i) The evolution of its phase, by expanding at first order in $\boldsymbol{dk}$, 
\begin{equation}
  \langle u_n(\vk)|u_n(\vk +\boldsymbol{dk})\rangle\approx  \langle u_n(\vk)|(1+\boldsymbol{dk} \cdot \boldsymbol{\nabla}_{\vk})|u_n(\vk)\rangle \approx e^{-i \boldsymbol{dk} \cdot \vA_n(\vk)}
\end{equation}
defines the Berry connection
\begin{equation}
 \mathbf{A}_n(\vk)=i\langle u_n(\vk)|\boldsymbol{\nabla}_{\vk} u_n(\vk)\rangle \,\, ,
\end{equation} which is the Bloch version of the TLS formula Eq.~(\ref{berryconnec}). This quantity is called $\boldsymbol{\mathfrak{X}}_{nn}(\vk)$ in~\cite{Blount:1962} and is related to the projected position operator. The Berry curvature is given by the curl of the Berry connection :
\begin{equation}
    F_{ij}^n(\vk)= i\langle \partial_i u_n|\partial_j u_n\rangle +\text{c.c.} \,\, ,
    \label{Curvat_u}
\end{equation}
which is the Bloch counterpart of Eq.~(\ref{berrycurvatureSPIN}) for the TLS. The Berry curvature $\mathbf{F}_n(\vk)=\text{curl } \mathbf{A}_n(\vk)$ is called $\boldsymbol{\Omega}_{n}(\vk)$ in~\cite{Blount:1962}. The geometry of the fiber bundle is described by the Berry connection, curvature and phase, while its topology is characterized by the Chern number (see below). In the context of band theory, this was first recognized by Thouless et al.~\cite{Thouless:1982,Thouless:1983} and Simon and coworkers~\cite{Simon:1983,Avron:1983} who underlined the relation with Berry's contribution.

\medskip

(ii) The evolution of the norm of this overlap (\ref{overlap}) defines another geometric quantity, known as the quantum metric, obtained by introducing a distance (squared) in the Hilbert space
\begin{equation}
ds^2=1-|\langle u_n(\vk)|u_n(\vk +\boldsymbol{dk})\rangle|^2\approx  \sum_{i,j} g_{ij}^n (\vk) dk_i dk_j \, .
\end{equation}
This distance in projective Hilbert space was introduced by Provost and Vallée \cite{ProvostVallee:1980}. Expanding at second order in $\boldsymbol{dk}$, the quantum metric is obtained as
\begin{equation}
    g_{ij}^n (\vk)= \text{Re} \langle \partial_i u_n|(\mathds{1}-|u_n \rangle\langle u_n|)|\partial_j u_n\rangle \, ,
\end{equation}
where $\partial_i$ is a short-hand notation for $\partial_{k_i}$ and $\mathds{1}$ is the identity. The quantum metric and the Berry curvature are the real and imaginary part of a more general object called the quantum geometric tensor:
\begin{equation}
    T_{ij}^n (\vk)= \langle \partial_i u_n|(\mathds{1}-|u_n \rangle\langle u_n|)|\partial_j u_n\rangle \, .
\end{equation}
Indeed $g_{ij}^n= \text{Re} T_{ij}^n$ and $F_{ij}^n=-2 \text{Im} T_{ij}^n$. For more information on the quantum metric, see the article by Berry in~\cite{WilczekShapere:1989}. The quantum metric appears in some physical quantities such as the magnetic orbital susceptibility~\cite{Piechon:2016} or the superfluid weight~\cite{Liang:2017}. It is also useful to define localization in an insulator~\cite{Resta:2011} and gives a measure of the minimal wavepacket spreading for a Bloch electron~\cite{Marzari:2012}.

\subsubsection{Berry gauge transformation}
There is a gauge freedom in the cell-periodic Bloch states seen as functions of the parameter $\vk$. Indeed, provided the phase $\varphi(\vk)$ is a smooth enough function of $\vk$, 
\begin{equation}
|\tilde{u}_n(\vk)\rangle=e^{i\varphi(\vk)}|u_n(\vk)\rangle
\end{equation}
is also a valid choice for the cell-periodic Bloch states. We refer to it as a Berry gauge transformation to distinguish it from a real-space electromagnetic gauge transformation. In this gauge transformation $|u_n(\vk)\rangle \to |\tilde{u}_n(\vk)\rangle$, the Berry connection gets modified as
\begin{equation}
\tilde{\mathbf{A}}_n=i\langle \tilde{u}_n |\boldsymbol{\nabla}_{\vk} \tilde{u}_n\rangle=\mathbf{A}_n-\boldsymbol{\nabla}_{\vk} \varphi
\end{equation} 
and is therefore not gauge-invariant, as the vector potential in electromagnetism. However, the Berry curvature is gauge-invariant since 
\begin{equation}
\tilde{\mathbf{F}}_n=\text{curl } \tilde{\mathbf{A}}_n = \text{curl } \mathbf{A}_n + 0 =\mathbf{F}_n
\end{equation}
just like the magnetic field. More generally, the quantum geometric tensor, i.e. both the Berry curvature and the quantum metric, is gauge-invariant and therefore measurable. A map of the Berry curvature and of the quantum metric in the whole BZ has been measured in artificial crystals, see e.g.~\cite{Flaeschner:2016,Gianfrate:2020}.

\subsubsection{Berry curvature and virtual transitions}
\begin{figure}
\begin{center}
\includegraphics[width=5cm]{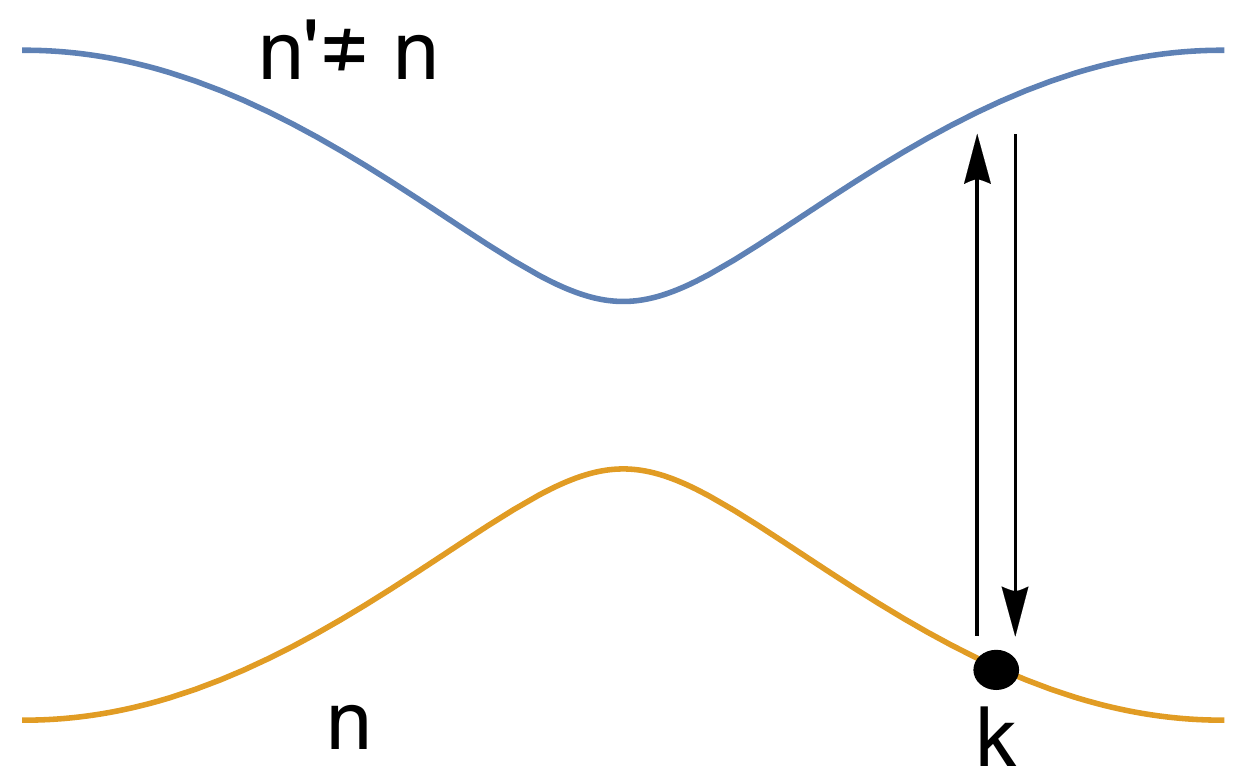}
\caption{\label{fig:virtualtrans} Virtual transitions (vertical arrows) between two bands at fixed wavevector $k$ leading to the appearance of an emergent gauge field (Berry phase effects) for an electron in the $n^\text{n}$ band.}
\end{center}
\end{figure}
The above expression may give the impression that the Berry curvature in the $n^\text{th}$ band only depends on a single band. This is actually not the case. To show that the Berry curvature is related to virtual transitions between bands (at fixed $\vk$), it is useful to rewrite it using perturbation theory to express $|\partial_j u_n\rangle$ as a function of $|u_{n'}\rangle$ with $n'\neq n$:
\begin{equation}
\label{BerryCurvatHamil}
    F_{ij}^n(\vk)= i\sum_{n'\neq n }\frac{\langle u_n|\partial_i H(\vk) | u_{n'}\rangle \langle u_{n'}|\partial_j H(\vk) | u_{n}\rangle}{(E_{n'}-E_n)^2} +\text{c.c.}
\end{equation}
This expression has the flavor of second-order perturbation theory and clearly shows that the Berry curvature in the $n^{\rm{th}}$ band is the result of virtual transitions (see Fig.~\ref{fig:virtualtrans}) to other bands and that the inter-band coupling is related to the velocity operator $\partial_i H(\vk)$. As a consequence, it is obvious that if there is a single band in the model, these effects are totally absent. Also, the Berry curvature is well-defined only when there is a gap between bands (at fixed $\vk$, i.e. $E_{n'}(\vk)\neq E_n(\vk)$) and becomes large when this gap becomes small. The above expression makes it obvious that the Berry curvature is gauge-independent (as for each bra $\langle u_{n'}|$, it involves the corresponding ket $| u_{n'}\rangle$), in contrast to the Berry connection which is gauge-dependent (as it involves a bra $\langle u_{n}|$ but a different ket $|\boldsymbol{\nabla}_\vk u_{n}\rangle$). The Berry curvature also has the periodicity of the reciprocal Bravais lattice $F_{ij}^n(\vk+\vG)=F_{ij}^n(\vk)$, although this is not the case of the $|u_n (\vk)\rangle$'s. A consequence of Eq. (\ref{BerryCurvatHamil}) is that the sum over all bands of the Berry curvature at a given $\vk$ point vanishes:
\begin{equation}
\label{sumrule}
    \sum_n F_{ij}^n(\vk)=0\, .
\end{equation}
Time-reversal symmetry implies
\begin{equation}
    F_{ij}^n(-\vk)= - F_{ij}^n(\vk)\, ,
    \label{trcurvature}
\end{equation}
while inversion symmetry imposes
\begin{equation}
    F_{ij}^n(-\vk)= F_{ij}^n(\vk)\, .
    \label{sicurvature}
\end{equation}
Hence, if both symmetries are present, the Berry curvature vanishes everywhere in the BZ.  

\subsubsection{Periodic versus canonical Bloch Hamiltonian (basis I/II issue)}

In general, the Bloch Hamiltonian $H(\vk)$ does not have the periodicity of the reciprocal lattice, even if the spectrum $E_n(\vk)$ always has it. The reason is that the definition of the Bloch Hamiltonian Eq. (\ref{eq:blochham}) involves the Hamiltonian $H$ \emph{and} the position operator $\vr$. The latter depends on the position of every site in the crystal, including sites within the unit cell (e.g. for a lattice with a basis). The distance between sites within the unit cell need not have a special relationship (i.e. need not be commensurable) with the Bravais lattice vectors. Therefore, in general, the Bloch Hamiltonian need not be a periodic function of $\vk$~\cite{Lim:2015}. In Sec.~\ref{section1D}, we will see on the example of the SSH chain that the Bloch Hamiltonian is periodic but with a double periodicity compared to the reciprocal lattice. In Sec.~\ref{section2D}, we will see another example: the honeycomb lattice, for which the Bloch Hamiltonian has a triple periodicity compared to the reciprocal lattice~\cite{Lim:2015}. 

An alternative ``periodic Bloch Hamiltonian'' is often defined as follows:
\begin{equation}
\mathcal{H}(\vk)=e^{- i \vk \cdot \vR} H e^{ i \vk \cdot \vR}\, .
\end{equation}
It depends only on the position operator for the unit cell $\vR$ (i.e. the position on the Bravais lattice) and not on the full position operator $\vr$. We will write $\vr=\vR+\boldsymbol{\delta}$, where $\boldsymbol{\delta}$ is the position operator within the unit cell. The canonical and the periodic Bloch Hamiltonian have the same energy bands $E_n(\vk)$. This alternative Bloch Hamiltonian is periodic with the reciprocal lattice $\mathcal{H}(\vk+\vG)=\mathcal{H}(\vk)$ unlike the Bloch Hamiltonian $H(\vk+\vG)=e^{-i\vk \cdot \boldsymbol{\delta}}H(\vk)e^{i\vk \cdot \boldsymbol{\delta}}$, which is not, in general. Note also that the canonical Bloch Hamiltonian is unique, while the periodic Bloch Hamiltonian depends on the choice of the unit cell. One should therefore better speak of \emph{a}, rather than \emph{the}, periodic Bloch Hamiltonian. This issue of canonical versus periodic Bloch Hamiltonian is crucial in the case of a lattice with a basis (e.g. the honeycomb lattice or the SSH chain). In the literature, it is sometimes known as basis I versus basis II, see Refs.~\cite{Bena:2009,Fuchs:2010,Fruchart:2014,Lim:2015}. The unique (canonical) Bloch Hamiltonian $H(\vk)$ being that in basis II and the various possible periodic Bloch Hamiltonians $\mathcal{H}(\vk)$ belonging to basis I.

The eigenvectors of a periodic Bloch Hamiltonian are not the cell-periodic Bloch eigenvectors $|u_n(\vk)\rangle$. In order to distinguish them, we call them $|v_n(\vk)\rangle=e^{-i\vk \cdot \vR}|\psi_{n \vk}\rangle=e^{i\vk \cdot \boldsymbol{\delta}}|u_n(\vk)\rangle$, where $\boldsymbol{\delta}=\vr-\vR$ is the position operator within the unit cell. They also satisfy
\begin{equation}
\mathcal{H}(\vk) |v_{n}(\vk)\rangle = E_n(\vk) |v_{n}(\vk)\rangle \, ,
\end{equation}
and the $v_{n, \vk}(\vr)$'s have the periodicity of the Bravais lattice. The $|v_{n}(\vk)\rangle$'s have the periodicity of the reciprocal lattice (at least under the ``periodic gauge choice''): $|v_{n}(\vk+\vG)\rangle=|v_{n}(\vk)\rangle$.

Let us restrict for a moment to a tight-binding model with Hamiltonian $H=\sum_{i,j} t_{ij} |i\rangle \langle j|$ and position operator $\vr=\sum_i \vr_i |i \rangle \langle i|$. The Hamiltonian stores the information about the connectivity between orbitals, that are assumed to form a complete orthogonal set: $\langle i |j \rangle = \delta_{i,j}$ and $\sum_i | i \rangle \langle i | = \mathds{1}$. The position operator contains the information on the position of the orbitals in space. The canonical Bloch Hamiltonian should be used in order to compute geometrical quantities that crucially depend on the spatial location (or spatial embedding, as Haldane calls it) of the orbitals used to define the model (i.e. Berry curvature, quantum metric, etc). The periodic Bloch Hamiltonian is more convenient in computing topological invariants such as winding numbers, Chern numbers, symmetry-based indicators (such as that of Fu and Kane~\cite{Fu:2007}) etc. To be on the safe side, it is always better to work with the canonical Bloch Hamiltonian. As the periodic Bloch Hamiltonian is blind to the exact location of orbitals within the unit cell, the two Bloch Hamiltonians are \emph{not equivalent} when there is a lattice with a basis (e.g. graphene or the SSH chain). The canonical Bloch Hamiltonian incorporates more information about the spatial location of orbitals than the periodic Bloch Hamiltonian. Colloquially speaking, the canonical Bloch Hamiltonian $H(\vk)$ knows the connectivity contained in the tight-binding Hamiltonian $H$ and the complete position operator $\vr=\vR+\boldsymbol{\delta}$. In contrast, the periodic Bloch Hamiltonian $\mathcal{H}(\vk)$ only knows $H$ and the Bravais lattice position operator $\vR$ but is unaware of the position operator within the unit cell $\boldsymbol{\delta}$.

\subsubsection{Conclusion}

In conclusion of this section, we wish to emphasize that there is \emph{no need} of introducing Berry phases. One could study the complete quantum mechanical problem of an electron in a band structure in the presence of external fields, without ever projecting on a given subset of bands. The appearance of Berry phase effects is only related to the approximate treatment of  restricting to a subset of bands and asking for an effective description within this subspace. In practice, solving the full quantum mechanical problem is often un-doable analytically (but may be done numerically) and one ends up asking for an analytically-tractable effective description restricted to a subset of bands. In such a case, Berry phase effects necessarily appear. For example, the orbital magnetic susceptibility or the electric polarization, which are usually discussed in terms of Berry phases, can be studied using only the energy spectrum numerically-computed in a finite magnetic field (Hofstadter butterfly)~\cite{Raoux:2014} or in a finite electric field (Wannier-Stark ladder)~\cite{Combes:2016}.

\subsection{Geometric phases and consequences}
\subsubsection{Berry phase}
If an electron in the $n^\text{th}$ band performs a closed (and contractible) orbit $\mathcal{C}$ in $\vk$-space under the influence of a force, it will acquire a geometric phase due to the encircled Berry flux in addition to the dynamical phase. This extra phase is known as the Berry phase~\cite{Berry:1984}: 
\begin{equation}
\Gamma_n(\mathcal{C}) = \oint_\mathcal{C} \boldsymbol{dk}\cdot\vA_n(\vk) = \int_\mathcal{S} d^2k \, F_{xy}^n(\vk) \,\,[2\pi], 
\end{equation}
where $\partial\mathcal{S}=\mathcal{C}$ (here we assumed that the electron is moving in the $xy$ plane). The Berry phase is defined modulo $2\pi$. Using Stokes' theorem in order to go from the expression involving the connection to that involving the curvature, one needs to assume that the connection is well-defined over the whole patch $\mathcal{S}$. When expressed in terms of the Berry curvature, it is obvious that the Berry phase is gauge-invariant. This Berry phase is a dual of the Aharonov-Bohm phase~\cite{AB:1959} in the sense that it is acquired in $\vk$-space (rather than real space) and due to the Berry curvature (rather than to the magnetic field). In the case of a closed cyclotron orbit performed under an applied magnetic field, the electron wave function actually acquires both an Aharonov-Bohm phase (in real space) and a Berry phase (in reciprocal space). See, for example, the discussion of semi-classical quantization of cyclotron orbits for Bloch electrons in~\cite{Fuchs:2010}. The Berry phase factor  $W(\mathcal{C})=\exp[i\Gamma_n(\mathcal{C})]$ is sometimes called an abelian Wilson loop (see Sec.~\ref{sec:nabp}).

\subsubsection{Zak phase\label{sec:zak}}
If the Berry phase is computed over a non-contractible loop $\mathcal{P}$ of the BZ torus, then it is known as a Zak phase~\cite{Zak:1989}:
\begin{equation}
Z_n(\mathcal{P}) = \int_\mathcal{P} \boldsymbol{dk}\cdot\vA_n(\vk) \,\,[2\pi]\, .    
\end{equation}
Stokes' theorem can no longer be used to relate it to a Berry curvature and it is not obvious that the Zak phase is gauge-invariant. Here there is no smooth gauge for the Berry connection over the complete path $\mathcal{P}$. Actually, the Zak phase is only gauge-invariant provided the ``periodic gauge choice'' condition is imposed~\cite{Zak:1989,Resta:2000}. This is an example of \emph{open-path geometrical phase}, as clearly discussed by Resta~\cite{Resta:2000}. Indeed, the final ket $|u_n(\vk_f)\rangle $ in the path is not the same as the initial one $|u_n(\vk_i)\rangle$ because the canonical Bloch Hamiltonian is not in general periodic with the first BZ (see the discussion about the canonical Bloch Hamiltonian versus a periodic Bloch Hamiltonian). It is possible to impose a definite phase relation between $|u_n(\vk_f)\rangle $ and $|u_n(\vk_i)\rangle$ in order to render the Zak phase gauge independent. This phase relation involves the position operator and is known as the ``periodic gauge choice'':
\begin{equation}
  |u_n(\vk_f)\rangle =|u_n(\vk_i+\vG)\rangle =e^{-i\vG\cdot \vr}|u_n(\vk_i)\rangle  \, .
\end{equation}
It does not completely fix the gauge, but only restricts possible gauge choices. As a consequence of this periodic gauge choice, the Zak phase depends explicitly on the position operator $\vr$ and therefore on the choice of position origin~\footnote{If the crystal has inversion symmetry, the position origin is often chosen on an inversion center, in which case the Zak phase is quantized: it is either 0 or $\pi$.}. The Zak phase is therefore best thought as being a particular position within the unit cell known as the Wannier center. The Zak phase is especially relevant in one dimension, where the BZ is a circle (see Sec.~\ref{section1D}). If the base space (here the BZ) was not a torus but a sphere or a manifold with a trivial first homotopy group (i.e. no non-contractible loop), then there would be no sense in defining a Zak phase and only the Berry phase would be defined. The Zak phase factor $W(\mathcal{P})=\exp([i Z_n(\mathcal{P})]$ is sometimes called an abelian Wilson-Zak loop. 

The Zak phase should be clearly distinguished from the Berry phase. The former can not be expressed in terms of the Berry curvature and is closely related to the position operator. A convenient habit is to think of the Zak phase as the Wannier center. In contrast, the Berry phase measures the Berry flux across a patch in BZ, has a minor dependence on the position operator (see the above discussion about basis I versus basis II) and does not depend on the choice of position origin. A word of caution to the reader: in many papers, a winding number is mistakenly called a Zak phase.

\subsubsection{Wannier functions\label{sec:wannier}}
Here we give a minimal introduction to Wannier functions, which is a whole subject in its own, see Refs.~\cite{Marzari:2012,Vanderbilt:2018}. They were introduced long ago~\cite{Wannier:1937} as the Fourier transform of Bloch states in a given band:
\begin{equation}
w_{n,\vR} (\vr) = \int_\text{BZ} \frac{d \vk}{(2\pi)^D} \, e^{-i\vk \cdot \vR} \psi_{n,\vk} (\vr) = \int_\text{BZ}  \frac{d \vk}{(2\pi)^D} \, e^{-i\vk \cdot (\vR-\vr)} u_{n,\vk} (\vr) \,.
\end{equation}
In 1D, they can also be defined as eigenvectors of the projected position operator onto a given band~\cite{Kivelson:1982}. There is one Wannier function per unit cell and per band. The Wannier states form an orthonormalized basis $\langle w_{n,\vR}|w_{n',\vR'}\rangle=\delta_{n,n'}\delta(\vR-\vR')$. In a given band, one may concentrate on $w_{n}(\vr) = w_{n,\mathbf{0}}(\vr)$ as other Wannier functions in the same band are obtained by translation by a Bravais lattice vector $w_{n,\vR} (\vr)=w_{n} (\vr-\vR)$. Physically, Wannier functions are the solid-state equivalent of atomic or molecular orbitals and are sometimes called Wannier orbitals. As compared to Bloch states, they are better localized in real space but they are not energy eigenvectors. There is a certain gauge-freedom in their definition (related to the Berry-gauge freedom in the definition of the cell-periodic Bloch states). Two important characterization of Wannier functions are:

\begin{itemize}

\item Their average position defined as
\begin{equation}
\langle \vr_{n,\vR} \rangle = \int d \vr \,  \vr \,  |w_{n,\vR}(\vr)|^2 = \langle \vr_n \rangle + \vR \, ,
\end{equation}
which is gauge-invariant and known as the Wannier (or band) center. One is usually mainly interested in $\langle \vr_{n} \rangle$ which is the Wannier center modulo a Bravais lattice vector $\vR$. The Wannier center is related to the electric polarization~\cite{KingSmith:1993,Vanderbilt:1993,Resta:1994} (see Sec.~\ref{sec:zakwannier}).

\item Their localization, which depends on the chosen gauge. The issue of exponential localization (or not) of Wannier functions has a long history starting with Kohn~\cite{Kohn:1959b} (see e.g. \cite{Strinati:1978} for a discussion of its relation to singularities in Bloch states). Roughly speaking, in a trivial insulator, Wannier functions can be exponentially localized~\cite{Brouder:2007}; in a Chern insulator, they are only algebraically localized~\cite{Thouless:1984,Thonhauser:2006}; and in a metal, they are delocalized. In particular, Thouless has shown that a non-zero Chern number implies an obstruction in finding an exponentially localized Wannier function~\cite{Thouless:1984}. One may characterize the localization by defining the spread or extension $\langle \vr_n^2 \rangle-\langle \vr_n \rangle^2$ of a Wannier function. Playing with the gauge freedom, it is possible to define so-called maximally localized Wannier functions (MLWF)~\cite{Marzari:2012}. The gauge-invariant part of the spread of a MLWF is related to the quantum metric~\cite{Resta:2011,Vanderbilt:2018}. The obstruction in finding exponential-localized Wannier functions respecting a given symmetry will turn out to play an important role in the definition of symmetry-protected topological insulators.

\end{itemize}

\subsubsection{Chern number\label{sec:chernnumber}}
In space dimension two, the integral of the Berry curvature over the whole BZ torus $T^2$ is quantized:
\begin{equation}
    C_n = \frac{1}{2\pi} \int_{T^2} d^2 k F_{xy}^n(\vk) \in \mathbb{Z}\, .
\end{equation}
This integer is called the Chern number and tells whether the corresponding Bloch bundle is twisted or not. For time-reversal-invariant materials, the Chern number is always zero, being the integral of an odd function of $\vk$ over the BZ. Therefore breaking time-reversal symmetry is a necessary condition to obtain a band with non-zero Chern number, but it is not sufficient, as we will see in the Haldane model. The Chern number can be seen as an obstruction to having a well-defined connection over the whole BZ. Indeed, if a well-defined Berry connection exists over the whole BZ, then the Berry phase computed over the null path should be equal to the total Berry flux across the BZ via Stokes' theorem and should therefore vanish, i.e. $C_n=0$. Therefore $C_n\neq 0$ means that there is no well-defined connection over the whole BZ.

An alternative view of the Chern number is the following. A band contact point (or degeneracy) acts as a magnetic monopole in parameter space (a Berry monopole)~\cite{Berry:1984}. Berry calls it a diabolical point because of its diabolo shape~\cite{Berry:2010}. It is also known as a conical intersection. The Chern number counts the number of such degeneracies which are enclosed by the BZ torus, i.e. which are \emph{inside} the torus. The precise meaning of \emph{inside} is the following. The band structure has no band degeneracy on the surface of the BZ torus, otherwise the bands would not be well separated, there would be no gap, and the Chern number would not be well-defined. Therefore the band degeneracies that we are talking about actually occur not on the surface of the BZ torus, but really inside a toroid or solid torus~\cite{Simon:1983}. It means that one should imagine extending the 2D $(k_x,k_y)$ model with a third dimension, that we call $k_z$ even it does not correspond to a spatial direction but to some parameter of the Bloch Hamiltonian (such as a hopping amplitude or an on-site energy) that upon tuning creates such a degeneracy. The torus spanned by $(k_x,k_y)$ is now filled inside into a toroid spanned by $(k_x,k_y,k_z)$. We will give a concrete example later in this review when discussing 3D Weyl points in Sec.~\ref{sec:weyl}. The latter is a contact point between two bands in 3D reciprocal space which is very analogous to the Dirac magnetic monopole.

An equivalent of the Chern number can also be defined at finite temperature~\cite{Viyuela:2014}. It requires extending the notion of a geometric phase to mixed (i.e. not pure) states and is known as the Uhlmann phase.

\subsection{Semi-classical equations of motion of a Bloch electron}

\subsubsection{Standard equations of motion}
In order to discuss the Berry phase effects in the semi-classical equations of motion for a Bloch electron, we first recall the standard equations (see, e.g., \cite{AshcroftMermin,Peierls:1955}), which were obtained in the 1930's by Bloch~\cite{Bloch:1928}, Peierls~\cite{Peierls:1929}, Jones and Zener~\cite{Jones:1934}. For simplicity, we consider the motion of a spinless electron restricted to a non-degenerate band and under the influence of external electromagnetic fields (they can be slightly inhomogeneous but here independent of time). The latter are responsible for the dynamics and also for possible transitions between bands. We assume that the electron stays in a given band (adiabatic following, semi-classical approximation): there are no inter-band transitions. This means that the external fields are sufficiently small and that the gap between the bands are sufficiently large. The coupled equations of motion for an electron with average (or center of mass) wave vector $\vk$ and position $\vr_c$ in the $n^\text{th}$ band are
\begin{eqnarray}
\hbar \dot{\vk}&=&-\boldsymbol{\nabla}_{\vr_c}\widetilde{E}_n -\dot{\vr}_c \times e\vB(\vr_c) 
= -e\left[\vE(\vr_c)+ \dot{\vr}_c \times \vB(\vr_c)\right]
\nonumber\\
\dot{\vr}_c&=&\frac{1}{\hbar}\boldsymbol{\nabla}_{\vk}\widetilde{E}_n = \frac{1}{\hbar}\boldsymbol{\nabla}_{\vk} E_n \, ,
\label{eq:smcp}
\end{eqnarray}
where the semi-classical energy 
\begin{equation}
    \widetilde{E}_n (\vk,\vr_c)=E_n(\vk)-e A_0 (\vr_c) \, 
    \label{eq:entildep}
\end{equation}
is the sum of the band energy and the potential energy of a charge $-e$ in the external electrostatic potential $A_0(\vr)$. The first equation in~(\ref{eq:smcp}) looks like Newton's equation for a particle with gauge-invariant momentum $\hbar \vk$ and electric charge $-e$ under the influence of the Coulomb and Lorentz forces in an electric field $\vE=-\boldsymbol{\nabla}_{\vr} A_0(\vr)$ and in a magnetic field $\vB=\boldsymbol{\nabla}_\vr\times \vA (\vr)$. The second equation in~(\ref{eq:smcp}) is the statement that the electron velocity $\dot{\vr}_c$ is given by the group velocity of the band dispersion relation $E_n(\vk)$.

It is possible to requantize the above equations and obtain an effective one-band quantum Hamiltonian describing the electron in the $n^\text{th}$ band
\begin{equation}
    H_n^\text{eff}= \widetilde{E}_n (\vk,\vr_c)=E_n(\vk)-e A_0(\vr_c)
\end{equation}
with
\begin{eqnarray}
\vk&=&\vq+\frac{e}{\hbar}\vA(\vr_c) \to -i\boldsymbol{\nabla}_{\vr_c}+\frac{e}{\hbar}\vA(\vr_c)\nonumber\\
\vr_c &=&\vx \, ,
\label{eq:peierls}
\end{eqnarray} 
where $\hbar \vq$ and $\vx$ are the canonical momentum and position operators, and $\hbar \vk$ is the (electromagnetic) gauge-invariant momentum operator. This is known as the Peierls substitution~\cite{Peierls:1933}. To summarize the ``Peierls strategy'', one diagonalizes the Bloch Hamiltonian $H(\vk)$ in the absence of external fields, projects on a given band to obtain an effective Hamiltonian $H_n^\text{eff}=E_n(\vk)$, then introduces external fields in the effective Hamiltonian $H_n^\text{eff}=E_n(\vq+\frac{e}{\hbar}\vA)-e A_0(\vr_c)$  and eventually requantizes $\vq\to -i\boldsymbol{\nabla}_{\vr_c}$. It is then obvious that inter-band transitions induced by the external fields are neglected in this process.

These equations were able to explain and predict many transport phenomena occurring in crystals (e.g. Bloch oscillations, Hall effect, conduction by holes, cyclotron motion). However, in the 1950 and 1960's, it became clear that these equations were not complete because the electron dynamics is actually influenced by the possibility of virtual transitions to other bands driven by the external fields~\cite{Blount:1962}. In other words, while it is possible to render real (Landau-Zener) inter-band transitions vanishingly small (by having small external fields and large gaps), it is not possible to forbid virtual inter-band transitions. In the ``Peierls strategy'', in order to describe the effective behavior in a given band, one relies \emph{only} on the band's dispersion relation $E_n(\vk)$ (obtained in the absence of external fields) and on no other information (e.g. such as the cell-periodic Bloch states $|u_{n'}(\vk)\rangle$). Berry phase effects are essentially the statement that cell-periodic Bloch states do play a role in the effective dynamics, as we now discuss. 

\subsubsection{Equations of motion including Berry phase effects}
The complete (at first order in the external fields) equations of motion were obtained in final form by Qian Niu and coworkers~\cite{Chang:1996,Sundaram:1999,XiaoBerryRMP:2010} using a wave packet approach. The derivation is quite tedious and we do not reproduce it here. For an alternative Hamiltonian approach (i.e. without wave packets), see~\cite{Gosselin:2006}. The semi-classical equations of motion for an electron wave packet built from states in the $n^\text{th}$ band and having average wave vector $\vk$ and position $\vr_c$ are
\begin{eqnarray}
\dot{\vk}&=&-\frac{1}{\hbar}\boldsymbol{\nabla}_{\vr_c}\widetilde{E}_n -\dot{\vr}_c \times \frac{e}{\hbar}\vB(\vr_c) \nonumber\\
\dot{\vr}_c&=&\frac{1}{\hbar}\boldsymbol{\nabla}_{\vk}\widetilde{E}_n -\dot{\vk} \times \boldsymbol{F}_n(\vk)
\label{eq:smc}
\end{eqnarray}
where the semi-classical energy is
\begin{equation}
    \widetilde{E}_n (\vk,\vr_c)=E_n(\vk)-e A_0(\vr_c)-\boldsymbol{m}_n(\vk)\cdot \vB(\vr_c) \, .
    \label{eq:entilde}
\end{equation}
At second order, extra terms appear, some of which are discussed in~\cite{Gao:2014}. 

Compared to the standard equations (\ref{eq:smcp},\ref{eq:entildep}), Eqs.~(\ref{eq:smc},\ref{eq:entilde}) contain two extra contributions (there is also a third hidden modification, related to phase-space measure and that we discuss below):

One $-\dot{\vk} \times \boldsymbol{\vF}_n$ is called the anomalous velocity and depends on the Berry curvature $\vF_n(\vk)=\boldsymbol{\nabla}_{\vk} \times \vA_n$. It was first found by Karplus and Luttinger in a particular case~\cite{Karplus:1954}. It is a dual of the Lorentz magnetic force (in $\vk$-space the Berry curvature is the analogous of a magnetic field and $\dot{\vk}$ is the analogous of a velocity) and is at the origin, for example, of the integer quantum Hall effect (see below).

The other extra term $-\boldsymbol{m}_n\cdot \vB$, first obtained by Kohn~\cite{Kohn:1959}, is a kind of Zeeman effect and involves another geometrical object (not previously introduced) called the orbital magnetic moment $\boldsymbol{m}_n(\vk)=-\frac{e}{2}\langle \vr \times (\boldsymbol{v} - \langle \boldsymbol{v} \rangle) \rangle$, where $\boldsymbol{v}$ is the velocity operator and the average is taken over a wave packet restricted to the $n^\text{th}$ band~\cite{XiaoBerryRMP:2010}. It has an expression similar to (\ref{BerryCurvatHamil}) for the Berry curvature:
\begin{equation}
\label{eq:omm}
    \vm_n(\vk)= i\frac{e}{2\hbar} \sum_{n'\neq n }\frac{\langle u_n|\boldsymbol{\nabla}_{\vk} H(\vk) | u_{n'}\rangle \times \langle u_{n'}|\boldsymbol{\nabla}_{\vk} H(\vk) | u_{n}\rangle}{E_{n'}-E_n}  \, .
\end{equation}
This Zeeman-like effect is emergent as the electron considered here is spinless. Its emergence is similar to that of the Zeeman effect in the Pauli equation (with the famous $g=2$ factor) when taking the low-energy limit of the 3D Dirac equation and projecting on the positive energy states~\cite{Blount:1962}. Niu and co-workers~\cite{XiaoBerryRMP:2010} provide the following picture of this emerging orbital magnetic moment. When restricted to a finite energy band, an electron wavepacket has a minimum size due to the uncertainty principle. This minimum size is similar to the Compton wavelength for the Dirac electron as discussed by Blount~\cite{Blount:1962}. This means that the electronic charge is now spread over a finite volume and there is the possibility of self-rotation as in the Uhlenbeck and Goudsmit picture of the rotating electron as a mechanism for the appearance of the intrinsic magnetic moment. The orbital magnetic moment is a crucial ingredient in the ``modern theory of orbital magnetism'', see~\cite{XiaoBerryRMP:2010,Thonhauser:2011} for review.

It is also possible to generalize the Peierls substitution in order to obtain an effective one-band quantum Hamiltonian including the Berry phase corrections~\cite{XiaoBerryRMP:2010}. As before the effective Hamiltonian is given by the semi-classical energy
\begin{equation}
    H_n^\text{eff}= \widetilde{E}_n (\vk,\vr_c)=E_n(\vk)-e A_0(\vr_c)-\boldsymbol{m}_n(\vk)\cdot \vB(\vr_c)
    \label{eq:hneff}
\end{equation}
upon identifying $\vk$ and $\vr_c$ with operators. However, the canonical quantization is not easy, because $\vr_c$ and $\hbar\vk$ are not canonical position and momentum: their Poisson bracket is not standard. We consider three cases. 

(i) In the particular case in which the Berry curvature vanishes, one uses the standard Peierls substitution (\ref{eq:peierls}). 

(ii) In the particular case where the magnetic field vanishes, these operators are given by a kind of dual of the Peierls substitution
\begin{eqnarray}
\vk &=&\vq \nonumber\\
\vr_c &=&\vx + \vA_n(\vk)\to  i \boldsymbol{\nabla}_{\vk}+\vA_n(\vk)\, ,
\label{eq:dualpeierls}
\end{eqnarray} 
where $\hbar \vq$ and $\vx$ are the canonical momentum and position operators, and $\vr_c$ is the (Berry) gauge-invariant position operator. It can be shown that $\vr_c= \vA_n(\vk) + \vx$ is also the position operator projected onto the $n^\text{th}$ band [In the crystal momentum representation~\cite{Blount:1962}, the complete position operator $\vr$ also has matrix elements between different bands, that are given by $\vA_{n,n'}(\vk)=\langle u_n|i \boldsymbol{\nabla}_\vk u_{n'}\rangle$ so that $\vr_{n,n'}=\delta_{n,n'} i \boldsymbol{\nabla}_\vk + \vA_{n,n'}(\vk)=\delta_{n,n'}\vr_c + (1-\delta_{n,n'})\vA_{n,n'}(\vk)$. This is the reason for distinguishing between the position operator $\vr$ and the projected position operator $\vr_c$.]. Its average over all wave vectors $\vk$ in the BZ $\langle\vr_c\rangle=\langle \vr_{n} \rangle +\vR$ is equal to the Wannier center $\langle \vr_{n} \rangle=\langle \vA_n\rangle$ modulo a Bravais lattice vector $\vR$. Roughly speaking, the Wannier center $\langle \vr_{n} \rangle$ is the electron position within the unit cell and $\vR$ is the position of the unit cell. The Wannier center plays an important role in the ``modern theory of electric polarization''~\cite{KingSmith:1993,Vanderbilt:1993,Resta:1994}, as we discuss below.

(iii) In the general case, when both the magnetic field and the Berry curvature are non-zero, it can be shown that the operator identification~\cite{Gosselin:2006,XiaoBerryRMP:2010} is
\begin{eqnarray}
\vk &=&\vq+\frac{e}{\hbar}\vA (\vr_c)+e\vB(\vr_c)\times \vA_n(\vk)\nonumber\\
\vr_c &=&\vx + \vA_n(\vk)\, ,
\label{eq:gps}
\end{eqnarray} 
where $(\hbar \vq,\vx)$ are the canonical momentum and position operators. Note that this is not simply (\ref{eq:peierls}) together with (\ref{eq:dualpeierls}).

There is a third modification of the electron dynamics that is not apparent in the above equations of motion (\ref{eq:smc},\ref{eq:entilde}), but which is important. It is a consequence of the fact that the wave packet momentum $\hbar\vk$ and position $\vr_c$ are gauge-invariant (both under an electromagnetic gauge transformation and under a Berry gauge transformation) but are not canonical. Their Poisson bracket (and the corresponding quantum commutator) is not the usual one but is modified by the Berry curvature and the magnetic field. Therefore the volume occupied by a state in the $(\vr_c,\vk)$ phase-space is no longer $(2\pi)^D$. In order to take this fact into account, one should modify the phase-space measure as follows~\cite{Xiao:2005,XiaoBerryRMP:2010}:
\begin{equation}
    \frac{d \vr_c d \vk}{(2\pi)^D} \to \frac{d \vr_c d \vk}{(2\pi)^D}(1+\frac{e}{\hbar}\vB\cdot \vF) \, .
\end{equation}
This modified phase-space density simply comes from the Jacobian in the transformation from the canonical ($\vx$,$\hbar\vq$) to the gauge-invariant ($\vr_c$,$\hbar\vk$) position and momentum variables~\cite{Gosselin:2006} as given in Eq.~(\ref{eq:gps}):
\begin{equation}
    d \vx d \vq =d \vr_c d \vk (1+\frac{e}{\hbar}\vB\cdot \vF) \, .
\end{equation}
The modified phase-space density affects thermodynamic quantities such as the orbital magnetization and susceptibility~\cite{XiaoBerryRMP:2010,Thonhauser:2011}.

As a side remark, it is easy to recover the behavior of the Berry curvature under time-reversal (\ref{trcurvature}) and space inversion (\ref{sicurvature}) from the anomalous velocity in Eq.~(\ref{eq:smc}). Under time reversal, the velocity, momentum and time-derivative change sign so that $\dot{\vk}$ does not change, which means that $\boldsymbol{F}_n$ must change sign. Therefore $\boldsymbol{F}_n(\vk)\to -\boldsymbol{F}_n(-\vk)$ under time-reversal and if it is a symmetry then $\boldsymbol{F}_n(\vk)=-\boldsymbol{F}_n(-\vk)$. Under space inversion, the velocity and momentum change sign so that $\dot{\vk}$ changes sign as well and the curvature must remain. Therefore $\boldsymbol{F}_n(\vk)\to \boldsymbol{F}_n(-\vk)$ under inversion and if it is a symmetry then $\boldsymbol{F}_n(\vk)=\boldsymbol{F}_n(-\vk)$. From the Zeeman-like term $-\vm_n \cdot \vB$, the same considerations apply to the orbital magnetic moment (\ref{eq:omm}): time-reversal symmetry makes it odd $\vm_n(-\vk)=-\vm_n(\vk)$ and inversion symmetry makes it even $\vm_n(-\vk)=\vm_n(\vk)$. If both symmetries are present, the Berry curvature and the orbital magnetic moment vanish at every $\vk$.

\medskip

In summary, at first order in the external fields, the equations of motion of a spinless Bloch electron restricted to a single band are given by Eqs.~(\ref{eq:smc}) with the effective energy~(\ref{eq:entilde}) and the relation~(\ref{eq:gps}) between gauge-invariant and canonical momentum and position. The extra terms (as compared to the standard equations) vanish in the presence of both time-reversal and inversion symmetries. Compared to the ``Peierls scheme'', here the restriction to an effective single-band description is done in the presence of external fields and virtual transitions to other bands are allowed, which lead to geometric forces of reaction~\cite{Berry:1989}. Only real (Landau-Zener) inter-band transitions are neglected.

\subsubsection{Quantized Hall conductivity of a filled band \label{sec:qhe}}
 
 Here, we specialize to space dimension $D=2$ and time-reversal breaking systems. In their landmark paper, Thouless-Kohmoto-Nightingale-den Nijs (TKNN) have shown that the Hall conductivity of a band insulator is quantized because it is related to a topological number: the sum of the Chern numbers of the occupied bands \cite{Thouless:1982}. They originally considered the case of a two-dimensional electron gas in an applied perpendicular magnetic field. Here, we will consider a slightly different version -- the so-called quantum anomalous Hall effect (QAHE) -- by assuming that time-reversal symmetry is broken but no external magnetic field is applied. We therefore consider both translation invariance under a Bravais lattice and time-reversal breaking. We will use the semi-classical equations of motion for an electron restricted to a given band and show that a filled band may nevertheless carry a quantized Hall current.

Consider the semi-classical equations of motion (\ref{eq:smc}) for a single electron in a given band $n$ in 2D in the presence of an applied electric field but no applied magnetic field. They read
\begin{eqnarray}
\dot{\vk}&=&-\frac{e}{\hbar}\vE \nonumber\\
\dot{\vr}_c&=&\frac{1}{\hbar}\boldsymbol{\nabla}_{\vk}E_n(\vk) -\dot{\vk} \times \vF_n = \frac{1}{\hbar}\boldsymbol{\nabla}_{\vk}E_n(\vk) + \frac{e}{\hbar}\vE \times \vF_n\, .
\end{eqnarray}
The first term in the above velocity is the familiar group velocity, while the second term is the anomalous velocity of Karplus and Luttinger~\cite{Karplus:1954}. We can easily obtain the average velocity of an electron and deduce the electric current carried by a filled band as
\begin{equation}
\vj_n = (-e) \int_{T^2}\frac{d^2 \vk}{(2\pi)^2} \dot{\vr}_c =  (-e) \int_{T^2}\frac{d^2 \vk}{(2\pi)^2}\left[\frac{1}{\hbar}\boldsymbol{\nabla}_{\vk}E_n(\vk) +\frac{e}{\hbar}\vE \times \vF_n\right]   \,  . 
\end{equation}
The first term (group velocity) vanishes after integration over a filled band as is well known (``a filled band does not conduct electricity''). However, the second term (anomalous velocity) need not vanish and involves the Chern number $C_n$ of the $n^\text{th}$ band:
\begin{equation}
\vj_n =  -\frac{e^2}{h}\vE  \times \int_{T^2}\frac{d^2 \vk}{2\pi} \vF_n =  -\frac{e^2}{h}C_n \vE  \times \ve_z .
\end{equation}
This means that, provided the Chern number is non-zero, a filled band can nevertheless have a Hall current, which is quantized in units of $e^2/h$, as the Chern number is an integer. For a band insulator with several filled bands (those with $n\leq n_F$), one finds that the Hall conductivity is given by:
\begin{equation}
\sigma_{xy}  =  -\frac{e^2}{h}  \sum_{n\leq n_F} C_n = -\frac{e^2}{h} n_H \, .
\label{conductivityHallChern}
\end{equation}
The topological invariant $n_H$ characterizing the band insulator is called the Hall (or TKNN) number and is given by the sum of the Chern numbers of the filled bands. The Chern number $C_n$ characterizes a band $n$, whereas the Hall number $n_H$ is attached to a band gap and characterizes a band insulator (gap labeling). At this point, we have not yet shown that a band insulator with a non-zero Hall number exists. From symmetry arguments, we know that if we do not break time-reversal symmetry, then $C_n=0$ for all bands and therefore $\sigma_{xy}  =0$. Below, we will discuss the Haldane model, which is precisely a band model that breaks time-reversal symmetry and is able to produce a non-zero Chern number for a band (note that breaking time-reversal is a necessary but not sufficient condition).

Thouless has also shown how the Chern number quantizes adiabatic pumping in a 1D band insulator~\cite{Thouless:1983}. Actually the 2D quantum Hall system can exactly be mapped on a time-periodic 1D system.

Another point to note is that from the perspective of the bulk, the Hall current is carried by all the filled bands. However, in a finite sample with edges, the quantized Hall current is carried by gapless and chiral edge modes as in the usual quantum Hall effect~\cite{Halperin:1982}. In the familiar quantum Hall effect, the bulk invariant $n_H$ is just the number of filled Landau levels. And the number of chiral gapless modes per edge is also equal to $n_H$. This is a first example of bulk-edge correspondence between a topological invariant in the bulk (number of filled Landau levels) and a topological invariant on an edge (the number of chiral gapless edge modes).

\begin{figure}
\begin{center}
\includegraphics[width=12cm]{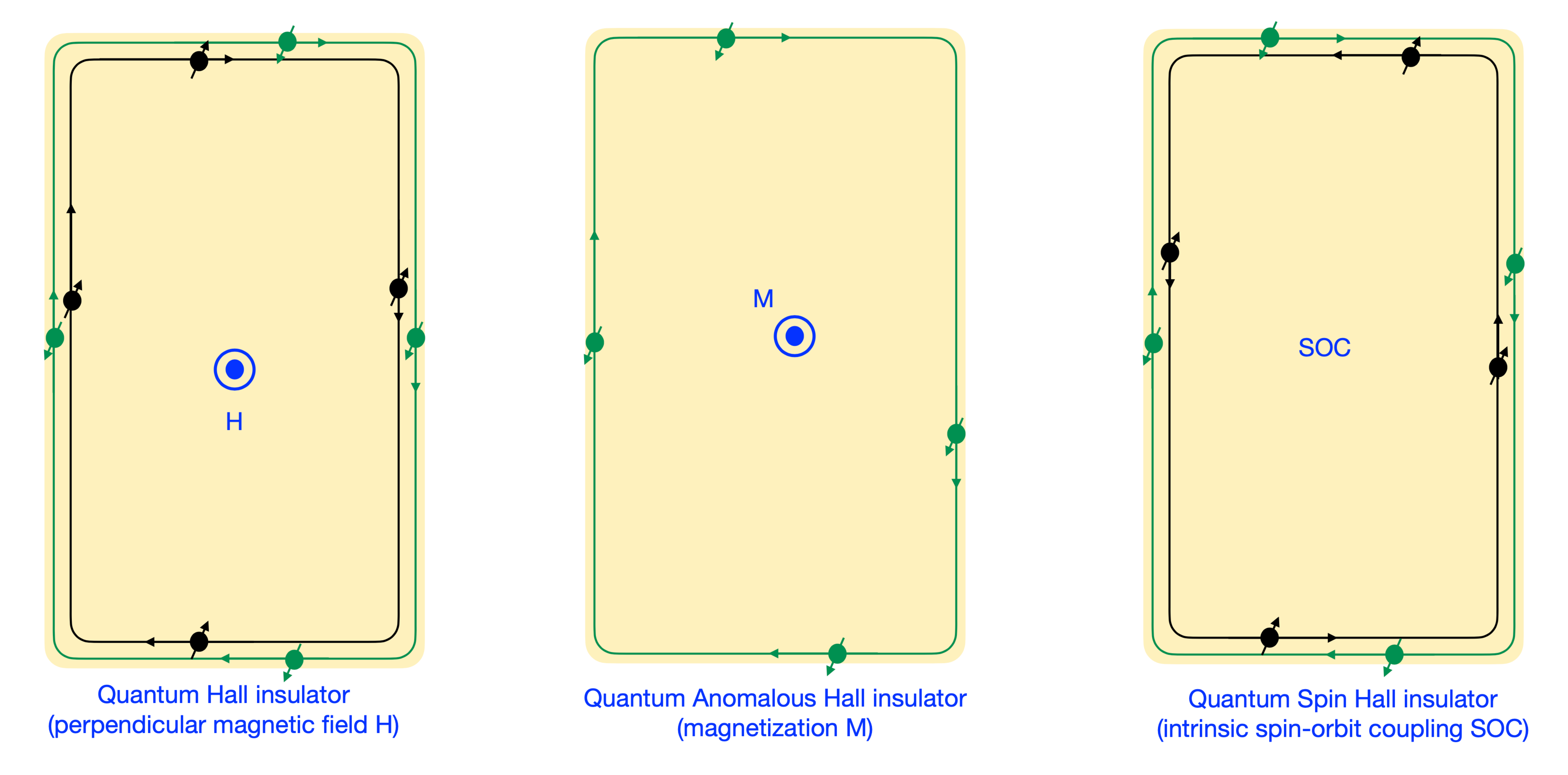}
\caption{Quantized Hall effects in insulators. H designates an externally applied magnetic field, M indicates a spontaneous magnetization and SOC stands for spin-orbit coupling. The chirality (clockwise or counter-clockwise) of edge states and their spin projections are also indicated. Time-reversal symmetry is present only in the quantum spin Hall insulator. For a comparison to the associated unquantized Hall effects in metals or semiconductors (Hall effect, anomalous Hall effect and spin Hall effect), see~\cite{Oh:2013}. \label{fig:qht}}
\end{center}
\end{figure}

When they occur in insulators, Berry curvature effects are quantized/topological because one integrates over the whole BZ (topological numbers). That is for example the case of the QAHE that occurs in Chern insulators. A typical representative is the Haldane model of graphene~\cite{Haldane:1988}, that we will treat in detail in Sec.~\ref{section2DHaldane}. Another example is the quantum spin Hall effect (QSHE) that occurs in time-reversal invariant topological insulators with strong intrinsic spin-orbit coupling, see e.g. Kane and Mele's model~\cite{Kane:2005a,Kane:2005b}. Figure~\ref{fig:qht} summarizes three kinds of quantized Hall effects.

\subsubsection{Berry phase effects in doped semiconductors and metals}
We have seen that Berry phase effects, such as the existence of the anomalous velocity, lead to robust quantized responses in insulators. However, the anomalous velocity also exists in doped semiconductors and metals where it induces observable but unquantized effects, such as the anomalous Hall effect~\cite{Nagaosa:2010} in time-reversal breaking compounds (ferromagnetic semiconductors or metals), or the spin Hall effect~\cite{Sinova:2015} in time-reversal invariant materials with strong-spin orbit coupling. These effects are still due to geometric quantities such as the Berry curvature but which are not integrated over the complete BZ, but over a finite portion of the BZ delimited by the Fermi surface, and are therefore not topological but merely geometrical. The anomalous Hall effect~\cite{Nagaosa:2010} and the spin Hall effect~\cite{Sinova:2015} are fundamental in the field of spintronics and have important applications. The inverse spin Hall effect is a very convenient method to measure a spin current by a charge signal. 

Even more recently the nonlinear electromagnetic responses of non-centrosymmetric crystals have been reinterpreted in the perspective of Berry curvature properties~\cite{Orenstein:2010,Fu:2015,Moore:2016,Moore:2019,Sodemann:2019prb,Sodemann:2019prl}. The photogalvanic effect (PGE), the second-harmonic generation or the frequency difference generation are captured by various frequency dependencies of the nonlinear second-order susceptibility tensor. At low frequency of the driving field, the rectified current is determined by the scattering time and the intrinsic Berry curvature dipole, which is a measure of the average gradient of Berry curvature of the occupied states~\cite{Fu:2015,Sodemann:2019prb,Sodemann:2019prl}. Provided inversion symmetry is broken, this intraband effect is present even in time-reversal invariant materials and has been coined nonlinear Hall effect (or nonlinear anomalous Hall effect depending on the authors). Nonlinear Hall or optical effects are also present in 3D materials, and are especially strong in recently discovered Weyl semimetals. Experimentally the nonlinear Hall effect has been measured in the few-layer TMDC WTe$_2$~\cite{Ma:2019,Kang:2019} and also in 3D Dirac or Weyl semimetals~\cite{Shvetsov:2019}. At higher frequencies, the DC rectified current is an inter-band effect known as the shift current. Interestingly, the frequency integral of the rectified second order conductivity is a purely geometrical quantity that depends neither on the scattering time nor on the band dispersion, but is solely determined by the Berry curvature dipole~\cite{Sodemann:2019prl}. In principle, those nonlinear response susceptibilities are not quantized in the metallic regime, but in some particular circumstances they also might be quantized \cite{deJuan:2017}. 

\subsection{Non-abelian Berry phases\label{sec:nabp}}
If instead of projecting on a single band, one projects on a group of $N$ bands (say $N=2$ bands) in a model containing more bands (for example because they are degenerate), then one ends up with an emergent $SU(2)$ static gauge field. The main difference is that the structure group is no longer the abelian $U(1)$ but becomes the non-abelian $SU(2)$. The main object in this description is the Wilczek-Zee (or non-abelian Berry) connection 
 \begin{equation}
     \vA_{nn'}(\vk)=i\langle u_n|\boldsymbol{\nabla}_\vk u_{n'}\rangle ,
 \end{equation}
 which is a matrix-valued version of the (abelian) Berry connection $\vA_{n}(\vk)=i\langle u_n|\boldsymbol{\nabla}_\vk u_n\rangle$~\cite{Wilczek:1984}. The same quantity is called $\boldsymbol{\mathfrak{X}}_{nn'}(\vk)$ by Blount~\cite{Blount:1962} and $\boldsymbol{R}_{nn'}(\vk)$ by Xiao et al.~\cite{Xiao:2005}. The matrix corresponding to $\vA_{nn'}(\vk)$ is noted $\boldsymbol{\mathcal{A}}$ which is both a $D$-component vector (hence in bold) and a $N\times N$ matrix (hence the notation with a curly letter). The corresponding curvature  \begin{equation}
     \boldsymbol{\mathcal{F}}=\boldsymbol{\nabla}_\vq \times \boldsymbol{\mathcal{A}} - i \boldsymbol{\mathcal{A}} \times \boldsymbol{\mathcal{A}}
 \end{equation}
 becomes a matrix. For a presentation of the semi-classical equations of motion in the case of degenerate bands see~\cite{XiaoBerryRMP:2010}.
 
The Berry phase factor becomes a matrix known as the Wilson loop~\cite{Alexandradinata:2014} 
 \begin{equation}
     \mathcal{W}(\mathcal{C})=\mathbb{P}\exp(i\int_\mathcal{C} \boldsymbol{dk}\cdot \boldsymbol{\mathcal{A}})\, ,
 \end{equation}
 where $\mathbb{P}$ is the path-ordering operator. Its eigenvalues are the non-abelian Berry phase factors.
 If the path is along a non-contractible loop $\mathcal{P}$ in the BZ, then the Zak phase factor of the abelian case gets promoted to a matrix known as the Wilson-Zak loop or large Wilson loop~\cite{Benalcazar:2017}:
  \begin{equation}
     \mathcal{W}(\mathcal{P})=\mathbb{P}\exp(i\int_\mathcal{P} \boldsymbol{dk}\cdot \boldsymbol{\mathcal{A}})\, .
 \end{equation}
 Its eigenvalues are the non-abelian Zak phase factors.

\subsection{Conclusion: Berry phase effects in solids, gauge fields and fiber bundles}

\begin{table}[h!]
    \centering
    \begin{tabular}{c|c|c}
    
        \textbf{Geometrical band theory} & \textbf{Electromagnetism (gauge fields)} & \textbf{Fiber bundles} \\
        \hline
         $\vk$-space (BZ torus $T^2$)& $\vr$-space (sphere $S^2$ around the monopole $\subset \mathbb{R}^3$) & base space\\
          \hline
          phase of $|u_n(\vk)\rangle$ & phase of $\psi(\vr)$ & fiber  \\
         \hline
          $U(1)_\text{Berry}$ &$U(1)_\text{electric charge}$ & structure group  \\
         \hline
         Berry connection $\vA_n(\vk)=\langle u_n|i\boldsymbol{\nabla}_\vk u_n\rangle$ & vector potential $\vA(\vr)$ & connection \\
         \hline 
         projected position $\vr_c=i \boldsymbol{\nabla}_{\vk} + \vA_n(\vk)$ & gauge-invariant momentum $\boldsymbol{\Pi}=-i \hbar \boldsymbol{\nabla}_{\vr}+e \vA(\vr)$ & covariant derivative \\
         \hline 
         Berry curvature $\vF^n(\vk)=\boldsymbol{\nabla}_{\vk}\times \vA_n(\vk)$& magnetic field strength $\vB(\vr)=\boldsymbol{\nabla}_{\vr}\times \vA(\vr)$ & curvature (Chern class) \\
         \hline
         Berry phase $\Gamma_n(\mathcal{C}) =\oint_\mathcal{C} \boldsymbol{dk}\cdot\vA_n(\vk)\, [2\pi]$ & Aharonov-Bohm phase $-e\oint \boldsymbol{dr}\cdot\vA(\vr)/\hbar$  $[2\pi]$ & (an)holonomy \\
          $=\int_{\mathcal{S}} \boldsymbol{dS}_\vk\cdot\vF_n(\vk)\, [2\pi]$ &  $=-e\int \boldsymbol{dS}\cdot\vB(\vr)/\hbar$  $[2\pi]$ & \\
         \hline
         Zak phase $Z_n=\int_\mathcal{P} \boldsymbol{dk}\cdot\vA_n(\vk)$ $[2\pi]$ & does not apply: no non-contractible loops  &  \\
         \hline
         Band contact point (Berry monopole)  & Dirac's magnetic monopole & singular curvature source \\
         \hline
         TKNN number $C_n = \int_{T^2} d \vk F_{xy}^n /(2\pi) \in \mathbb{Z}$ & monopole charge $-e\int_{S^2} d\vS \cdot  \vB/h \in \mathbb{Z}$ & first Chern number \\
         \hline
          & generalized wave function (Dirac) & section \\
         \hline
         orbital magnetic moment $\boldsymbol{m}_n(\vk)$ & spin magnetic moment (Dirac eq. $\to$ Pauli eq.)& \\
         \hline
        quantum metric $g_{ij}^n(\vk)$ & ? & ?
    \end{tabular}
    \caption{Analogy between geometrical band theory (projection on a single band) and electromagnetism in the case of a two-dimensional parameter space (a torus in the band theory case and a sphere in the Dirac monopole case). The last column makes a relation to the mathematical language of fiber bundles. In the left column, $\mathcal{P}$ refers to a non-contractible loop over the BZ torus and $\mathcal{C}$ to a contractible loop (the surface $\mathcal{S}$ is such that $\mathcal{C}=\partial \mathcal{S}$).}
    \label{tab:analogy}
\end{table}

To conclude this section, it is interesting to draw an analogy between geometrical band theory, gauge theories such as electromagnetism, and fiber bundles. A useful analogy exists between the geometry of a single band (i.e. of the restriction of band theory to a single band by projection) on the one hand and electromagnetism (gauge theory) on the other hand, see the two first columns in Table~\ref{tab:analogy}. Furthermore, as understood by Wu and Yang~\cite{Wu:1975}, what physicists call gauge structure or gauge theory is what mathematicians call fiber bundles, see the last two columns in Table \ref{tab:analogy}. In the present case, the gauge structure is emergent and exists in parameter space. Its origin comes from projecting on a sub-set of bands (typically one band) and reflects the effect of virtual transitions to the other bands. The general mechanism for the appearance of such emergent gauge structure lies in the separation between slow (heavy) and fast (light) degrees of freedom. In the context of band structure, this is the separation between the ``external'' dynamics of the electron (wave vector $\vk$) within a band, i.e. the slow motion from unit cell to unit cell, and the ``internal'' dynamics between bands (band index $n$) and corresponding to the intra-unit cell fast motion. When one wishes to have an effective description for the heavy system only, the reaction of the light system on the heavy system happens through an emergent gauge field. See the general discussion in~\cite{WilczekShapere:1989}. In the above effective description, the Bloch wave-vector $\vk$ plays the role of a parameter, whereas the band index $n$ that of a quantum number. 

Inter-band effects are known in the literature under various names, such as Berry phase effects, band coupling, emergent gauge field, Bloch bundle, etc. An important message of the present section is that they give rise not only to global (in parameter space) topological effects but also to interesting local geometrical effects. In other words, there can be interesting geometrical effects already in insulators that are topologically trivial, such as boron nitride. Another important message is that these effects are only present if the bands are coupled: they are absent in a single-band model, and also in models containing several bands that are totally decoupled. Coupling means hopping terms between the corresponding orbitals in a tight-binding model.

\section{Lattice Dirac fermions in 1D \label{section1D}}
In the previous section, we have introduced the key concepts of modern band structure theory, including geometrical (Berry curvature) and topological aspects (Chern number) from a rather general point of view. In this section and the following ones, we provide examples of such effects and emphasize the specificity of each space dimension, starting with 1D systems (Sec. \ref{section1D}) before going on with 2D honeycomb lattices (Sec. \ref{section2D}), topological insulators (Sec. \ref{sec:topinsulators}) and 3D Weyl semimetals (Sec. \ref{sec:topmetal}). Here we discuss the famous SSH model for polyacetylene chains \cite{Su:1979,Heeger:1988}, and its generalization to diatomic polymers, the RM model \cite{Rice:1982}, which proved also to be of interest in the study of ferroelectricity \cite{Vanderbilt:1993,Onoda:2004}. The SSH and RM models describe  spinless fermions on a bipartite 1D lattice in the presence of several types of symmetries. Interestingly, the polyacetylene chain realizes a solid state implementation of the Jackiw-Rebbi mechanism for the generation of fractionalized excitations \cite{Jackiw:1976,Goldstone:1981}. 

Here we will use the SSH and the RM models as pedagogical guidelines to understand the emergence of Dirac fermions and topology in 1D electronic band structures. Many features of these models enable to learn a lot about Dirac fermions and topological effects in higher dimensions. For instance the zero energy states of the SSH model generalizes into 1D edge states of topological insulators in 2D, and surface states of 3D topological insulators, although the bulk edge correspondence properties may differ. Moreoever, the SSH model has also been extended to superconducting systems by Kitaev which led to the discovery of Majorana end states~\cite{Kitaev:2001}. Recently, the SSH model was experimentally realized on platforms, like cold atoms, photonic crystals or graphene nanoribbons, which allows a much higher tunability of the parameters, thereby allowing an exploration of the phase diagram and the observation of the interesting features of the SSH model. Among reviews on SSH model, we recommend the historic one~\cite{Heeger:1988} and the more recent~\cite{CooperRMP:2019} with illustrations in cold atom systems and photonic systems.

\subsection{Su-Schrieffer-Heeger model}
\begin{figure}[h!]
\begin{center}
\includegraphics[width=14cm]{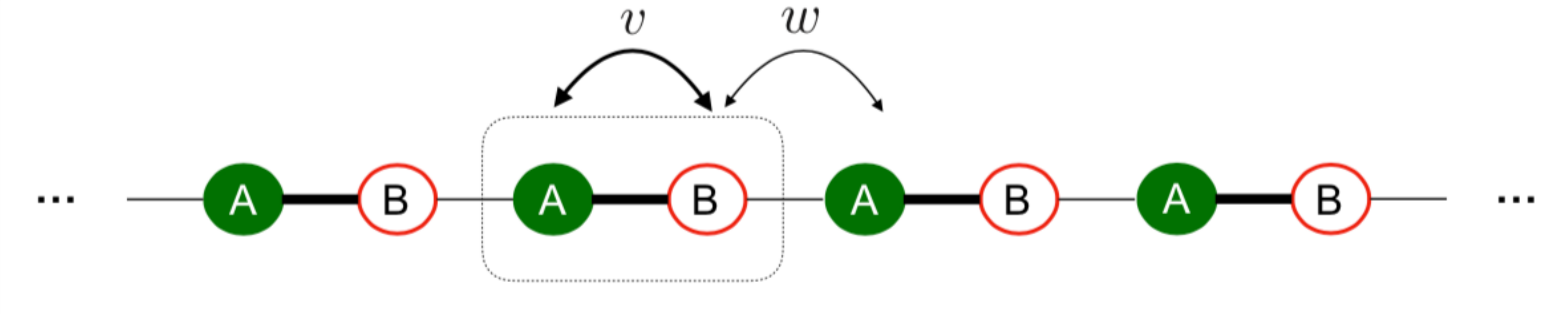}
\caption{\label{fig:ssh_chain} Schematic representation of the SSH model. The SSH chain is represented as a linear chain with alternating strong and weak bonds. In real polyacetylene, the bonds are not aligned and differ in length, but the alternating bond model captures the essential features for the electronic states. Due to the fact that the hopping integrals $v$ and $w$ are distinct, the unit cell has to contain two atomic sites thereby defining $A$ and $B$ sublattices. The hopping amplitudes $v=t+\delta/2$ and $w=t-\delta/2$ are called intra-cell and inter-cell respectively. Note that sites $A$ and $B$ are occupied by 
the same carbon $p_z$ orbitals, and that the choice of unit cell is arbitrary.}
\end{center}
\end{figure}
Polyacetylene is a 1D polymer consisting in a large number of -CH- monomers. The $2s$, $2p_x$, $2p_y$ orbitals of the C atom, arranged in $sp^2$ hybridization, and the $1s$ orbitals of H atom form the covalent C-C and C-H bonds and provides its planar structure to the 1D polymer. There is a single remaining $2p_z$ orbital on each C atom whose hybridization with neighboring $2p_z$ orbitals leads to $\pi-$bands. Those $\pi-$bands are the highest occupied and lowest unoccupied molecular orbitals of the polymer, and may also be called conduction and valence bands in the language of solid-state physics. Polyacetylene undergoes a dimerization transition where the gain in electronic energy overcompensates the elastic energy cost for creating the lattice distorsion (Peierls instability)~\cite{Peierls:1955}. Due to this Peierls instability, polyacetylene is stable in a dimerized form consisting in alternating shorter and longer C-C bonds (see Fig.~\ref{fig:ssh_chain}). The unit cells are labelled by the integer $n$, and each unit cell contains two atomic sites, labelled $A$ and $B$ respectively. The SSH tight-binding Hamiltonian for the $\pi-$bands reads~\cite{Su:1979}  :
\begin{equation}
\label{SSH_Hamiltonian}
H_{SSH}  =  v \sum_{n}    c_B^\dagger(n) c_A(n) + w  \sum_{n}    c_B^\dagger(n) c_A(n+1) + {\rm H.c.}   \,\, ,
\end{equation}
where H.c. means hermitian conjugation, and the operator $c_l^\dagger(n)$ creates a spinless fermion in the $p_z$ orbital at site $(n,l)$, with $l=A,B$. The parameter $v$ is the intra-cell hopping and $w$ the inter-cell hopping amplitude. The definition of the unit cell is a matter of convention, and therefore which coupling is called intra-cell or inter-cell is arbitrary. The hopping parameters $v$ and $w$ are real because there is no external magnetic field or internal magnetic flux. Physically, the largest (resp. smallest) coupling among $v$ and $w$, in absolute values, corresponds to the shortest (resp. longest) bond. In polyacetylene, the hopping amplitudes $v$ and $w$ are fixed and close to $3.5$~eV (strong bond) and $2.5$~eV (weak bond)~\cite{Heeger:1988}. The average hopping amplitude $(v+w)/2=t\sim 3$~eV in polyacetylene is close to that in graphene.

\subsection{Band structure and Berryology of the SSH model} 

We now turn to the band structure of the SSH model which can be handled using two distinct representations, a periodic Bloch Hamiltonian $\mathcal{H}(k)$ and the (canonical) Bloch Hamiltonian $H(k)$. Both are presented because they are both useful, the periodic $\mathcal{H}(k)$ being convenient for some purposes, but the standard formula for Berry quantities are only correct in their forms given in Sec. \ref{sec:bandtheory}, when using the canonical Bloch Hamiltonian $H(k)$.  

\subsubsection{Bloch Hamiltonians (two bases), band structure and spinors}

In the absence of disorder, translation invariance allows one to diagonalize the Hamiltonian with respect to the cell index $n$. The field operators $c_l(n)$ in real space are expanded over operators in reciprocal space $c_{l}(k)$ as : 
\begin{equation}
\label{Fourier1D}
c_l (n)= \frac{1}{\sqrt{N}} \sum_{k} e^{i k  n} c_{l}(k) \, , 
\end{equation}
where $l=A,B$ is the sublattice index, $k$ the crystal momentum, and $N$ the number of unit cells. The unit cell is chosen to have a unit length $a=1$ and we choose units such that $\hbar=1$. After substitution of Eq. (\ref{Fourier1D}) in Eq. (\ref{SSH_Hamiltonian}), the Hamiltonian becomes :
\begin{equation}
\label{RMHamiltonianFourier}
H  = \sum_k    c_l^\dagger(k) \mathcal{H}_{lm}(k) \, c_m (k),
\end{equation}
where momentum $k$ is restricted to the 1D BZ $[-\pi,\pi[$ and the Einstein summation over repeated sublattice indexes $(l,m)$ is implied. The SSH Hamiltonian which is a  $2N \times 2N$ matrix in real space, becomes a $2 \times 2$ matrix $\mathcal{H}(k)$ for each value of $k$. Note that the exponential factors in Eq. (\ref{Fourier1D}) contain only the cell number $n$, without any mention about the real position of the sites within each cell. As a result, the Bloch Hamiltonian $\mathcal{H}(k)$ is $2\pi-$periodic. This \emph{periodic} Bloch Hamiltonian  can be written as a linear combination of the Pauli matrices $\sigma_x$ and $\sigma_y$ only :
\begin{equation}
\mathcal{H}(k)= d_x(k) \sigma_x + d_y(k) \sigma_y   \, \, ,
\label{hRM}
\end{equation}
with real coefficients 
\begin{equation}
 d_x(k) = v + w \cos{k }  \hspace{5mm} {\rm }    \hspace{5mm} {\rm and}  \hspace{5mm}  \hspace{5mm}  d_y(k)=w \sin{k }   \, \, .
\label{dxdydzRM}
\end{equation}
The Pauli matrices above act on the sublattice index $(A,B)$ and not on real electronic spin (we assume spinless electrons here). We call $t=(v+w)/2$ the average hopping amplitude, take units such that $t=1$ and call $\delta=v-w$ the difference in hopping amplitudes. As already discussed, a periodic Bloch Hamiltonian depends on the choice of unit cell. Making the other choice is equivalent to $\delta\to -\delta$.

\medskip

For comparison and later use, one defines also the \emph{canonical} Bloch Hamiltonian (or simply \emph{the} Bloch Hamiltonian) $H(k)$. It is obtained by using the following unitary transformation for diagonalizing $H$ :
\begin{equation}
\label{Fourier1Dcano}
c_l (n)= \frac{1}{\sqrt{N}} \sum_{k} e^{i k \, x_{nl}} c_{l}(k) \, , 
\end{equation}
where $x_{nl}$ ($x_{nl}=n+0$ if $l=A$ and $x_{nl}=n+1/2$ if $l=B$) represents the exact position of the type-$l$ site of the cell with number $n$. Note that one should have used a different notation for the operators $c_{l}(k)$ in order to distinguish (\ref{Fourier1D}) and (\ref{Fourier1Dcano}). After substitution of Eq.~(\ref{Fourier1Dcano}) in Eq.~(\ref{SSH_Hamiltonian}), one obtains the Bloch Hamiltonian :
\begin{equation}
H(k)=(v+w)\cos(k/2) \sigma_x + (v-w) \sin(k/2) \sigma_y =2\cos(k/2) \sigma_x + \delta \sin(k/2) \sigma_y \, .
\label{sshcanonical}
\end{equation}
Since the exact position of sites is involved in the unitary transformation, the Bloch Hamiltonian $H(k)$ does not depend on the choice of unit cell, but its periodicity in reciprocal space is altered. Indeed, note that $\mathcal{H}(k+2\pi)=\mathcal{H}(k)$ but $H(k+2\pi)=-H(k)$ such that $H(k+4\pi)=H(k)$. The Bloch Hamiltonian has a doubled periodicity related to the fact that the distance between the two sites within the unit cell is half of the unit cell size. Later, we will encounter a similar effect in graphene. The Bloch Hamiltonian seems to depend on the sign of $\delta$. Actually $H_{-\delta}(k)$ can be mapped back to $H_{\delta}(k)$ by a unitary transform which exchanges $A$ and $B$ sites: $\sigma_x H_{-\delta}(k) \sigma_x = H_{\delta}(x)$. This relabeling transformation is a do-nothing transformation. One could therefore restrict the study of the infinite SSH chain to $\delta \geq 0$.
\begin{figure}
\begin{center}
\includegraphics[width=5.5cm]{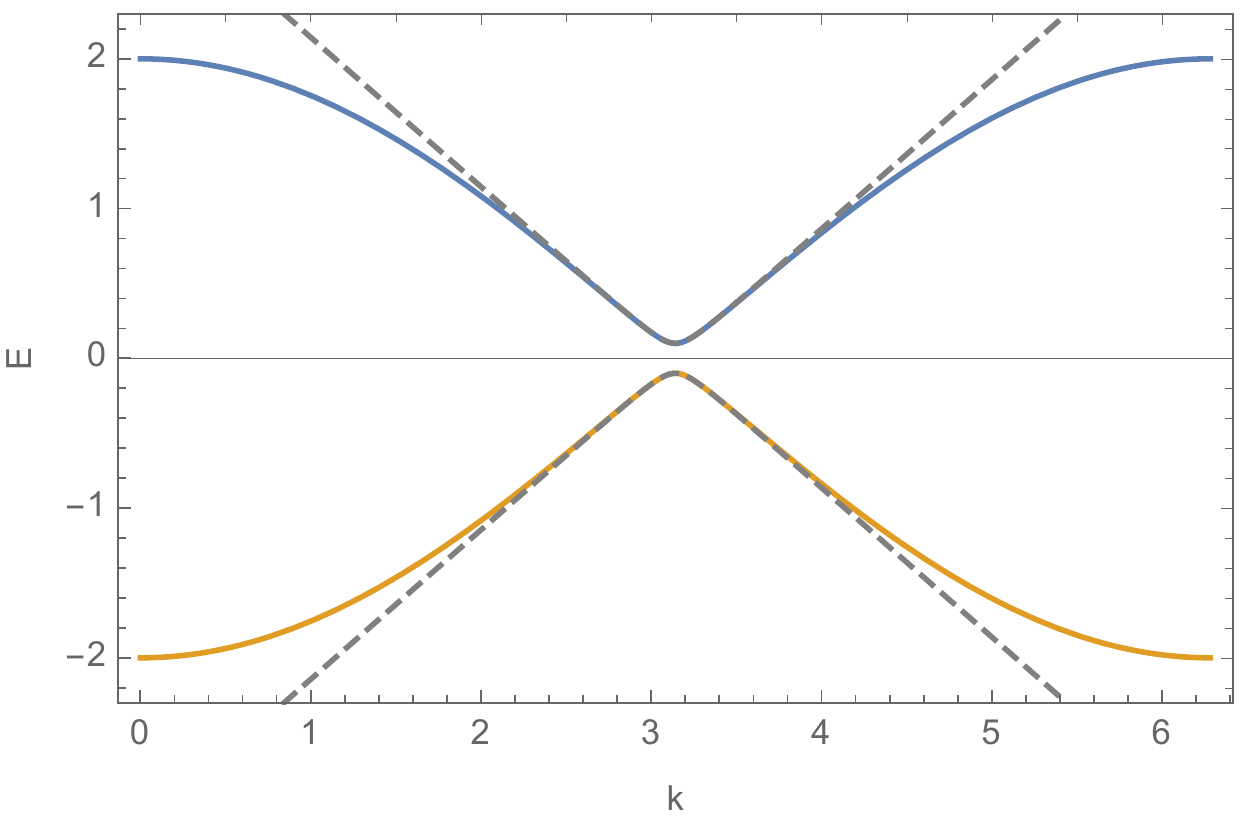}
\includegraphics[width=5.5cm]{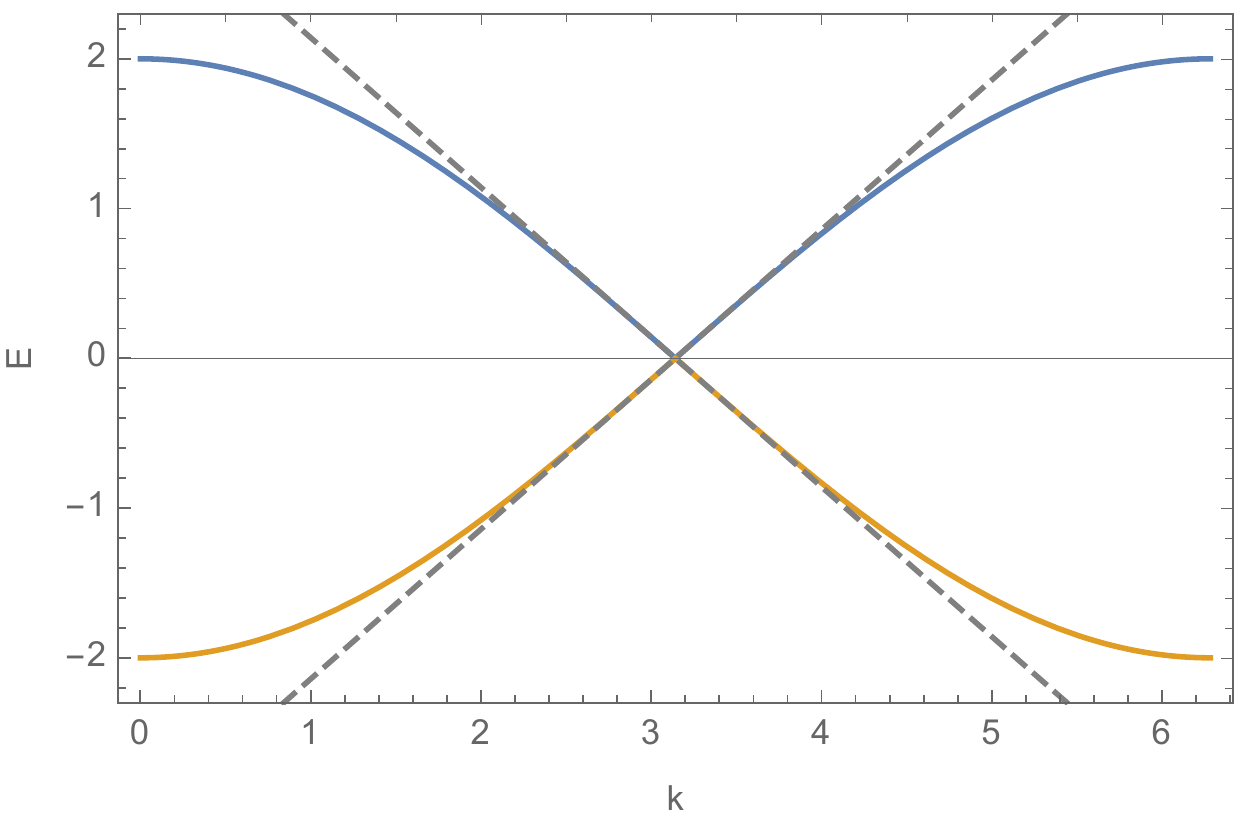}
\includegraphics[width=5.5cm]{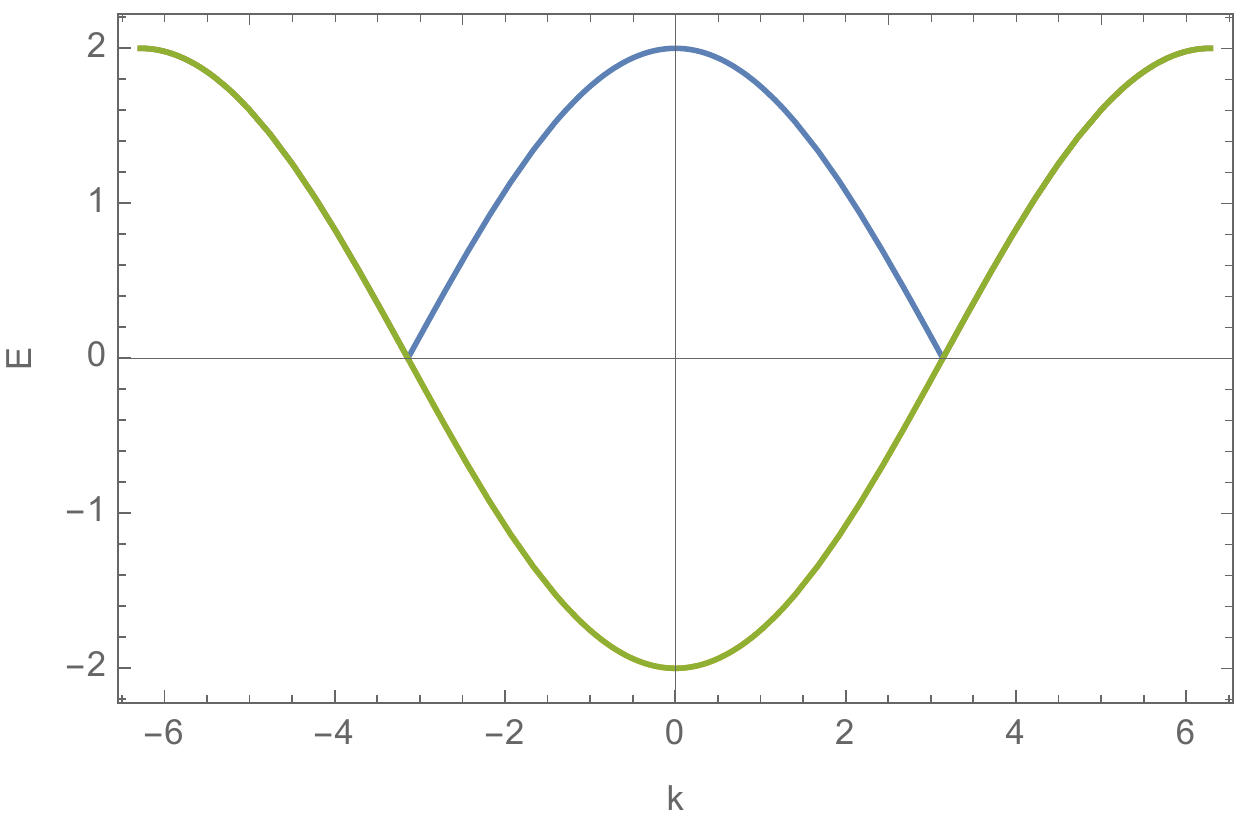}
\caption{\label{fig:dispssh} (Left) Dispersion relation of the SSH chain in the first BZ $k\in [0,2\pi]$ exhibiting the massive Dirac cone in the vicinity of $k=\pi$ (in dashed gray). (Middle) Dispersion relation of the gapless chain ($\delta=0$) with $k\in [0,2\pi]$ (folded scheme, not the first BZ) exhibiting the massless Dirac cone in the vicinity of $k=\pi$ (in dashed gray). (Right) Dispersion relation of the gapless chain showing the folding from $k\in [-2\pi,2\pi]$ (first BZ, single band in green) to $k\in [-\pi,\pi]$ (two bands in blue, the one at negative energy is overlaid by the green curve).}
\end{center}
\end{figure}
 
In both representations, we have obtained that $d_x$ is an even and $d_y$ an odd function of $k$ by explicit calculation, but this is also a consequence of the symmetries of the SSH model. Like in any two-band model, the physics is encoded in the functions $d_x(k)$,  and $d_y(k)$, which are similar to the components of the external magnetic field in the TLS case (section \ref{section0D}), and determine the geometry of the spinors. Note the absence of a $d_z(k) \sigma_z$ term in the SSH model. We will study the effect of such a term within the RM model in Sec.~\ref{sec:rm}.

\begin{figure}[ht]
\begin{center}
\includegraphics[width=8.25cm]{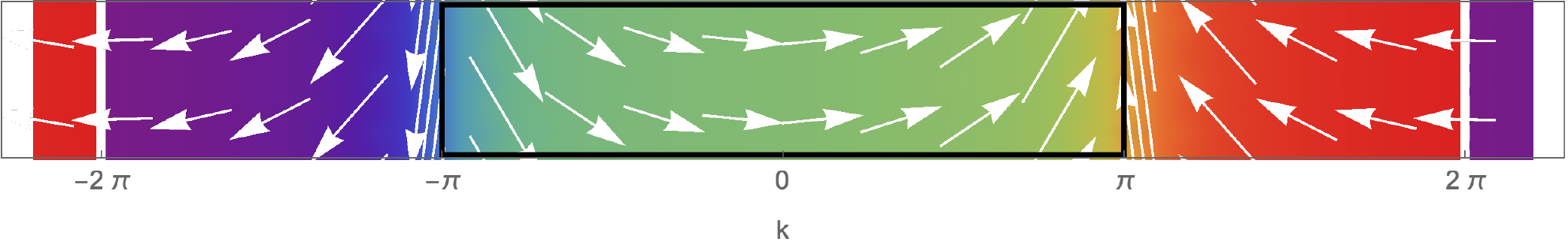}
\hspace{1cm}
\includegraphics[width=8.25cm]{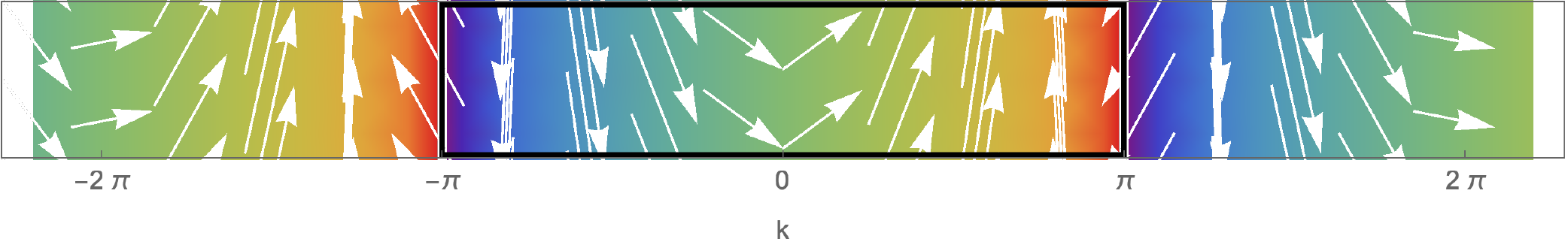}
\caption{Azimuthal angle for the SSH model with $\delta=0.3$ versus $k$ in reciprocal space (the first BZ $[-\pi,\pi[$ is indicated by a black line). The vertical axis has no meaning. (Left) Phase $\varphi_k$ showing the enlarged periodicity of the (canonical) Bloch Hamiltonian $H(k)$. (Right) Phase $\phi_k$ obtained from a periodic Bloch Hamiltonian $\mathcal{H}(k)$.}
\label{FigAzimuthalPhaseSSH}
\end{center}
\end{figure}

The electronic band structure of the infinite chain is obtained by diagonalizing Eq. (\ref{hRM}) or Eq. (\ref{sshcanonical}), which leads to the dispersion :
\begin{equation}
E_{\pm}(k) = \pm |\vd(k)| =  \pm \sqrt{v^{2} + w^{2} + 2 v w \cos(k)}= \pm \sqrt{4  \cos^2(k/2)+\delta^2 \sin^2(k/2)}\, .
\label{Essh}
\end{equation}
In the general case $\delta\neq 0$, the conduction band ($E_{+}(k)>0$) and the valence band ($E_{-}(k)<0$) are separated by a gap, and the chain is therefore insulating at half-filling, see Fig.~\ref{fig:dispssh} (Left). One may also define an azimuthal angle (along the equator of the Bloch sphere) $\varphi_k$ reflecting the geometrical/topological properties of the Bloch Hamiltonian beyond the energy spectrum:
\begin{equation}
H(k)=E_+(k)(\cos \varphi_k \sigma_x + \sin \varphi_k \sigma_y)\, \text{ and } \mathcal{H}(k)=E_+(k)(\cos \phi_k \sigma_x + \sin \phi_k \sigma_y)\, .    
\end{equation}
In order to clearly distinguish them, we call $\varphi_k$ the azimuthal angle for the canonical Bloch Hamiltonian $H(k)$ and $\phi_k$ that for a periodic Bloch Hamiltonian $\mathcal{H}(k)$. They are both plotted in Fig.~\ref{FigAzimuthalPhaseSSH}.

One may wonder if there are exceptional situations where the spectrum Eq. (\ref{Essh}) becomes gapless. The condition for a gap closing between these two bands requires all coefficients $d_i(k)$ (for $i=x,y$) in Eq.(\ref{dxdydzRM}) to vanish simultaneously at the same point $k$ of the BZ. One has two constraints and only one variable $k$, which is a typical level crossing/repulsion situation in quantum mechanics. Nevertheless, here it leads to $d_x = v + w \cos(k) = 0$ and $d_y = w \sin(k) = 0$. Therefore, considering solely positive $v$ and $w$, the gap closing arises for :
\begin{equation}
k= \pi  \hspace{5mm}  ,  \hspace{5mm}  \delta=v-w = 0    \, .
\end{equation}
The SSH model becomes gapless when $\delta=0$, which in fact means that the chain is no longer dimerized. Indeed a monoatomic chain is known to lead to a single metallic band $E(k)=2\cos(k/2)$. It can be described by a single band in its natural BZ $[-2 \pi,2\pi[$. Here the two bands in Fig.~\ref{fig:dispssh} (Middle) correspond to the folding of this metallic band in the BZ of the SSH model $[-\pi,\pi[ $ [see Fig.~\ref{fig:dispssh} (Right)].

\medskip

Going beyond the energy level description, one considers now the evolution of the stationary states of $H(k)$ when $k$ runs over the circular BZ. For each $k$, the Bloch Hamiltonian has exactly the structure of the two-level Hamiltonian of Section \ref{subsec:BerryTLS} with $d_z =0$ and therefore $\theta_k = \pi/2$. The spinors (cell-periodic Bloch states) are located along the equator of the Bloch sphere and read :
 \begin{equation}
\ket{u_+(k)} =\frac{1}{\sqrt{2}}  \begin{pmatrix}  1  \\  e^{i \varphi_k} \end{pmatrix}  \hspace{5mm} {\rm et}  \hspace{5mm} \ket{u_-(k)}= \frac{1}{\sqrt{2}}  \begin{pmatrix} 1 \\ - e^{i \varphi_k} \end{pmatrix} \, ,
\end{equation}
which are well defined for all $k$. The ket $\ket{u_+(k)} $ represents the excited state associated with the positive energy $E(k)$ (conduction band), while the eigenstate $\ket{u_-(k)} $ is the ground state with negative energy $-E(k)$ (valence band). 

\subsubsection{Symmetries}

The symmetries of the SSH model translate into constraints fulfilled by its Bloch Hamiltonian. It turns out that, for the SSH model (but this is not a general property), these constraints are the same whether expressed on the canonical or a periodic Bloch Hamiltonian.

The time-reversal symmetry $\mathcal{T}$ for spinless fermions is expressed by 
\begin{equation}
\mathcal{T} : \, \, \mathcal{K} \mathcal{H}(k) \mathcal{K} =\mathcal{H}(k)^*= \mathcal{H}(-k) \, ,
\end{equation}
where $\mathcal{K}$ means complex conjugation. This invariance is valid because the hopping parameters $v$ and $w$ are real.
The chiral symmetry reads :
\begin{equation}
\mathcal{S} : \, \, \sigma_z \mathcal{H}(k) \sigma_z = - \mathcal{H}(k) \, .
\end{equation}
The chiral symmetry is equivalent to the absence of $\sigma_z$ in $\mathcal{H}(k)$.  The 
chiral symmetry, also called sublattice symmetry, $\mathcal{S}$ involves only $\mathcal{H}(k)$, and not a relation between periodic Bloch Hamiltonians $\mathcal{H}(k)$ and $\mathcal{H}(-k)$ at opposite quasimomenta $k$ and $-k$. For each state $\ket{\Psi_k} $ with energy $E$, there is a state $\sigma_z \ket{\Psi_k} $ with opposite energy $-E$, which explains the electron/hole symmetry between the conduction and valence bands Eq.(\ref{Essh}).   
The RM term $\Delta \sigma_z$ would break this chiral symmetry as it does not respect bipartiteness. The charge conjugation $\mathcal{C}=\mathcal{S}\mathcal{T}$ appears to be a combination of time-reversal, which involves complex conjugation, and sublattice symmetry which involves flipping the sign of energy:  
\begin{equation}
\mathcal{C} : \, \, \sigma_z \mathcal{H}(k)^* \sigma_z = - \mathcal{H}(-k) \, .
\end{equation}
The SSH model is also a condensed matter realization of the quantum field theory concept of charge conjugation. In particle physics, charge conjugation is the interchange of particles and antiparticles. In polyacetylene, there are no genuine positrons of course but the electronic states splits into electron and hole states. In a half-filled chain, hole excitations can propagate with the same parameters of the electron excitation, except for the opposite electric charge. Formally it corresponds to a map between the positive energy solutions and the negative energy solutions of the model. 

\medskip
Finally, the SSH model has an additional symmetry which is the invariance under space inversion, which implies the exchange of $A$ and $B$ isospin index, $\mathcal{I}:c_A \rightarrow c_B$ and $\mathcal{I} :c_B \rightarrow c_A$, as the inversion center is mid-bond and not on-site. The inversion symmetry reads :
\begin{equation}
\mathcal{I} : \, \, \sigma_x \mathcal{H}(k) \sigma_x =  \mathcal{H}(-k) \, .
\end{equation}

In the ten-fold classification of topological insulators~\cite{Schnyder:2008,Kitaev:2009,Qi:2008,Chiu:2016}, it would therefore appear that the SSH model belongs to the BDI class of symmetry (i.e. time-reversal symmetry TRS with $\mathcal{T}^2=+1$, particle-hole symmetry PHS with $\mathcal{C}^2=+1$ and therefore chiral or sublattice symmetry SLS $\mathcal{S}=1$) which is characterized by a $\mathbb{Z}$ topological index (winding number). Actually, we will see that the bulk winding number has no precise meaning here and that the SSH model is best described as an inversion-symmetric 1D band insulator characterized by a $\mathbb{Z}_2$ invariant.

\subsubsection{Winding number and edge modes \label{sec:winding}}
\begin{figure}[h!]
\begin{center}
\includegraphics[width=5.5cm]{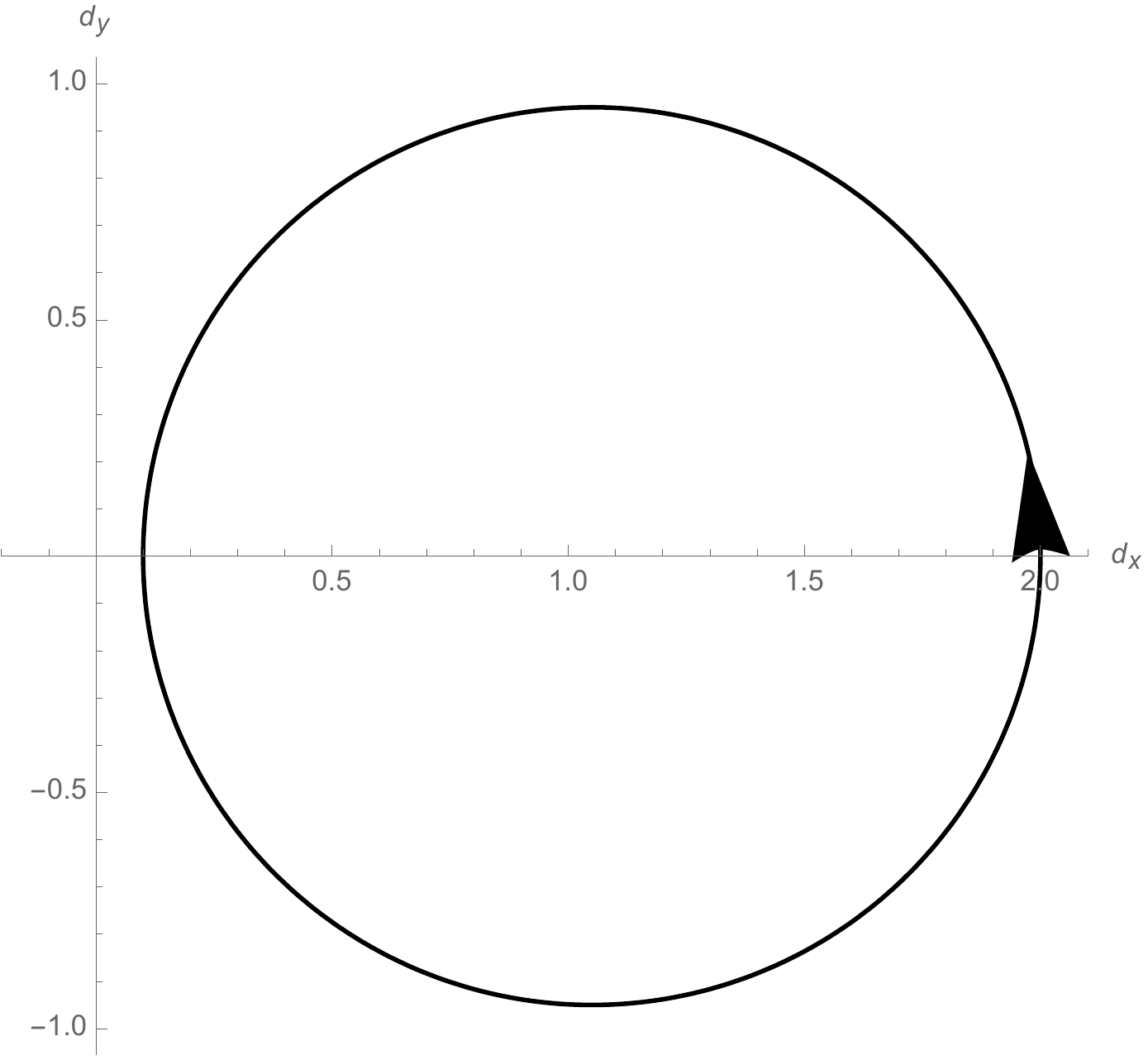}
\includegraphics[width=5cm]{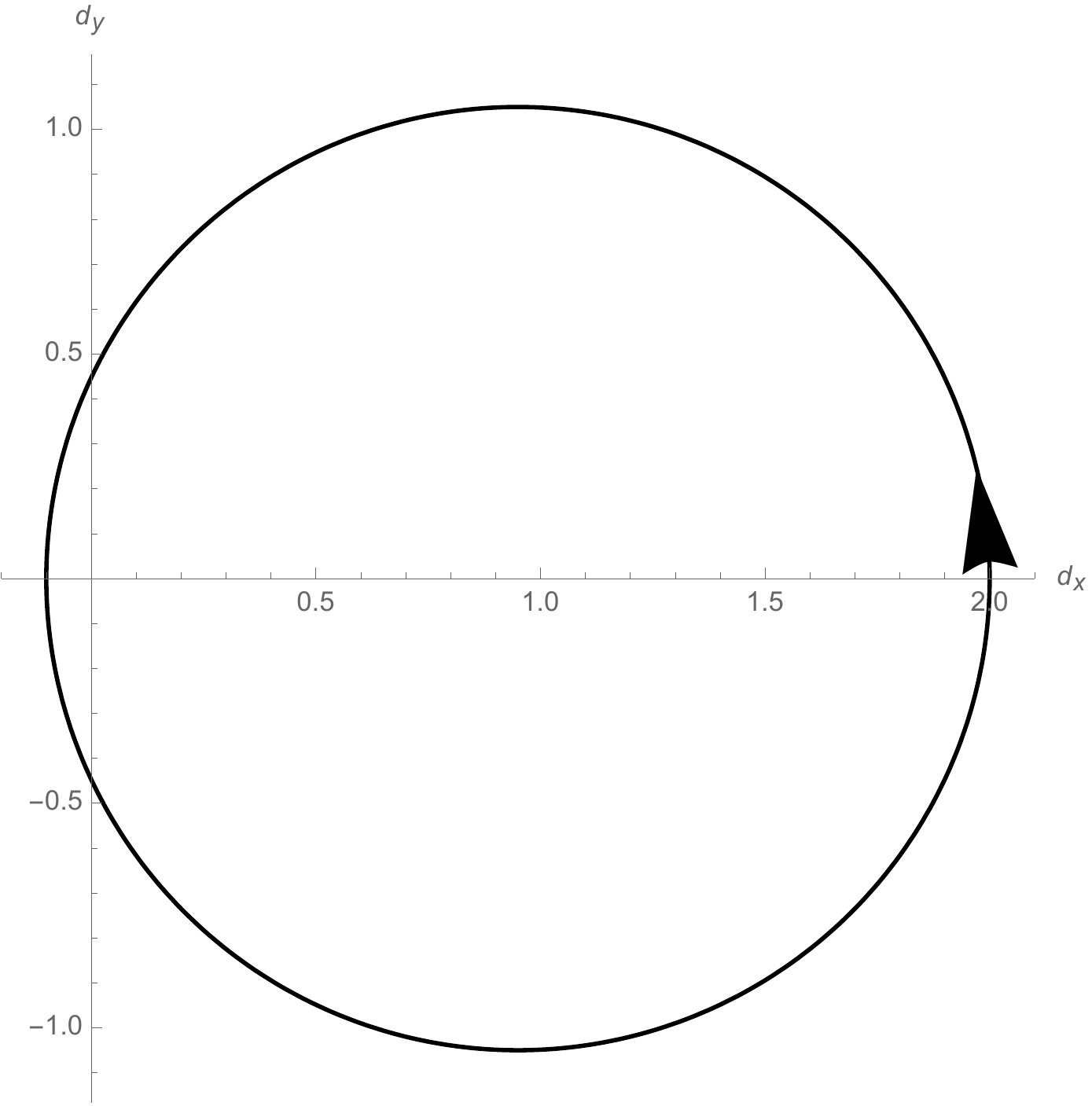}
\caption{\label{fig:windingssh} Parametric plot $(d_x(k),d_y(k))$ giving the winding number $n_w$ for the SSH model computed with a periodic Bloch Hamiltonian. (Left) When $\delta=v-w>0$, the curve does not enclose the origin and $n_w=0$. (Right) When $\delta<0$, the curve encloses the origin once in the anti-clockwise orientation and $n_w=+1$.}
\end{center}
\end{figure}
For each $k$, the \emph{periodic} Bloch Hamiltonian $\mathcal{H}(k)$ (and hence its two eigenstates) is represented by the unit vector $\mathbf{\hat{d}}(k)=\mathbf{d}(k)/|\mathbf{d}(k)|$ on the Bloch sphere $S^2$. Besides explaining the electron/hole symmetry between the conduction and valence bands Eq.(\ref{Essh}), the chiral symmetry also constrains the tip of this unit vector $\mathbf{\hat{d}}(k)$ to stay along the equator of the Bloch sphere when $k$ runs over the BZ. The 1D BZ has the structure of the circle $S^1$ since its extremities $k=\pm \pi$ represent the same state. The mapping from the circle (BZ) to the circle (equator) allows one to define a winding number $n_w$, which is related to the first homotopy group of the circle $\Pi_1(S^1)=\mathbb{Z}$. The relative integer $n_w$ is the winding number of $\mathbf{d}(k)$ around the origin. For our initial choice of unit-cell convention, when $w<v$, the origin is outside the circle traced by $\mathbf{d}(k)$, so $n_w = 0$ [see Fig.~\ref{fig:windingssh} (Left)]. In contrast, when $w>v$, the origin is inside the circle and $\mathbf{d}(k)$ winds exactly once around the origin : $n_w=1$ [see Fig.~\ref{fig:windingssh} (Right)]. This winding number $n_w$ is protected by the chiral symmetry of the SSH chain, namely $\sigma_z \mathcal{H}(k) \sigma_z = -\mathcal{H}(k)$ (this periodic Bloch Hamiltonian $\mathcal{H}(k)$ anticommutes with $\sigma_z$).  

There is an ambiguity in the assignment of $n_w =0$ or $n_w=1$ to one phase or the other, namely to a definite sign of the parameter $v-w$. For a genuinely infinite system, one cannot determine whether the system is in a trivial or non-trivial phase based on $n_w$ if one is not aware of the chosen unit cell. The only unambiguous statement is that this winding number changes by $\pm 1$ when going from one dimerization to the other. This ambiguity is lifted for finite chains, because it is then possible to state whether the chain starts (or ends) by a weak or a strong bond and there is a natural choice of unit cell. There is a relation between this winding property and the presence of protected zero modes at an edge or a domain wall~\cite{Ryu:2002,Delplace:2011}. It is an example of the Jackiw and Rebbi mechanism (see below). 

In contrast, the staggered on-site potential $\Delta \sigma_z$ of the RM model breaks explicitly the chiral symmetry, and the tip of $\mathbf{\hat{d}}(k)$ is now free to follow any loop on the whole Bloch sphere. Since all such unconstrained loops can be smoothly deformed and reduced to a point, there is no possibility to define a winding number for the RM model (at least when inversion symmetry is not imposed, see below).  

In summary, the winding number is useful in discussing open chains with edges or domain walls and the presence of protected edge modes. However, as a bulk invariant, it is meaningless as it depends on the unit cell choice.


\subsubsection{Zak phase, position operator, Wannier center and electric polarization\label{sec:zakwannier}}
A more interesting quantity is the Zak phase~\cite{Zak:1989}. It is a kind of Berry phase defined for a given band but computed along a \emph{non-contractible} path in the BZ thanks to its torus shape. In 1D, it reads
\begin{equation}
    Z_n = \int_{-\pi}^\pi dk \,  A_n(k)= \int_{-\pi}^\pi dk \,  \langle u_n(k)|i\partial_k u_n(k)\rangle, 
\end{equation}
for the $n^\text{th}$ band. This bulk quantity is measurable (and was actually measured~\cite{Atala:2013}) and related to the electronic contribution to the polarization~\cite{Resta:2007}. The Zak phase should be computed using the \emph{canonical} Bloch Hamiltonian $H(k)$ (and not the periodic one $\mathcal{H}(k)$), as it is related to the projected position operator $x_c =  i\nabla_k + A_n(k)$. It is best thought as being an average electron position (known as the Wannier or band center) within the unit cell~\cite{Zak:1989}
 \begin{equation}
 \langle x_-\rangle = \int_0^1 dx |w_n(x)|^2 x = \frac{Z_-}{2\pi}
 \end{equation}
 modulo $a=1$, where 
 \begin{equation}
 w_n(x)=\int_{-\pi}^\pi dk e\, ^{i k x} u_{n k}(x)=\int_{-\pi}^\pi dk \, \psi_{n k}(x)     
 \end{equation}
is the Wannier function of the $n^\text{th}$ band (here positioned in the unit cell at $R=0$)~\cite{Marzari:2012}. 
 Because the Wannier center is a position, it continuously depends on the choice of position origin. Therefore the Zak phase continuously depends on the choice of position origin. Also, it is actually an \emph{open-path} geometrical phase and therefore becomes gauge-invariant only upon imposing a definite phase relation between initial and final states~\cite{Zak:1989,Resta:2000}. Indeed, on the non-contractible path from $k=-\pi$ to $k=+\pi$, the final state $|u_n (\pi)\rangle$ is not the same as the initial state $|u_n (-\pi)\rangle$: the path is open in Hilbert space, although it is closed in parameter space, i.e. in the BZ. In order to give a gauge-invariant quantity, the Zak phase should be computed under the following restriction called the ``periodic gauge choice''~\cite{Zak:1989,Resta:2000} $|u_n (k+G)\rangle =e^{-iG x} |u_n (k)\rangle$, where $x$ is the position operator, i.e. $|u_n (\pi)\rangle =e^{-i 2\pi x} |u_n (-\pi)\rangle$. The Zak phase being position origin-dependent, computed along a non-contractible path and an open-path geometrical phase subject to a periodic gauge condition, it is truly different from a Berry phase. The two should be carefully distinguished.

The 1D Berry connection defined from the Bloch spinor $\ket{u_-(k)}$ (ground state or valence band) reads :
 \begin{equation} 
 A_- (k)=i \braket{u_-(k)   \mid  \partial_k u_-(k)}=\frac{1}{2} \frac{d \varphi_k}{dk} \, ,
 \end{equation} 
 and can be easily integrated along a non-contractible closed path in the BZ to give the Zak phase :
  \begin{equation} 
 Z_-=\int_\text{BZ} A_- (k) dk =\frac{1}{2} \int_{k=-\pi}^{k=\pi} d \varphi_k\, =\frac{1}{2} \left( \varphi_\pi - \varphi_{-\pi}  \right) = \frac{\pi}{2} \text{sign}\frac{w-v}{w+v}=\pm \frac{\pi}{2}\, .
 \end{equation} 
The Zak phase is also related to the electronic contribution to the polarization 
 \begin{equation}
 P_\text{el}= -e \langle x_-\rangle = -e \frac{Z_-}{2\pi} =\mp \frac{e}{4}
 \end{equation}
 modulo the electron charge $e$~\cite{Resta:2007}. In the following, we take units such that $e=1$ and it therefore seems that the polarization quantum~\footnote{The polarization ``quantum'' has nothing to do with quantum mechanics.}, i.e. the modulo in the electric polarization, is $P_q=1$. But this is not correct. Actually, the only meaningful electric polarization is defined for a charge-neutral crystal and should therefore include the contribution of the ions: 
 \begin{equation}
     P_\text{tot}=P_\text{el}+P_\text{ions}\, .
 \end{equation}
In order to have a charge-neutral crystal, we assume that each ion ($A$ or $B$) carries a $+1/2$ charge. As another consequence of the presence of the ions (or of the rigid lattice of sites), it will turn out (see Sec.~\ref{sec:rm}) that the polarization is actually defined modulo $P_q=1/2$ and not $P_q=1$.

In conclusion of these short sections on the winding number and the Zak phase, we remark the following. The winding number is very often mistaken for the Zak phase. They should be clearly distinguished. The winding number is computed using a periodic Bloch Hamiltonian, whereas the Zak phase should be computed using the canonical Bloch Hamiltonian, as it is related to the position operator (see the above discussion). On the one hand, the winding number depends on the choice of unit cell and is therefore only well-defined when the choice of unit cell is fixed, e.g. because of a boundary~\cite{Delplace:2011} or at a domain wall between two dimerizations. On the other hand, the Zak phase requires using the periodic gauge choice, it continuously depends on the position origin (it is best thought as being proportional to the Wannier center) and it is evaluated along a non-contractible loop in BZ. 

\subsection{Linearization and effective Dirac Hamiltonian} 
In the generic case, $\delta\neq 0$, the SSH model is gapped. The gap is direct, located at $k=\pi$, and its magnitude is $2|\delta|$. Therefore the sign of $\delta$ is unimportant for the energy spectrum. In contrast, this sign, and the related band inversion, is important for the wave functions, and the topological properties. 

A first way to understand this consists in analyzing the SSH chain in the continuous limit, namely at length scales exceeding the lattice spacing between cells $a=1$. We consider the SSH chain  with a narrow gap $|\delta|\ll 1$, close to the semi-metallic regime. We can expand this periodic Bloch Hamiltonian $\mathcal{H}(k)$ around $k=\pi$ by writing $k = \pi + q$, with $q \ll 1$ :
\begin{eqnarray}
\mathcal{H}(k=\pi + q)= \mathcal{H}_{D}(q) =[v+w\cos(\pi+q)] \sigma_x + w \sin(\pi+q) \sigma_y
\approx -q \sigma_y + \delta \sigma_x \, ,
\label{JRq}
\end{eqnarray}
with $\delta=v-w$ and $w=1-\delta/2\approx 1$. This has the form of a 1D Dirac Hamiltonian. 

The Dirac equation was originally invented to describe relativistic electrons in three dimensional space~\cite{Dirac:1928} but it can be generalized to any space dimension $D$. Generally speaking, the Dirac Hamiltonian reads 
\begin{equation}
H_D=c\vp\cdot \boldsymbol{\alpha}+mc^2 \beta \, ,
\end{equation}
and involves $D+1$ anti-commuting matrices $\beta, \alpha_1,...,\alpha_D$ that square to one (the so-called Clifford algebra). Here $\vp$ is the momentum operator and $\boldsymbol{\alpha}=(\alpha_1,\alpha_2,...)=(\alpha_x,\alpha_y,...)$. In 3D, the matrices are $4\times 4$, but in 2D and 1D, $2\times 2$ matrices are possible. The Dirac Hamiltonian depends on two parameters, which are the velocity of light $c$ and the electron mass $m$. When it emerges in the low-energy limit as an effective description of a lattice model in solid-state physics, the velocity $c$ and mass $m$ are effective parameters that have no simple relation to the velocity of light or the electron mass. In the present context, the two Dirac matrices are $\alpha_x=-\sigma_y$ and $\beta=\sigma_x$, and they satisfy a Clifford algebra as the Pauli matrices anti-commute and square to one. The velocity is $c=t a=1$ (it is usually called Fermi velocity $v_F$ in solid-state physics) and the mass is $m=\delta$. The band structure is that of a massive Dirac cone :
\begin{equation}
    E(q)=\pm\sqrt{q^2+\delta^2}\, ,
\end{equation}
that becomes massless $E(q)=\pm |q|$ when $\delta\to 0$ (see Fig.~\ref{fig:dispssh}).

This effective Dirac Hamiltonian is useful to analyze the band inversion at $k=\pi$. If we sit exactly at $q=0$, the Hamiltonian is simply the mass term $\delta \sigma_x$, and the stationary states reduce to the eigenstates of the Pauli matrix $\sigma_x$, denoted $\ket{\sigma _x =1} $ and $\ket{\sigma _x =-1} $ respectively. The ground state is $\ket{\sigma _x =-1} $ if $\delta>0$, and switches to $\ket{\sigma _x =1} $ when $\delta<0$. Therefore this local analysis shows that the parameter $\delta$ drives a band inversion. 

In the limit $\delta=0$, the system is gapless and described by a 1D massless Dirac equation. As both chiralities are present (left and right movers), this is not a 1D Weyl equation despite the masslessness.

\subsection{Jackiw-Rebbi mechanism and charge fractionalization in Dirac insulators \label{sec:jr}}
In the low-energy effective description in terms of a Dirac Hamiltonian (\ref{JRq}), let us consider that the mass $m=\delta$ is $x-$dependent and changes sign at $x=0$. This is a domain wall in the Dirac mass. In the seminal paper by Jackiw and Rebbi~\cite{Jackiw:1976}, this mass was provided by the coupling of massless fermions to a bosonic background, and such a bosonic field was assumed to have a topological defect, called a kink. In solid-state physics, polyacetylene provides such a situation at any domain wall junction between a $\delta<0$ and $\delta>0$ chain (the two possible dimerizations)~\cite{Su:1979,Su:1980}. This is a topological defect in the dimerization. The mass kink $m(x)$ breaks translation invariance and one should substitute $q$ in Eq.(\ref{JRq}) by $-i \partial_x$, yielding the first-order differential equation (obtained after multiplication by $i\sigma_y$)
\begin{equation}
i \sigma_y \frac{\partial \Psi}{\partial x} + m(x) \sigma_x \Psi = E \Psi  \, ,
\label{JRx}
\end{equation}
where $m(x)$ is a mass profile varying from $m(x=-\infty)=-m_0$ to $m(x=\infty)=m_0>0$. Since, the gap changes sign between $-\infty$ and $\infty$ it has to close somewhere and it is natural to investigate the possibility of zero energy states. Such a zero energy state should satisfy the equation :  
\begin{equation}
 \frac{\partial \Psi}{\partial x} = m(x) \sigma_z \Psi  \, ,
\label{JRzero}
\end{equation}
which is equivalent to two uncoupled first order differential equations for the $A$ and $B$ components of the wave function respectively. The fact that the mass has opposite signs at $x=\pm \infty$ allows one of these equations to have bounded solutions. For a step in $m(x)$ with no sign inversion, there is no zero-energy bound state because on both sublattices the solution would grow exponentially on one side of the line or the other.  Let us first consider a sharp mass profile $m(x) =  m_0 \Theta(x) - m_0 \Theta(-x)$ where $\Theta(x)$ is the Heaviside step function. For positive $m_0$, the zero energy bound state reads : 
\begin{equation}
\Psi(x) =\left( \Theta(x) e^{-m_0 x}  + \Theta(-x) e^{m_0 x} \right)  \ket{\sigma_z=-1}  \, ,
\end{equation}
which is a state completely localized (or polarized) on the $B$ sublattice. For the choice $m_0 < 0$, the bounded solution would be polarized on the $A$ sublattice. Upon increasing $m_0$, the bound state gets more and more localized around $x=0$ over a typical length given by $ \xi = 1/ m_0$. For a general kink (smooth kink), the general solution of Eq. (\ref{JRx}) reads :
\begin{equation}
\Psi(x) =\left( e^{-\int_0^x \mathrm{d}x' m(x') \sigma_z} \right) \ket{\sigma_z=\sigma}  \qquad \qquad \forall x \in \mathbb{R} \, ,
\end{equation}
where the sublattice polarization is determined by $\sigma = - \text{sign}(m_0)$. The zero energy state is localized near the location where the mass $m(x)$ changes sign, and it is stuck between two insulators on both sides. 

The zero-energy states are protected by the chiral symmetry, which means that any perturbation anticommuting with $\sigma_z$ will preserve the zero energy level. For instance, disorder in the hopping parameters $v$ and $w$ will preserve the zero-energy states provided the gap remains. In contrast, switching on a staggered on-site energy $\pm \Delta$ breaks the chiral symmetry and splits the zero-energy level in two levels at finite energies $\pm \Delta$. 

If the SSH chain is cut on a strong bond, then the edge hosts a dangling bond. This mid-gap state sits exactly at zero energy and can be empty or occupied by an electron. If it is empty, then the edge has an excess charge $+1/2$ (that of the isolated cation at the edge). If it is occupied, then the charge turns negative as cation + electron = $+1/2-1=-1/2$. This is the simplest example of a very general mechanism for charge fractionalization discovered by Jackiw and Rebbi~\cite{Jackiw:1976}. Here, starting with spinless electrons, one obtains emergent excitations at zero energy and with charge $\pm 1/2$. The general recipe is: take a Dirac equation, make its mass spatially inhomogeneous and change its sign by inserting a topological defect (a domain wall in 1D, a vortex in 2D, etc.) so that it changes sign and the result is a trapped zero-energy mode localised on the defect. See the short review by Jackiw~\cite{Jackiw:2012}.

In this section, we have seen that the SSH chain has localized zero-energy modes either on a domain wall (mass kink between two dimerizations) or at an edge provided the cut occurs on a strong bond. In solid-state physics, it means that any insulator that has a low-energy description in terms of a Dirac equation (i.e. we could call it a \emph{Dirac insulator}) may host such a topologically-protected zero-mode trapped on a topological defect in the Dirac mass. Note that this is independent of the fact that this Dirac insulator is, in addition, a trivial or a topological insulator. In other words: topologically protected zero-mode on a defect is not the same thing as topological band. The former is a local property in the vicinity of a real-space defect, the latter is a property involving the whole band, the whole BZ and not only the low effective description in the vicinity of the gap. For example, boron nitride is a Dirac insulator but it is topologically trivial (the two valleys carry opposite Berry curvature, see Sec.~\ref{section2Dbn}). This is also the case of the SSH chain (see Sec.~\ref{sec:rm}). In contrast, the Haldane model in graphene is also a Dirac insulator but this time it can be a topological (Chern) or trivial insulator depending on its parameters (see Sec.~\ref{section2DHaldane}).

\subsection{Inversion-protected topological insulator \label{sec:rm}}
\begin{figure}[h!]
\begin{center}
\includegraphics[width=10cm]{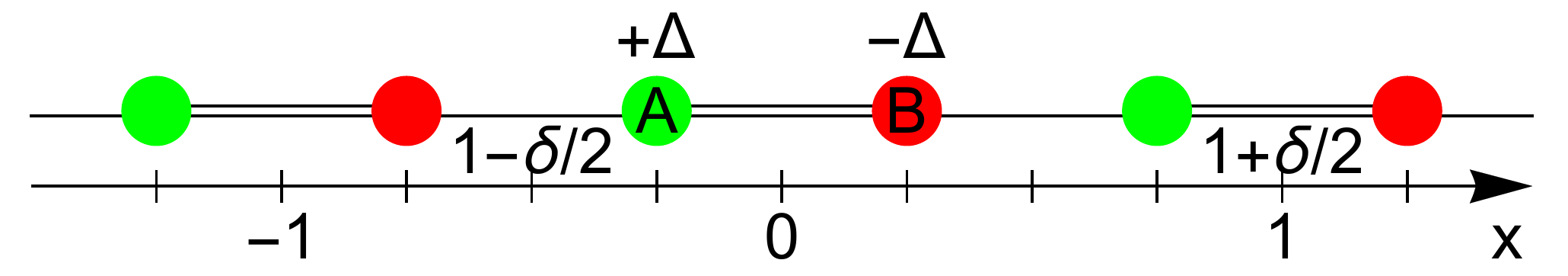}
\caption{\label{fig:rmchain} Rice-Mele chain showing the dimerized hopping amplitudes $v=1+ \delta/2$, $w=1-\delta/2$ and the staggered on-site energies $E_{A,B}=\pm \Delta$. The position origin $x=0$ is chosen on a mid-bond and corresponds to an inversion center for the SSH chain ($\Delta=0$). An inversion center for the charge density wave case (CDW, $\delta=0$) is on site, e.g. at $x=-1/4$.}
\end{center}
\end{figure}
The SSH chain is often taken as the simplest example of a one-dimensional topological insulator. Indeed, it is a two-band insulator when half-filled. However, its two ``phases'' characterized by the sign of $\delta$ are actually the same, as is obvious by realizing that a shift of the infinite chain by half a unit cell makes $\delta \to - \delta$. Therefore, in the bulk, there is not a trivial phase (say $\delta>0$) and a topological phase (say $\delta<0$) as a naive analysis of the winding number $n_w$ would suggest (see Fig.~\ref{fig:windingssh}). In addition, zero-energy edge states are not present for any cut of the chain but only when the cut occurs on a strong bond. Also it is impossible to define a genuine bulk topological invariant that would distinguish the two phases (for example, neither the winding number $n_w$ nor the Zak phase qualify as a well-defined topological invariant). These three facts make the SSH model a poor example of a topological insulator. Nonetheless, it is a good example of a Dirac insulator featuring zero-modes trapped on a topological defect (see Sec.~\ref{sec:jr} on the Jackiw-Rebbi mechanism).

\subsubsection{Rice-Mele model with inversion symmetry}
However, there is one way in defining a proper 1D topological crystalline insulator~\cite{Fu:2011}, i.e. a band insulator whose topology is protected by a point group symmetry. We have already noted that the SSH chain has an inversion symmetry and that the inversion center is mid-bond (at equal distance between two sites $A$ and $B$) and not on-site. Another type of 1D two-band insulator with inversion symmetry is a regular tight-binding chain (as the SSH chain with $\delta=0$) but with staggered on-site energies $E_A=\Delta$ and $E_B=-\Delta$. When half-filled, this gapped two-band model realizes a band insulator of the charge density wave (CDW) type. It has inversion symmetry but the inversion center is now on-site (rather than mid-bond). 

In order to realize a phase transition between these two types of inversion-symmetric insulators, we consider the RM model~\cite{Rice:1982} having both the SSH dimerization $\delta$ and the CDW staggered on-site energy $\Delta$. The (canonical) Bloch Hamiltonian reads
\begin{equation}
H(k)=2\cos(k/2) \sigma_x + \delta \sin(k/2) \sigma_y + \Delta \sigma_z .
\label{rmcanonical}
\end{equation}
Following~\cite{Vanderbilt:1993,XiaoBerryRMP:2010,Combes:2016}, we introduce an angle $\theta_\text{RM}$ such that $\cos \theta_\text{RM} = \Delta/M$ and $\sin \theta_\text{RM} = \delta /M$, where $M=\sqrt{\Delta^2+\delta^2}\geq 0$, in order to parametrize the RM model. However, instead of considering the full range of parameters, we restrict them such that the model has inversion symmetry. On-site inversion symmetry means
\begin{equation}
H(k)\to H(-k)=2\cos(k/2) \sigma_x - \delta \sin(k/2) \sigma_y + \Delta \sigma_z = H(k),
\end{equation}
which only occurs if $\delta=0$ and mid-bond inversion symmetry means
\begin{equation}
H(k)\to \sigma_x H(-k)\sigma_x =2\cos(k/2) \sigma_x + \delta \sin(k/2) \sigma_y - \Delta \sigma_z = H(k),
\end{equation}
which is possible only if $\Delta=0$. Therefore, we choose $(\delta,\Delta)$ to be $(\neq 0,0)$ i.e. $\theta_\text{RM}=\pi/2$ or $3\pi/2$ (SSH); or $(0,0)$ (gapless); or $(0,\neq 0)$ i.e. $\theta_\text{RM}=0$ or $\pi$ (CDW). Inversion symmetry imposes that only two out of the three Pauli matrices appear simultaneously in the Bloch Hamiltonian, which means that this model actually has a chiral symmetry. Indeed, when $\Delta=0$, $H(k)$ anticommutes with $\sigma_z$ and when $\delta=0$, it anticommutes with $\sigma_y$ (in the latter case, chiral symmetry is no longer related to the two sublattices and bipartiteness). This means that when $k$ spans the BZ, the Bloch Hamiltonian is restricted to move on a great circle of the Bloch sphere. When $\Delta=0$, this great circle is the equator ($xy$ plane), and when $\delta=0$, it is a meridian ($xz$ plane). We also introduce the dimensionless parameter $\lambda$ such that $\Delta=\lambda M \Theta(\lambda)$ and $\delta=-\lambda M \Theta(-\lambda)$, where $\Theta(x)$ is the Heaviside step function. When $\lambda<0$, the model is an SSH chain with positive $\delta$ and when $\lambda>0$ it is a CDW chain with positive $\Delta$. When $\lambda=0$ it is gapless. When $\lambda$ varies from negative to positive, a phase transition occurs between two different types of band insulators having inversion symmetry. At the transition, the gap closes and the low-energy physics is described by a massless Dirac equation (see Fig.~\ref{fig:phasediagrm}).
\begin{figure}[h!]
\begin{center}
\includegraphics[width=8cm]{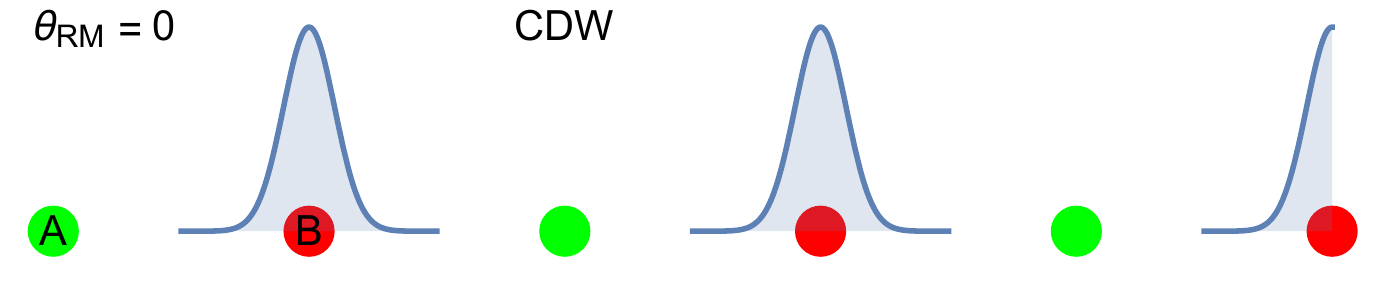}
\includegraphics[width=8cm]{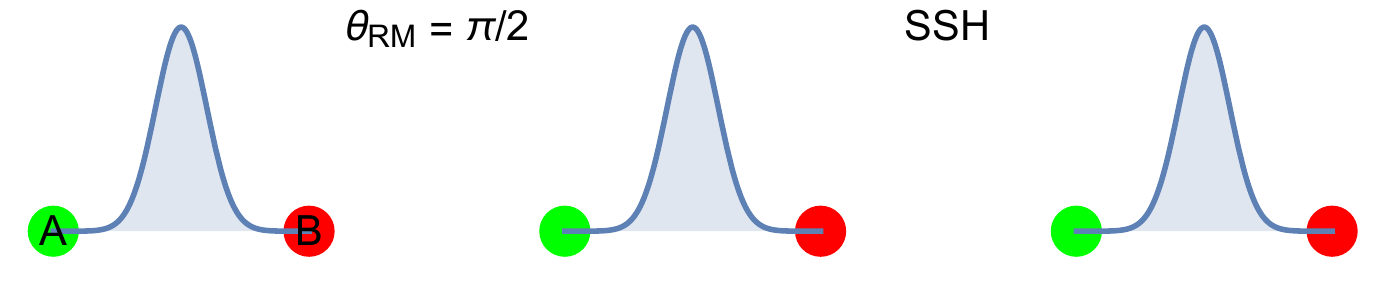}
\includegraphics[width=8cm]{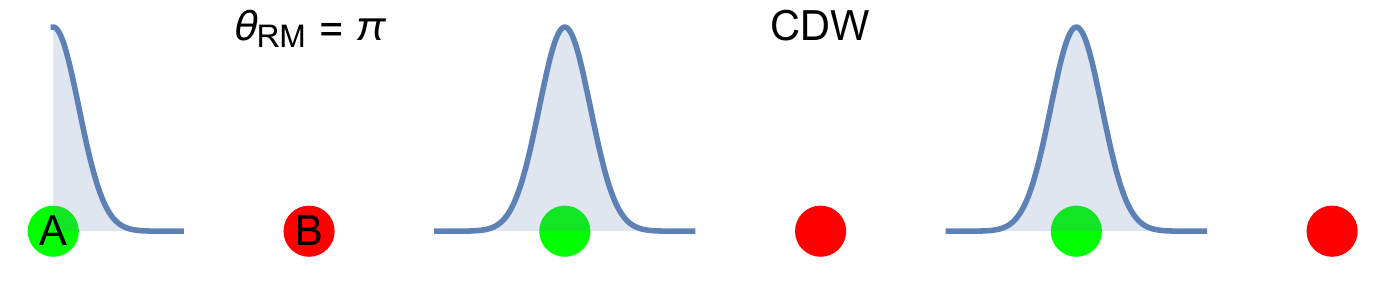}
\includegraphics[width=8cm]{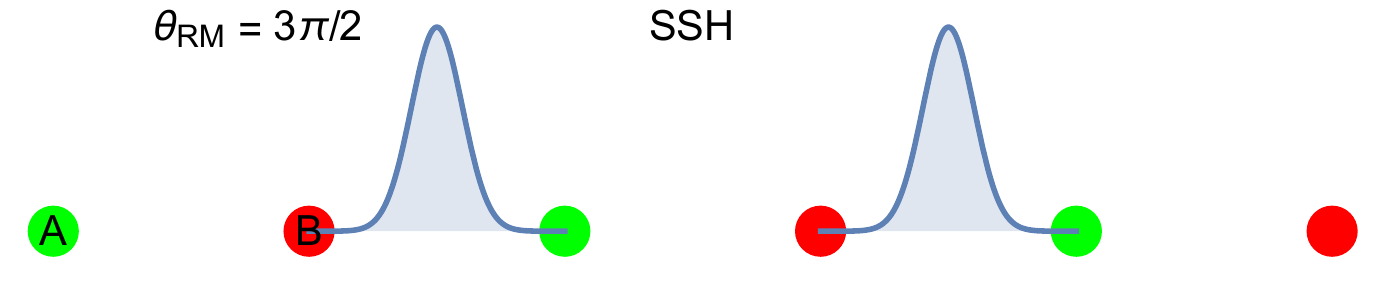}
\caption{\label{fig:wannier} The Rice-Mele chain with $\theta_\text{RM}=0$ (CDW), $\pi/2$ (SSH), $\pi$ (CDW) and $3\pi/2$ (SSH) has inversion symmetry. The $A$ (green) and $B$ (red) ions carry a $+1/2$ charge. The Wannier function is schematized as a Gaussian wavepacket that shows the position of the electron (of charge -1) in the unit cell. In a classical description, the CDW appears as an alternation of positive and negative charges $+1/2,-1/2,+1/2,-1/2,...$ (an ionic crystal, similar to rock salt), whereas the SSH chain $+1/2-1+1/2=0,+1/2-1+1/2=0,+1/2-1+1/2=0,...$ is similar to a chain of neutral dimers (a molecular crystal made of non-polar dimers).}
\end{center}
\end{figure}

Both phases (SSH and CDW) are Dirac insulators. Indeed at low energy, they are both described by a massive Dirac equation. In the SSH case, the Dirac Hamiltonian is given in (\ref{JRq}). In the CDW case, it is
\begin{eqnarray}
\mathcal{H}(k=\pi + q) \approx -q \sigma_y + \Delta \sigma_z \, .
\label{CDWD}
\end{eqnarray}
Beware that the Dirac Hamiltonian is here obtained from linearizing the periodic (and not the canonical) Bloch Hamiltonian $\mathcal{H}(k)$. The Dirac mass is not the same in the two cases: $\delta \sigma_x$ versus $\Delta \sigma_z$. However, a crucial point is that, in each phase, only two Pauli matrices appear, otherwise this would not be Dirac matrices satisfying the Clifford algebra in 1D. It is interesting to note that the Jackiw-Rebbi mechanism therefore applies to both phases as well with either a domain wall in $\delta(x)$ or in $\Delta(x)$. There is a subtlety involved here in the fact that the Jackiw-Rebbi mechanism strictly applies only in the continuum limit in which $a\to 0$ and $t\to \infty$ such that the velocity $ta$ remains finite. In the tight-binding model, there may be surprises as found in the corresponding 2D case (boron nitride)~\cite{Semenoff:2008}.

\subsubsection{Bulk electric polarization}
Let us now discuss the different phases by computing the electric polarization in the bulk. We first recall that, as we are considering spinless electrons and a half filled two-band chain, it means that there is a single electron per unit cell carrying a charge $-e=-1$. In order to ensure electric neutrality, we assume that the ions $A$ and $B$ carry each a charge $+1/2$ (they are cations). It is crucial to consider a neutral system, otherwise the electric polarization has no meaning. In addition, the complete polarization $P_\text{tot}$ is defined modulo $P_q=1/2$ (see e.g., the pedagogical discussion about the polarization quantum $P_q$ in Ref.~\cite{Spaldin:2012}). This can be seen as follows. Consider the CDW situation with the electron localized on $A$ sites. The total charge is therefore $+1/2-1=-1/2$ on $A$ sites and $+1/2$ on $B$ sites. Now, if we move every electron by a distance of $1/2$ to the right we obtain a state in which every $A$ site has a charge $+1/2$ and every $B$ site a charge $-1/2$. But up to a translation by half a unit cell, these two states are identical (they correspond to the RM model with $\theta_\text{RM}=0$ and $\pi$). Because we have translated a single electron of charge $-1$ by a distance of $1/2$ in order to be back on the same state, it means that the polarization is defined modulo a quantum of $1\times 1/2=1/2$. A subtle issue is the fact that the electronic contribution to the polarization reads $P_\text{el}=-e \langle x_- \rangle = -\frac{Z_-}{2\pi}$ (see below), which is obviously defined modulo $1$ as $Z_-$ is a phase. However, the complete polarization $P_\text{tot}$ is defined modulo $P_q=1/2$. Next, we remark that under space inversion, the total polarization $P_\text{tot}\to -P_\text{tot}$ and because inversion is a symmetry it means that $-P_\text{tot}=P_\text{tot}$ modulo $P_q$. Therefore 
\begin{equation}
    P_\text{tot}=0 \text{ or } P_q/2 \text{ modulo } P_q
\end{equation}
in a inversion-symmetric band insulator. This is a $\mathbb{Z}_2$ invariant.

\begin{figure}
\begin{center}
\includegraphics[width=7cm]{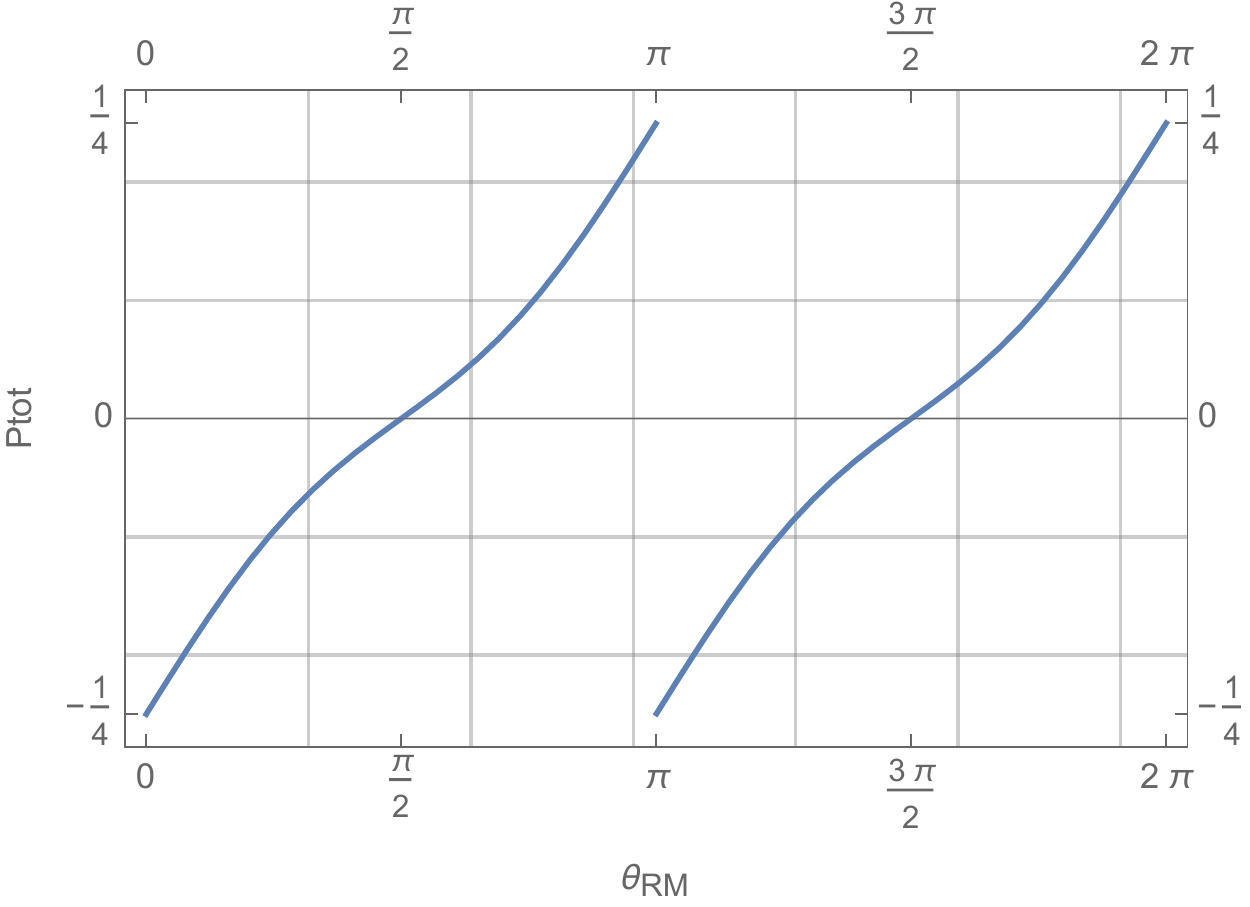}
\caption{\label{fig:polarization} Bulk electric polarization for the Rice-Mele model as a function of $\theta_\text{RM}$ (see Fig.~3 in \cite{Combes:2016}. In this reference, the polarization quantum was wrongly assumed to be $P_q=1$ instead of $P_q=1/2$ for a spinless RM chain.). The polarization quantum $P_q=1/2$ and the polarization is therefore plotted between $-1/4$ and $+1/4$. The SSH phase corresponds to $\theta_\text{RM}=\pi/2$ and $3\pi/2$ with $P_\text{tot}=0$ and the CDW phase to $\theta_\text{RM}=0$ and $\pi$ with $P_\text{tot}=1/4$. Other values of $\theta_\text{RM}$ correspond to the RM model when it breaks inversion symmetry.}
\end{center}
\end{figure}

The total electric polarization $P_\text{tot}$ has two contributions: that of ions and that of electrons. The former is computed classically as $P_\text{ions}=\sum_j q_j x_j =(x_A+x_B)/2=\bar{x}$, where $q_j=+1/2$ is the ion charge. $P_\text{ions}$ depends on the average position $\bar{x}$ of ions within the unit cell. This contribution can be made to vanish by properly choosing the position origin on a bond, half way between $A$ and $B$ (this is an inversion center for the SSH chain), see Fig.~\ref{fig:rmchain}. The electronic contribution can be easily computed thanks to the prescription given by King-Smith, Vanderbilt and Resta~\cite{Resta:2007}: do as if the electron charge was entirely localized at the Wannier center of the occupied band $P_\text{el}=-\langle x_-\rangle = -Z_- /(2\pi)$. In the SSH phase, the Wannier centers are localized on the bonds, half-way between $A$ and $B$ sites (see Fig.~\ref{fig:wannier}). The sign of $\delta$ only decides whether these Wannier centers are to the left or to the right of $A$ sites (they are on the strong bonds). For such a chain, the total electric polarization (including the electrons and the ions) is $P_\text{tot}=0$ modulo $1/2$. The Zak phase is $0$, i.e. $\langle x_-\rangle=0$, or $\pi$, i.e. $\langle x_-\rangle=1/2$, (depending on the sign of $\delta$ and when computed from the inversion center, which is on a bond, can be either chosen to the left or to the right of an $A$ site). In the CDW phase, the Wannier centers are located on-sites. The sign of $\Delta$ only decides whether they are located on an $A$ site (if $\Delta<0$) or a $B$ site (if $\Delta>0$). The total electric polarization here is $P_\text{tot}=1/4$ modulo $1/2$. The Zak phase is also $0$ or $\pi$ when computed from the inversion center which is now on-site. The value $0$ or $\pi$ depends on the sign of $\Delta$ but also on whether one has chosen the inversion center to be on an $A$ or a $B$ site. If one fixes the same position origin to compute the Zak phase in both the SSH and the CDW phases, then it is $0$ or $\pi$ in the SSH phase and $\pm \pi/2$ in the CDW phase. Figure~\ref{fig:polarization} presents the electric polarization for the complete RM model. When $\theta_\text{RM}$ is not $0,\pi/2,\pi$ or $3\pi/2$, this model breaks inversion symmetry and the polarization is not constrained to $0$ or $1/4$.

In a one-dimensional crystal with inversion symmetry, there are two special points in each unit cell that are known as Wyckoff positions. In our case, they are either on-site ($A$ or $B$) or at mid-distance between two sites (to the right of $A$ or to its left). We find that the Wannier center can only occupy one of these two positions and as a consequence the electric polarization is quantized: either $P_q/2$ or $0$ modulo $P_q$. This is the fundamental reason for having two classes of 1D insulators protected by inversion symmetry: CDW has $P_\text{tot}=P_q/2$ and SSH has $P_\text{tot}=0$ modulo $P_q$.

\subsubsection{Which phase is trivial and which is topological?}
We give two different answers before concluding.

\emph{First answer:}  The vacuum is a band insulator with inversion symmetry and vanishing polarization. We take it as a definition of a trivial insulator. On the one hand, the SSH chain has a vanishing bulk polarization and is smoothly connected to a molecular insulator (this is obvious in the limit $|\delta|=2$ in which it describes uncoupled non-polar dimers). It is therefore a trivial insulator. 
On the other hand, the CDW chain has a non-vanishing bulk polarization $P_\text{tot}=P_q/2\, [P_q]$ and is smoothly connected to an ionic insulator (a one-dimensional version of rock salt Na$^+$ Cl$^-$ obtained in the limit $t=(v+w)/2\to 0$). It is therefore a symmetry-protected topological insulator. The $\mathbb{Z}_2$ invariant $\nu$ can be taken to be:
\begin{equation}
\nu = -\frac{P_\text{tot}}{P_q/2} = 4(\langle x_-\rangle-\bar{x}) = \frac{Z_- -2\pi \bar{x}}{\pi/2} \,\, [2]\, .
\label{z2ssh}
\end{equation}

This characterization by a $\mathbb{Z}_2$ invariant is consistent with that based on a topological $\theta$-term in the effective gauge theory~\cite{Qi:2008,Chen:2011}, that we here recall. The long-wavelength electromagnetic response of a dielectric material in one spatial dimension is described by the following Lagrangian density
\begin{equation}
\mathcal{L}=-\frac{\epsilon}{4}F^{\mu \nu} F_{\mu \nu} -e\frac{\theta}{4\pi}\epsilon^{\mu\nu}F_{\mu \nu} = \epsilon\frac{E_x^2}{2}+P_\text{tot} E_x    \, ,
\label{eq:thetaterm}
\end{equation}
where $E_x$ is the electric field and $F^{\mu\nu}$ is the field tensor with $\mu=0,1=t,x$. There are two material parameters that distinguish the dielectric material (here a 1D band insulator) from the vacuum: the familiar dielectric constant $\epsilon$ -- related to the electric susceptibility $\chi$ by $\epsilon=1+\chi$ -- and the less familiar angle $\theta$. The latter is related to the total (spontaneous) electric polarization by $P_\text{tot}=-P_q \theta/(2\pi)$ modulo $P_q$. The $\theta$ angle has only two possible values $\theta=0$ (trivial) or $\pi$ (non-trivial) when the band insulator respects both time-reversal and inversion symmetries, which is the case here. It appears that $P_\text{tot}(\text{CDW})=P_q/2\neq 0\, [P_q]$ (topological, $\theta=\pi$) and $P_\text{tot}(\text{SSH})=0\, [P_q]$ (trivial, $\theta=0$). The $\mathbb{Z}_2$ invariant is 
\begin{equation}
    \nu = \frac{\theta}{\pi} \, [2]
\end{equation}.

\medskip 

\emph{Second answer:} We now take a different definition of a trivial insulator. It is an atomic-like insulator: it should have exponentially-localized Wannier orbitals and the Wannier centers should be on the ions. In this case, we would call CDW a trivial insulator because the Wannier centers are on the ions, whereas SSH would be a topological insulator because the Wannier centers are exactly in between two nearest-neighbor ions (see Fig.~\ref{fig:wannier}). 

Although we do not prove it here, in both phases (SSH and CDW), there exists exponentially-localized Wannier functions (as is always the case in 1D) that respect inversion symmetry. The existence of an exponentially-localized Wannier function respecting some symmetry is often taken as a definition of an atomic limit~\cite{Bradlyn:2017}. In the present case, it would therefore seem that there are two inequivalent atomic limits. A naive atomic limit with the Wannier centers on the ions (the CDW case) and an \emph{obstructed atomic limit}~\cite{Bradlyn:2017} with the Wannier centers in between the ions (SSH case). Such an obstructed atomic limit occurs because of bonding and the fact that the natural objects in this insulator are molecules (here non-polar dimers) rather than atoms.

\medskip 

\emph{Conclusion:} In the end, there are two different classes of inversion-symmetric band insulators in one dimension that are distinguished by their electric polarization. We disagree with the commonly-found statement that the two phases of the SSH model with positive or negative $\delta$ would correspond to the two different values of the polarization (see e.g. Ref.~\cite{Benalcazar:2017}). We think that these two values of polarization actually correspond to SSH and to CDW. The difficulty in this interpretation is in the fact that what seems trivial from the polarization perspective seems non-trivial from the Wannier center perspective and vice-versa. We believe that the first answer is better motivated as it corresponds to a physically measurable quantity, the electric polarization. We therefore call SSH trivial and CDW topological, but this may be a matter of convention.

The above peculiarities of the RM model is related to its having two sites per unit cell. Another model of 1D inversion-symmetric two-band insulator but with a single site per unit cell -- the coupled $s$ and $p$ bands -- is studied in~\cite{Vanderbilt:1993} and was originally introduced by Shockley~\cite{Shockley:1939}. It has a simpler behavior than SSH and CDW (e.g. there is no difference between the periodic and the canonical Bloch Hamiltonian). In this model, the non-trivial (covalent $sp$ hybridized) insulator corresponds to $P_\text{tot}=P_q/2$ $[P_q]$ and having Wannier centers in between ions, whereas the trivial (atomic $s$ band) insulator has a vanishing polarization and Wannier centers on the ions~\cite{Vanderbilt:1993}.

\subsubsection{Topological phase transition}
\begin{figure}[h!]
\begin{center}
\includegraphics[width=7cm]{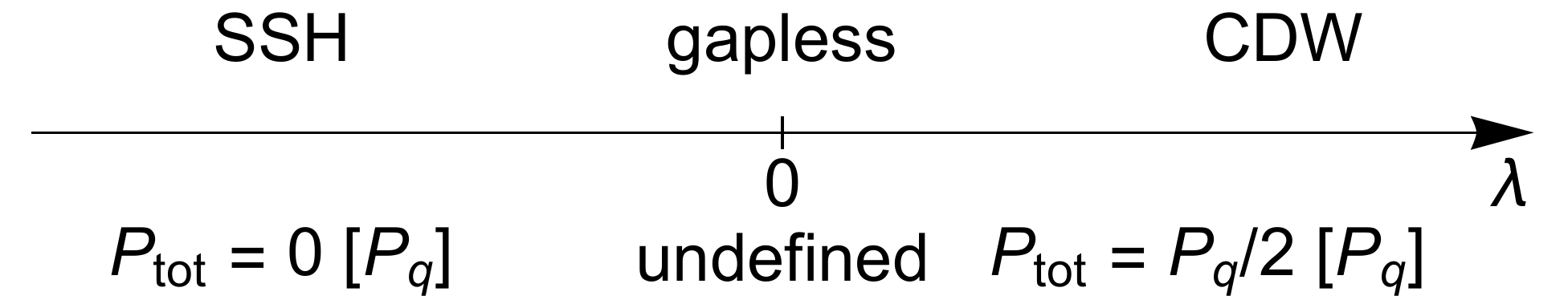}
\hspace{0.5cm}
\includegraphics[width=6cm]{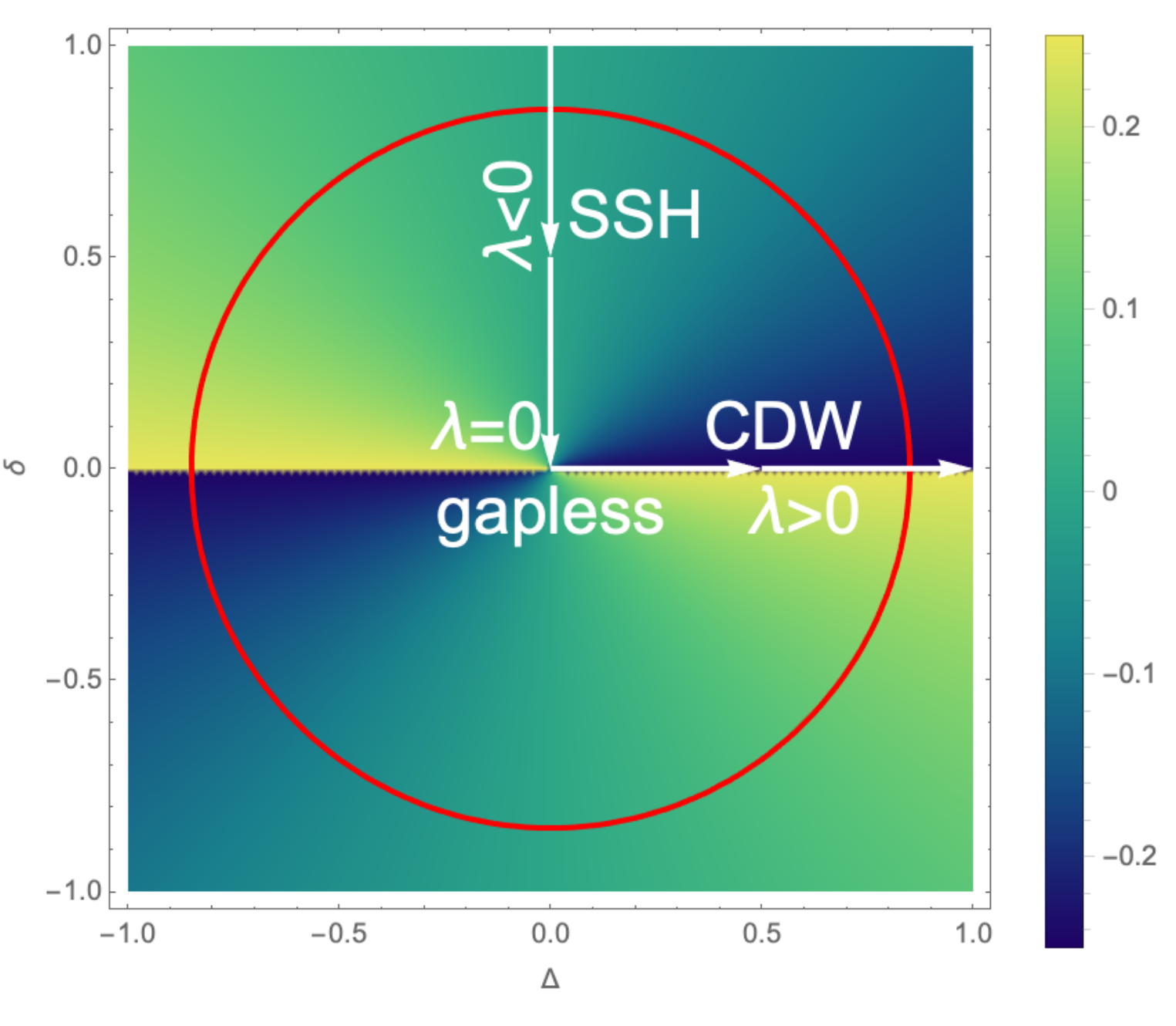}
\caption{\label{fig:phasediagrm} (Left) Phase diagram of the Rice-Mele model with inversion symmetry. A gapless point separates two topologically distinct band insulators protected by inversion symmetry and characterized by their electric polarization that can only take two values: $2 P_\text{tot}/P_q=0$ or $1$, where $P_q$ is the polarization quantum. For spinless (spinful) electrons $P_q=e/2$ ($P_q=e$). (Right) Bulk polarization as a function of the angle $\theta_{RM}$. Compare with Fig.~2 in \cite{XiaoBerryRMP:2010}. The difference comes from taking $P_q=1/2$ instead of $1$.}
\end{center}
\end{figure}

Imagine tuning a transition between the SSH and the CDW phases. We start with $\delta>0$ and $\Delta=0$ (i.e. $\lambda<0$), then diminish the dimerization until we reach $\delta=0$ maintaining $\Delta=0$ (i.e. $\lambda=0$) and then we increase $\Delta$ while maintaining $\delta=0$ (i.e. $\lambda>0$). In such a case, the inversion symmetry is preserved all the way. Then, we necessarily have to close the gap at $\lambda=0$, a point at which the electron is delocalized over the whole crystal and the Wannier center is ill defined. Actually, in the process the Wannier center jumps from on-site to mid-bond exactly when $\lambda=0$. See the phase diagram in Fig~\ref{fig:phasediagrm}.


\subsubsection{Conclusion}
In summary, the RM model with inversion symmetry connects the SSH model to a CDW model. It allows one to study the phase transition between the two phases of a 1D band insulator with inversion symmetry. The non-trivial phase is a symmetry-protected topological (SPT) insulator -- rather than an intrinsic topological insulator. As the symmetry in question is the space inversion (and not an external symmetry such as time-reversal or particle-hole), it is known as a topological crystalline insulator~\cite{Fu:2011}. It is characterized by a $\mathbb{Z}_2$ topological invariant given in (\ref{z2ssh}) and
built from the bulk electric polarization. The CDW chain is an ionic insulator (inversion-protected topological insulator), whereas the SSH chain is a molecular insulator (trivial). 

If space inversion is broken (which happens in the RM model whenever $\theta_\text{RM}\neq 0$ modulo $\pi/2$), the system has a single gapped phase and is a trivial band insulator. Indeed, one can continuously go from the CDW to the SSH phase without closing the gap by breaking inversion symmetry. Its electric polarization is unquantized (neither 0 nor $P_q/2$, see Fig.~\ref{fig:polarization}) and the Wannier center is not at a special position.

\section{Dirac fermions on 2D honeycomb lattice \label{section2D}}

\begin{figure}[ht]
\begin{center}
\includegraphics[width=7cm]{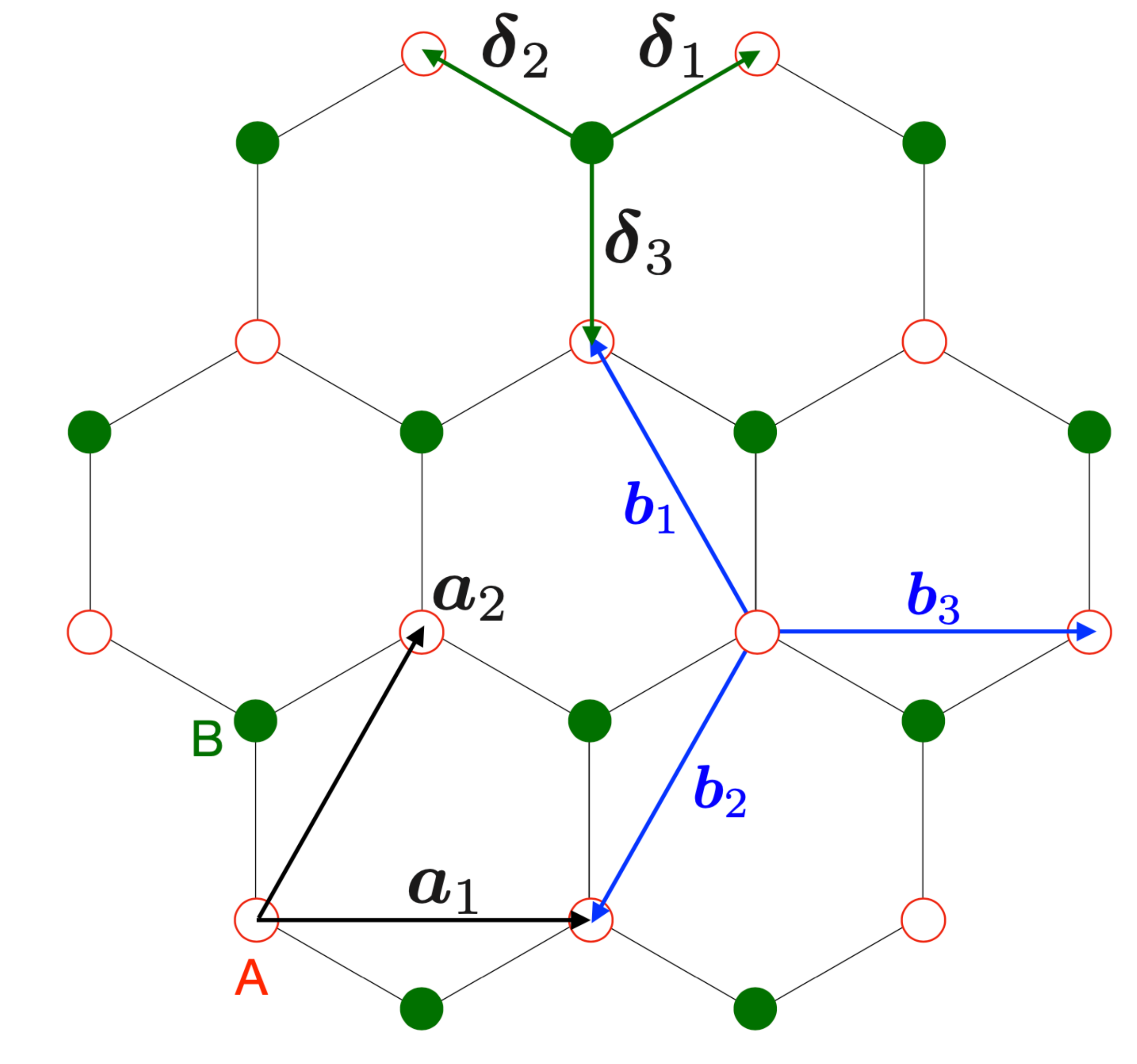}
\hspace{15mm}
\includegraphics[width=8.5cm]{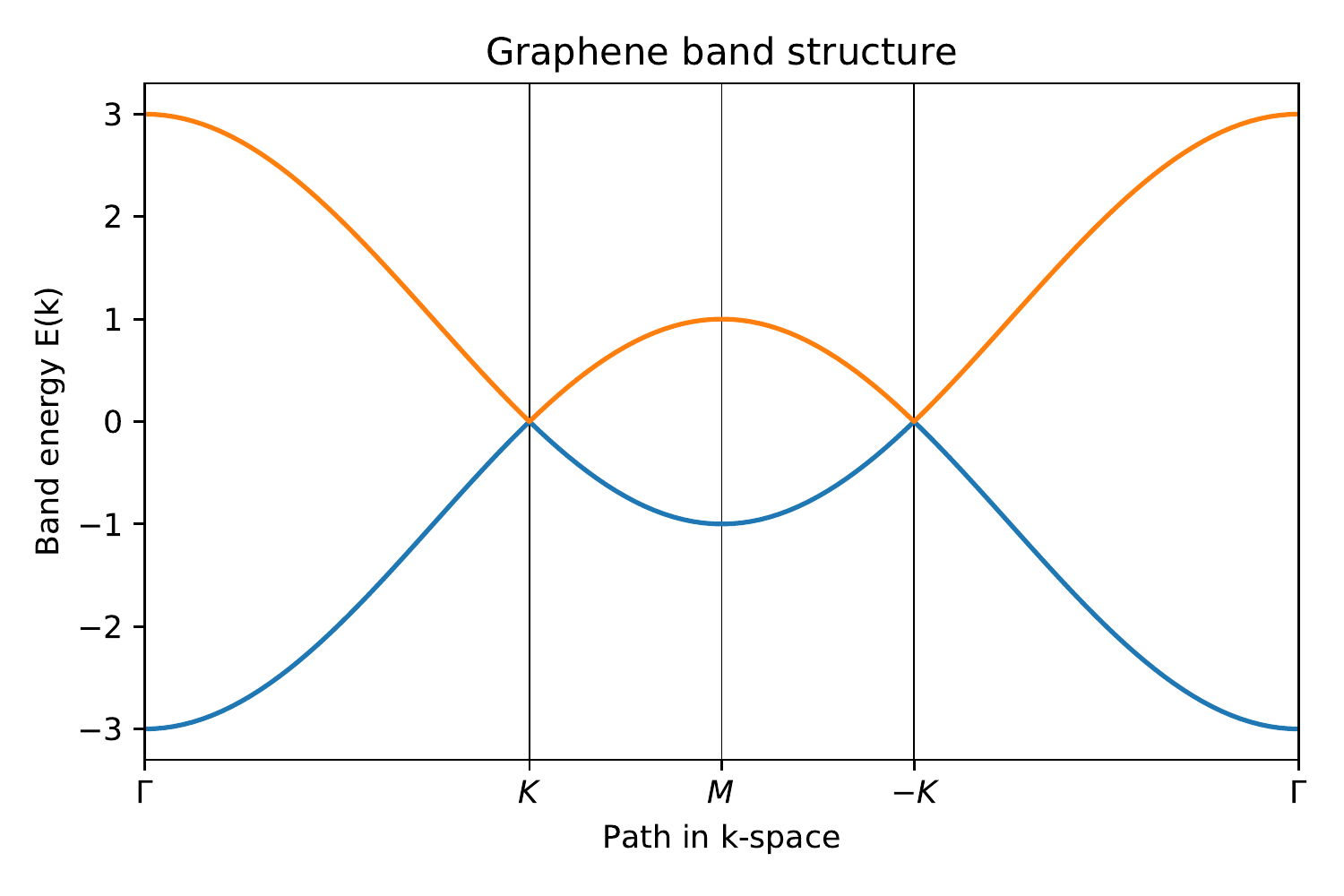}
\caption{(Left) Graphene honeycomb lattice structure. Red open (green filled) dots for A (B) sublattice. The basis vectors $\va_{1}$ and $\va_{2}$ generate the Bravais lattice. The vectors $\boldsymbol{\delta}_\alpha$ ($\alpha=1,2,3$) are connecting a given site to its three nearest neighbors. The vectors $\vb_i$ and their opposite $-\vb_i$ (i=1,2,3) connect a given site to its six second-nearest neighbors. The distance between two sites is $a=0.142$ nm and the surface of the unit cell is $A_{cell}=3 \sqrt{3} a^2 /2$. (Right) Section of the electronic energy dispersion $E(\vk)=\pm |\vd(\vk)|$ of graphene for $k_y=0$, showing the two Dirac points at $\vk=\pm \vK$. The first BZ has an hexagonal shape. High symmetry points in this BZ are: the center $\Gamma$, two inequivalent corners of the hexagon $K$ and $K'$ and the mid-point on the boundary called $M$.}
\label{FigHoneycomb}
\end{center}
\end{figure}

Graphene is the archetype of a strictly 2D crystal whose electronic excitations obey a Dirac-(or Weyl-)like equation for multicomponent wave functions. The spinor components correspond to the amplitudes of the Bloch waves on the two inequivalent triangular sublattices of the honeycomb structure \cite{CastroRMP:2009,GoerbigRMP:2011}, instead of the real spin projections involved in the historical Dirac equation of high-energy physics \cite{Dirac:1928,Dirac:1930}. Since graphene was isolated in 2004 \cite{Novoselov:2004,Novoselov:2005,Zhang:2005}, the family of 2D crystalline solids has gained many new members exhibiting striking electrical and optical properties, e.g. transition metal dichalcogenides (TMDC) \cite{RMP_TMDC:2018}, twisted graphene bilayers, van der Waals heterostructures. Single layer graphene is still remarkable by the simplicity of its band structure around the Fermi level \cite{Wallace:1947}. The relevant excitations near the Fermi level form two $\pi$-bands that are very well isolated from the other energy bands because carbon is a light element. In comparison, the band structure of MoS$_2$, and more generally of any TMDC, involves much more complicated orbital combinations. 

Graphene has been a fertile playground for the development of topological concepts from the Haldane model \cite{Haldane:1988} to the Kane-Mele model \cite{Kane:2005a,Kane:2005b} to cite only two major milestones. The Haldane model is the first representative of the Chern insulator class which are the time-reversal breaking topological insulators, namely band insulators exhibiting the QHE without the Landau level structure. Like the QHE, Chern insulators are also characterized by chiral edge states. The Kane-Mele model describes graphene with a finite intrinsic spin-orbit (SO) coupling, which turns out to be a time-reversal invariant topological insulator associated to a robust $\mathbb{Z}_2$ index, and helical edge states. In graphene, the SO coupling is low and therefore the real spin of electrons in graphene is almost uncoupled to the motion. Nevertheless in TMDCs, the SO coupling is stronger and may lead to Quantum Spin Hall (QSH) topological insulators. The compound WTe$_2$ has been predicted \cite{Qian:2014} and experimentaly demonstrated \cite{Fei:2017,Tang:2017,Wu:2018} to be a QSH insulator. We will not cover the time-reversal invariant TIs in this review, see \cite{KaneRMP:2011,QiRMP:2011,KonigJPSJ:2008}. 

In this section, we use graphene to exemplify the concepts of geometrical and topological band theory, introduced in Sec. \ref{sec:bandtheory}. We start with pristine graphene (without any SO coupling) which has a gapless band structure consisting of two bands that cross each other at two isolated points of the BZ. Pristine graphene being both centrosymmetric and time-reversal invariant, its Berry curvature is zero everywhere in the BZ except at the Dirac band touching points, which are local singularities and where it is ill-defined. In the following, we present the spectral and Berry curvature properties of the Semenov insulator obtained by gapping graphene via a staggered on-site potential \cite{Semenoff:1984}. This model describes hexagonal boron nitride ($h$-BN) and has a well-defined finite Berry curvature, but its total flux through the BZ is zero. It is therefore the typical example of a material whose band structure has non trivial local geometrical properties (we call it a Dirac insulator), but still a trivial global topology. Then we introduce the Haldane model where a periodic pattern of magnetic fluxes breaks time-reversal symmetry without breaking the translational invariance of the Bravais lattice. The resulting Haldane insulator has bands that each carries a finite Chern number, thereby leading to QHE in the absence of any net magnetic flux through the sample. We conclude by discussing the band inversion mechanism and topological transitions, using a simple method to evaluate Chern numbers for any two-band model.

The Haldane insulator has been experimentally observed in Bi$_2$Se$_3$ or Bi$_2$Te$_3$ films doped with magnetic impurities \cite{Chang:2013,Bestwick:2015}, and more recently in the intrinsic magnetic insulator MnBi$_2$Te$_4$ \cite{Deng:2020}. It was also realized with cold atoms trapped in optical lattices using dynamical methods to induce complex hopping amplitudes \cite{Jotzu2014}. Pristine graphene and the Semenov model were also implemented using cold atom vapors \cite{tarruell:2011}.

\subsection{Pristine graphene and massless Dirac fermion \label{section2Dgraphene}}

We introduce here the tight-binding model of graphene~\cite{Wallace:1947} and its symmetries. We review its full band structure and the emergence of Dirac fermions. We put an emphasis on the properties of the spinor wave functions.

\subsubsection{Hamiltonian and band structure in the whole BZ}
Graphene is the one-atom thick layer of carbon atoms arranged with the honeycomb lattice structure, made of two interpenetrating triangular sublattices, respectively denoted $A$ and $B$ (Fig. \ref{FigHoneycomb}). Each carbon atom has six electrons: five core electrons filling the inner shells (2 electrons in the $1s$ orbital and 3 electrons in the covalent $sp^2$ bonds) while a single valence electron fills the $p_z$ orbital perpendicular to the plane. As in polyacetylene, the $p_z$ orbitals lead to $\pi$ bands that are well ``isolated" from filled lower bands and empty higher bands. Much of the physics of graphene is related to those two-dimensional $\pi$ bands that are accurately described by the following tight-binding Hamiltonian~\cite{Wallace:1947}:
\begin{equation}
\label{GrapheneHamiltonian}
H_0  =  t \sum_{\vr_A} \sum_{\alpha=1}^{3}   c_B^\dagger(\vr_A+\boldsymbol{\delta}_\alpha) c_A(\vr_A) + {\rm h.c.}  \,\,  ,
\end{equation}
where $t \simeq -2.7$ eV is the hopping amplitude between the $p_z$ orbitals of two adjacent carbon atoms. The operator $c_l(\vr)$ destroys an electron in the $p_z$ orbital at site $\vr$, with $l=A,B$ indicating the sublattice. The sum over $\vr_A$ runs over the $A$-sites which form a triangular Bravais lattice generated by the basis vectors (Fig. \ref{FigHoneycomb}) :
\begin{equation}
\boldsymbol{a}_{1}=   \sqrt{3}  a \,  \boldsymbol{e}_x ,     \,   \hspace{1cm} \boldsymbol{a}_2=  \frac{a}{2} \left(    \sqrt{3} \boldsymbol{e}_x + 3 \boldsymbol{e}_y \right)    ,
\end{equation}
where $a=0.142$ {\rm nm} is the length of the carbon-carbon bond. The vectors $\boldsymbol{\delta}_{\alpha}$ defined by 
\begin{equation}
\boldsymbol{\delta}_{1,2}=   \frac{a}{2} \left(   \pm \sqrt{3} \boldsymbol{e}_x +  \boldsymbol{e}_y \right)   ,     \,  \hspace{1cm}  \boldsymbol{\delta}_{3}=   -  a \, \boldsymbol{e}_y  \, ,
\end{equation}
connect any $A$-site to its three $B$-type nearest neighbors (Fig. \ref{FigHoneycomb}). The hopping matrix elements between next-nearest neighbors can be safely neglected, being roughly ten times smaller than the main hopping $t$ \cite{CastroRMP:2009,GoerbigRMP:2011}. Sites from distinct sublattices, $A$ and $B$, are crystallographically differents by the orientations of the three attached bonds. In graphene, all sites are occupied by identical carbon atoms, while in hexagonal boron nitride ($h$-BN) the boron atoms are distributed on one sublattice while the nitrogen atoms lie on the other one, see Sec. \ref{section2Dbn}.

The graphene lattice can be seen as a 2D version of the polyacetylene chain obtained by replacing the C-H bonds by C-C bonds to another chain. Although the tight-binding models of graphene and polyacetylene look quite similar, there are also important differences. First, in graphene each $p_z$ orbital is coupled to 3 neighboring $p_z$ orbitals, instead of two neighbors for polyacelylene. Second, there is no Peierls instability in graphene, although dimerization effects have been studied: for a magnetic-field induced Peierls instability see~\cite{Fuchs:2007} and for a K\'ekul\'e type of distortion see~\cite{Hou:2007}. Finally, and more fundamentally, the quasi-momentum spans a 2D BZ in graphene, which is crucial for topology, since a 2D compact manifold is a necessary condition for a Chern number to be defined. 
  
\medskip  
  
\begin{figure}[ht]
\begin{center}
\includegraphics[width=7cm]{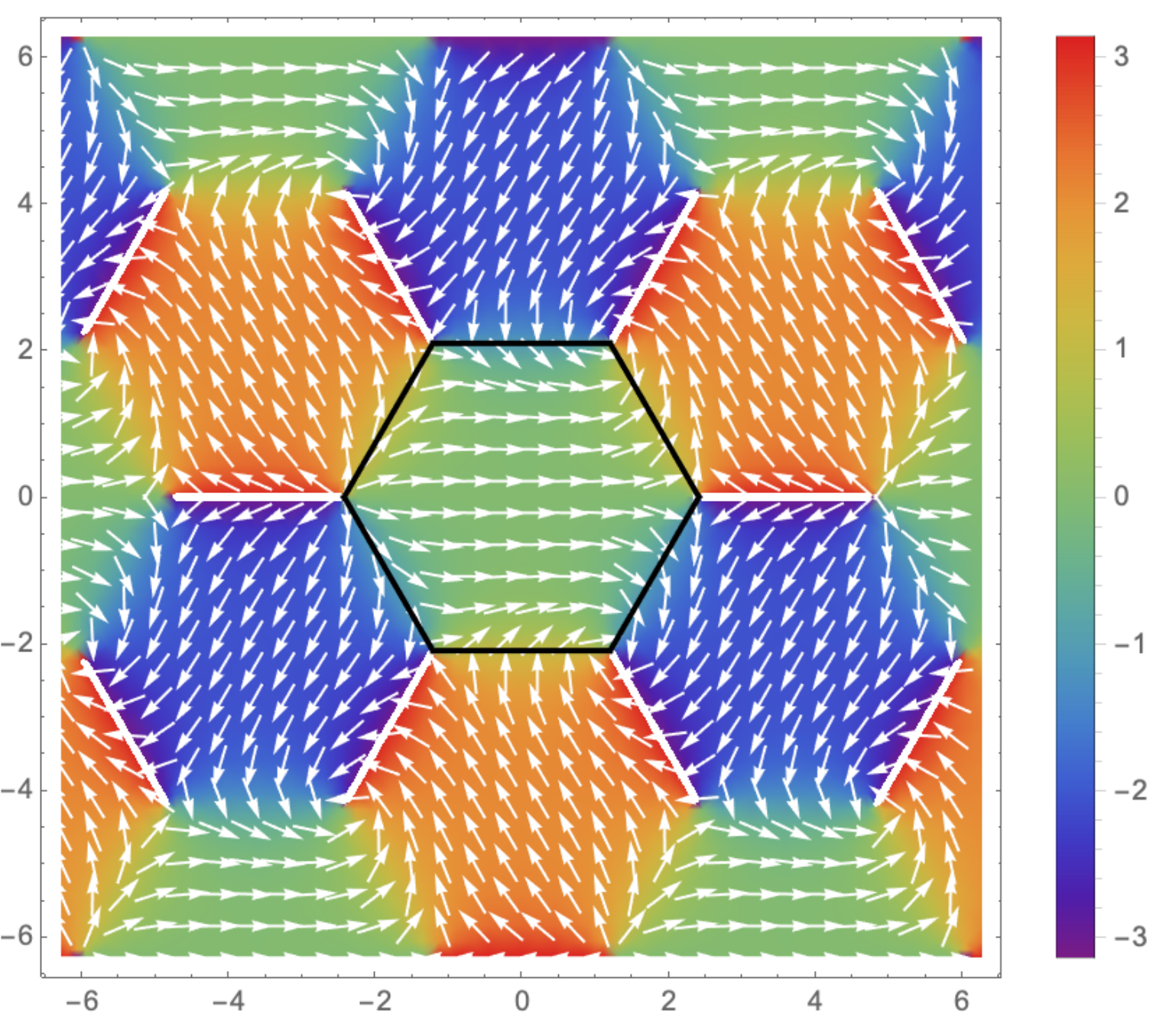}
\hspace{1cm}
\includegraphics[width=7cm]{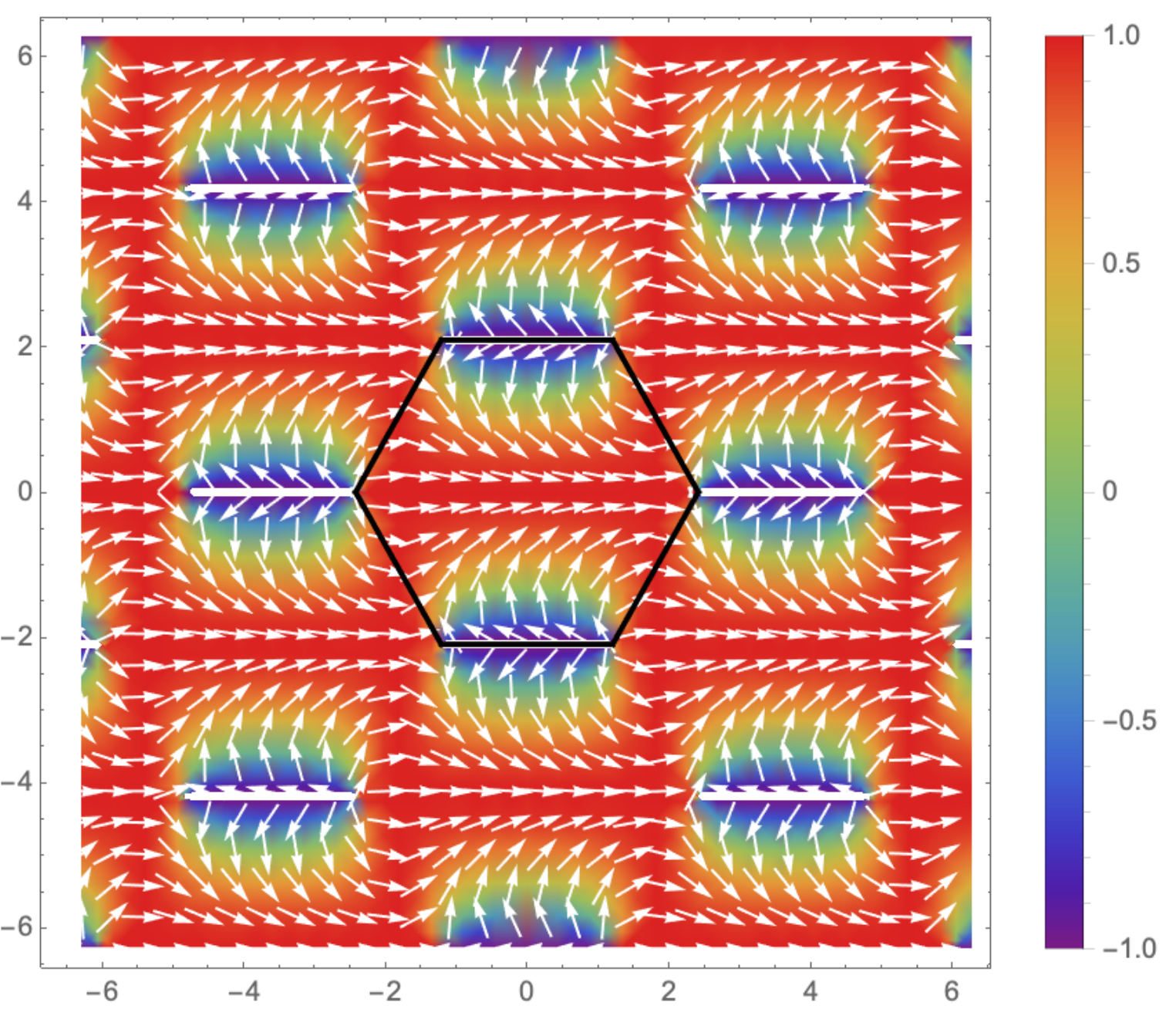}
\caption{(Left) Phase $\varphi_\vk$ in reciprocal space (the hexagonal BZ is indicated by a black line) showing the enlarged periodicity of the (canonical) Bloch Hamiltonian $H_0(\vk)$. (Right) Phase $\phi_\vk$ obtained from a periodic Bloch Hamiltonian $\mathcal{H}_0(\vk)$.}
\label{FigAzimuthalPhase}
\end{center}
\end{figure}

Owing to translation invariance, the two-dimensional quasi-momentum $\vk =(k_x,k_y)$ is a good quantum number. In order to diagonalize the Hamiltonian Eq. (\ref{GrapheneHamiltonian}), we expand the field operator $c_l(\vr)$ as a sum of Fourier modes :  
\begin{equation}
\label{Fourier}
c_l(\vr)= \frac{1}{\sqrt{N}} \sum_{\vk} e^{i \vk \cdot \vr} c_l(\vk) \, , 
\end{equation}
where $l=A,B$ is the sublattice index and $N$ is the total number of unit cells. Note that the exponential phase factors contain the exact location of the atomic sites. After substitution of Eq.(\ref{Fourier}), the Hamiltonian Eq. (\ref{GrapheneHamiltonian}) becomes diagonal in momentum and reads : 
\begin{equation}
\label{AGrapheneHamiltonianFourier}
H_0  = \sum_\vk    c_l^\dagger(\vk) [H_0(\vk)]_{lm} \, c_m (\vk) \, ,
\end{equation}
where $\vk$ is restricted to the first BZ. The (canonical) Bloch Hamiltonian $H_0(\vk)$, which acts on the sublattice isospin, is given by 
\begin{equation}
H_0(\vk)=d_x(\vk) \sigma_x + d_y(\vk) \sigma_y = |\vd(\vk)|(\cos\varphi_\vk \sigma_x +\sin\varphi_\vk \sigma_y) \, ,
\label{hzerographene}
\end{equation}
since only off-diagonal hopping amplitudes are included in the model defined by Eq.~(\ref{GrapheneHamiltonian}). The phase $\varphi_\vk$ is the azimuthal angle along the equator of the Bloch sphere. The real functions $d_x(\vk)$ and $d_y(\vk)$ are defined by :
\begin{equation}
d_x (\vk)   =  t \sum_{\alpha=1}^{3} \cos( \vk \cdot \boldsymbol{\delta}_\alpha)   \hspace{5mm} {\rm and}  \hspace{5mm}   d_y (\vk)   =   t \sum_{\alpha=1}^{3} \sin( \vk \cdot \boldsymbol{\delta}_\alpha)  \,  ,
\label{dxdyGraphene}
\end{equation}
over the whole BZ. The functions $d_x (\vk)$ and $d_y(\vk)$ are respectively even and odd in momentum reversal $\vk \rightarrow -\vk$, which is related to time-reversal and inversion symmetries of pristine graphene. 
The electronic energy spectrum is given by the length of the vector $\vd =(d_x,d_y)$ :
\begin{equation}
\label{SpectrumTBM}
E(\vk)= \pm |\vd(\vk)| = \pm \sqrt{d_x^2 (\vk)+d_y^2 (\vk)} \, ,
\end{equation}
which describes a valence band (minus sign) and a conduction band (plus sign) that are symmetric with respect to $E=0$. The zero energy corresponds to the common energy of the $p_z$ atomic orbitals on sublattices $A$ and $B$. The corresponding wave functions are the spinors:
\begin{equation}
\ket{u_+(\vk)} =\frac{1}{\sqrt{2}}\begin{pmatrix}
1 \\ 
 e^{i\varphi_\vk}
\end{pmatrix},        \hspace{5mm} {\rm and}  \hspace{5mm}  \ket{u_- (\vk)}=\frac{1}{\sqrt{2}}\begin{pmatrix}
1 \\ 
- e^{i\varphi_\vk}
\end{pmatrix},   
\label{SpinorGraPsiPlus}
\end{equation}
in the upper and lower bands respectively. These are the same as for the SSH model albeit the angle $\varphi_\vk$ is now a function of a 2D quasimomentum given by $\varphi_\vk=\text{arg}(k_x+i k_y)$. The Bloch Hamiltonian does not have the periodicity of the reciprocal lattice. The reason is that its definition involves the distance between sites within the unit cell. For the honeycomb lattice, the distance between nearest-neighbor $A$ and $B$ sites is actually one-third of a lattice vector. For example, $\boldsymbol{\delta}_3=(\va_1-2\va_2)/3$. This translates into the fact that the Bloch Hamiltonian has a triple periodicity, as can be seen by plotting the phase $\varphi_\vk$ [see Fig.~\ref{FigAzimuthalPhase}(Left)]~\cite{Lim:2015}. This fact has measurable consequences which have been observed by quantum tomography using cold atoms in an optical lattice, see~\cite{Li:2015}. If instead of using the canonical Bloch Hamiltonian (\ref{hzerographene}), one uses a periodic Bloch Hamiltonian
\begin{equation}
\mathcal{H}_0(\vk)= |\vd(\vk)|(\cos\phi_\vk \sigma_x +\sin\phi_\vk \sigma_y) \, ,
\end{equation}
then the azimuthal angle is different and called $\phi_\vk$. The latter has the periodicity of the hexagonal BZ and is related to $\varphi_\vk$ by $\phi_\vk=\varphi_\vk-\vk\cdot \boldsymbol{\delta}_{3}$ [see Fig.~\ref{FigAzimuthalPhase}(Right)], where $-\boldsymbol{\delta}_{3}$ is the position of a $B$ site with respect to an $A$ site within the unit cell [see Fig.~\ref{FigHoneycomb}(Left)].

The valence and conduction bands touch at isolated points of the BZ obtained by solving  the equations $d_x(\vk)=d_y(\vk)=0$. There are only two inequivalent solutions located at :
\begin{equation}
\vk = \pm \vK=\pm \frac{4 \pi}{3\sqrt{3}a} \, \boldsymbol{e}_x \, ,
\label{DiracPoints}
\end{equation}
and called the Dirac points. Any other solutions of the equation $\vd(\vk)=0$ can be linked by a reciprocal lattice vector to one of these two solutions, and therefore would describe the same physical state. In the SSH chain, the band touching conditions were also obtained by cancelling two functions $d_x$ and $d_y$, but there was only a single momentum component $k$, so it needed a fine tuning of hopping parameters to close the gap, namely $\delta=v-w=0$. In graphene, the momentum $\vk$ runs over a 2D manifold for graphene, so there is no need for fine tuning of the Hamiltonian parameters and one generically gets Dirac points. Even for anisotropic graphene, where the hoppings $t_\alpha$ along the bonding directions $\boldsymbol{\delta}_\alpha$ might differ, there will be always some isolated points $\vk =(k_x,k_y)$ satisfying the 2 equations : $d_x(\vk )=d_y(\vk )=0$ \cite{Goerbig:2008}.

\subsubsection{Effective Hamiltonian near the Dirac points and symmetries}

We consider now the low-energy theory for the single-particle states near the Dirac points~\cite{DiVincenzo:1984}, see Eq.~(\ref{DiracPoints}). The momenta are written as $\vk =\pm  \vK + \vq$ where $\vq = q_x \ve_x +  q_y \ve_y$ is a small momentum deviation from the Dirac points, namely $|\vq|a \ll 1$. The annihilation operators for these states are relabelled as $c_{A\pm\vK} (\vq)=c_{A}(\pm\vK+\vq)$ to indicate the valley they belong to. Expanding to first order in momenta, the Bloch Hamiltonian describing the low-energy excitations in the valley near $\xi \vK$ ($\xi=\pm 1$) reads:
\begin{equation}
H_0(\xi \vK + \vq)=H_0^{(\xi)}(\vq) =v_F \begin{pmatrix} 
0 & \xi q_x -i q_y \\
 \xi q_x  + i q_y & 0   
\end{pmatrix} = v_F(q_x \xi \sigma_x+q_y \sigma_y)
\, ,
\end{equation}
where $v_F = -3 at/2 \simeq 10^6$ m.s$^{-1} \simeq c/300$ is the Fermi velocity. The Fermi velocity is roughly the bandwidth $t$ divided by the BZ size $1/a$ (when $\hbar=1$). The Hamiltonian can also be written $H_0^{(\xi)}(\vq)=v_{ij} q_i \sigma_j$ with $i,j=x,y=1,2$ here. Therefore, near each Dirac point, one obtains a 2D Weyl Hamiltonian describing massless fermions carrying a sublattice isospin coupled to their momentum. The two species of massless Dirac fermions are attached to a given valley, labelled by $\xi$. The chirality (or winding) of each Dirac point is given by $\chi_\xi=\text{sign } \text{det} (v_{ij}) = \xi$. The low-energy dispersion of those fermions is valley independent and reads :
\begin{equation}
E_\pm (\vq) = \pm v_F  |\vq| ,
\label{DispersionMasslessDirac}
\end{equation}
which is typical of a relativistic massless particle with velocity of light replaced by $v_F$. This linearized dispersion is reminiscent 
of the 1D fermions obtained in the SSH chain near $k=\pi$ at the gap closure point $m=v-w=0$. It also has a 3D counterpart in Weyl and Dirac semimetals~\cite{Armitage:2018}. As a general reference on the distinction between Dirac (i.e. complex and having both chiralities), Weyl (i.e. complex and chiral) and Majorana (i.e. real and achiral) fermions, we recommend~\cite{Pal:2011}.

Nevertheless we would like to emphasize the differences between the Dirac (and Weyl) equation in the contexts of graphene and particle physics, respectively. In high-energy physics, the Dirac equation comes from Lorentz-invariance and very general considerations related to special relativity and quantum mechanics. Then, in 3+1 space-time dimensions, the minimal objects to satisfy Dirac equation are bispinors combining the spin and particle/antiparticle degrees of freedom. In graphene, the origin of the Dirac physics is totally different. As we have seen, the spinors originate from a $\vk .\vp$ expansion around special points of a particular band structure. Hence in graphene, there is no fundamental issue with the negative energy states that are just the valence band states (these states are in fact bounded from below by the bottom of the valence band). Finally the emergent Lorentz invariance of Eq. (\ref{DispersionMasslessDirac}) is only valid near the Dirac point, namely for wave vectors $\vq$ located in a disk whose radius is far smaller than the inverse lattice spacing $1/a$, whereas Lorentz invariance applies in the whole Minkowski space-time. Finally the 4 components of the spinors are associated to the sublattice isospin (instead of real spin), and to the valley index (instead of particle/antiparticle label).

\medskip

\subsubsection{Protection of the Dirac points by symmetry}  

Graphene is invariant under space inversions with respect to particular points of the lattice, which are the centers of the hexagons of carbon atoms and the centers of the carbon-carbon bonds. In the absence of any magnetic field or impurities, graphene is also time-reversal invariant and therefore all hopping parameters are real numbers.

It is rather difficult to gap out the Dirac points between two bands of spinless fermions. Indeed, there are only four possible types of perturbations which mathematically corresponds to the identity and the three Pauli matrices. A scalar perturbation (proportional to identity in sublattice isospin space) will just shift both bands in energy without separating them. Moreover perturbations in $\sigma_x$ or $\sigma_y$ would shift the position of the band touching to another location in the BZ without removing the degeneracy. Only a perturbation acting as $\sigma_z$ could gap out the Dirac points, but we are now going to explain that such a term is forbidden by the time-reversal and inversion symmetry of graphene, and more generally in any two-band model which is invariant under those two symmetries.  
Let us consider a generic two-band model in 2D, described by the Bloch Hamiltonian $H(\vk)=\vd(\vk)\cdot \boldsymbol{\sigma}$, and with sublattice isospin ($\mathcal{T}^2=+1$).
The time-reversal $\mathcal{T}$ symmetry condition reads :
\begin{equation}
 H(\vk) = H^* (-\vk)  \implies  \hspace{2mm}  d_x(\vk) = d_x(-\vk)   ,  \hspace{2mm}   d_y(\vk) = -d_y(-\vk)    \hspace{2mm} {\rm  and } \hspace{2mm}  d_z(\vk) = d_z(-\vk)    \, ,
\end{equation}
which is indeed satisfied by graphene and $h$-BN.

The inversion $\mathcal{I}$ symmetry condition can be written :
\begin{equation}
\sigma_x H(\vk) \sigma_x =  H(-\vk)  \implies  \hspace{2mm}   d_x(\vk) = d_x(-\vk)   , \hspace{2mm}  d_y(\vk) = -d_y(-\vk)    \hspace{2mm} {\rm  and } \hspace{2mm}  d_z(\vk) = -d_z(-\vk)    \, ,
\end{equation}
which is satisfied by graphene but violated by $h$-BN. 

Finally the combined $\mathcal{I}\mathcal{T}$ symmetry leads to : 
\begin{equation}
\sigma_x H^*(\vk) \sigma_x =  H(\vk)  \implies  \hspace{2mm}   d_x(\vk) = d_x(\vk)   ,\hspace{2mm}   d_y(\vk) = d_y(\vk)    \hspace{2mm} {\rm  and } \hspace{2mm}  d_z(\vk) = -d_z(\vk)    \, .
\end{equation}

Therefore, the presence of both $\mathcal{T}$ and $\mathcal{I}$ enforces the function $d_z(\vk)$ to be odd and even in $\vk$, which means $d_z(\vk)=0$ for all $\vk$. The third line shows that the symmetry $\mathcal{I}\mathcal{T}$ is enough to protect the existence of a robust contact, even if  $\mathcal{T}$ and $\mathcal{I}$ were separately broken. Then a gap closing requires only the simultaneous cancellation of two functions $d_x(\vk)=d_y(\vk)=0$ at some $\vk$ which can be varied in a 2D manifold. This is why isolated solutions are expected as it happens in graphene. The existence of such isolated solutions of  $\vd(\vk)=0$, preventing the system to become gapped, is robust even if some crystal symmetries are lost and more hopping amplitudes are added. For instance additional second-neighbor hopping will break the electron/hole symmetry discussed above, but will not affect the existence of Dirac points. Other perturbations, like an anisotropic deformations on one type of bond, only shift the Dirac points and modify the conical dispersion around them \cite{Goerbig:2008,guinea:2010,Ghaemi:2012}. The touching points are protected by more fundamental symmetries, namely space inversion and time-reversal symmetries. Breaking at least one of these symmetries usually results in a gap opening at the Dirac points, as we will see below with noncentrosymmetric $h$-BN, and in the time-reversal breaking Haldane model.  

\subsection{Hexagonal boron nitride and massive Dirac fermions \label{section2Dbn}}  

Like graphene, hexagonal boron nitride ($h$-BN) crystallizes in the honeycomb structure, but two different elements (boron and nitrogen) occupy the $A$ and $B$ inequivalent sublattices respectively. The corresponding tight-binding model is given by $H_0$ in Eq. (\ref{GrapheneHamiltonian}) plus a staggered potential $H_1$ that takes into account the difference in energy between the boron and nitrogen orbitals. The on-site staggered potential simply reads :
\begin{equation}
\label{HS}
H_{1} = M \sum_{\vr_A}  \left(  c_A^\dagger(\vr_A) c_A(\vr_A) -  c_B^\dagger(\vr_A+\boldsymbol{\delta}_3) c_B(\vr_A+\boldsymbol{\delta}_3)  \right)   \, ,
\end{equation}
which is very similar to the asymmetry term in the Rice-Mele model, albeit defined here on the honeycomb lattice. This perturbation $H_1$ breaks the inversion symmetry $\mathcal{I}$ between $A$ and $B$ sublattices while still keeping time-reversal invariance. The effect of this staggered potential, first studied by Semenov \cite{Semenoff:1984}, is to gap out the Dirac points, thereby turning graphene into an insulator, at half filling. If we take the limit $M \rightarrow \infty$, thereby making the nearest neighbour hopping amplitude negligible, the electrons are forced to sit on one of the triangular sublattice (the $B$ one) and one obtains an ionic insulator. 

The corresponding Bloch Hamiltonian can be written :
\begin{equation}
H(\vk)=d_x(\vk) \sigma_x + d_y(\vk) \sigma_y +d_z(\vk) \sigma_z     \hspace{3mm}  {\rm with }   \hspace{3mm}     d_z(\vk) =M           \, ,
\label{HBlochGeneric}
\end{equation}
the functions $d_x(\vk) $ and $d_y(\vk) $ being still given by Eqs. (\ref{dxdyGraphene}). Hence at $\vk=\pm \vK$, the Bloch Hamiltonian reduces to $H(\pm \vK) = M \sigma_z$ and 
the Dirac points are gapped because $H^{(\xi)}(\vq) = v_F(q_x \xi \sigma_x+q_y \sigma_y)+M\sigma_z$. The spectrum is :
\begin{equation}
\label{SpectrumTBMSemenov}
E_\pm (\vk)= \pm |\vd(\vk)|\approx \pm \sqrt{v_F^2\, \vq^2+M^2} \, ,
\end{equation}
which corresponds to a massive Dirac fermion. The spinor eigenfunctions of Eq. (\ref{HBlochGeneric}), expressed in the ``north gauge'', read:
\begin{equation}
\ket{u_+(\vk)} =\begin{pmatrix}
\cos \frac{\theta_\vk}{2}  \\ 
 \sin \frac{\theta_\vk}{2} \, e^{i\varphi_\vk}
\end{pmatrix},      \hspace{5mm} {\rm and}  \hspace{5mm} \ket{u_- (\vk)}=\begin{pmatrix}
\sin \frac{\theta_\vk}{2}  \, e^{-i\varphi_\vk}\\ 
- \cos \frac{\theta_\vk}{2} 
\end{pmatrix}  \, , 
\label{uspinors}
\end{equation}
in the upper and lower bands respectively, and depends on the parameter $M$ via the angle $\theta_\vk$. The $\vk$-dependent quantities $\theta_\vk$ and $\varphi_\vk$ are the spherical coordinate angles of the unit vector :
\begin{equation}
\dhat (\vk) = \frac{\vd(\vk)}{|\vd(\vk)|}=
\begin{pmatrix}
\cos \varphi_\vk \sin \theta_\vk \\ 
\sin \varphi_\vk \sin \theta_\vk \\
\cos \theta_\vk
\end{pmatrix} \, .
\label{dspherique}
\end{equation}
At each quasi-momentum $\vk$, these spinors are exactly the same as those of a spin one-half in a fictitious magnetic field $\vd(\vk)$, given by Eq. (\ref{spinors0D}). The mapping $\vk \rightarrow \dhat (\vk) = \vd(\vk)/|\vd(\vk)|$ is essential and captures the global topological properties of the Hamiltonian $H(\vk)=\vd(\vk) . \boldsymbol{\sigma}$, as we will see in Sec \ref{subsec:chernpontrya}. The image manifold of the $2$-dimensional BZ $T^2$ is a sub-set of the unit sphere $S^2$, that is included in the north-hemisphere (resp. south hemisphere) for $M>0$ (resp. $M<0$).

\begin{figure}[ht]
\begin{center}
\includegraphics[width=6cm]{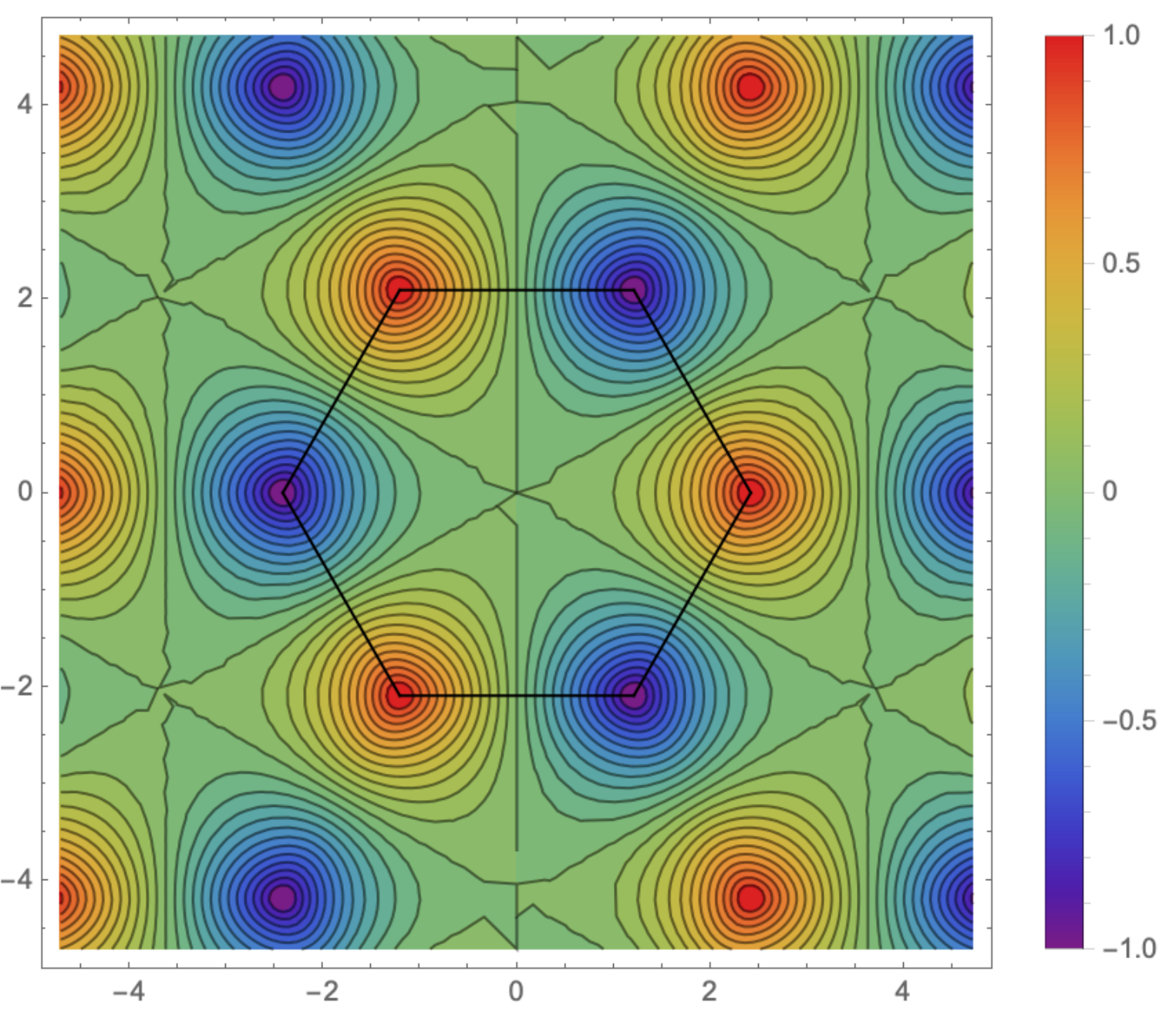}
\hspace{1cm}
\includegraphics[width=6cm]{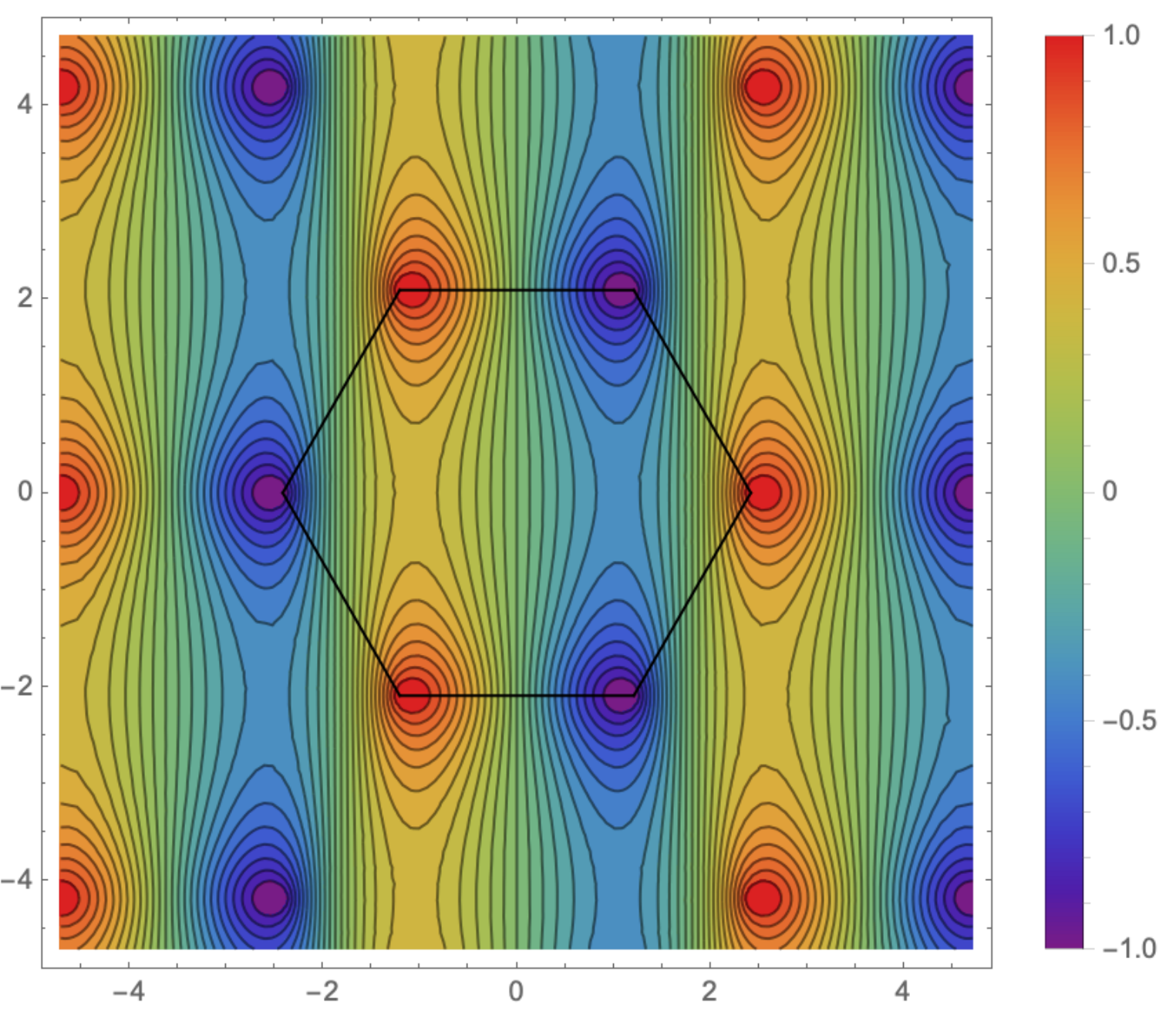}
\caption{(Left) Berry curvature for boron nitride as a function of $(k_x,k_y)$ (the hexagonal BZ is indicated by a black line). (Right) Same but computed from a periodic Bloch Hamiltonian instead of the canonical one. The latter is therefore not correct. See Refs.~\cite{Fuchs:2010,Fruchart:2014,Lim:2015}. \label{FigBerryCurvature}}
\end{center}
\end{figure}
Locally in $\vk$-space, $h$-BN has interesting and non trivial geometrical properties. This can be seen by  computing the Berry curvature of boron nitride~\cite{Fuchs:2010} :
\begin{equation}
F_{xy}^\pm (\vk) = \pm a^2  \frac{\sqrt{3} t^2 M}{\mid E(\vk)\mid^3 }  \, \sin \left(    \vk \, . \, \frac{\boldsymbol{\delta}_2 - \boldsymbol{\delta}_3}{2}   \right)\,  \sin \left(    \vk \, . \, \frac{\boldsymbol{\delta}_3 - \boldsymbol{\delta}_1}{2}   \right)\,  \sin \left(    \vk \, . \, \frac{\boldsymbol{\delta}_1 - \boldsymbol{\delta}_2}{2}   \right) \, ,
\label{berryCurvBN}
\end{equation}
a formula which is valid within the whole BZ (here $\pm$ refers to the band index). Near a Dirac point (in a given valley $\xi=\pm 1$), this formula simplifies as: 
\begin{equation}
F_{xy}^\pm (\xi \vK + \vq) = \pm \frac{\xi v_F^2 M  }{2(M^2+v_F^2 q^2)^{3/2}} \, \, ,
\label{berryCurvBNlocal}
\end{equation}
thereby recovering rotational invariance around the Dirac points. In the limit $M \rightarrow 0$, we recover that the Berry curvature is vanishing everywhere except at $\vq=0$. In the limit first $\vq=0$, and then $M\rightarrow 0$, the sign (even for a given valley) is not well defined.  The sign depends on the valley (time-reversal symmetry). So the total flux is zero. The Berry curvature is plotted in Fig.~\ref{FigBerryCurvature} (Left). If, instead of using the (canonical) Bloch Hamiltonian, one wrongly uses a periodic Bloch Hamiltonian to compute the Berry curvature with the usual formula, it would give the result plotted Fig.~\ref{FigBerryCurvature} (Right). This obviously lacks the correct rotational symmetry~\cite{Fuchs:2010,Fruchart:2014,Lim:2015}.

Locally in reciprocal space, near each valley, boron nitride behaves as an interesting Dirac insulator. It therefore features the physics related to the Jackiw-Rebbi mechanism (see Sec.~\ref{sec:jr}). For example, if there is a domain wall (a grain boundary) between two different staggered on-site potentials $M>0$ on one side and $M<0$ on the other side, there should be a gapless chiral one-dimensional edge mode running along the boundary between the two domains. The Jackiw-Rebbi mechanism occurs in each valley. Therefore the corresponding gapless edge modes are actually valley-filtered: one valley has a given chirality and the other has the opposite chirality. These were studied in~\cite{Semenoff:2008}. Such valley-filtered gapless edge modes are not very robust and may be gapped by any inter-valley scattering process. For example, they depend on whether the boundary is of zig-zag or armchair type~\cite{Semenoff:2008}. If inter-valley scattering is negligible then these modes give rise to a quantized valley Hall effect~\cite{Xiao:2007}.

Even if the Berry curvature is finite and exhibits pronounced peaks near low gaps in the spectrum, its total flux must vanish in $h-$BN due to the time-reversal invariance of the model. This is because time-reversal invariance implies that the Berry curvature is an odd function of quasimomentum. Here, the Berry curvature contributions from the two valleys compensate each other. This cancellation of the Chern number also confirms the absence of quantized Hall effect in this model and boron nitride is considered as a trivial insulator. In the next section, we explain how Haldane devised a band insulator model on the honeycomb lattice in which the total flux of the Berry curvature does not vanish, and is even quantized when the Fermi level lies in the band gap. The expressions above suggest that the Berry flux cancellations between valleys is related to the fact that the sign of $M$ is identical in both valleys. Therefore having a finite Chern number requires to have opposite signs of those masses through inverting the bands in a single valley (and not in the other). That's indeed what the Haldane model does.

\subsection{Haldane model of graphene and topology \label{section2DHaldane}}

\begin{figure}[ht]
\begin{center}
\includegraphics[width=7cm]{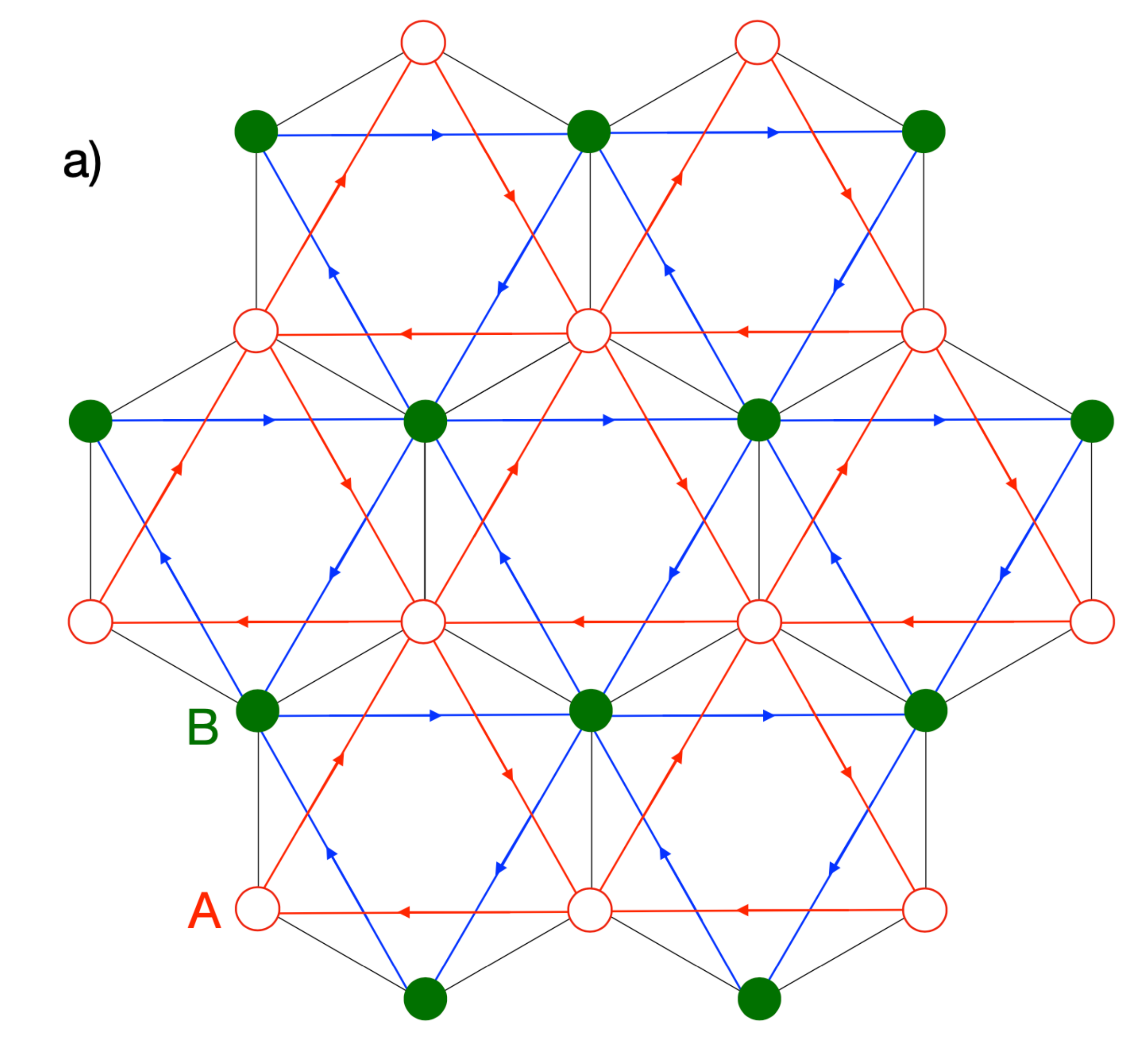}
\hspace{15mm}
\includegraphics[width=6cm]{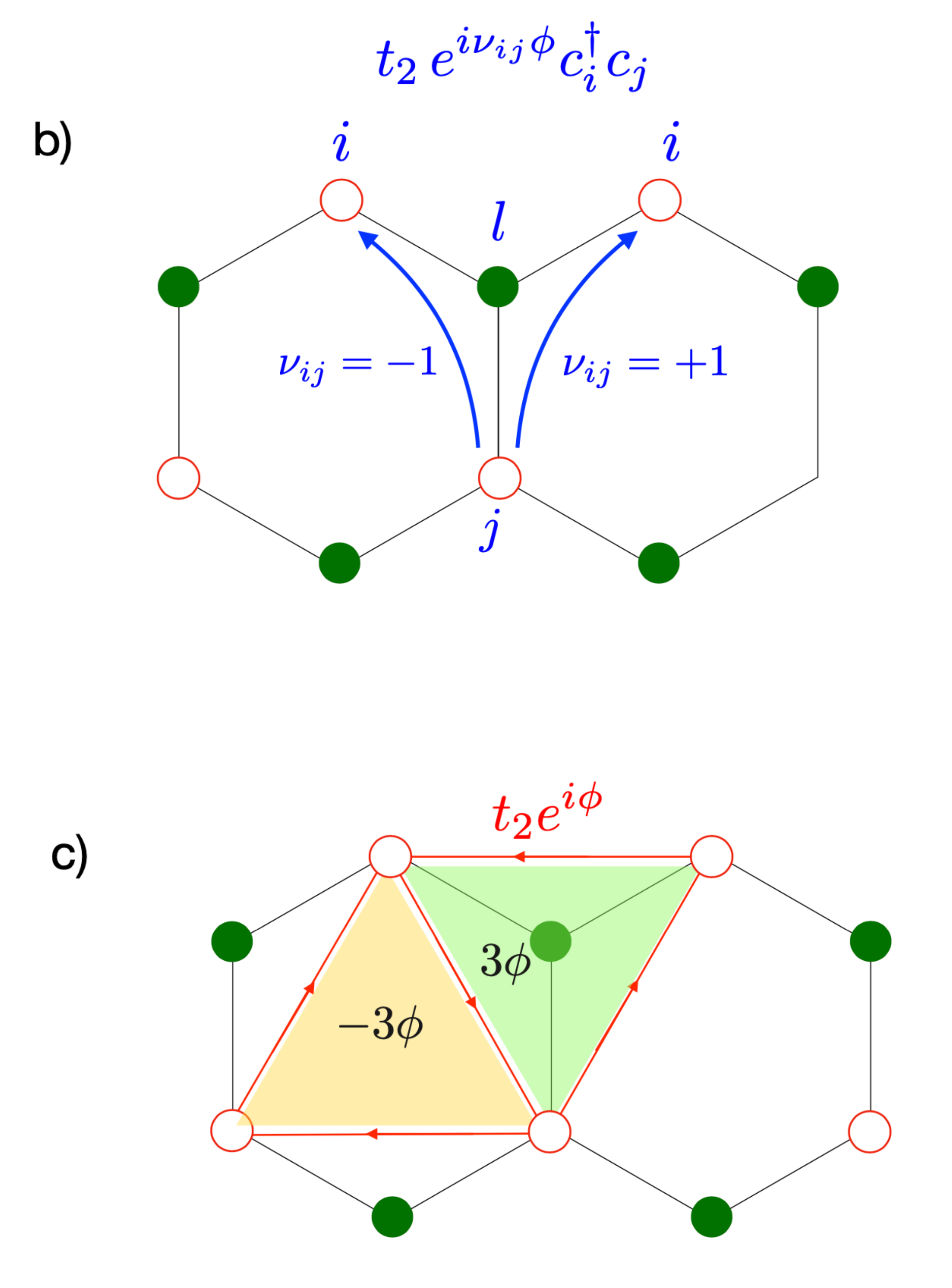}
\caption{(a) Flux pattern defining the Haldane model. The arrows (blue and red) stand for $t_2 e^{i\phi}$, and $t_2$ is real. Those arrows circulate clockwise around the center of each carbon atom hexagon. Therefore the reversed arrows (not represented) would correspond to the hopping amplitude $t_2 e^{-i\phi}$.  (b) The definition of the $\nu_{ij}$ for the phase signs of a NNN hopping term $t_2 e^{i \nu_{ij} \phi} c^\dagger_i c_j$ in the Haldane Hamiltonian. (c) A typical unit cell (parallelogram) is represented with the NNN complex hoppings (red arrows). The overall flux through such a unit cell is zero, resulting from the cancellation between the opposite flux piercing each half-unit cell (shaded triangle).}
\label{FigHaldaneModel}
\end{center}
\end{figure}
In 1982, TKNN had considered the quantum Hall effect (QHE) in the presence of a periodic lattice \cite{Thouless:1982}, where the interplay of the magnetic field and lattice generates the famous Hofstadter spectrum \cite{Hofstadter:1976}.
In his seminal work Ref.\cite{Haldane:1988}, Haldane made a step forward by realizing that the mandatory condition for QHE was time-reversal symmetry breaking, while the homogeneous magnetic field is not necessary. To prove it, he devised a toy model based on a 2D single sheet of graphite (although at that time graphene was far from being an experimental reality) showing quantum Hall effect without Landau levels. The recipe consists in breaking time-reversal symmetry while preserving a zero net magnetic flux per unit cell and therefore Bloch states. It is also important to have an insulator to get quantized Hall conductivity. Haldane realized that a specific pattern of complex second neighbor hoppings can bring all these ingredients together \cite{Haldane:1988}. The Hamiltonian of the Haldane model reads :

\begin{equation}
    H   = H_0 +    H_1 + H_2   \, ,
\label{Hhaldane}
\end{equation}
where the Hamiltonian of pristine graphene $H_0$, Eqs. (\ref{hzerographene},\ref{dxdyGraphene}), and the on-site staggered potential $H_1$, Eq. (\ref{HS}), have already been encountered to describe $h$-BN \cite{Semenoff:1984}. In order to break time-reversal symmetry, Haldane introduced the term $H_{2}$ which contains the complex next-nearest neighbour (NNN) hoppings, equal either to $t_2 e^{i \phi}$ or $t_2 e^{-i \phi}$ depending on the NNN bond \cite{Haldane:1988}. Those two values are not distributed randomly over the bonds but are organized in a way related to the orbital motion of electrons, see Fig. \ref{FigHaldaneModel} a). The value $t_2 e^{i \phi}$ iq associated to any term $c^\dagger_i c_j$ where $i$ and $j$ are second neighbors that have a common intermediate first neighbor $l$ on the left of the vector joining site $j$ to site $i$, see Fig. \ref{FigHaldaneModel} b). The Haldane term can be written in a compact form as
\begin{equation}
\label{GrapheneH2ij}
H_{2} =  t_2   \sum_{((i \leftarrow j))}     c_i^\dagger c_j e^{i \nu_{ij }\phi}     \, ,
\end{equation}
where the sum is over all oriented second-neighbor pairs $((i\leftarrow j))$. Each non-oriented pair $((i,j))$ contributes to two conjugated terms $((i\leftarrow j))$ and $((j\leftarrow i))$, thereby making $H_2$ Hermitian. The index $\nu_{ij}$ of the oriented pair $((i \leftarrow j))$ is defined by :
\begin{equation}
\label{Nuij}
\nu_{ij} =  \ve_z . \left( \boldsymbol{\delta}_{il} \times \boldsymbol{\delta}_{lj} \right) /\mid\mid  \boldsymbol{\delta}_{il} \times \boldsymbol{\delta}_{lj}  \mid\mid      \, ,
\end{equation}
where $l$ is the common first-neighbor shared by the sites $i$ and $j$, see Fig. \ref{FigHaldaneModel} b).

Going back to our notations with explicit mention to the $A$ and $B$ sublattices, this leads to :
\begin{equation}
\label{GrapheneH2}
H_{2} =  t_2   \sum_{i=1}^{3}   \left( \sum_{\vr_A} c_A^\dagger(\vr_A) c_A(\vr_A +\vb_i) e^{i \phi} +\sum_{\vr_B}   c_B^\dagger(\vr_B) c_B(\vr_B + \vb_i) e^{-i\phi} \right) + {\rm H.c.} \, ,
\end{equation}
where $\vb_1=\boldsymbol{\delta}_{2}-\boldsymbol{\delta}_{3}$, $\vb_2=\boldsymbol{\delta}_{3}-\boldsymbol{\delta}_{1}$, and $\vb_3=\boldsymbol{\delta}_{1}-\boldsymbol{\delta}_{2}$, are the vectors connecting next-nearest neighbor sites (see Fig. \ref{FigHoneycomb}). 
The Haldane term $H_2$ breaks time-reversal symmetry $\mathcal{T}$ because the hoppings $t_2 e^{i \phi}$ are complex when $\phi\neq 0$ or $\pi$, but respects inversion symmetry.
Both the Semenov term $H_1$ and the Haldane term $H_2$ break the chiral symmetry because they do not change sign under the transformation : $c_A \rightarrow c_A$ and $c_B \rightarrow - c_B$. For the topological symmetry classes~\cite{Schnyder:2008,Kitaev:2009}, only $\mathcal{T}$ and $\mathcal{S}$ matter, so the Haldane model belongs to the $A$-class of the ten-fold periodic table in two dimension, just like the quantum Hall effect.

\medskip

\begin{figure}[ht]
\begin{center}
\includegraphics[width=7cm]{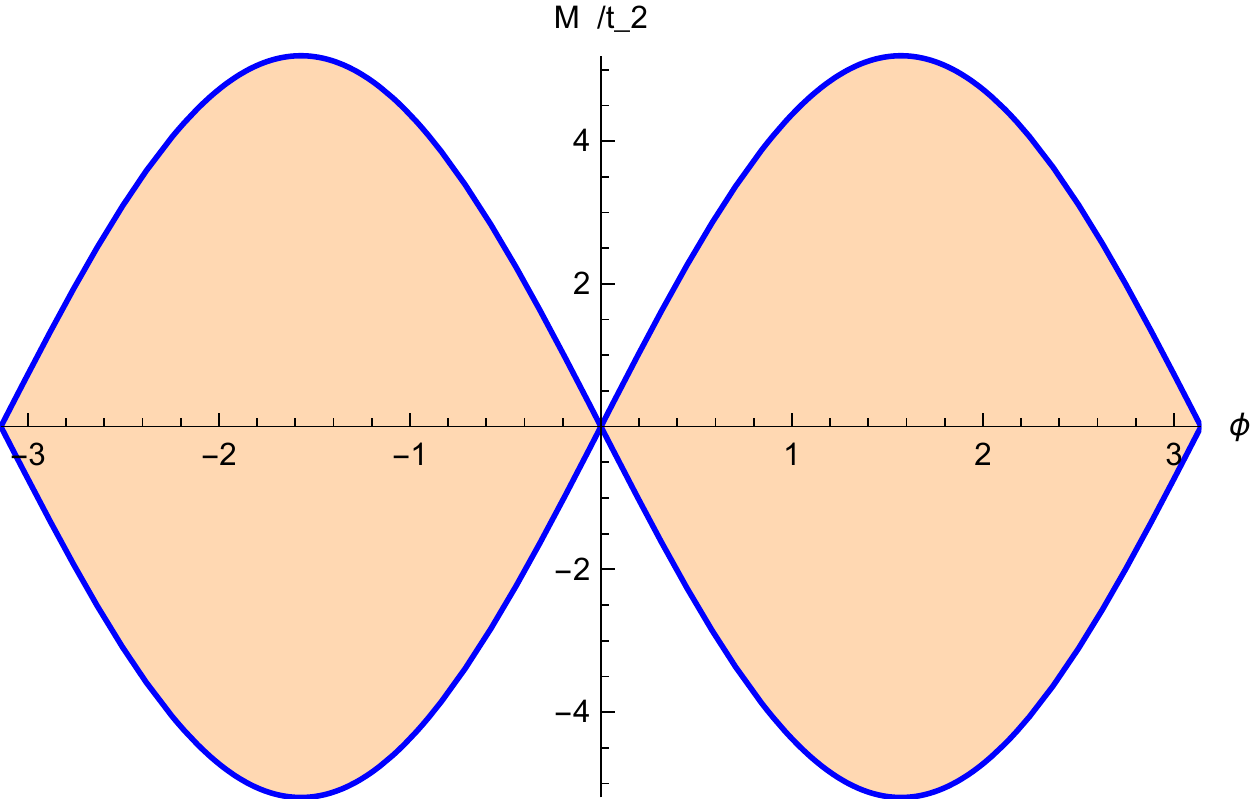}
\hspace{15mm}
\includegraphics[width=7cm]{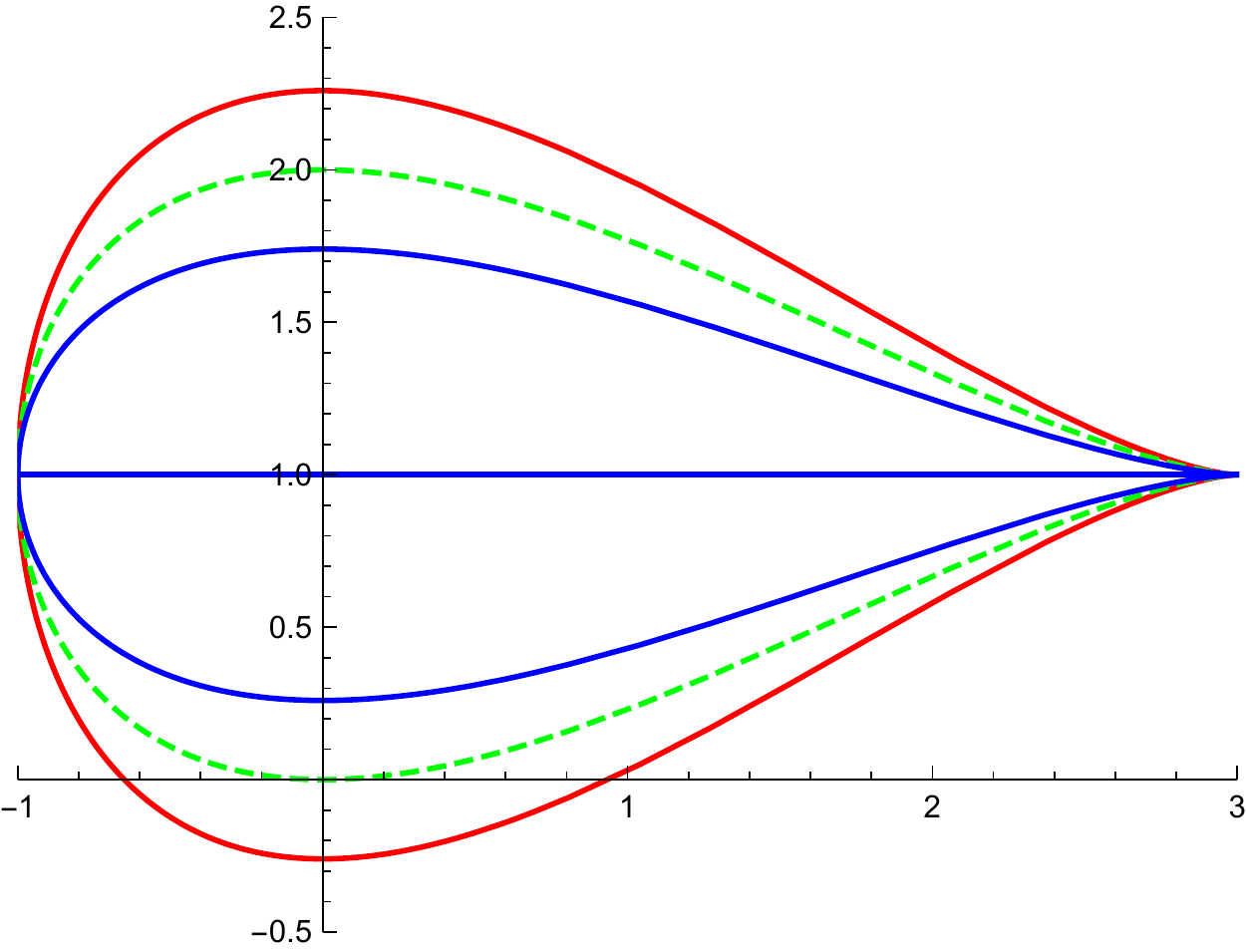}
\caption{(Left) Phase diagram of the Haldane model in the $(\phi,M/t_2)$ plane. The blue lines are the transition lines between the topological phase (coloured in orange) and the trivial phase (white).  (Right) A representation of the manifold $\mathcal{M}$ -- i.e. the image of the whole BZ through the map $\vk \to \vd(\vk)$ -- in the $(d_x,d_z)$ plane, defined by $d_y=0$, for various values of $t_2$ (with $M$ set to $1$) and at $\phi=\pi/2$. The blue lines do not enclose the origin and corresponds to values of $M/t_2$ in the trivial phases, the flat horizontal segment being $t_2=0$ (Semenov insulator). The red curve encloses the origin and is therefore in the topological phase. The green dashed line corresponds to $M=3 \sqrt{3}t_2$.}
\label{FigHaldanePhaseDiagram}
\end{center}
\end{figure}

Going to reciprocal space, the Bloch Hamiltonian for the Haldane model reads :
\begin{equation}
H(\vk)=\epsilon_0 (\vk) \sigma_0 + d_x(\vk) \sigma_x + d_y(\vk) \sigma_y +d_z(\vk) \sigma_z     \hspace{3mm}          \, ,
\label{HBlochHaldane}
\end{equation}
and the corresponding band structure is given by :
\begin{equation}
E_{\pm}(\vk)=\epsilon_0(\vk) \pm |\vd(\vk)| \, , 
\label{Spectrum2BandsBis}
\end{equation}
where the functions $d_x(\vk) $ and $d_y(\vk) $ are still given by the same Eqs. (\ref{dxdyGraphene}) as for pristine graphene. The specificity of the Haldane model lies in the following $\vk$-dependent terms : 

\begin{equation}
\epsilon_0(\vk) =2 t_2 \cos (\phi)   \sum_{i=1}^{3}       \cos(\vk . \vb_i)   \hspace{5mm} {\rm and}   \hspace{5mm}  d_z (\vk)   =M +    2 t_2 \sin (\phi)   \sum_{i=1}^{3}    \sin(\vk . \vb_i) \,  .
\label{dzHaldane}
\end{equation}
The NNN perturbation exhibits spatial dispersion ($\vk-$dependence) because it is nonlocal in real space. The part of $H(\vk)$ which is proportional to the identity just shifts the energies and breaks the electron-hole symmetry of the purely NN model. Nevertheless if $M=0$, the system remains gapless (if $\phi=0,\pi$) under introduction of a real NNN hoppings $t_2 \neq 0$, because both $\mathcal{I}$ and $\mathcal{T}$ are preserved for real NNN hoppings. In contrast, for complex hoppings (namely $\phi \neq 0,\pi$), the term proportional to $\sigma_z$ opens gaps at the Dirac points. The mass term $d_z(\vk)$ is the sum of a constant $M$ and a contribution proportionnal to $t_2$ which is odd in momentum and in particular changes sign in different valleys. Near the Dirac points, $d_x$ and $d_y$ vanish and one simply has to substitute $\vk=\vK$ (or $\vk=-\vK$) in $d_z$ as a zero order approximation :
\begin{equation}
d_z (\vk = \xi \vK)  = M - 3 \sqrt{3} t_2 \, \xi \, \sin (\phi)   ,
\label{HaldaneTerm}
\end{equation}
where $\xi=\pm 1$ is the valley index. The term in $t_2$ depends on the flux and vanishes for $\phi=0$ and $\phi=\pi$ where the time-reversal symmetry is restored. Therefore there is a competition between $M$ and $t_2 \sin(\phi)$, and three different scenarios are possible depending on parameters. First, when $|M|$ is very large, the Semenov mass $M$ dominates and the masses $d_z (\vk = \xi \vK)$ have the same sign at both Dirac points. This situation can be connected continuously to an atomic insulator without closing any gap, so this corresponds to the trivial insulating phase. Second, when $|M|$ is small (eventually zero), the $\xi t_2$ term dominates and induces opposite masses at the Dirac points $\xi=\pm 1$. This is also an insulating phase and we will see that it is a non-trivial phase in the sense that the valence band carries a non zero Chern number $C_{-}=\pm 1$ (orange regions in Fig. \ref{FigHaldanePhaseDiagram}). Finally, the transition between the trivial insulator and the non-trivial insulator occurs when one of the mass or the gap closes and changes sign at one of the Dirac point. This situation corresponds to the blue lines in Fig.~\ref{FigHaldanePhaseDiagram} that are given by :    
\begin{equation}
\frac{M}{t_2} = 3 \sqrt{3} \, \xi \, \sin (\phi)   ,
\end{equation}
which are transition lines between the two classes of band structures.

This seminal paper by Haldane was not fully nor widely appreciated at that time, because graphene was not experimentally available and even considered as a toy model for theorists. Moreover the time-reversal breaking terms involved in the Haldane model seemed (and are still) very difficult to realize : it requires a pattern of alternating magnetic fluxes at the scale of the unit cell. The works of Kane and Mele around 2005 \cite{Kane:2005a,Kane:2005b} lead to a revival of interest in the Haldane model physics and attempts to realize it experimentally. This has been done in some condensed matter materials and in cold atom systems. That was the first example of a topological band insulator: a Chern or QAH insulator.

\subsection{Chern number of two-band models in 2D \label{subsec:chernpontrya}}

We consider the very important case of a crossing between two non-degenerate bands. This is the most elementary case in topological band theory where it plays a role similar to the TLS in atomic physics. In such a case, the Berry curvature in one band is only originating by virtual transitions to the only other band.  The general Bloch Hamiltonian that captures the two-band system is :
\begin{equation}
H(\vk)=\vd(\vk)\cdot \boldsymbol{\sigma} \, .
\end{equation}
We have seen examples with the models of spinless fermions on the honeycomb lattice described above.

\subsubsection{Chern number as a wrapping number}

\begin{figure}
\begin{center}
\includegraphics[width=13cm]{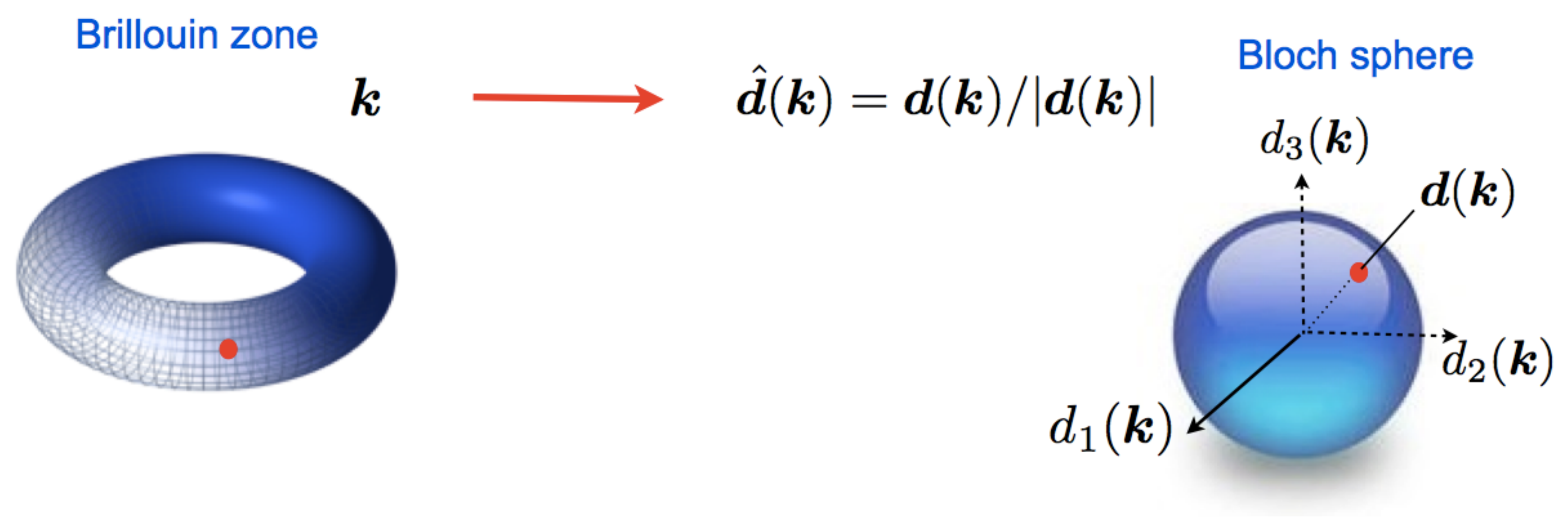}
\caption{Mapping $\vk \rightarrow \dhat (\vk) = \vd(\vk)/|\vd(\vk)|$ between the BZ (torus) and the Bloch sphere. Each point on the unit sphere represents a spinor parametrized by the angles $\theta_\vk$ and $\phi_\vk$.}
\label{FigMapping}
\end{center}
\end{figure}

To compute the Chern number, one can use Eq.~(\ref{BerryCurvatHamil}) and inject 
\begin{equation}
   \partial_i H(\vk) = \sigma_a \partial_i d_a(\vk)   \hspace{5mm} {\rm and}  \hspace{5mm}    (E_{n'} -E_n)^2=4 d^2  \, ,
\end{equation}
where summation over repeated index $a=x,y,z$ is implied. The Berry curvature is given by :
\begin{eqnarray}
F_{xy}^n(\vk)&=& \frac{i}{4d^2}  \langle u_n|\partial_x H(\vk) \partial_y H(\vk) | u_{n}\rangle +\text{c.c.} = -\frac{1}{2d^2} \epsilon_{abc} \langle u_n| \sigma_c  | u_{n}\rangle \partial_x d_a(\vk)  \partial_y d_b(\vk) \, , \\
&=&\frac{1}{2d^3} \epsilon_{abc} d_c   \partial_x d_a(\vk)  \partial_y d_b(\vk) \, ,
\end{eqnarray} 
where the last line holds for the valence band because then $\langle u_n| \sigma_c  | u_{n}\rangle=-d_c/d$. The Berry curvature is opposite for the conduction band. Then we can express the Chern number of the valence band $n=-$ under the following form :
\begin{equation}
    C_- = \frac{1}{2\pi} \int_{T^2} d^2 \vk \, \, F_{xy}^-(\vk) =  \frac{1}{4 \pi}  \int d^2 \vk   \left(  \frac{ \partial \vd }{\partial k_x}     \times  \frac{ \partial \vd }{\partial k_y}   \right) . \, \, \frac{ \dhat }{d^2} = \frac{1}{4 \pi}  \int d\Omega \,\,  \in \mathbb{Z}\, ,
    \label{CmoinsFormula}
\end{equation}
which is the Chern number of the fiber bundle. This formula is useful because it expresses the Chern number of a band as an integral over the parameter vector components with no explicit mention of spinors, unlike Eq. (\ref{Curvat_u}), nor velocity operator averages, unlike Eq.~(\ref{BerryCurvatHamil}). It has a simple geometric interpretation, which explains why it has to be an integer. Let us consider the image $\mathcal{M}$ of the whole BZ through the map $\vk \rightarrow \vd(\vk)$  (Fig.~\ref{FigMapping}). The number $C_n$ counts how many times the closed manifold $\mathcal{M}$ wraps around the origin, because it is the global solid angle $\Omega$ from which this image manifold is seen from the origin divided by $4 \pi$. If the origin is outside the image manifold $\mathcal{M}$ then the integral vanishes and $C_n=0$. If the origin is inside the image manifold, then the integrated is a multiple of $4\pi$, and therefore $C_n$ is a finite integer (see Fig.~\ref{FigHaldanePhaseDiagram} right). To change $C_n$ it is necessary to change the parameter of the bulk Hamiltonian $\dhat(\vk)$ in such a way that the bulk gap closes. Below, we will see that an alternative view on a Chern insulator is to consider it as a topological texture in reciprocal space. The corresponding invariant will be shown to be the skyrmion number (which is still another name for the Chern number, the Pontryagin index, the wrapping number, etc). An alternative formula using only the unit vector $ \dhat$ is :
\begin{equation}
C_n =\frac{1}{4 \pi}  \int d^2 \vk   \left(  \frac{ \partial \dhat (\vk)}{\partial k_x}     \times  \frac{ \partial \dhat(\vk) }{\partial k_y}   \right) . \,\,  \dhat 
\label{Winding2}
\end{equation}
which is the wrapping number of the map $\vk \rightarrow \dhat (\vk) = \vd(\vk)/|\vd(\vk)|$ between the BZ (torus $T^2$) and the unit sphere ($S^2$). In accordance with general classifications, $C_n$ is a single number that characterizes the general structure of wave functions globally in $\vk$-space. In Sec.~\ref{sec:qhe}, we have seen  that this topological number is observable and measures the charge Hall conductance of insulators in units of $e^2/h$. In the following subsection, we evaluate $C_n$ for the Haldane model in its different insulating phases.

Finally, we can relate those Chern number of Bloch bands to expressions for Chern numbers in the TLS case. By integrating over the BZ and injecting the spinor components Eq. (\ref{uspinors}), one gets the Chern number :
\begin{equation}
C =  \frac{1}{2 \pi} \int_{\rm BZ} d\vk \,\,\,  \frac{\sin \theta_\vk}{2} \left(  \frac{\partial \theta_\vk}{\partial k_x} \frac{\partial \phi_\vk}{\partial k_y} -  \frac{\partial \phi_\vk}{\partial k_x} \frac{\partial \theta_\vk}{\partial k_y}  \right) =  \frac{1}{2 \pi}  \int_{\rm BZ'} d\theta d\varphi \,\,\,  \frac{\sin \theta}{2} \, \, ,
\label{ChernNumberJacob}
\end{equation}
where $\text{BZ}'$ is the image of the BZ on the Bloch sphere, or equivalently the central projection of the closed manifold $\mathcal{M}$ onto the unit sphere. The last equality simply follows from the change of integration variables between the $\vk$ of the BZ and the polar angles on the Bloch sphere, the quantity between comes being the Jacobian of this transformation. This is the flux of a radial Berry curvature through the unit sphere.

\subsubsection{Practical evaluation of the Chern number with a degree formula}

Equation~(\ref{Winding2}) expresses the Chern number as a winding number of the map $\vk \rightarrow \dhat (\vk) = \vd(\vk)/|\vd(\vk)|$, which is the composition of the map $\vk \rightarrow \vd(\vk)$ followed by the central projection $\pi$ to the unit sphere. It is possible to transform this integral expression as a discrete sum by using the Brouwer degree of the map. The Brouwer degree reads \cite{Sticlet:2012} :
\begin{equation}
C =\sum_{y \in Y_z} \, \, \, \sum_{\vk \in d^{-1}(y)}   \text{sign} \left[  \left( \partial_{k_x} \vd \times \partial_{k_y}  \vd \right) \cdot \vn  \right]  \,\, ,
\label{Brouwer}
\end{equation} 
where $z$ is an arbitrary point on $S^2$, $\vn$ the unit vector towards $z$, and $Y_z=\mathcal{M} \cap \pi^{-1}(z)$. The procedure for evaluating $C$ consists in choosing a point $z$ on $S^2$ and then determining the intersections of $\mathcal{M}$ with the ray originating from $(0,0,0)$ to $z$. These intersections typically form a discrete set $Y_z$ of isolated points of $\mathbb{R}^3$. For our purpose, it is very useful to choose $z$ to be on one of the $\sigma$ axis, let us say $z$ is the north pole of the Bloch sphere. Then we have to sum in Eq. (\ref{Brouwer}) over points that satisfy $d_x(\vk)=d_y(\vk)=0$. 

Let us treat the case of the Haldane model to be more specific. Then $d_x(\vk)=d_y(\vk)$ leads to the Dirac points of pristine graphene $\vk=\pm \vK$, but note that we are considering the gapped Haldane model, so that importantly $d_z(\pm \vK)$ is non zero. Near those points the Bloch Hamiltonians read :  
\begin{equation}
H(\xi \vK + \vq)=H^{(\xi)}(\vq) = v_F \left( \xi q_x \sigma_x + q_y \sigma_y \right) + d_z(\xi \vK) \sigma_z \,\,  , 
\end{equation}
$\vd_\xi = (\xi v_F q_x , v_F q_y,d_z(\xi \vK)) $. At this point, one needs to consider different cases depending on the signs of $d_z(\pm \vK)$. Let us first assume that $d_z$ is positive at both Dirac points $\pm \vK$. Then the set $Y_z$ consists in two points and the Chern number is : 
\begin{equation}
C = \sum_{\xi=\pm 1}   \text{sign} \left[  \left( \partial_{q_x} \vd_\xi \times \partial_{q_y}  \vd_\xi \right) \cdot \ve_z  \right]  = \sum_{\xi=\pm 1}   \text{sign}(\xi v_F^2) =0 \,\, ,
\label{BrouwerHaldanePP}
\end{equation} 
reflecting that the local orientations of the two sides of the surface $\mathcal{M}$ are opposite. If $d_z$ is negative at both Dirac points $\pm \vK$, then set $Y_z=$ and immediately $C=0$. Therefore one finds that the valence band is trivial in the $h-$BN model and also in regions of the Haldane phase diagram where the masses have the same sign at the Dirac point.   

We now consider the case with opposite signs of the masses, for instance $d_z(\vK)>0$ and $d_z(-\vK)<0$. In such a case, the ray $z$ intersects the surface $\mathcal{M}$ only once and the Chern number is simply : 
\begin{equation}
C = \sum_{\xi=+ 1}   \text{sign} \left[  \left( \partial_{q_x} \vd_\xi \times \partial_{q_y}  \vd_\xi \right) \cdot \ve_z  \right]  =  \text{sign}(v_F^2) =1 \,\, .
\label{BrouwerHaldanePN}
\end{equation} 
For the opposite choice of signs, $d_z(\vK)<0$ and $d_z(-\vK)>0$, the intersection $Y_z$ also reduces to a single point but which maps back to valley $\xi=-1$ in the BZ. Hence $C=-1$.  
The above formulas can be symmetrized by considering the whole axis $z$ instaed of the ray $z$, leading to : 
\begin{equation}
C =- \frac{1}{2} \sum_{\xi=\pm 1}  \xi\, \text{sign} (M_\xi)=\frac{1}{2} \left[  \text{sign}(M_-) - \text{sign}(M_+) \right] \, ,
\end{equation} 
where $M_\xi = d_z(\xi \vK)$. 

Finally, the above formula has been obtained in a more general way in Ref. \cite{Sticlet:2012} which is valid for any two-band insulator and reads :
\begin{equation}
    C = \frac{1}{2}\sum_{\vk_D} \, \chi (\vk_D)\, \text{sign } [d_z(\vk_D)],
\end{equation}
where $\chi(\vk_D)=\pm 1$ denotes the chirality or winding number of the Dirac point located at $\vk_D$ [for example, in graphene, this chirality is $\chi_\xi=\text{sign det}(v_{ij})=\xi$] and $\text{sign } [d_z(\vk_D)]=\pm 1$ is the sign of the Dirac fermion mass. Lattice Dirac fermions always occur in pairs (fermion doubling theorem), which makes the Chern number an integer. Each massive Dirac fermion contributes $\pm 1/2$ to the Chern number. However, it is not always easy to identify all massive Dirac fermions from the energy spectrum alone. Indeed, there are often cases with \textit{spectator} Dirac fermions of very large mass that are totally undetectable in the energy spectrum. Let us illustrate that in the case of two bands with $H(\vk)=\vd(\vk) \cdot \boldsymbol{\sigma}$. The way to identify Dirac fermions correctly is to find points in reciprocal space at which two (out of three) components of $\vd(\vk)$ vanish (e.g. $d_x(\vk_D)=d_y(\vk_D)=0$). The third component $d_z(\vk_D)$ giving the mass of the Dirac fermion (it can not be zero otherwise the system would be gapless at $\vk_D$). The choice of components is arbitrary: it does not matter which two components vanish and which plays the role of the finite mass. Spectator fermions are discussed in \cite{Haldane:1988} and also in section 8.2 of \cite{Bernevig:2013}.

\medskip

Let us now consider the example of graphene in presence of some inversion breaking and time-reversal breaking terms. So the mass matrix is $(M_S - 3 \sqrt{3} t_2 \sin (\phi) \xi ) \sigma_z$ implying that the gap can close for $M_S =\xi 3 \sqrt{3} t_2 \sin \phi$ in one valley (labelled by $\xi=\pm 1$). This equality signals a one-electron topological quantum transition separating an QAH insulator and a trivial atomic insulator. Finally we would like to make a comment on the terminology. This type of phase transition is purely a change between two one-electron Hamiltonians. It has in particular nothing to do with topological order defined by Wen~\cite{Wen:2004,Wen:2019}. In particular the transition discussed here is not a transition between two topological orders. It is rather a transition between two band-insulators having distinct topological invariants (which characterize the winding of one-electron wave functions). More on this distinction in the conclusion, see Sec.~\ref{sec:conclusion}.

When proposing his model, Haldane had a two-fold goal~\cite{Haldane:1988}. One was to find a quantum Hall effect without Landau levels, i.e. as a band effect, breaking time-reversal symmetry but preserving translation symmetry. This goal was achieved. His other goal was to have a condensed matter realization of the parity anomaly known to occur in the 2D massless Dirac equation. One manifestation of this anomaly (i.e. a symmetry which exists classically but does not survive quantization) is the fact that, in a perpendicular magnetic field, there should be a half-quantized quantum Hall effect. However, in the Haldane model on the honeycomb lattice, Dirac fermions always occur in pairs even if one member of the pair has a zero mass and the other has a very large mass (of the order of the bandwidth). Therefore, no half-quantized Hall effect occurs in the Haldane model and the parity anomaly of one Dirac fermion (say the massless one in one valley) is compensated by the other Dirac fermion (say the spectator one in the other valley). Hence, from this perspective the second goal was not achieved. Achieving this parity anomaly requires a further trick, namely to produce a single 2D massless Dirac equation at the surface of a 3D topological insulator~\cite{Fu:2007,Moore:2007,Roy:2009}.

\section{Topological insulators: a bigger picture}
\label{sec:topinsulators}

We have already discussed in detail a few examples of topological insulators, including the Haldane model of spinless fermions on the honeycomb lattice (an intrinsic topological insulator), and the Rice-Mele chain with inversion symmetry (a topological crystalline insulator). We have also alluded to the Kane-Mele model as an example of time-reversal invariant topological insulator (i.e. a topological insulator protected by an extraordinary symmetry such as time-reversal, charge conjugation or chiral). In this section, we build upon those representative examples and enlarge our point of view to describe the current ``standard model" of insulating topological phases. Below, we give two definitions of a topological insulator. The first (see Sec.~\ref{sec:firstdef}) mainly involves cell-periodic Bloch states in reciprocal space and focuses on bulk invariants and quantized electromagnetic responses. The second (see Sec.~\ref{sec:seconddef}) mainly involves Wannier functions in real space and focuses on their localization. Topological metals will be the subject of the next section~\ref{sec:topmetal}. 


\subsection{Bulk quantized electromagnetic response and gapless boundary states \label{sec:firstdef}}

In this section, we use a characterization of topological insulators in reciprocal space focusing on two ingredients: a bulk topological invariant (usually associated to a quantized electromagnetic response) and the related existence of gapless surface states. We first discuss intrinsic topological insulators and then symmetry-protected topological insulators. In the latter case, we further distinguish between extraordinary symmetries (time-reversal, charge conjugation or chiral) and crystalline symmetries (such as rotations, mirrors, inversion, etc). These three symmetries are called extraordinary because either they are anti-unitary (time-reversal and charge conjugation) or/and they anti-commute with the Bloch Hamiltonian (charge conjugation and chiral symmetry). We refer to the first case simply as symmetry-protected topological insulators and to the second case as topological crystalline insulators.

\subsubsection{Intrinsic topological insulators}
A well-known topological insulator, the Haldane model~\cite{Haldane:1988}, has been discussed in detail in this review (see Sec.~\ref{section2DHaldane}). This is the telltale example and the representative of a larger class of topological insulators, called Chern insulators or QAH insulators. Its phase diagram shows several gapped phases separated by lines where the gap closes and identified by a specific value of the Chern number $C_-$ of the occupied valence band  (Fig. \ref{FigHaldanePhaseDiagram}, left). The Chern number of a band is a topological invariant which measures a global property of the Bloch wave functions of this band over the whole BZ. It can only be defined in the presence of a finite bulk gap. Each gapped phase is robust in the sense that modifying continuously the parameters of the model does not change the Chern number. The only way to jump from an integer value of the Chern number to another one is to cross a gap closing.  At the gap closure, $C_-$ is ill-defined as can be seen from the formula Eq.~(\ref{CmoinsFormula}) where the denominator $d$ vanishes, or in more geometric terms, the manifold $\mathcal M$ crosses the origin which is the source of the Berry curvature flux (Fig.~\ref{FigHaldanePhaseDiagram}, right). Note that the total Chern number, including the two bands of the Haldane model (the occupied valence and the empty conduction band) is always zero : $C=C_- + C_+ =0$. Each band can separately carries both zero Chern number (trivial phase), or opposite $C_n=\pm 1$ (topological phase). More generally, a Chern insulator is a two-dimensional band insulator that breaks time-reversal symmetry and that has a non-vanishing total Chern number of the occupied bands at zero temperature (it may have much more bands than the Haldane model). The QH insulator is a close relative where the translational invariance is broken by the magnetic field and replaced by the Landau level structure in the continuum, and eventually magnetic subbands and Hofstadter spectrum in presence of a periodic potential. 

\bigskip

We now turn to the characterization of topological insulating phases in terms of macroscopic observables, like the transverse Hall conductance which is a bulk response function odd with respect to time-reversal. Via the TKNN formula (\ref{conductivityHallChern}), the Hall conductance is directly proportionnal to the TKNN number with a prefactor $e^2/h$ depending only on fundamental constants. There is a subtle distinction here between the TKNN number and Chern numbers. A Chern number characterizes a band (or a group of bands) whereas a TKNN number characterizes a gap. The TKNN number is simply the sum of the Chern numbers of the occupied bands below the gap in which the Fermi level is assumed to be. This explains why each insulating phase is distinguished by a robust and quantized Hall conductance $\sigma_{xy}=0$ or $\sigma_{xy}=\pm e^2/h$. Translation symmetry is very useful in order to relate the robust observable (here the Hall conductance) and the Bloch wave functions. Nevertheless it is not essential and the Hall quantization survives disorder. One may even envision the existence of topological invariants that do not rely on Bloch states, as in quasicrystals (see for example~\cite{Tran:2015, Fuchs:2016, Akkermans:2017} and \cite{Zilberberg:2020} for review) or in amorphous solids~\cite{Agarwala:2017,Grushin:2020}. In order to study such cases, one needs generalizations of the Chern number beyond reciprocal space. This was pioneered by Niu et al.~\cite{Niu:1985} with the sensitivity of the many-body ground-state to twisted boundary conditions. There are other possibilities such as the Bott index~\cite{Loring:2019} or the local Chern marker~\cite{Bianco:2011}.

\bigskip

So far, we have only discussed a bulk characterization of the gapped phases and stated that topological phases are separated from trivial phases by a gap closing transition, where the Chern number and the Hall conductance encounter a finite jump. This is also true for a transition between two topological phases with different Chern numbers. Note that the vacuum is considered to be a trivial band insulator. This fact has very important consequences for the physics at the boundary of a topological insulator, or at the interface between distinct topological phases. At such an interface, the parameters of the model have to be tuned in such a way that the Chern number jumps by an integer number, which implies a region where the gap closes. The gap closing makes possible the presence of gapless conducting states at the interface between topologically-distinct phases. In addition, the existence of chiral gapless modes is guaranteed by the bulk-edge correspondence and can also be understood as following from the Jackiw-Rebbi mechanism (see Sec.~\ref{sec:jr}). Bulk-edge correspondence states that the number of gapless chiral modes per edge is equal to the jump of the bulk topological invariant~\cite{Hatsugai:1993}. These boundary modes have unique properties, meaning that they typically cannot be realized in a lattice system. For Chern insulators, the 1D edge states are chiral, i.e. they circulate only in one direction around the insulating bulk (see Fig.~\ref{fig:qht}). The chiral nature of motion indicates the time-reversal breaking. Such chiral fermions can only emerge at the boundary at a non trivial insulator : there are no 1D chain or wire with unidirectional electrons. In the QHE, the direction of motion is determined by the orientation of the magnetic field, while in the Haldane model it is determined by the internal inhomogeneous magnetic field. These modes are protected from backscattering by impurities or disorder because there is no available state for the carrier to back scatter on the same edge and tunneling to the opposite edge is exponentially forbidden. 

\medskip

From the above discussion, we can formulate a first definition of a topological insulator. It is a bulk band insulator, that is characterized by a non-zero integer, called a (bulk) topological invariant, and a robust quantized bulk response. To change this bulk topological invariant it is necessary to close the bulk gap. The edges of a topological insulator necessarily host gapless modes that are robust because they are chiral and therefore immune to disorder. In addition there is a relation between the number of these edge modes and the bulk topological invariant: this is the so-called bulk-edge correspondence. The case of the Chern insulator is particular in that it requires no symmetry for its protection and is truly robust to any perturbation, as long as it does not close the bulk gap. It is intrinsically topological.

\subsubsection{Symmetry-protected topological insulators}

It was long believed that broken time-reversal symmetry and space dimension two were a mandatory setting for the appearance of topological phases. In 2005, Kane and Mele~\cite{Kane:2005a,Kane:2005b} uncovered a new type of topological insulator where time-reversal plays the role of a protective symmetry for metallic edges, and allows one to draw a topological distinction between two insulating states. They introduced a model of spinful fermions in graphene consisting in two independant copies of the Haldane model respectively attached to each spin projection, each copy being the time-reversed of the other, e.g. with Chern number $C_\uparrow=+1$ for spin-up fermions and $C_\downarrow =-1$ for spin down fermions. As a result the overall Kane-Mele model is time-reversal invariant and describes graphene with a very particular intrinsic spin-orbit (SO) coupling conserving the spin projection $S_z$. The total Chern number of the occupied bands is $C=C_\uparrow+C_\downarrow =0$ (no quantum Hall effect), but there is a $\mathbb{Z}_2$ topological invariant $\nu = (C_\uparrow-C_\downarrow)/2 \, \, [2]$, which distinguishes only two classes: trivial ($\nu=0$) and non-trivial ($\nu=1$).

\bigskip

Because the $S_z$-conserving Kane-Mele model consists in two independent copies of a Haldane model with opposite Chern numbers, it has counter-propagating edge states with a full spin-momentum locking property : spin up propagates in only one direction, and spin-down propagates in the opposite direction. This spin-momentum locking insures that backscattering by a non-magnetic impurity (i.e. an impurity that respects time-reversal symmetry) is not possible. One could expect that the gapless helical edge states are an artefact of this $S_z$ conserving juxtaposition of two chiral edge states, and that any spin mixing perturbation would mix the two spins and gap the edge states. In fact the gapless character of the helical edge states is robust again both staggered potential perturbations and Rashba spin-orbit perturbations (mixing spin up and down projections) provided those perturbations do not close completely the bulk gap. With those time-reversal symmetric perturbations, the only way to get rid of the helical edge state is to close the bulk gap. In contrast, a time-reversal breaking perturbation (like a Zeeman effect due to transverse magnetic field in $x$ or $y$ directions) would immediately spoil the metallic character of the helical edge. The crossing of the left and right movers branches is protected by Kramers' theorem as $\mathcal{T}^2=-1$. In principle, it is possible to construct models with a larger number of pairs of counter-propagating modes. If the number of Kramers' pairs is even, then disorder may eventually gap out all the edge states. In contrast, for an odd number of Kramers' pairs there will always be a gapless pair left as long a the bulk gap is finite and time-reversal symmetry is obeyed. This expresses that the bulk $\mathbb{Z}_2$ index is the parity of the number of Kramers' pairs of edge states. Intrinsic spin-orbit coupling, at the heart of the Kane and Mele proposal, is too small in graphene for the effect to be measurable. However, quantum spin Hall insulators have been realized experimentally in HgTe/CdTe quantum wells following the theoretical proposal by Bernevig, Hughes and Zhang, see Ref.~\cite{Bernevig:2013} for review.

\medskip

In contrast to the Chern insulator (quantum Hall effect), that only exists in two space dimension~\footnote{There are also so-called weak topological insulators protected by time-reversal symmetry in 3D and characterized by three integers. They are best seen as layered versions of 2D time-reversal-symmetric topological insulator, just like there is a layered 3D version of the quantum Hall effect~\cite{Kohmoto:1992}.}, this type of time-reversal symmetric insulator has a three-dimensional counterpart known as a strong topological insulator~\cite{Fu:2007,Moore:2007,Roy:2009}. The strong topological insulator is also classified by a $\mathbb{Z}_2$ index and also has a topological metal on its surface. It is a single massless two-dimensional Dirac fermions that is spin-momentum locked. It appears similar to graphene, except that it has a single (instead of four) Dirac cone. Indeed, the real spin is actually the one appearing in the Dirac equation (and not a sublattice pseudo-spin as in graphene) and there is also no valley degeneracy. Such a strange metal can only exist as a surface mode of a higher dimensional system and has a true parity anomaly and a half-quantized quantum Hall effect.

\medskip

In conclusion, the $\mathbb{Z}_2$ topological index and the helical edge states are both protected by time-reversal symmetry. There is no smooth path (in the space of Hamiltonians) connecting the $\nu=1$ and $\nu=0$ phase while simultaneously keeping the bulk gap open and respecting the time-reversal invariance. To change the $\mathbb{Z}_2$ topological index, it is necessary either to use time-reversal breaking perturbations or to close the bulk gap.  In 2D, the odd and even phases of the Kane-Mele model can be characterized by a bulk quantized response which is a spin Hall conductance. Nevertheless, this only holds for the $S_z$ conserving models, and it is difficult to build a macroscopic observable that could play the role of the Hall conductance for Chern insulators. The bulk characterizations of the 2D or 3D time-reversal invariant phases relies on $\mathbb{Z}_2$ pumps~\cite{Fu:2006,Essin:2007}. The 3D strong topological insulators have a specific orbital magnetoelectric polarizability described by axion electrodynamics [with a topological $\theta$-term similar to the one discussed in the 1D context, see Eq.~(\ref{eq:thetaterm})] that can play such a role~\cite{Essin:2009}. 

\medskip

The ideas of Kane and Mele have been extended to extraordinary symmetries such as time-reversal, particle-hole (charge conjugation) and chiral and to the concept of symmetry-protected topological insulators. There is a general classification of topological insulators (and superconductors) for non-interacting fermions that is known as the ten-fold way periodic table~\cite{Schnyder:2008,Qi:2008,Kitaev:2009,Chiu:2016}. Depending on space dimension and on three extraordinary symmetries resulting in ten Altland-Zirnbauer classes, it indicates whether there is a single type of band insulator (only trivial), two types of band insulators ($\mathbb{Z}_2$ index separating trivial from non-trivial band insulators) or as many band insulators as relative integers $\mathbb{Z}$. Kitaev has realized that, as a function of space dimension $D$, there is a form of regularity or pattern in this table related to real and to complex K-theory and known as Bott periodicity~\cite{Kitaev:2009}. Such topological insulators that are protected by a extraordinary symmetry are also called strong topological insulators~\cite{Cano:2020} as they resist the absence of crystalline symmetries that we study next.

\medskip   

\subsubsection{Topological crystalline insulators}

So far we have discussed topological band insulators protected by time-reversal symmetry, which is a fundamental and generic symmetry of space-time. As initiated by Liang Fu~\cite{Fu:2011}, it is also possible to use spatial symmetries (e.g. rotations or reflections) of the crystal itself to protect and maintain topological distinctions between various insulating phases. Such phases have been coined topological crystalline insulators (TCI) and have also been classified (see e.g. Refs.~\cite{Slager:2013,Kruthoff:2017}). 

We have already discussed a very simple example, namely an inversion-symmetric insulator in 1D. In Sec.~\ref{sec:rm}, we have seen that provided inversion symmetry is maintained, it is impossible to connect the SSH insulating chain (at $\delta \neq 0$ and $\Delta =0 $) to the CDW insulator (at $\delta=0$ and $\Delta \neq 0$). Both phases are characterized by a distinct robust observable, which is the electric polarization. It is vanishing in the SSH phase and quantized in the CDW phase.

This simple example of a quantized electric polarization in 1D has been generalized to quantized electric multipole insulators~\cite{Benalcazar:2017} leading to the concept of higher order topological insulators (HOTI)~\cite{Schindler:2018}. These latter systems are insulators that do not possess gapless surface states but rather gapless hinge states, which are gapless states living on the boundary of a boundary. For example, a 2D second order topological insulator known as the quadrupole insulator does not host 1D gapless edge states but possesses 0D gapless corner states for a square patch protected by mirror symmetries. In other words, the surface of this second order topological insulator is not a metal but is itself a 1D topological insulator. HOTI escape the simple bulk-boundary correspondence between a $D$-dimensional insulator and $D-1$ metal but rather feature a correspondence to $D-2$ or $D-3$ gapless states. They only exist under the protection of point-group symmetries.

\subsubsection{Conclusion: first definition of a topological insulator}
A topological insulator is a band insulator (it has a bulk gap) characterized by a bulk topological invariant computed from its cell-periodic Bloch states (e.g. a Chern number or a $\mathbb{Z}_2$ invariant). This invariant is typically related to a quantized electromagnetic response (e.g. a quantized Hall conductivity or a quantized electric polarization). In order to change this bulk topological invariant it is necessary to close the bulk gap or to break the protecting symmetry. In a finite sample with a surface that does not break the protecting symmetry, a topological insulator host gapless surface (or hinge) states that have some form of robustness towards perturbations that respect the corresponding symmetry.

\subsection{Obstruction to exponentially-localized and symmetric Wannier functions \label{sec:seconddef}}

The picture of the previous section focuses first on the topology of Bloch states in the bulk and second on the eventual gapless boundary states (and their dispersion in momentum space) as a hallmark signature of a topological insulator. Now we present an alternative point of view based on the real-space localization of the Wannier functions in the bulk (see Sec.~\ref{sec:wannier}). 

According to Kohn~\cite{Kohn:1964}, a (trivial) insulator is exponentially insensitive to boundary conditions. This is similar to an atomic insulator, which can be seen as a periodic solid made from bringing closer initially isolated atoms. As in isolated atoms, valence electrons remain close to their nuclei (or ions). In the language of solid-state physics, the Wannier functions are exponentially-localised on the ions. In the case of an atomic insulator, the Wannier functions are simply the atomic orbitals. As a consequence, the many-body ground-state (for non-interacting electrons) is barely sensitive to what happens near the boundary of a macroscopic sample. In contrast, a topological insulator is a band insulator that is not smoothly or adiabatically connected to an atomic insulator. Smoothly means without closing the gap for intrinsic topological insulators, and without closing the gap while keeping some protective symmetry for SPT insulators. 

Such a viewpoint of a topological insulator featuring an obstruction to exponentially-localized and symmetric Wannier functions has been put forward recently~\cite{Bradlyn:2017,Po:2017} and is sometimes called topological quantum chemistry (see~\cite{Cano:2020} for review).

\subsubsection{Intrinsic topological insulators}

Bloch functions and Wannier functions are two dual possible representations of the electronic states of a crystal. Intrinsic topological insulators (Chern insulator and QH states) are characterized by obstructions in the realization of either of these representations. First, we explain the obstruction for the Bloch states in reciprocal space, and second the corresponding veto for exponentially-localized Wannier functions in real space. 

\medskip

A band with a finite Chern number implies an obstruction in finding a unique gauge for the cell-periodic Bloch states which would be smooth over the whole BZ. Let us consider the explicit case of the Haldane model and use the Bloch sphere representation to visualize the spinors (cell-periodic Bloch states) of the valence band. In the trivial phase, $d_z(\vk)$ being always positive it is possible to pick the $(n)$-gauge which has no singularity except at the south pole. In the topological phase, $d_z(\vk)$ changes sign somewhere in the BZ, and the spinor explores both north and south poles : it is not possible to use a single gauge and one has to use the Wu-Yang construction with overlapping $(n)$ and $(s)$ regions. From this perspective, going from a trivial to a topological phase requires a band inversion (in one of the valley for the Haldane model) which necessarily occurs via a gap closing. In other words, there is an obstruction to smoothly or adiabatically connect a topological insulator to an atomic insulator. Note that the vacuum is considered to be a trivial (atomic) band insulator. This obstruction is also a well-known property of the QH states~\cite{Kohmoto:1985}. Reciprocally, without this obstruction, it would be possible to apply Stokes' theorem to the whole BZ and to obtain that the flux of the Berry curvature vanishes in contradiction with the hypothesis of finite Chern number.  

\medskip

Let us now turn to the Wannier function representation. Wannier functions are the Fourier transforms of Bloch states that, in conventional insulators, are usually exponentially-localized around specific locations of the solid, the Wannier centers. The Wannier functions are not unique, because there is a gauge freedom to define the Bloch functions. Changing, locally in $\vk$, the $U(1)$ phases of the Bloch functions results in changing the spread of each Wannier function, while keeping the Wannier center of the band (or isolated group of bands) invariant. Vanderbilt and Marzari~\cite{Marzari:1997} devised a procedure to compute the maximally localized Wannier functions (MLWF) of a given lattice model, see Ref.~\cite{Marzari:2012} for a review. Going back to the Haldane model, Thonhauser and Vanderbilt have shown that this procedure fails at the topological transition, where the spread of the Wannier functions diverges~\cite{Thonhauser:2006}. Indeed, at the transition, Wannier functions are completely delocalized, as in any metallic gapless state. For a band with finite Chern number, it is not possible to build a basis of exponentially-localized Wannier functions, meaning that they are only algebraically localized~\cite{Thouless:1984,Thonhauser:2006}. This is in fact a general property shared by all Chern insulators~\cite{Brouder:2007} and by QH insulators~\cite{Thouless:1984}. 

\medskip

Finally, this impossibility to represent a topological phase in terms of localized basis functions shows that the quantum degrees of freedom are not stored locally in such phases. As a plausible consequence, the non-triviality of such phases should be evidenced by their sensitivity to twists in the boundary conditions. For the QHE, Niu, Thouless and Wu~\cite{Niu:1985} gave a real-space expression for the Chern number. This requires putting the system on a real space torus and twisting the boundary conditions by inserting fluxes in the inequivalent non-contractible loops of the torus. The space of boundary conditions replaces the BZ. The beauty of this formulation is that translation invariance is no longer needed. The topological invariant can be defined directly for the many-body ground-state (instead of ``band by band'') and is valid also for a disordered (and in some cases interacting) system.

\subsubsection{Symmetry-protected topological insulators}

The idea of obstruction is also relevant to investigate SPT phases, but here the obstruction is based on the fact that one enforces a protective symmetry. Without such a constraint, the obstruction would generally be lifted. 

For concreteness, we consider the case where the symmetry is time-reversal and first discusse real space. According to Ref.~\cite{Brouder:2007}, it is always possible to construct a basis of exponentially-localized Wannier functions for time-reversal invariant insulators both in the even (trivial) and odd (topological) phases. This is true but it comes always at a price, which is that one cannot maintain the time-reversal invariance of the exponentially-localized Wannier functions in the topological phase~\cite{Soluyanov:2011,Soluyanov:2012}. In a topological insulator, there is an obstruction to finding symmetric and exponentially-localized Wannier functions~\cite{Po:2017,Bradlyn:2017}.

In reciprocal space, the obstruction is manifest in the fact that it becomes impossible to find a smooth gauge for the whole BZ that respects time-reversal symmetry. Although it is possible to find a smooth gauge over the whole BZ (because the Chern number vanishes), it necessarily breaks time-reversal symmetry. This is nicely reviewed in~\cite{Fruchart:2013}.

\subsubsection{Topological crystalline insulators}

We start by considering the simplest possible 1D monoatomic chain and search for topological insulator phases protected by inversion symmetry (the simplest point-group symmetry). The two possible configurations that respects inversion symmetry correspond to either having electrons wave functions localized at every site, or alternatively having them localized at the center of each bond. Those two situations have respectively electric polarization $0$ and $P_q/2$ modulo the quantum of polarization $P_q$. Of course it seems very simple to go continuously from one to the other situation by simply shifting all electrons by half the lattice constant. But this way breaks inversion symmetry at every steps between the two configurations. Alternatively one may keep inversion symmetry at all stage if one first spreads the wave functions in a symmetric way before reconcentrating them around the new localization centers. Nevertheless this latter protocol goes through a fully delocalized state that is gapless. In conclusion, it seems impossible to go continuously from the configuration $P_\text{tot}=0$ to the configuration $P_\text{tot}=P_q/2$ while simultaneously keeping the whole system gapped and inversion symmetric at each intermediate steps. Such a topological transition can be implemented in a 1D model with coupled $s$ and $p$ bands~\cite{Vanderbilt:1993,Shockley:1939}. The trivial phase is the atomic insulator (a filled $s$ band) with vanishing polarization and the topological phase is a covalent insulator (a filled $sp$ hybridized bonding band) with quantized polarization.

Actually, these two 1D insulating phases have exponentially-localized and inversion-symmetric Wannier functions. They are therefore both eligible for being atomic limits~\cite{Po:2017}. One is a naive atomic limit (the atomic insulator with Wannier center on the ions), while the other has been called an obstructed atomic limit (the covalent insulator with Wannier center on the bonds)~\cite{Bradlyn:2017}. What distinguish the two is whether the Wannier center is located on the ions or mid-bonds. The conclusion is that in this example, there is no obstruction in finding symmetric and localized Wannier functions. Still, in a finite chain, as shown by Shockley, the covalent insulator hosts edge states, but not the atomic insulator~\cite{Shockley:1939}.

A general theory based on the obstruction in finding symmetric localized Wannier functions has recently been developed~\cite{Po:2017,Bradlyn:2017} and is reviewed in~\cite{Cano:2020} (it is sometimes called topological quantum chemistry). It is based on the key idea of a mismatch or a compatibility between the symmetry of atomic orbitals in real-space and the topology of bands in the BZ. First, starting from well-localized atomic orbitals, the notion of elementary band representations is used to build all possible atomic limits. These are essentially the cases where symmetric localized Wannier functions exist. Second, starting from the opposite view of delocalized electrons in Bloch states, one considers all possible band structures, which defines quasi-band representations. Third, one analyses the compatibility of those two views of band theory. Topological semimetals correspond to the case where compatibility relations can not be satisfied. When they can be satisfied, as there are more quasi-band representations than elementary band representations, one defines atomic limits (i.e. trivial insulators) as being the quasi-band representations that are elementary band representations. There exists several such limits and hence the notion of obstructed atomic limits. Quasi-band representations that are not elementary band representations are called topological insulators. In order to detect such topological insulators it is possible to use symmetry-based indicators~\cite{Po:2017} following the initial idea of Fu and Kane for inversion-symmetric topological insulators~\cite{Fu:2007}. Another interesting notion that emerges from this framework is that of fragile topological bands~\cite{Po:2018}. Using these ideas, several catalogues of topological materials based on the 230 space groups appeared recently~\cite{Zhang:2019,Vergniory:2019,Tang:2019}.

\subsubsection{Conclusion: second definition of a topological insulator}
An atomic limit corresponds to exponentially-localized and symmetric Wannier functions. A trivial insulator is such that it can be smoothly continued to an atomic limit. Due to the different possible protecting symmetries, there may be several distinct atomic limits and therefore several types of trivial insulators (see the notion of an obstructed atomic limit). A topological insulator is such that it can not be continued to an atomic limit without closing a gap or breaking a protecting symmetry. In a symmetry-protected topological insulator there is an obstruction to finding symmetric and exponentially-localized Wannier functions.

This second definition of a topological insulator is actually more stringent than the first one. For example, according to the first definition, a 1D inversion-symmetric band insulator with quantized electric polarization (e.g. the CDW chain realizing an ionic insulator, or the Shockley model of coupled $s$ and $p$ bands in the covalent insulator phase) is a topological insulator. But according to the second definition, a 1D inversion-symmetric band insulator is always trivial: it is either a naive atomic limit if the Wannier centers are on the ions (e.g. the CDW chain, or the Shockley model in the atomic insulator limit) or an obstructed atomic limit if the Wannier center are on the bonds (e.g. the SSH chain realizing a molecular insulator, or the Shockley model in the covalent insulator limit).

\section{Topological metals: Fermi surface as a topological defect}
\label{sec:topmetal}

Up to now, we have mainly dealt with band insulators having non-trivial topology, i.e. topological insulators. Here, we want to expand the picture and consider also metals, i.e. fermionic systems having a Fermi surface, and their topological characterization. For example, graphene is a peculiar 2D semi-metal, in which the Fermi surface is limited to two points corresponding to the contacts between two bands. The idea of classifying Fermi surfaces as topological defects goes back to Volovik (see Ref.~\cite{Volovik:2003} for review, the original idea dates from the 1980's), following the pioneering work of Lifshitz \cite{Lifshitz:1960} in the 1960's. The latter understood that there could be a phase transition in a metal that is not related to a change in symmetry but rather to a change in the topology of the Fermi surface. Is this surface made of a single piece or of two disconnected pieces, for example? Volovik later took the step of considering Fermi surfaces as topological defects in reciprocal space and of treating either the inverse Green function (or single-electron propagator) or the Bloch Hamiltonian as a kind of order parameter. These systems are sometimes called topological (semi-)metals. An extension of the idea of a topological defect (namely a topological texture) will eventually lead us to an alternative point of view on topological insulators.

\subsection{Introduction to homotopy groups for topological defects and textures \label{homotopy}}
These ideas followed from earlier developments in the classification of topological defects in real space using homotopy groups, that we now review. This subject flourished in the 1970's mainly under the impulsion of Mineeev, Volovik, Toulouse, Kl\'eman and Mermin (see Ref.~\cite{Mermin:1979} for review, see also the very accessible lectures notes by Sethna \cite{Sethna:1992}). These authors have shown that, in the context of phase transitions resulting from spontaneous symmetry breaking and characterized by an order parameter, defects (i.e. zeroes) in the order parameter could be systematically classified using homotopy theory. We will use similar ideas to discuss Fermi surfaces as if they were topological defects. 

Consider the ordered phase of a 3D ferromagnet. The order parameter is the magnetization $\vM$, its direction $\vn$ is uniform at equilibrium and its norm $|\vM|\approx M_0$ is almost constant at low temperature. Spatially varying the direction of the magnetization corresponds to excitations in the system. If these variations are small perturbations around equilibrium, these are known as spin waves (or generically as soft or Goldtsone modes). But there are other type of excitations of the order parameter, which are not smooth and small perturbations. These corresponds to situations in which the order parameter vanishes on points or lines or planes and are known as defects. Some of these defects are remarkably stable due to a topological protection. A topological defect is a singularity of the order parameter that can not be patched or repaired by any local rearrangement of the microscopic spins. In order to characterize a singularity as a topological object, one proceeds in four steps:

First, one needs to trap the defect inside a cage $C$. What is the dimensionality of this cage? If the defect is point-like and space is 3D, then the cage is a sphere $S^2$ of dimensionality 2. If the defect is line-like, then the cage is a circle $S^1$ of dimensionality 1. The general formula for the corresponding dimensions is known as the hunter's rule and reads~\cite{Toulouse:1976}:
\begin{equation}
    \text{space} = \text{defect} + \text{cage} + 1\, .
    \label{eq:hunter}
\end{equation}
It is often written $d=d'+r+1$, where $d$ is the space dimension, $d'$ the defect dimension (0 for a point, 1 for a line, 2 for a plane, etc), $r$ the cage dimension (2 for a sphere $S^2$, 1 for a circle $S^1$, 0 for two points $S^0$).

Second, one identifies the order parameter space $V$, which is the remaining freedom for the order parameter once deep in the ordered phase. For the ferromagnet, the magnetization is a 3-dimensional vector. In the ordered phase at low temperature, its norm is essentially fixed, only its direction has some freedom and therefore $V=S^2$.

Third, in order to represent the configuration of the magnetization around the defect, one needs to consider maps from the cage space $C$ to the target space $V$. For example, for a point-like defect in a 3D ferromagnet where the magnetization $\vM \approx M_0 \vn$:
\begin{eqnarray}
\vn:   C=S^2 &\to& V=S^2\nonumber \\
    \vr &\longrightarrow& \vn(\vr) \, .
\end{eqnarray}

Fourth, we need to use a mathematical tool from topology known as homotopy groups. Homotopy groups answer the following question: how many inequivalent classes of maps exist from $C$ (where $C$ is a sphere $S^r$) to $V$? By inequivalent, we mean in a topological sense: we are allowed to smoothly deform the cage space that we are applying onto the target space but we are not allowed to cut it or paste it. Homotopy groups are noted $\Pi_r(V)$. In the present case, we are interested in the second homotopy group of the sphere $S^2$. This is $\Pi_2(S^2)=\mathbb{Z}$ as can be found in tables. This means that there are as many inequivalent classes as there are integers. And therefore that point-like defects in a 3D ferromagnet can be attributed an integer topological charge. In the present context, it is known as the wrapping number. 
\begin{figure}
\begin{center}
\includegraphics[width=2.5cm]{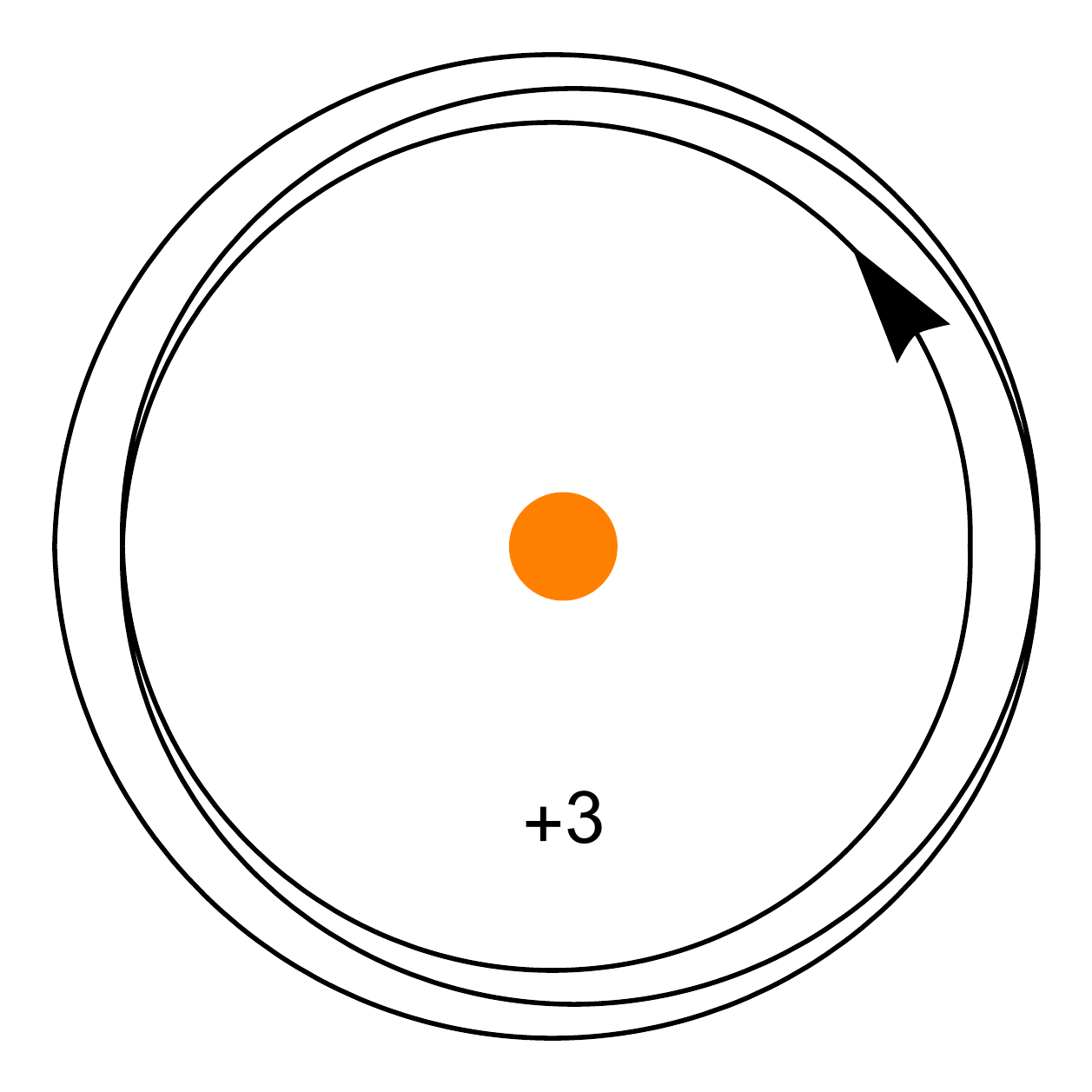}
\includegraphics[width=2.5cm]{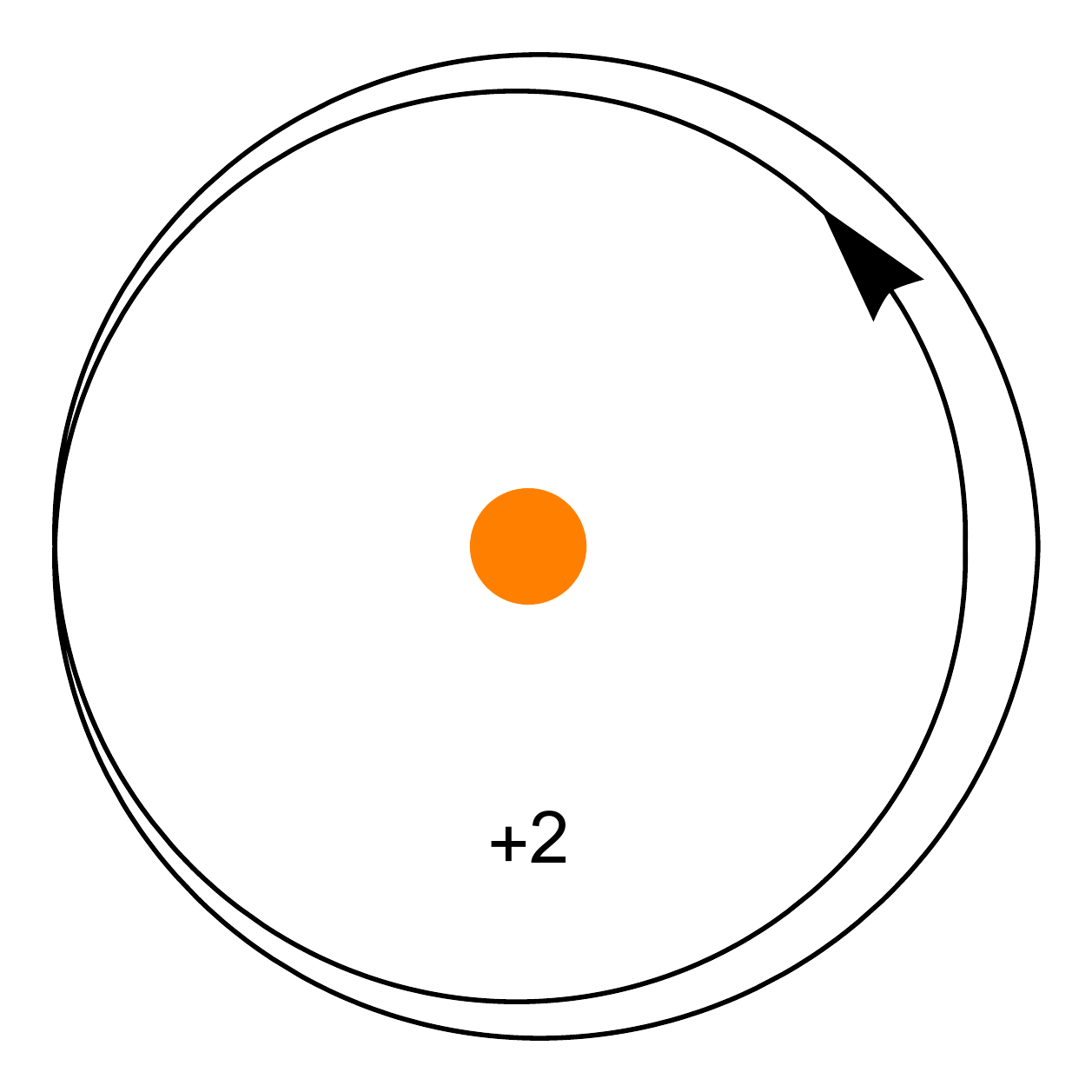}
\includegraphics[width=2.5cm]{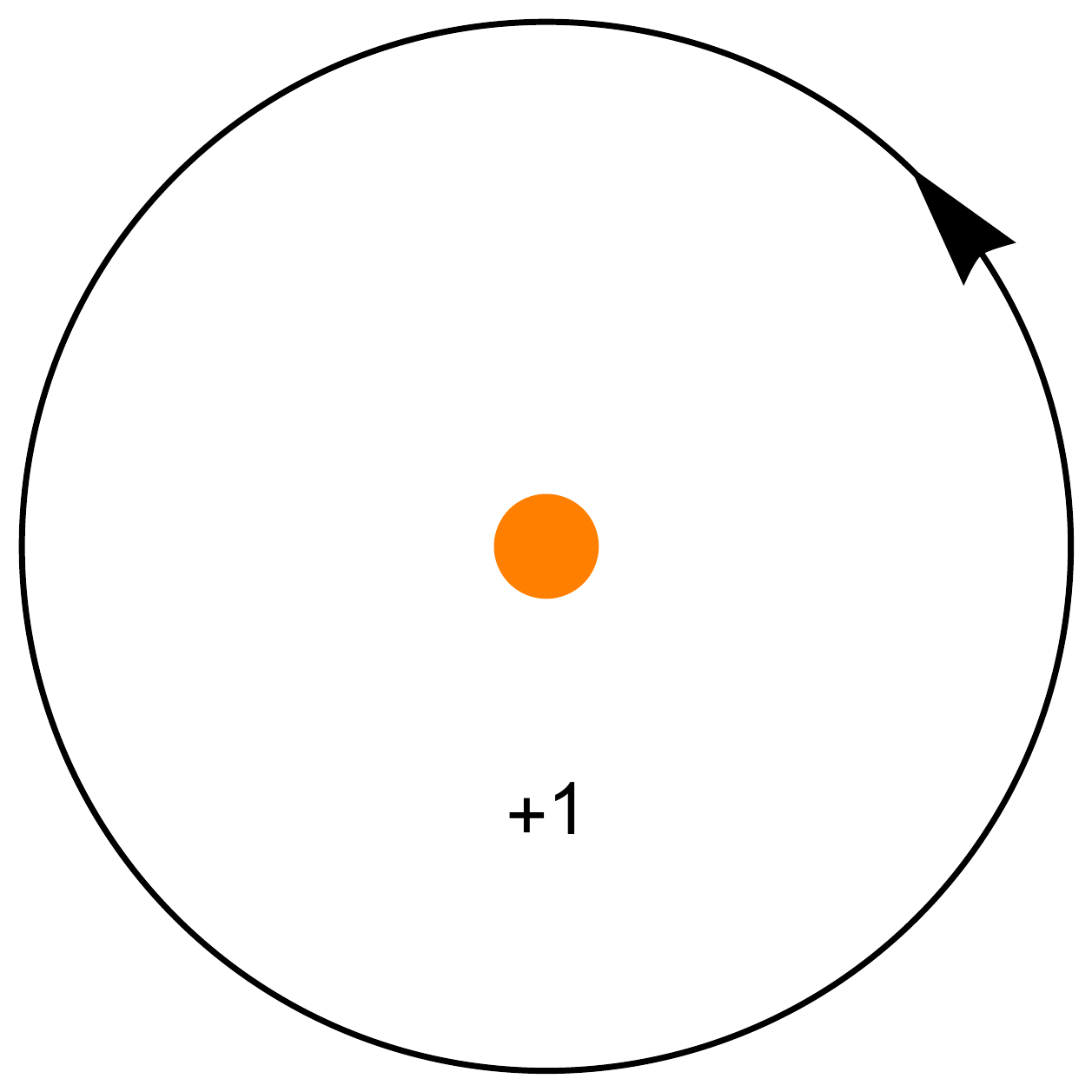}
\includegraphics[width=3cm]{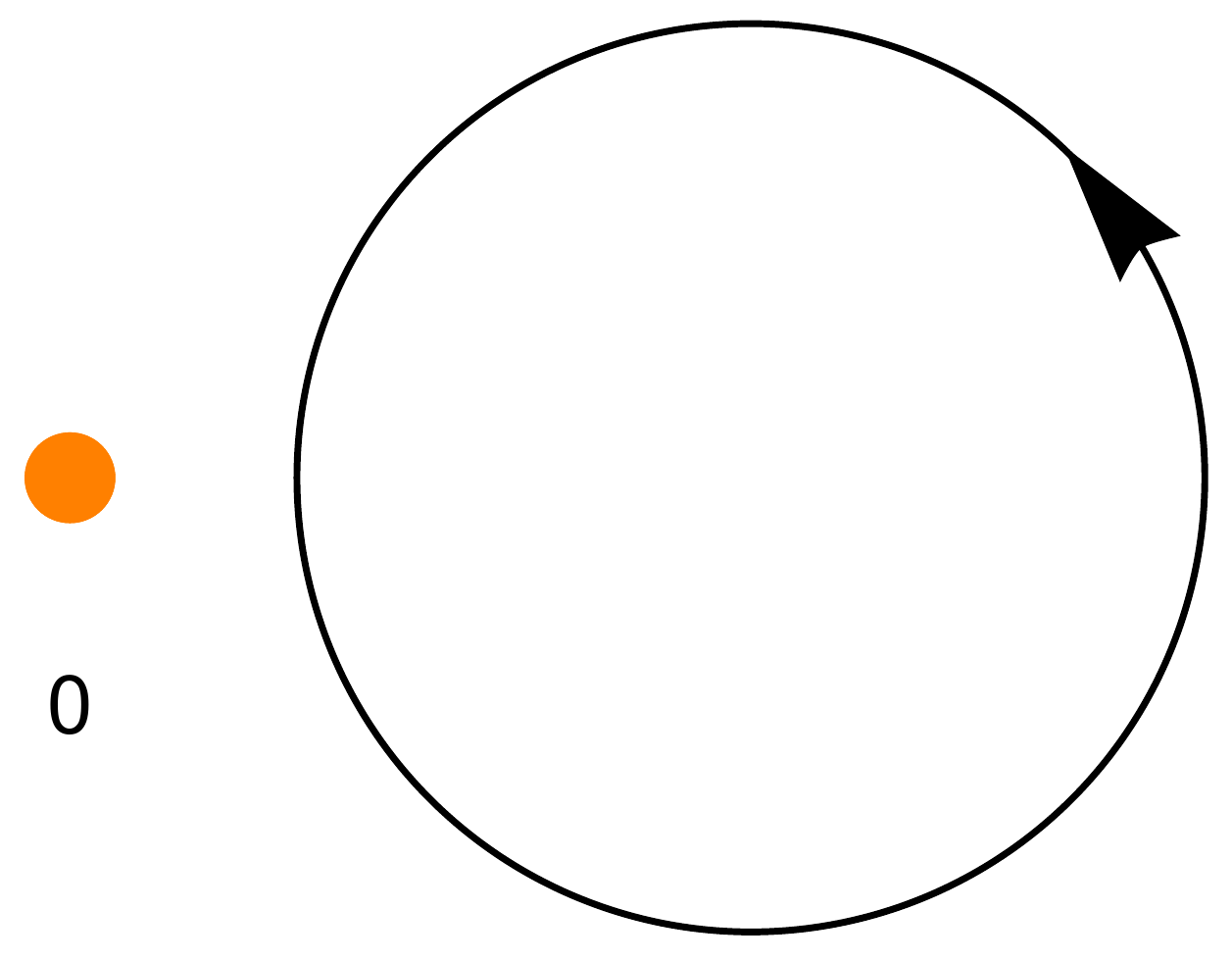}
\includegraphics[width=2.5cm]{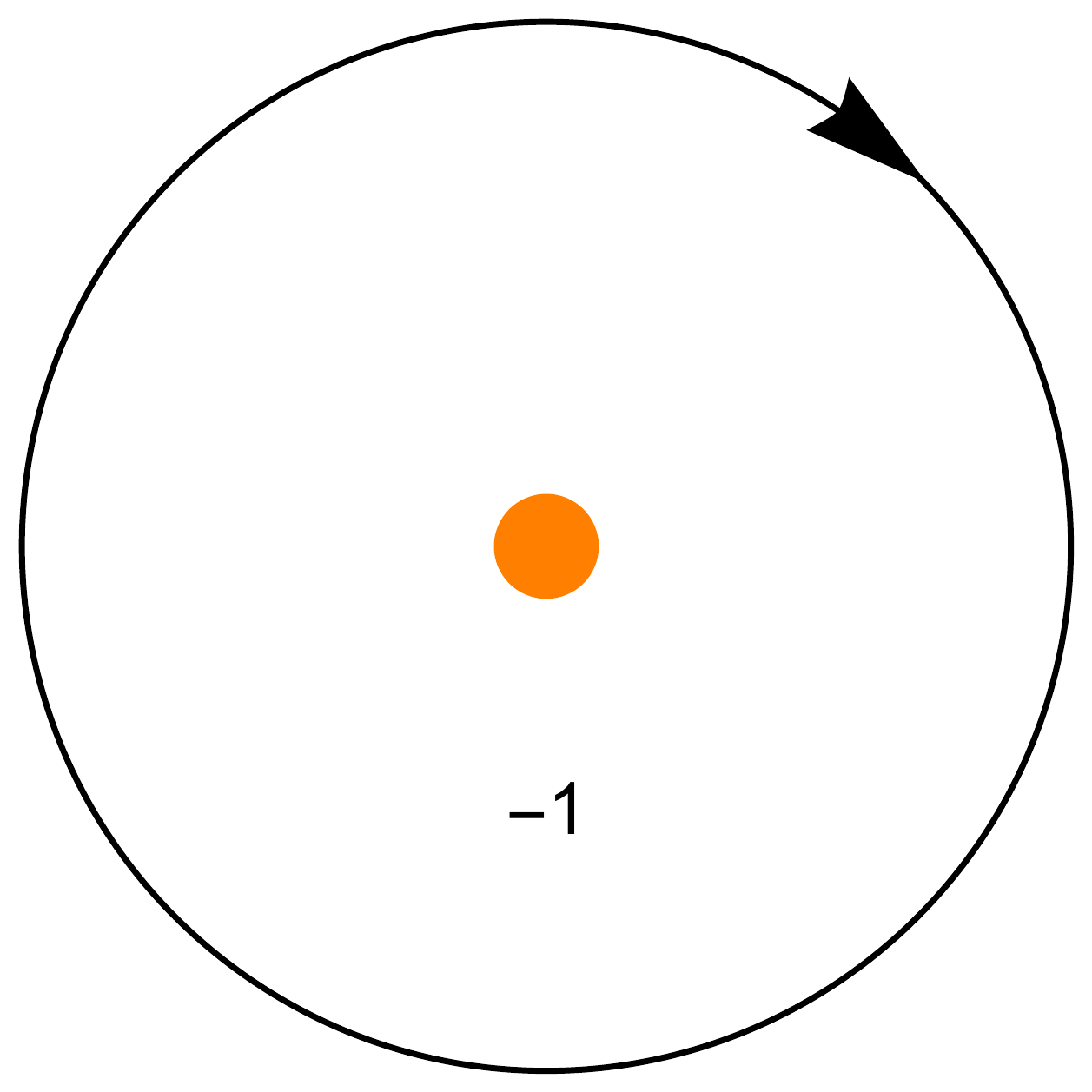}
\includegraphics[width=2.5cm]{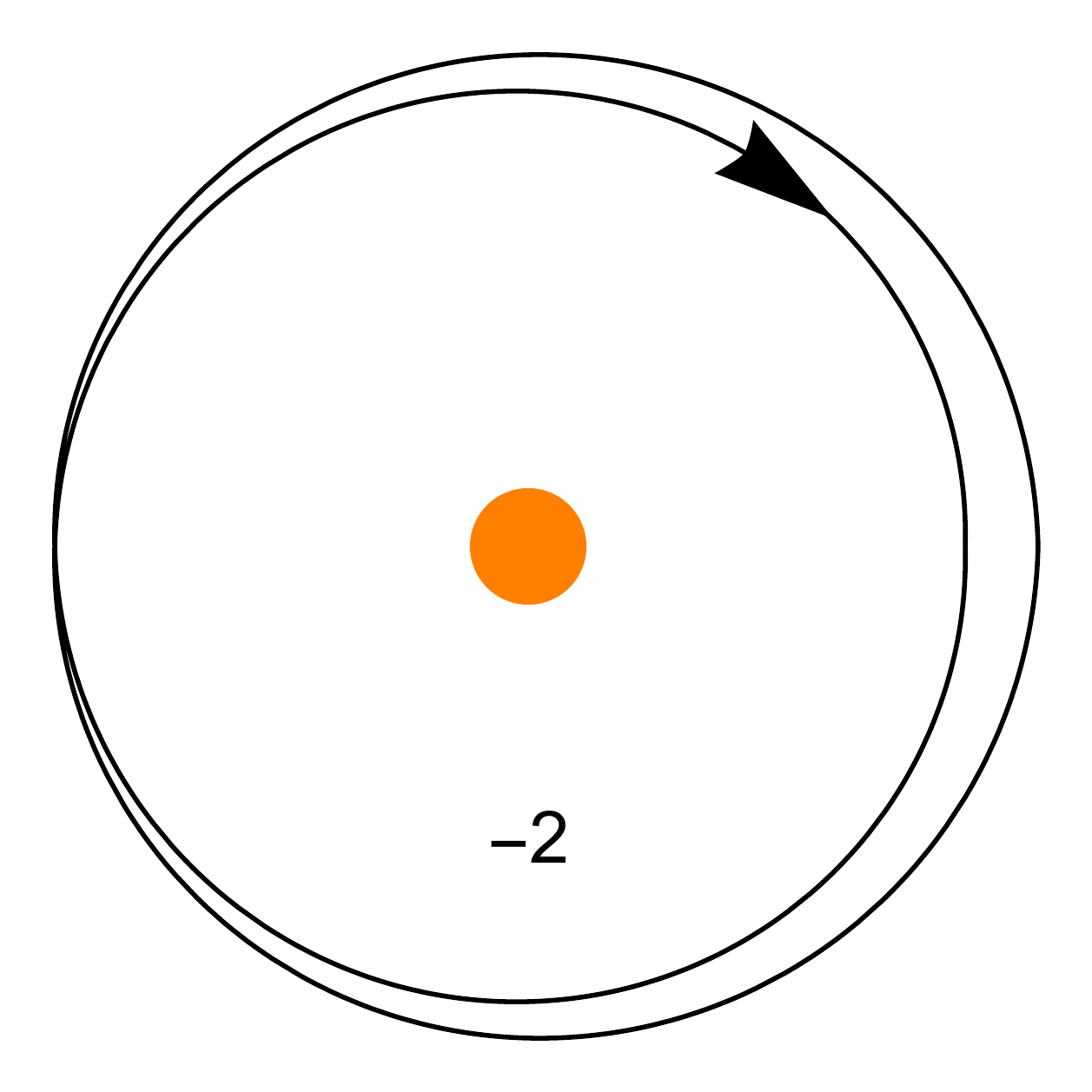}
\caption{First homotopy group of the circle and winding number $W$. Examples from $W=+3$ to $-2$. \label{FigWinding}}
\end{center}
\end{figure}

In order to better understand the notion of a homotopy group, we give several simple examples before going back to the wrapping number. Consider maps going from a cage $C=S^1$ to a target space $V=S^1$. We want to know how many inequivalent ways there are to map a circle onto a circle or, to put it differently, to draw a closed path on a circle. This is given by the first homotopy group (a.k.a. the fundamental group) of the circle $\Pi_1(S^1)=\mathbb{Z}$ and is known as the winding number $W$ (see Fig.~\ref{FigWinding}). Another example concerns maps from the circle $C=S^1$ to the sphere $V=S^2$. This is actually the trivial group $\Pi_1(S^2)=0$, which means that there is a single class of closed paths on a sphere. All paths can be smoothly deformed to a null path, or to paraphrase Sethna~\cite{Sethna:1992}: ``one cannot lasso a basketball''. What about closed path on the torus? We are interested in maps from the circle $C=S^1$ to the torus $V=T^2$. As we have seen, the torus is the direct product of two circles $T^2=S^1\times S^1$. There are therefore two fundamentally distinct non-contractible loops on a torus. The first homotopy group is therefore $\Pi_1(T^2)=\mathbb{Z}\times \mathbb{Z}$. And classes of closed paths on the torus are characterized by a pair of integers.

As we have seen, homotopy groups can be generalized from the first homotopy group by considering maps that start from other spheres then $S^1$. For example starting from $S^2$ or from $S^0$. In the latter case, the zeroth homotopy group, by convention, gives the number of connected components of the target space. For example, $\Pi_0(S^0)=\mathbb{Z}_2$ because $S^0$ consists of two points, whereas $\Pi_0(S^r)=0$ when $r\geq 1$. 

The wrapping number corresponding to the above $\Pi_2(S^2)=\mathbb{Z}$ answers to the question: how many inequivalent ways of wrapping a sphere with a sphere? A pictorial way is to imagine that the target space is a basketball and the cage space is a specially designed spherical bag made of rubber and with a zipper that is used to carry the ball. A dummy would leave the ball outside the closed bag, corresponding to a wrapping of 0. When the ball is inside the bag it corresponds to a wrapping  of +1. Wrapping the ball twice with the bag (harder to imagine!), would correspond to +2. One could also turn the bag inside-out before enclosing the ball, which would correspond to a wrapping of -1. 

\begin{figure}
\begin{center}
\includegraphics[width=5cm]{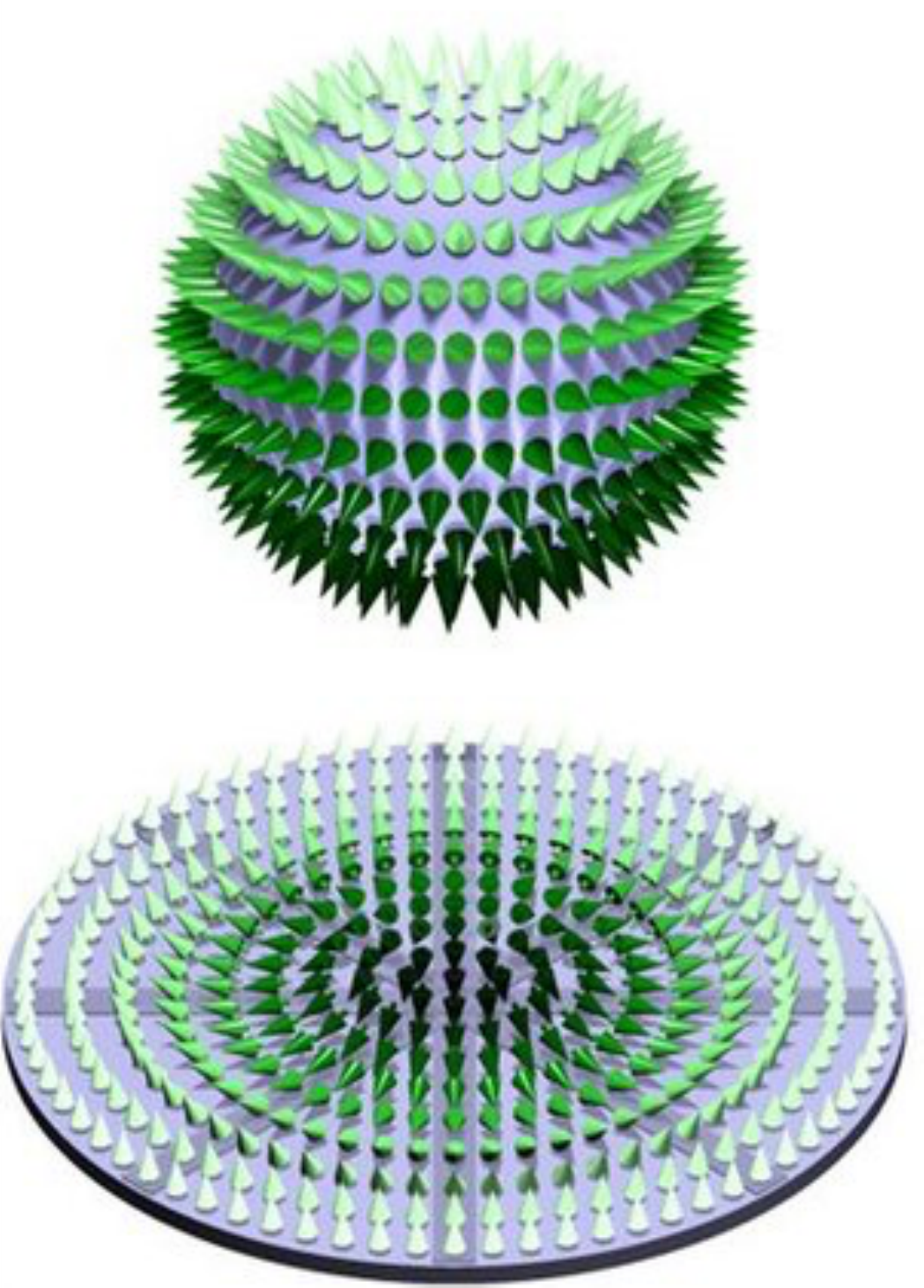}
\caption{\label{FigSkyrmion} Hedgehog (topological defect of a 3D ferromagnet) above and its stereographic projection into a skyrmion (topological texture in a 2D ferromagnet) below (adapted from~\cite{Hoffmann:2017}). In reciprocal space, the same figure would be an illustration of the relation between a 3D Weyl semi-metal (above) and a 2D Chern insulator (below).}
\end{center}
\end{figure}

Going back to our initial question, we have seen that point-like defects in a 3D ferromagnet carry an integer topological charge known as the wrapping number. The elementary topological defect with wrapping +1 is known as the hedgehog (or Bloch point) in the context of ferromagnets. One can picture it as a sphere, the surface of which carries outward pointing arrows (see Fig.~\ref{FigSkyrmion}).

The stereographic projection of the hedgehog on the 2D plane, gives rise to another interesting topological object known as a texture (see Fig.~\ref{FigSkyrmion}). In the present case the texture is called a (baby or 2D) skyrmion. A texture is not a defect because there is no singularity (no zero) in the order parameter on the 2D plane. In addition, it is related to special boundary conditions. Here, because of the stereographic projection sending the north pole of the sphere to the points at infinity on the plane, we see that on the boundary on the plane at infinity, all vectors $\vn$ point up. Note that the hedgehog is a topological point-like defect for a 3D ferromagnet, whereas the skyrmion is a topological texture (not a defect) for a 2D ferromagnet. Topological textures (also known as topological solitons or topological configurations) are discussed in~\cite{ChaikinLubensky} and are classified by relative homotopy groups. Topological textures are interesting in that they extend the topological ideas of homotopy groups to cases that have no defects. The price to pay is that the boundary conditions should be fixed.

\subsection{3D Weyl semi-metal as topological defect\label{sec:weyl}}
\begin{figure}
\begin{center}
\includegraphics[width=6cm]{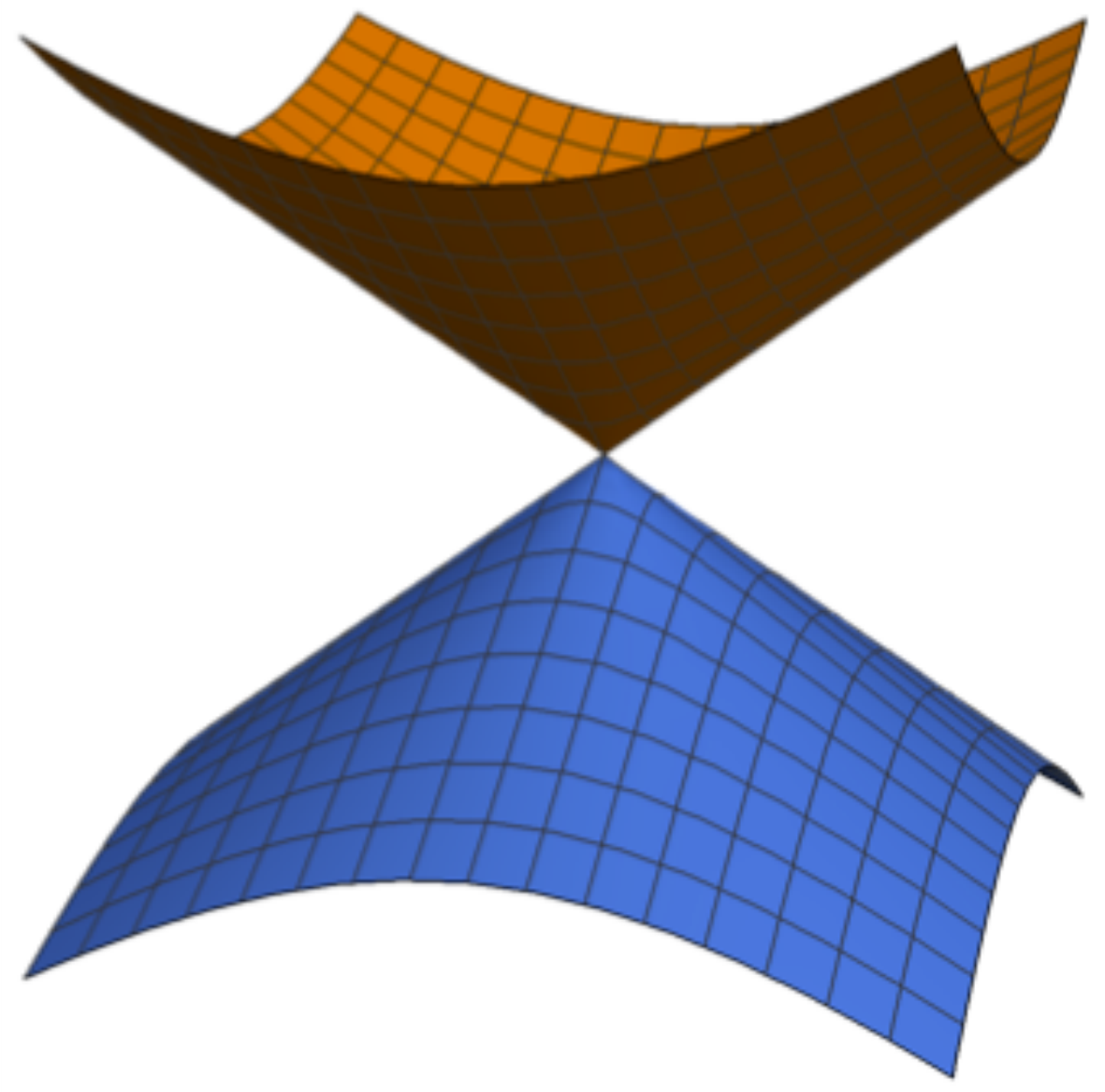}
\hspace{1cm}
\includegraphics[width=6cm]{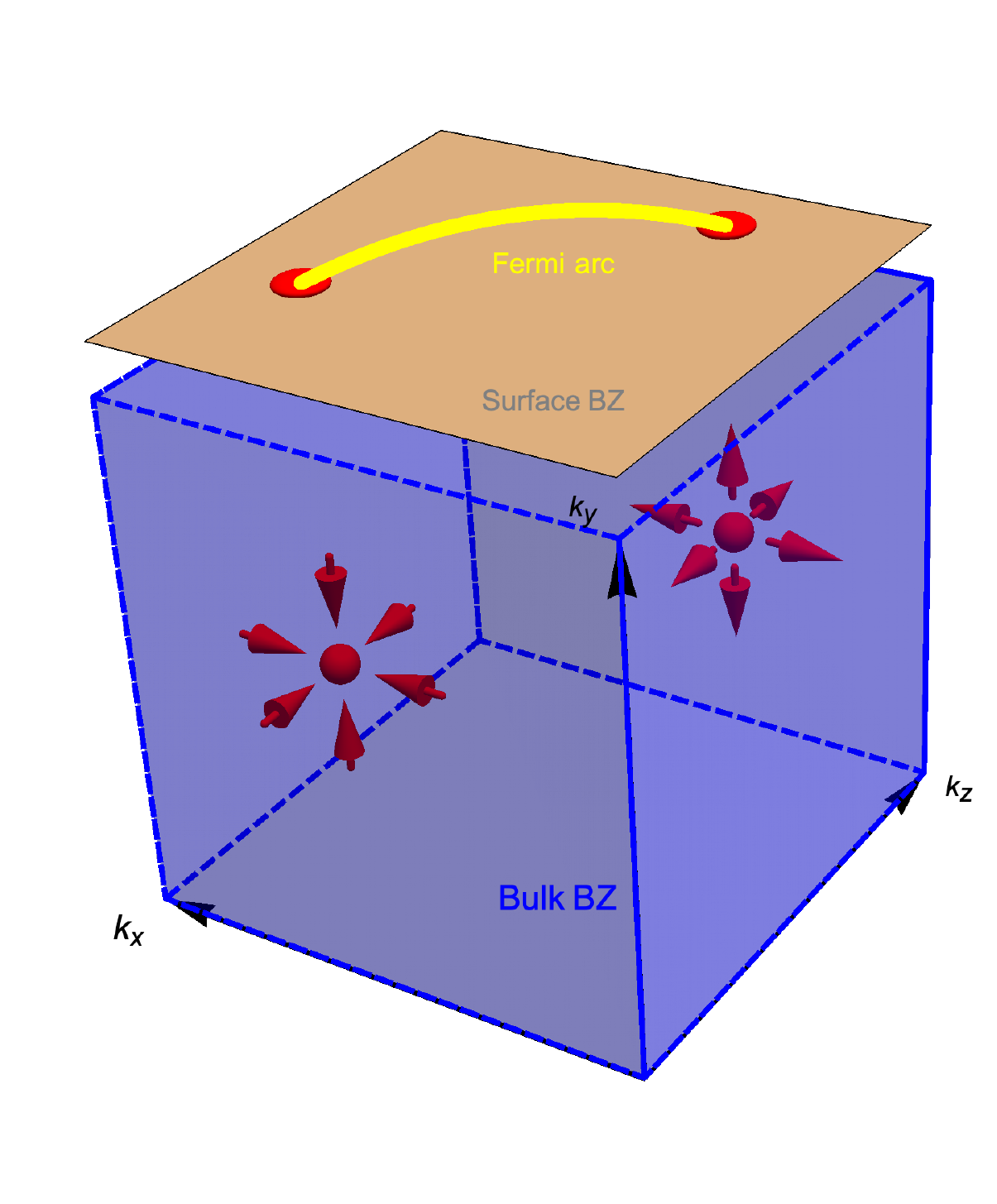}
\caption{\label{FigFermiArcs} (Left) Diabolo-shape dispersion relation in the vicinity of a Weyl point: energy $E$ as a function of $q_x$ and $q_y=q_z$. (Right) Two Weyl points in the bulk BZ. One acts as a Berry flux source and the other one as a sink. The surface BZ contains a Fermi arc that connects to the surface projection of the two Weyl points. Figure adapted from~\cite{Balents:2011}.}
\end{center}
\end{figure}

We are now back to band theory. Consider a contact point between two bands in 3 space dimensions. This can be modeled using a $2\times 2$ Bloch Hamiltonian  
\begin{equation}
    H(\vk) = \vd (\vk)\cdot \boldsymbol{\sigma},
\end{equation}
where $\vk=(k_x,k_y,k_z)$. The contact point correspond to $\vk_c$ such that $\vd (\vk_c)=0$. When considered as a topological defect, it should be characterized as follows. We first need to put this defect into a cage $C$. According to the hunter's rule (\ref{eq:hunter}), the cage has dimension 2 as the defect is 0-dimensional. Here the cage can be taken as a 2-dimensional sphere $C=S^2$ surrounding the contact point. The target space $V$, which in the context of phase transitions is known as the order parameter space in the ordered phase, is here the Bloch sphere $S^2$ spanned by $\hat{\vd} = \vd/|\vd|$. The defects are characterized by the inequivalent classes of maps going from $C=S^2 \to V=S^2$, i.e. by the second homotopy group $\Pi_2(S^2)=\mathbb{Z}$. The corresponding topological invariant is the wrapping number $N_3$ ($3$ refers to the co-dimension of the defect, which is codim = space - defect = 3-0). Here we are using the notation of Volovik $N_\text{codim}$ for the topological indices~\cite{Volovik:2003}. The wrapping number is actually also a Chern number in the present context and reads
\begin{equation}
    N_3 = \frac{1}{2\pi}\int_{C=S^2} \boldsymbol{dS} \cdot \boldsymbol{F}_n(\vk) .
\end{equation}
A Weyl point corresponds to an elementary such topological defect also known as a hedgehog and having a wrapping number or chirality $N_3=\pm 1$. In the vicinity of the contact point, the Bloch Hamiltonian is
\begin{equation}
    H(\vk_c+\vq)=H_W(\vq) = \sum_{i,j}v_{ij}\,  q_i \sigma_j,
\end{equation}
where $\vq=\vk-\vk_c$ and $i,j=x,y,z$. In that case, the wrapping number (or chirality) is given by $N_3=\text{sign} \det (v_{ij})$. Such a Hamiltonian was proposed by Weyl shortly after the discovery of the Dirac equation in order to describe hypothetical chiral fermions~\cite{Weyl:1929}. A Weyl fermion can be roughly thought as being half a Dirac fermion in 3D, in the sense that its Hamiltonian is $2\times 2$ rather than $4\times 4$. In addition chirality implies masslessness. For a nice discussion of Dirac versus Weyl versus Majorana fermions, see~\cite{Pal:2011}. In the original version, the Weyl Hamiltonian reads
\begin{equation}
    H_W = \pm c \, \vp \cdot \boldsymbol{\sigma}\, ,
\end{equation}
where $c$ is the velocity of light and the $\pm$ indicates the chirality (i.e. left or right Weyl fermion). The dispersion relation is $E=\pm c |\vp|$ (see Fig.~\ref{FigFermiArcs} Left).

An alternative way to understand that a contact point between two bands in 3D is a stable (topological) defect is to realize that the Bloch Hamiltonian being a $2\times 2$ matrix can be decomposed onto the three Pauli matrices (the identity matrix plays no role) and that the three corresponding coefficients $\vd=(d_x,d_y,d_z)$ are each analytical functions of three variables $\vk=(k_x,k_y,k_z)$. A contact point is a triplet $\vk_c$ such that $\vd(\vk_c)=0$. As there are three linear equations ($d_j=0$) and three unknowns ($k_j$), there generically exists a solution in reciprocal space. This type of reasoning belongs to von Neumann and Wigner~\cite{vonNeumann:1929}. It was applied to band theory by Herring~\cite{Herring:1937}. Such an accidental degeneracy is not required by symmetry but is topologically robust. To get rid of Weyl points, one needs to merge two of them with opposite chirality~\cite{Volovik:2007}. A Weyl point corresponds to a singular source (or sink) of Berry flux. It is the reciprocal-space analog of a Dirac magnetic monopole, which Berry calls a diabolical point and which we call a Berry monopole. Weyl points can also be seen as the 3D analogs of the Dirac cones of graphene.

Nielsen and Ninomiya have shown that in a lattice realization, chiral (Weyl) fermions occur in pairs~\cite{Nielsen:1983}. This is known as the fermion doubling theorem. It is related to the fact that a Weyl point carries a chirality of $N_3=\pm 1$, which is equivalent to an elementary Berry monopole, and that the total Berry flux across the BZ should vanish. In order to have a Weyl point, one needs a non-vanishing Berry curvature, and therefore either inversion or time-reversal symmetry must be broken, which leads to two types of Weyl semi-metals: inversion symmetric or time-reversal symmetric. 

An inversion-symmetric Weyl semi-metal necessarily breaks time-reversal symmetry: when there is a Weyl point at $\vk$ with chirality $N_3$, inversion symmetry implies that there is another one at $-\vk$ with chirality $-N_3$ (see Fig.~\ref{FigFermiArcs}). 
Therefore the minimal number of Weyl points is two in that case as $N_3+(-N_3)=0$. 

A time-reversal-symmetric Weyl semimetal necessarily breaks inversion symmetry and when there is a Weyl point at $\vk$ with chirality $N_3$, time-reversal symmetry guarantees that there is another one at $-\vk$ with the same chirality. In order for the total chirality in the BZ to vanish, one therefore needs a minimum of four Weyl points in that case.

Early on, Nielsen and Ninomiya have proposed a lattice realization of a Weyl semi-metal in order to study the associated chiral anomaly~\cite{Nielsen:1983}. A nice and simple tight-binding model is discussed in~\cite{Delplace:2012}. These author study a model for 3D spinless electrons on a lattice with two sites per unit cell, that breaks time-reversal symmetry and preserves inversion. In that way, they are able to find a situation in which there are only two Weyl points, which they study in detail including the resulting surface behavior. Indeed, a remarkable feature of Weyl semi-metals is that they host open Fermi arcs on their surface~\cite{Wan:2011}. Open Fermi arcs are two-dimensional Fermi surfaces (i.e. lines) that are opened instead of closed. 
Actually, a Fermi arc connects to the projection of the Weyl points on the surface. The Fermi arcs are the 3D equivalent of the zero-energy edge mode present in 2D graphene nanoribbons~\cite{CastroRMP:2009}. To put it differently, some of the surfaces of a 3D Weyl semi-metal host a peculiar two-dimensional metal, which is chiral and has an opened Fermi line.

For a general review of theoretical and experimental aspects of Weyl semimetals, see~\cite{Armitage:2018}.

\subsection{2D Chern insulator as topological texture\label{sec:skyrmion}}
From the perspective of topological defects, there is also an interesting relation between the Weyl point in 3D and the Chern insulator in 2D. A band insulator has no Fermi surface, it is gapped at every point in the BZ, and therefore it can not be characterized as a topological defect. However, it may still contain some kind of twist. This is related to the notion of a topological texture \cite{Michel:1980,ChaikinLubensky}. A topological texture is not a defect, as there is no vanishing of the ``order parameter'', no defect, but it can nevertheless be characterized by a topological invariant. Topological textures are classified by \textit{relative} homotopy groups \cite{Michel:1980}. They are best explained on the example of the 2D skyrmion, which is a simple topological texture. Let us consider the 2D reciprocal infinite plane $\vk=(k_x,k_y)\in \mathbb{R}^2$ (for the moment, we forget about the BZ torus $T^2$ relevant to the 2D Chern insulator and replace it by $\mathbb{R}^2$), in which at each point $\vk$ the vector field $\vd=(d_x,d_y,d_z)$ is defined. Imagine that at infinity $k\to \infty$, the vector $\vd$ points in the $z$ direction and that it smoothly evolves towards the center $k\to 0$, where it points in the $-z$ direction (see Fig.~\ref{FigSkyrmion}). Note that, in particular the vector $\vd$ never vanishes (one could consider that its norm is constant), so that there is no defect. The 2D plane with such a boundary condition at infinity $\mathbb{R}^2+\{\infty\}$ can be compactified into a sphere $S^2$, that will play a role similar to that of the cage in the case of a defect (remember that a point-like defect in 3D is trapped in an $S^2$ cage). The relevant mapping that we now consider are from this compactified complete space $\mathbb{R}^2+\{\infty\}\sim S^2$ to the Bloch sphere $S^2$. We already know that the relevant homotopy group is $\Pi_2(S^2)=\mathbb{Z}$. This means that topological textures in 2D are characterized by a wrapping number, just like topological defects in 3D. The elementary non-trivial topological texture is known as the 2D skyrmion and has a wrapping (or skyrmion) number $\tilde{N}_3=\pm 1$. The notation here is again that of Volovik~\cite{Volovik:2003}: the tilda means that it is a topological index related to a texture and not to a defect. In other words, there is a deep relation between the 2D skyrmion $\tilde{N}_3=\pm 1$ and the 3D hedgehog $N_3=\pm 1$. The former can be seen as the stereographic projection of the latter (see Figure~\ref{FigSkyrmion} bottom). In the above discussion, the parameter space, which is the BZ torus $T^2$, was replaced by a sphere $S^2$ (compactified plane). As explained in~\cite{Moore:2008} e.g., maps from $T^2$ to $S^2$ are actually topologically equivalent to maps from $S^2$ to $S^2$ (despite the fact that the sphere and the torus do not have the same genus). This means that topological textures also exist when the parameter space is a torus and are classified by a wrapping number.

In the context of band structures, the relation between the skyrmion and the hedgehog means that there is a relation between the 2D Chern insulator and the 3D Weyl semi-metal. In order to clearly expose this relation, we consider the phase transition between two phases of a 2D Chern insulators driven by a parameter $\lambda$. One may think of the Haldane model and of tuning the inversion-breaking mass $M$ (i.e. $\lambda=M$ for example) so as to go from a phase with Chern number $\tilde{N}_3^{(1)}$ to a phase with Chern number $\tilde{N}_3^{(2)}\neq \tilde{N}_3^{(1)}$ (here we use the notation of Volovik for a topological texture, but the Chern number $\tilde{N}_3$ is really the same thing as what we called $C_-$ before). The corresponding phase transition can be thought of as being a Weyl point in the space $(k_x,k_y,\lambda)$. The wrapping number $N_3$ of the corresponding Weyl point is related to the change in the Chern number as follows:
\begin{equation}
    \tilde{N}_3^{(2)}-\tilde{N}_3^{(1)} = N_3
\end{equation}
For a detailed proof see \cite{Belissard:1995}. Here, we recall our discussion in Sec.~\ref{sec:chernnumber}: a Chern insulator is a 2D band insulator that has a non-zero number of Berry monopoles (or Weyl points) \emph{inside} (i.e. enclosed by) its BZ.


In conclusion, gapped non-interacting fermionic systems can therefore be classified using ideas from topology. Two alternative viewpoints on topological insulators (or superconductors/superfluids) are: (1) twisted fiber bundles and (2) topological textures. The first viewpoint goes back to Thouless et al.~\cite{Thouless:1982}, the second to Volovik~\cite{Volovik:1988}.

\subsection{2D Dirac point as symmetry-protected topological defect}
In contrast to the 3D Weyl semi-metal, in 2D, a contact point between two bands is unlikely (unstable) because it requires the vanishing of three functions $\vd=(d_x,d_y,d_z)$ that only depend on two variables $\vk=(k_x,k_y)$. The corresponding homotopy analysis involves a cage $C=S^1$, a target space $V=S^2$ and a homotopy group $\Pi_1(S^2)=0$, which means that no topological invariant exists that protects a 2D contact point between two bands. Then, one may wonder about the case of graphene: why would Dirac points be stable? The contact points in graphene are actually not topologically stable but protected only as long as a certain symmetry is preserved. This symmetry is actually $\mathcal{I} \mathcal{T}$, i.e. the product of inversion and time-reversal transformations, and ensures that the target space is restricted to a great circle $V=S^1$ of the Bloch sphere (instead of the whole Bloch sphere $S^2$). This is known as a symmetry-protected topological defect. Indeed, in that case the relevant homotopy group is $\Pi_1(S^1)=\mathbb{Z}$ and the corresponding topological invariant is known as the winding number $N_2(\mathcal{I} \mathcal{T})$ using Volovik's notation (the subscript $2$ indicates the codimension and $\mathcal{I} \mathcal{T}$ indicates that this topological invariant requires the existence of a symmetry for its existence). In graphene, the Dirac points have winding number $N_2(\mathcal{I} \mathcal{T})=\pm 1$. Breaking the $\mathcal{I} \mathcal{T}$ symmetry, the Dirac points can be immediately gapped, which shows that they are not stable against any perturbation (see e.g., boron nitride that breaks inversion symmetry or the Haldane model that breaks time-reversal symmetry). However, if this symmetry is maintained, the Dirac points can only disappear via a merging (Lifshitz) transition \cite{Montambaux:2009}. This topological invariant $N_2(\mathcal{I} \mathcal{T})$ is known as the winding number or the chirality $\chi=\text{sign } \det(v_{ij})$. Symmetries are actually able to protect degeneracies between more than two bands (i.e. pseudo-spin 1/2). For a discussion on how site-permutation symmetries can be used to obtain the equivalent of pseudo-spin 1 or 2 fermionic quasiparticles, see e.g.~\cite{CrastoDeLima:2020}.

\subsection{Classification of topological metals}
The general idea of Volovik is to classify Fermi surfaces as topological defects~\cite{Volovik:2003,Volovik:2007}. Using elementary ideas from homotopy groups, he realized that in 3D, the Fermi surface $d'=2$ is stable (and characterized by a winding number), the Fermi line $d'=1$ is unstable and the Fermi point $d'=0$ is stable (also with an integer invariant). The latter is now known as a 3D Weyl semi-metal. In 2D, he found that the Fermi line is stable (with integer invariant) and the Fermi point is unstable (unless protected by a symmetry as in graphene). And in 1D, the Fermi point being the natural ``Fermi surface'' is stable (also with an integer invariant).

As in the case of gapped systems (topological insulators or superconductors), one may generalize the above ideas and classify all types of Fermi surfaces as topological defects. Important concepts are space dimension and the nature of the Fermi surface (does it involve complex (Dirac) fermions or real (Majorana) fermions?). As in the ten-fold periodic table, the classes are of three types ($0$, $\mathbb{Z}_2$ or $\mathbb{Z}$) and there is a form of regularity known as Bott periodicity and inherent to K-theory. This issue is beyond the scope of the present review. The interested reader will find more information in~\cite{Horava:2005,Zhao:2013}. 

\section{Conclusion \label{sec:conclusion}}
To conclude this review, we would like to highlight some take-home messages :
\begin{itemize}
    \item In band theory, the physical properties are not only determined by the energy level properties, but also by the Bloch wave functions. For instance the electrical polarization of crystals and the quantum Hall effect cannot be understood solely from the energy bands. In the central section of this review (Sec. \ref{sec:bandtheory}), we explain how the geometrical concepts (Berry connection and curvature, quantum metrics) and the topological invariants (Chern number) are useful to build a complete description of crystal band structures, going beyond the energy level spectrum. Then these latter concepts are applied to several systems : one-dimensional lattices (Sec.~\ref{section1D}), electrons on honeycomb lattices (Sec.~\ref{section2D}), and finally 3D Weyl/Dirac semimetals (Sec.~\ref{sec:topmetal}).
    
    \item Interesting physics emerges when bands are not independent but are rather coupled by virtual transitions due to external fields (e.g. electric and magnetic fields). These virtual transitions lead to geometrical effects locally in reciprocal $\vk$-space and to topological effects globally in reciprocal space (Sec.~\ref{sec:bandtheory}). Topologically trivial insulators (e.g. boron nitride) may still contain interesting geometrical effects. For example, Dirac insulators host zero-modes trapped on topological defects of the Dirac mass (see the Jackiw-Rebbi mechanism, Sec.~\ref{sec:jr}). 
    
    \item Berry phase concepts (including Berry connection and curvature) are not mandatory to describe the inter-band effects. In principle, if one could solve completely the energy spectrum in the presence of external fields all the physical properties could be calculated without resorting to Berry phase concepts. The Berry connection appears when one projects onto a single isolated band, or an isolated subset of bands. This is already clear in the context of the $0$D two-level system (Sec.~\ref{section0D}) and pertains for $D$-dimensional lattice systems (Sec.~\ref{sec:bandtheory}). The alternative when studying a crystal in an external field is therefore: either use the \emph{zero-field} energy bands and Bloch states (and therefore Berry phase effects); or compute the energy spectrum \emph{in the presence of external fields}.
    
    \item In practise, it is very important to distinguish a periodic Bloch Hamiltonian and the canonical Bloch Hamiltonian in the case of a crystal structure with several atoms per unit cell (i.e. lattice with a basis). A periodic Hamiltonian contains less information than the canonical Hamiltonian. The usual formula for the Berry connection or curvature, Zak phases are typically written in terms of the canonical Bloch Hamiltonian and the corresponding cell-periodic Bloch states.
    
    \item A definition for a topological insulator consists in a bulk insulator with a quantized bulk response, and a topologically protected metal at its boundary. Within this picture, there is only one way to be a trivial insulator (Sec.~\ref{sec:firstdef}).
    
    \item An alternative paradigm consists in defining as topological an insulator that cannot be continuously deformed (while keeping a protective symmetry) into an atomic insulator characterized by localized symmetric Wannier functions. In this framework, there can be several trivial atomic limits (Sec.~\ref{sec:seconddef}). 
    
    \item It is good to keep in mind a distinction between topological insulators (discussed in the present review) and the notion of topological order (not discussed). See Appendix~\ref{sec:to} for a short comparison between the two notions.
    
\end{itemize}

\medskip

\section*{Acknowledgments}
We acknowledge useful discussions with J\'anos Asb\'oth, Jens Bardarson, H\'el\`ene Bouchiat, Alexandre Buzdin, David Carpentier, Fr\'ed\'eric Combes, Pierre Delplace, Bal\'asz Dor\'a, Beno\^it Dou\c{c}ot, Cl\'ement Dutreix, Pierre Gosselin, Adolfo Grushin, Rony Ilan, Pavel Kalugin, Lih-King Lim, Andrej Mesaros, Roderich Moessner, Herv\'e Mohrbach, Joel Moore, R\'emy Mosseri, Louis Nouri, Hugo Perrin, Arnaud Raoux, Pascal Simon, Doru Sticlet, Andr\'e Thiaville, Max Trescher, Pierre Vallet, Julien Vidal and Ashvin Vishwanath. Special thanks to the gang of four in Orsay -- Mark Goerbig, Gilles Montambaux and Fr\'ed\'eric Pi\'echon (like the musketeers) -- for the enlightening discussions over the years since the first experimental papers on graphene that sparked our interest in these matters. We also thank Christophe Brun for organizing lectures at INSP in Jussieu that were a strong motivation in putting all this material together. J. Cayssol acknowleges support from the network Quantum Matter at Bordeaux University under project TaQuaMaUC.

\appendix

\section{Topological insulator versus topological order \label{sec:to}}
In this Appendix and in order to recap what we learned on \emph{topological insulators}, we wish to compare and distinguish them as clearly as possible from another type of gapped systems featuring some topological property and called \emph{topological order}. The following discussion is essentially summarized in Table~\ref{tab:sptvsto}. A good reference for this distinction is the commentary written by Fisher~\cite{Fisher:2013}. Topological insulators are discussed in the books by Bernevig and Hugues~\cite{Bernevig:2013}, Asb\'oth et al.~\cite{Asboth:2016} and Vanderbilt~\cite{Vanderbilt:2018}, whereas topological order is discussed in detail in the book by Wen~\cite{Wen:2004} and in the review~\cite{Wen:2019}.

We restrict the discussion to two-dimensional gapped systems. On the one hand, topological insulators are essentially non-interacting fermionic band insulators. Their topological nature comes from the fact that they can not be adiabatically  -- i.e. without closing the bulk gap -- deformed into an atomic insulator. This obstruction is typically characterized by a topological invariant such as a Chern number. Usually, the obstruction in doing so only occurs provided a symmetry condition is imposed. These are therefore also known as symmetry-protected topological (SPT) insulators. A good example is the Kane-Mele model for a quantum spin Hall insulator (QSHI), which is protected by time-reversal symmetry. As a consequence of this obstruction, a boundary with the vacuum (that behaves as a trivial insulator) necessarily hosts gapless edge modes, that have some form of robustness. In the case of the QSHI, the edge modes are spin-momentum locked and known as helical edge states, their direction of motion being tied to their spin projection. Therefore an impurity or a defect that does not act on spin can not backscatter. The groundstate of a SPT insulator is unique and is a Slater determinant. It has only short-range quantum entanglement. Bulk excitations are similar to that in a trivial band insulator: they are electronic quasiparticles. Topological insulators in this sense are a refinement of band insulators.

On the other hand, systems featuring topological order (TO) are typically strongly interacting. Examples are fractional quantum Hall (FQH) states and quantum spin liquids (QSL). In the FQH effect, microscopics is defined in terms of interacting electrons in a perpendicular magnetic field. In the QSL, the microscopics typically involves localized spins with exchange interaction and some form of frustration that prevents spontaneous symmetry breaking and long-range order. The topological nature of TO phases is revealed by their ground-state degeneracy depending on the genus of the space manifold: it is different for the system placed on a sphere or on a torus, for example. This property is intrinsic (the topological degeneracy is robust to any perturbation which is sufficiently small) and does not rely on the protection by a specific symmetry. The ground-state features long-range quantum entanglement despite the fact that correlation functions decay exponentially on a length scale set by the bulk gap. Furthermore, some of the excitations are exotic: they are topological excitations, created in pairs, related by an unobservable string and having fractionalized quantum numbers and exchange statistics. The latter properties means that they are anyons. A defining property of TO is the presence of such fractionalized excitations.

\begin{table}
    \centering
    \begin{tabular}{c|c|c}
        &\textbf{Topological insulators (SPT)} & \textbf{Topological order (TO)}\\
        \hline
         Full name & symmetry-protected topological phases  & intrinsic topological order \\
         \hline
         Space dimension & any D  & mainly 2D \\
          \hline
          Gapped bulk & yes  & yes \\
         \hline
         Any small perturbation & not robust  & robust \\
          \hline
         Effective description & topological band theory  & topological quantum field theory  \\
         \hline
         Interactions & not needed & required  \\
         \hline
         Ground-state & non-degenerate & degenerate on manifolds with non-trivial topology \\
         \hline 
         Bulk excitations & non-exotic & fractionalized, anyons, topological,\\ 
         & & created in pairs related by a string \\
         \hline 
         Robust gapless edge states & always & not always (chiral versus achiral TO) \\
         \hline
         Ground-state entanglement & short-range & long-range\\
         \hline
         Related to  & topological textures & lattice gauge theories, tensor categories, CFT \\
         \hline
         Historical example & integer quantum Hall effect (IQHE) & fractional quantum Hall effect (FQHE)\\
          \hline
         Further examples & Chern insulators, $\mathbb{Z}_2$ TI in 2D and 3D & quantum spin liquids\\
         \hline
         Models & Haldane, Kane-Mele, Bernevig-Hughes-Zhang, & toric code (Kitaev), string nets (Levin-Wen),\\
                & Volovik $p_x+ip_y$ superfluid, Majorana chain & Kitaev honeycomb \\
         \hline
         Historical roots & Thouless-Kohmoto-Nightingale-den Nijs 1982 & Wen's topological order 1990
    \end{tabular}
    \caption{Symmetry-protected topological phases (topological insulators) versus intrinsic topological order. \label{tab:sptvsto}}
\end{table}

In this dichotomy between SPT phases and TO phases, there are some cases which are not so clear cut. 
For example, the integer quantum Hall effect seems to be the historical example of a topological insulator but it exists in the absence of any protecting symmetry (it is intrinsically topological). However its excitations are not fractionalized. 
Also, superconductors are usually described as resulting from the spontaneous symmetry breaking of a gauge symmetry, despite the fact that a gauge symmetry can not be broken. Actually, they have been shown to be topologically ordered~\cite{Hansson:2004}. 
Furthermore, topological superconductors such as chiral $p_x+ip_y$ superconductors, when treated at the mean-field level so that they effectively appear as being non-interacting, belong to the $D$-class of the ten-fold periodic table and are characterized by a Chern invariant~\cite{Schnyder:2008,Kitaev:2009}. As such they are a form of SPT. But they are also known to host fractionalized excitations trapped in vortex cores and known as Majorana zero modes (or Majorana bound states). The latter behave as non-Abelian (Ising) anyons.


%

\end{document}